%% file: PhD-Thesis.tex
\documentclass{ucetd}
\usepackage{subfigure,epsfig,amsfonts}
\usepackage{amsmath}
\usepackage{amssymb}
\usepackage{amsthm}
\usepackage{booktabs}
\usepackage{graphicx}
\usepackage[latin1]{inputenc}
\usepackage{tikz}
\usepackage{nicefrac}
\usepackage{wrapfig}
\usepackage{acro}
\usepackage{wasysym}
\usepackage[load=addn,binary-units=true]{siunitx}
\usepackage[labelfont=bf]{caption}
\usepackage{mathtools}
\usepackage{url}
\usepackage{multirow}
\usepackage{placeins}
\usepackage{afterpage}

\newcommand{\isotope}[2]{\ensuremath{^{#2}\text{#1}}}
\newcommand{\Exp}[1]{\ensuremath{\text{Exp}\hspace{-4pt}\left[#1\right]}}

\def\cogent/{CoGeNT}
\def\coherent/{COHERENT}
\def\mcnp/{MCNPX-PoliMi ver.\ 2.0}
\def\NI/{National Instruments}
\def\csi/{{CsI[Na]}}
\def\nai/{{NaI[Tl]}}
\def\brillance/{BrilLanCe\texttrademark}
\def\ej299/{{EJ299-33A}}
\def\xs/{{cross-section}}
\def\geant/{{Geant4}}
\def\amcrys/{{AMCRYS-H}}
\def\labview/{{LabVIEW}}
\def\agc/{{\ensuremath{N^\text{max}_\text{pt}}}}
\def\chc/{{\ensuremath{N^\text{min}_\text{iw}}}}
\def\cDS/{{\ensuremath{\mathcal{C}}}}
\def\acDS/{{\ensuremath{\mathcal{AC}}}}
\def\onDS/{{\ensuremath{\mathcal{ON}}}}
\def\offDS/{{\ensuremath{\mathcal{OFF}}}}
\def\rDS/{{\ensuremath{\mathcal{R}}}}
\def\qspe/{{\ensuremath{Q_\text{spe}}}}
\def\npe/{{\ensuremath{N_\text{pe}}}}

\DeclareSIUnit{\micro}{\ensuremath{\mu\!}} 
\DeclareSIUnit{\keVnr}{\text{keV}\ensuremath{_{\text{nr}}}}
\DeclareSIUnit{\eVnr}{\text{eV}\ensuremath{_{\text{nr}}}}
\DeclareSIUnit{\keVee}{\text{keV}\ensuremath{_{\text{ee}}}}
\DeclareSIUnit{\sample}{S}
\DeclareSIUnit{\erg}{erg}
\DeclareSIUnit{\pe}{PE}
\DeclareSIUnit{\year}{yr}
\DeclareSIUnit{\inch}{\text{inch}}
\DeclareSIUnit{\atom}{\text{atom}}
\DeclareSIUnit{\mwe}{\text{m.w.e.}}
\DeclareSIUnit{\Vpp}{\text{V}_\text{pp}}
\DeclareSIUnit{\adc}{\text{ADC counts}}
\DeclareSIUnit{\adcq}{\text{ADC counts}\ensuremath{\,\times\,2\,}\text{ns}}
\DeclareSIUnit{\MWh}{\text{MWh}}
\DeclareSIUnit{\GWh}{\text{GWh}}

\DeclareAcronym{spe}{short = SPE, long = single photoelectron}
\DeclareAcronym{fwhm}{short = FWHM, long = full width at half maximum}
\DeclareAcronym{cenns}{short = CE$\nu$NS, long = coherent elastic neutrino-nucleus scattering}
\DeclareAcronym{tunl}{short = TUNL, long = Triangle Universities Nuclear Laboratory}
\DeclareAcronym{psd}{short = PSD, long = pulse shape discrimination}
\DeclareAcronym{cfd}{short = CFD, long = constant fraction discriminator}
\DeclareAcronym{sns}{short = SNS, long = Spallation Neutron Source}
\DeclareAcronym{pmt}{short = PMT, long = photomultiplier tube}
\DeclareAcronym{pe}{short = PE, long = photoelectron}
\DeclareAcronym{pot}{short = POT, long = protons-on-target}
\DeclareAcronym{tof}{short = ToF, long = time-of-flight}
\DeclareAcronym{roi}{short = ROI, long = region of interest}
\DeclareAcronym{ssa}{short = SSA, long = shielded source area}
\DeclareAcronym{ccd}{short = CCD, long = charge coupled device}
\DeclareAcronym{sm}{short = SM, long = Standard Model}
\DeclareAcronym{wimp}{short = WIMP, long = Weakly Interacting Massive Particle}
\DeclareAcronym{dsnb}{short = DSNB, long = diffuse supernova neutrino background}
\DeclareAcronym{lar}{short = LAr, long = liquid argon}
\DeclareAcronym{nin}{short = NIN, long = neutrino-induced neutron}
\DeclareAcronym{sba}{short = SBA, long = super-bialkali}
\DeclareAcronym{qe}{short = QE, long = quantum efficiency}
\DeclareAcronym{hdpe}{short = HDPE, long = high-density polyethylene}
\DeclareAcronym{ps}{short = PS, long=Phillips Scientific}
\DeclareAcronym{ni}{short = NI, long=National Instruments}
\DeclareAcronym{mots}{short = MOTS, long=Mercury Off-Gas Treatment System}
\DeclareAcronym{pnnl}{short = PNNL, long = Pacific Northwest National Laboratory}
\DeclareAcronym{ornl}{short = ORNL, long = Oak Ridge National Laboratory}
\DeclareAcronym{pt}{short = PT, long = pretrace}
\DeclareAcronym{sca}{short = SCA, long = single-channel analyzer}
\DeclareAcronym{mc}{short = MC, long = Monte Carlo}
\DeclareAcronym{fom}{short = FOM, long = figure of merit}
\DeclareAcronym{pdf}{short = PDF, long = probability density function}
\title{First observation of coherent elastic neutrino-nucleus scattering}
\author{Bjorn Jorg Scholz}
\department{Physics}
\division{Physical Sciences}
\degree{Doctor of Philosophy} 
\date{December 2017}

\dedication{To my parents,\\who made me who I am today.}
\epigraph{\textit{Gutta cavat lapidem non vi,\\ sed saepe cadendo.}}

\begin{document}
\maketitle

\makecopyright
\makededication
\makeepigraph

\tableofcontents
\listoffigures
\listoftables

\acknowledgments
\input{acknowledgments.tex}

\abstract
Coherent elastic neutrino-nucleus scattering (CE$\nu$NS) has the largest predicted cross-section of all low-energy neutrino couplings. However, as a neutral-current interaction, the only experimental signature of CE$\nu$NS is a low-energy nuclear recoil, which made its detection challenging. CE$\nu$NS remained unobserved for over four decades. I will describe the experiment that resulted in a CE$\nu$NS observation at a 6.7-sigma confidence level, which was performed in the framework of the COHERENT collaboration. A low-background, 14.6-kg CsI[Na] scintillator was exposed to the neutrino emissions from the Spallation Neutron Source at Oak Ridge National Laboratory. Characteristic CE$\nu$NS signatures in energy and time, compatible with predictions from the Standard Model, were observed in high signal-to-background conditions. CE$\nu$NS provides new opportunities to study neutrino properties, and enables the miniaturization of detectors.


\mainmatter

\input{introduction.tex}

\input{cenns-theory.tex}

\input{coherent-at-sns.tex}

\input{background-studies.tex}

\input{csi-setup.tex}
\input{am-calibration.tex}
\input{ba-calibration.tex}

\input{qf-measurements.tex}

\input{sns-analysis.tex}

\input{future-outlook.tex}

\newpage
\addcontentsline{toc}{chapter}{Abbreviations}
\printacronyms[name=Abbreviations]

\newpage
\addcontentsline{toc}{chapter}{References}
{\singlespace{
\bibliographystyle{bjsThesis}
\bibliography{PhD-Thesis}}}

\end{document}

%% file: acknowledgments.tex
First and foremost thank my advisor, Juan Collar. Juan has given me a nearly endless number of opportunities to learn new things and to grow as a scientist. He gave me ample space and time to work on my own, but also provided guidance whenever I needed it. He supported me and my research with a passion I have rarely seen. His physics (and Jazz) knowledge is unrivaled. In short, I could have never wished for a better advisor.\par

Next I have to deeply thank Philipp Barbeau and Grayson Rich for all the help and input they provided for my analysis. Phil has never failed to miss even subtle flaws in my reasoning, but also always provided ample input on how to address potential issues. I will eternally be grateful for Grayson's help during the quenching factor measurements at TUNL. Without his help I would have never made it through the misery of that measurement.\par

I also want to thank Nicole Fields for all the work she has put into the characterization of the CsI detector. Without her work this thesis would have been impossible. Everyone else from the COHERENT collaboration also deserves my deepest gratitude. Especially Jason Newby and Yuri Efremenko, for all the help they provided at the SNS. Whenever an issue with the CsI detector arose, they promptly fixed it. Without their help many hours of good beam time would have been lost. As a result the significance of the observation presented in this thesis would have been much lower. I also want to thank Alexey Konovalov for all the work he has put into the parallel analysis pipeline of the CE$\nu$NS search data. Having two independent analyses come to the same conclusion puts credibility into both.\par

I am extremely grateful to everyone from AMCRYS-H. Despite the ongoing Ukraine crisis they were able to grow a perfect CsI[Na] crystal that met all the specs we required. I am also eternally grateful to everyone involved in the transportation of the crystal from Kharkiv to the United States. A safe passage to Chicago was far from guaranteed, due to the close proximity of Kharkiv to the contested borderlands.\par

Finally I'd like to thank my committee Paolo Privitera, Carlos Wagner and Philippe Guyot-Sionnest for their advice.

%% file: introduction.tex
%
%
\chapter{Introduction}
\label{chapter:introduction}

\begin{figure}[htbp]
\begin{center}
\includegraphics[width=4in]{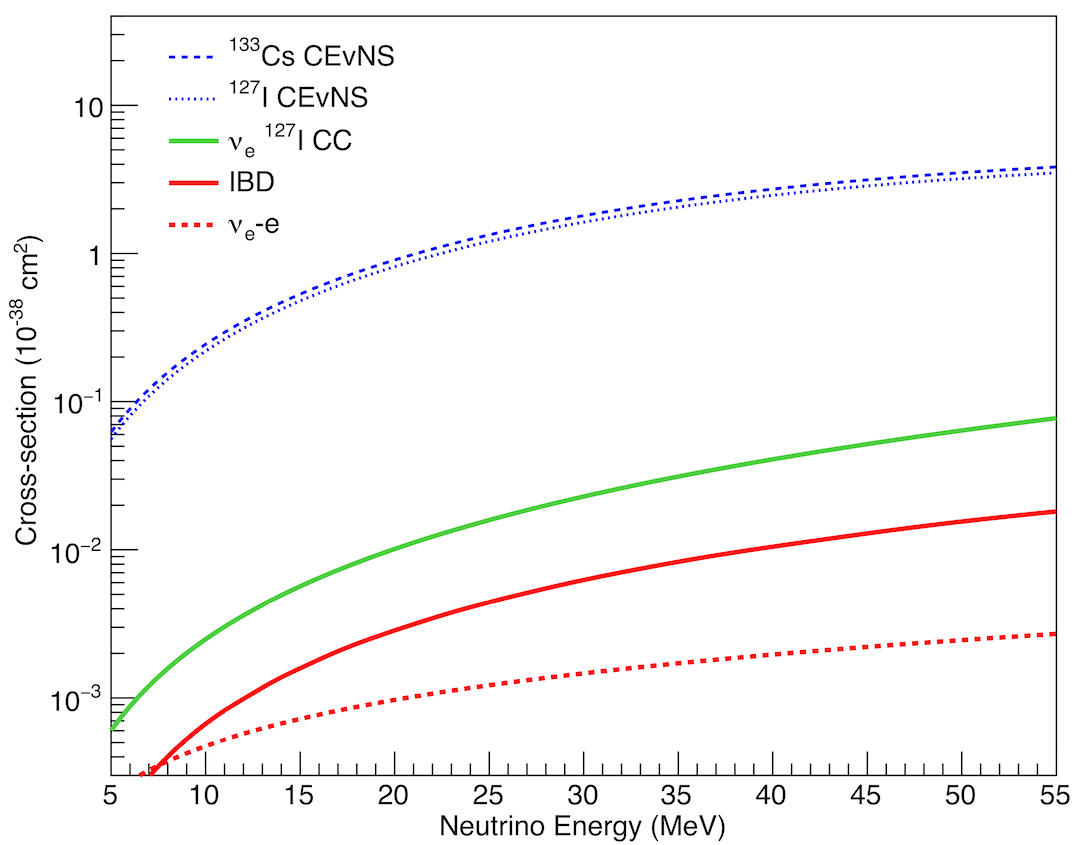}
\end{center}
\caption[Scattering \xs/s for different neutrino couplings]{Total \xs/ for \acs*{cenns} (blue) and other neutrino couplings. Shown are the \xs/s from charged-current (CC) interaction with iodine (green), inverse beta decay (red) and neutrino-electron scattering (dotted red). It is readily visible that \acs*{cenns} provides the largest \xs/, dominating over any charged-current interaction for incoming neutrino energies of less than \SI{55}{\MeV}. Plot adapted from \cite{collar-04}.}
\label{fig:cenns-theory:neutrino-cross-sections}
\end{figure}

In 1974, shortly after the first observation of a possible weak neutral current in neutrino-nucleus interactions \cite{hasert-01}, Daniel Z. Freedman provided the first theoretical description of \ac{cenns} as a \ac{sm} process~\cite{freedman-01}. In general, when a neutrino scatters off a nucleus the interaction depends on the non-trivial interplay between the neutrino and the individual nucleons. However, if the neutrino-nucleus momentum transfer is small enough, it does not resolve the internal structure of the nucleus, and as a consequence the neutrino scatters off the nucleus as a whole. This purely quantum mechanical effect gives rise to a coherent enhancement of the scattering \xs/, which scales approximately with the square of the target nucleus' neutron number. The resulting scattering \xs/ is therefore several orders of magnitude larger than any other neutrino-nucleus coupling (Fig.~\ref{fig:cenns-theory:neutrino-cross-sections}).\par

It might seem surprising that \ac{cenns} had eluded detection for over four decades given its large \xs/. However, being a neutral-current process, the only experimental signature of \ac{cenns} consists of difficult-to-detect nuclear recoils with energies of only a few \SI{}{\eVnr} to \SI{}{\keVnr}. In what follows the subscript `nr' emphasizes that the energy quoted is that of a nuclear recoil. In conventional radiation detectors only a small fraction of the total energy carried by such a nuclear recoil is converted into a detectable scintillation and ionization signal. The fraction between the detectable and the total energy is usually referred to as quenching factor~\cite{scholz-02}, and is typically on the order of a few to a few ten percent. To distinguish the quenched, detectable energy from the total energy of a nuclear recoil a subscript `ee' (electron equivalent) is used for the former throughout this thesis.\par

The low energy carried by \ac{cenns}-induced nuclear recoils combined with the aforementioned quenching factor had put a potential \ac{cenns} detection out of reach for any conventional neutrino detector. These detectors typically achieve an energy threshold of a couple of \SI{}{\MeV}~\cite{avanzini-01}, which is much larger than the threshold required to detect \ac{cenns}-induced recoils.\par

\begin{figure}[tb]
\begin{center}
\includegraphics[scale=1]{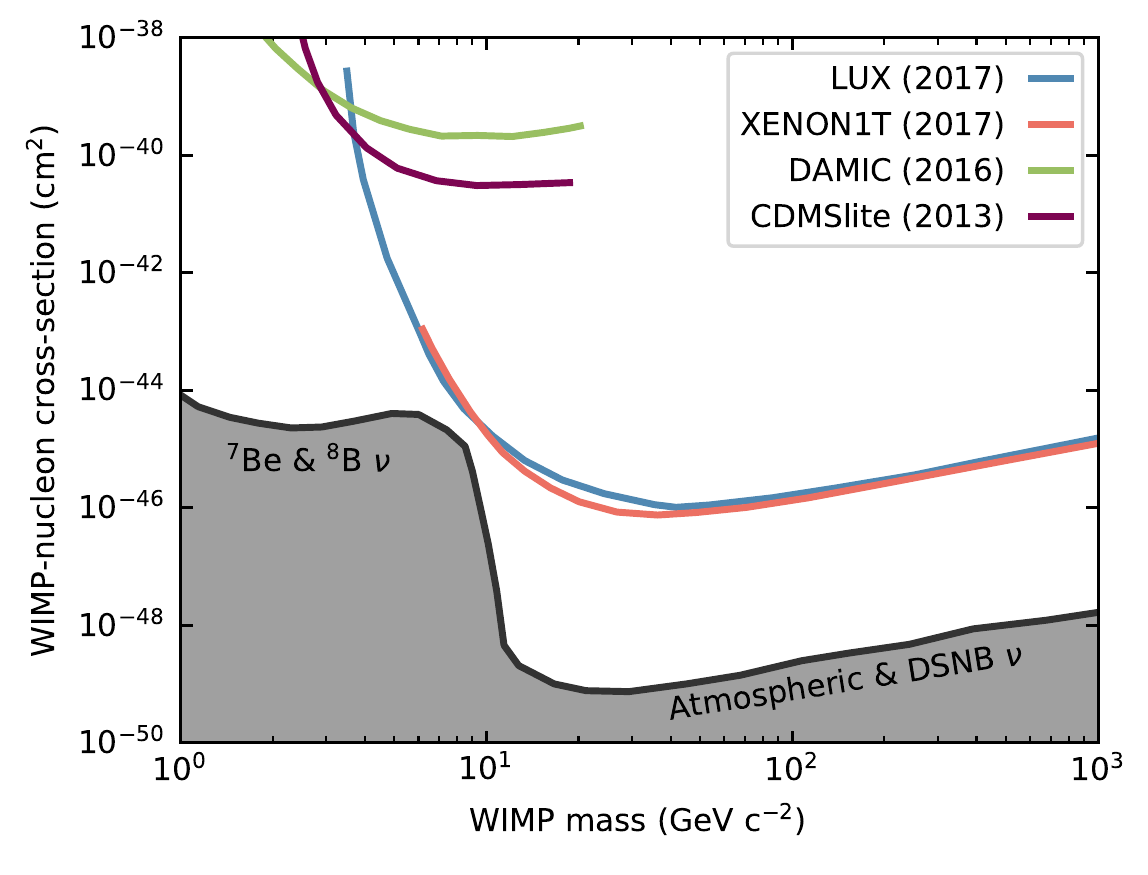}
\end{center}
\caption[\acs*{wimp} discovery limit due to the expected \acs*{cenns} neutrino floor]{The \acs*{wimp} discovery limit due to the expected \acs*{cenns} of neutrinos from different astrophysical sources is shown in black. Future sensitivity to low-mass \acsp*{wimp} will be limited by the neutrino emission from the sun, whereas the high-mass sensitivity is limited by atmospheric neutrinos and the \acs*{dsnb}. An increase in exposure would not significantly increase the sensitivity of dark matter experiments beyond this point. Yet there are still multiple ways to improve \acs*{wimp} sensitivity besides increasing exposure, such as the detection of an annual modulation or directional recoil measurements \cite{ohare-01}. Neutrino floor data is taken from \cite{billard-01}, CDMSlite \cite{agnese-01}, LUX \cite{akerib-01}, XENON1T \cite{aprile-01}, DAMIC \cite{aguilar-01}. Limits from other experiments were omitted for clarity. Additional information on current and planned dark matter experiments is provided in \cite{liu-01}.}
\label{fig:cenns-theory:neutrino-floor}
\end{figure}

A measurement of the \ac{cenns} \xs/ directly tests the description of neutrino-matter interactions as governed by the \ac{sm}. In addition it also substantiates the coherent \xs/ enhancement, which is a crucial component included in the calculation of limits on the \ac{wimp}-nucleus scattering \xs/. For a vanishing momentum transfer between the incoming \ac{wimp} and the target nucleus an analogous coherent enhancement of the scattering \xs/ is expected. As the neutrino coupling to protons is negligible for \ac{cenns}, the \ac{cenns} \xs/ scales with the total neutron number $N^2$. In contrast, the assumed dark matter coupling to both protons and neutrons is non-negligible and approximately equal~\cite{brice-01}. As a consequence, the \xs/ for \ac{wimp}-nucleus scattering scales with the square of the total nucleon number $A^2$ instead. Thus the expected enhancement of the \xs/ would be even larger.\par

A detection of \ac{cenns} also requires a shift in the modus operandi of the design of dark matter experiments. In the past, the response to an absent \ac{wimp} scattering signal usually was to increase the total exposure by building ever bigger detectors. This approach will no longer yield a significant increase in sensitivity, once these dark matter searches run into an irreducible \ac{cenns} background. This background is induced by neutrinos from astrophysical sources, such as the sun, or the \ac{dsnb}~\cite{billard-01}. Since the sole detectable signal for both \ac{cenns} and \ac{wimp} scattering is a low energy nuclear recoil it is impossible to distinguish the two. This inevitably requires a potential \ac{wimp} discovery to rely on additional information gained through new approaches, e.g. directional recoil measurements \cite{ohare-01}. Fig.~\ref{fig:cenns-theory:neutrino-floor} shows the so called \emph{neutrino floor}, i.e. the hard \ac{wimp} discovery limit for future dark matter experiments \cite{billard-01,billard-02} assuming a \ac{sm} \ac{cenns} \xs/. It also highlights current best limits on the \ac{wimp}-nucleus scattering \xs/ by several different dark matter experiments.\par

Another reason to measure \ac{cenns} stems from the desire to properly understand and model supernovae. When a star undergoes core collapse most of its gravitational energy is converted into energy driving the explosion. Among others, neutrino and nuclear physics are particularly important for describing this process. During the stellar core collapse approximately \SI{99}{\percent} of the gravitational energy, i.e. $E\approx\SI{3e53}{\erg}$, is emitted in \SI{}{\MeV} neutrinos compared to \SI{0.01}{\percent} as photons~\cite{ott-01}. About \SI{0.1}{\second} after the onset of the core collapse the density within the core reaches $\rho_c\,\sim\,\SI{e12}{\gram\per\cubic\cm}$~\cite{janka-01}. At this point the neutrinos no longer escape the core freely but remain trapped, coherently scattering off the heavy nuclei produced in the core. Given the vast amount of energy released via neutrinos it is thus crucial to have confidence in the \ac{cenns} \xs/s invoked in supernova calculations to properly describe this phase of the collapse.\par

Given the advances in ultra-sensitive detector technologies over the last decades, which was mainly driven by rare event searches such as dark matter or $0\nu\beta\beta$-decay experiments, it is now feasible to achieve energy thresholds low enough to directly measure \ac{cenns}-induced nuclear recoils and test the \ac{sm} prediction. There is a large variety of experiments currently being proposed and/or being built to measure \ac{cenns} using different neutrino sources as well as detector technologies.\par

One group of experiments aims to measure \ac{cenns} using a nuclear reactor as neutrino source \cite{ricochet-01, connie-01, miner-01, nucleus-01}, whereas another group aims to detect \ac{cenns} of neutrinos produced by a stopped pion source \cite{collar-02}. In a stopped pion source a proton beam impinges on a heavy target, which produces mesons. If the target is large enough the produced mesons are stopped and produce several neutrinos in their subsequent decay.\par

The flux emitted by a nuclear power plant is several orders of magnitude higher than what is to be expected from even the most intense stopped pion sources. However, the downside is the very low energy carried by these reactor neutrinos. Reactor experiments therefore have to achieve a sub-\SI{}{\keVee} energy threshold in order to measure the tiny recoil energies induced in the target material.\par

A broad array of detector technologies are used in reactor experiments to achieve this sub-\si{keVee} energy threshold. Among others there is Ricochet~\cite{ricochet-01}, which proposes a \SI{10}{\kg} array of low temperature Ge- or Zn-based bolometers to achieve an energy threshold of \SI{100}{\eV}. In contrast, CONNIE~\cite{connie-01} uses an array of \acp{ccd} to achieve an effective threshold of only \SI{28}{\eV}. MINER \cite{miner-01} is planning to use cryogenic germanium and silicon detectors in close proximity to a low power research reactor and {\Large{$\nu$}}-CLEUS~\cite{nucleus-01} further proposes to use cryogenic CaWO$_4$ and Al$_2$O$_3$ calorimeters to measure \ac{cenns}. All of these experiments build on the expertise gathered from previous efforts to measure \ac{cenns}, such as \cogent/ \cite{barbeau-02} and should in principle be able to measure \ac{cenns} if all design specifications are met.\par

The advantage of a stopped pion source lies in the higher energy of neutrinos produced, which is approximately one order of magnitude larger than the energy of reactor neutrinos. The \ac{cenns}-induced nuclear recoils at a stopped pion source therefore carry energies of up to tens of \SI{}{\keVnr}. Even after accounting for the quenching factor this corresponds to detectable energies of up to a few \SI{}{\keVee}, removing the requirement of a sub-\SI{}{\keVee} energy threshold. The neutrinos produced at a stopped pion source include three different flavors $\nu_\mu$, $\bar{\nu_\mu}$ and $\nu_e$, whereas the neutrino emission from a nuclear power plant only consists of $\bar{\nu_e}$. However, \ac{cenns} is insensitive to the neutrino flavor and as such all three flavors produced at a stopped pion source contribute to the nuclear recoil spectrum.\par

The experiment presented in this thesis is located at the \ac{sns}, a stopped pion neutrino source at \ac{ornl}. A low-background, \SI{14.57}{\kg} \csi/ detector was deployed $\SI{19.3}{\meter}$ away from the \ac{sns} mercury target, which isotropically emits three different neutrino flavors with well known emission spectra. The detector is located in a sub-basement corridor, which provides at least \SI{19}{\meter} of continuous shielding against beam-related backgrounds and a modest overburden of \SI{8}{\mwe} which reduces backgrounds induced by cosmic-rays. \csi/ provides an ideal target material for a \ac{cenns} search \cite{collar-02}. First, the large neutron number of both cesium and iodine result in a large enhancement of the \ac{cenns} \xs/ for both elements. Second, its excellent light yield, i.e. the number of \ac{spe} produced per unit energy, makes it possible to achieve an energy threshold of $\sim\SI{4.5}{\keVnr}$. After approximately two years of continuous data acquisition this thesis presents the first observation of \ac{cenns} at a $6.7$-$\sigma$ confidence level.\par

The theory of \ac{cenns} and its experimental detection is discussed in chapter~\ref{chapter:cenns-theory}. Chapter~\ref{chapter:coherent-at-the-sns} provides detailed information on the \ac{sns} and also provides an overview of the activities of the \coherent/ collaboration. Several different background studies were performed prior to the deployment of the \csi/ detector and are discussed in chapter~\ref{chapter:background-studies}. Three different \csi/ detector calibrations are covered in chapters~\ref{chapter:am-calibration},~\ref{chapter:ba-calibration} and ~\ref{chapter:quenching-calibration}. The full analysis of \ac{cenns} search data taken at the \ac{sns} as well as the first observation of \ac{cenns} is presented in chapter~\ref{chapter:sns-analysis}. Chapter~\ref{chapter:future-outlook} provides a brief summary of the findings presented in this thesis and shortly discusses future efforts by \coherent/ aiming to decrease the uncertainty on the \ac{cenns} cross-section, test the $N^2$ dependance, and to put more stringent limits on non-standard interactions.

%% file: cenns-theory.tex
%
%
\chapter{Coherent Elastic Neutrino-Nucleus Scattering}
\label{chapter:cenns-theory}

\begin{figure}[htbp]
\begin{center}
\includegraphics[width=2.3in]{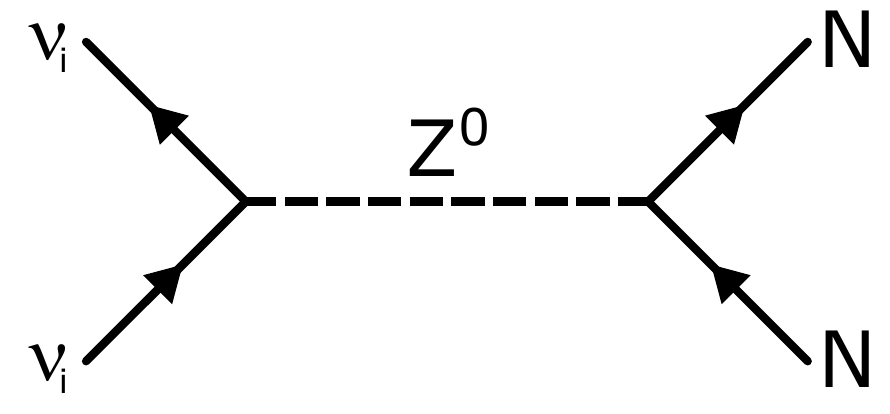}
\end{center}
\caption[Feynman diagram for \acs*{cenns}]{Feynman diagram for \acs*{cenns}. Here $\nu_i$ describes both neutrinos and anti-neutrinos of any flavor and N denotes any nucleus.}
\label{fig:cenns-theory:feynman-diagram}
\end{figure}

In the \ac{sm} \ac{cenns} is mediated via $Z^0$ exchange (Fig.\ \ref{fig:cenns-theory:feynman-diagram}). Being a neutral current process \ac{cenns} is insensitive to the flavor of the incoming neutrino resulting in an identical scattering \xs/ for all neutrino types apart from some minor corrections. Following Drukier and Stodolsky \cite{drukier-01}, the differential, spin-independent \ac{cenns} \xs/ assuming a negligible momentum transfer (i.e. $q \equiv \left|\vec{q}\right|\rightarrow 0$) between neutrino and nucleus can be written as
\begin{align}
\frac{\text{d}\sigma_0}{\text{d}\cos\phi} = \frac{G_f^2}{8\pi}\left[Z\left(4\sin^2\Theta_\text{W}-1\right)+N\right]^2E_\nu^2\left(1+\cos\phi\right), \label{eq:cenns-theory:diff-cross-section-1}
\end{align}
where $G_f$ is the Fermi coupling constant, $Z$ and $N$ are the number of protons and neutrons in the target nucleus respectively, $\sin^2\Theta_\text{W}$ is the weak mixing angle, $E_\nu$ is the incoming neutrino energy, and $\phi$ is the scattering angle. Any contribution from axial vector currents and any radiative corrections above tree level were neglected in Eq.\ (\ref{eq:cenns-theory:diff-cross-section-1}). Including these contributions at this point would not lead to a deeper understanding of \ac{cenns} but would rather only distract from the core concepts driving \ac{cenns}.\\

Since for low momentum transfers $\sin^2\Theta_\text{W}\,=\,0.23867\pm0.00016\,\approx\,\frac{1}{4}$~\cite{erler-01}, Eq.~(\ref{eq:cenns-theory:diff-cross-section-1}) can be simplified by eliminating any contribution from proton coupling in a first order approximation. It is further beneficial to express Eq.\ (\ref{eq:cenns-theory:diff-cross-section-1}) in terms of the three momentum transfer $q$ between neutrino and nucleus, which yields
\begin{align}
\frac{\text{d}\sigma_0}{\text{d}q^2} = \frac{G_f^2}{8\pi}N^2\left(1-\frac{q^2}{q_\text{max}^2}\right)\quad\text{with}\quad q^2=2E_\nu^2\left(1-\cos\phi\right).
\end{align}
Here $q_\text{max}$ denotes the maximum momentum transfer at $\phi=\SI{180}{\degree}$. As the only visible outcome of a \ac{cenns} event is an energy deposition in the detector in the form of a nuclear recoil it is advantageous from an experimental point of view to express the differential \xs/ in terms of the recoil energy $E_r$ carried by the target nucleus. It is
\begin{align}
\frac{\text{d}\sigma_0}{\text{d}E_r} = \frac{G_f^2}{4\pi} m_a N^2\left(1-\frac{m_a E_r}{2 E_\nu^2}\right)\quad\text{as}\quad E_r = \frac{q^2}{2 m_a}, \label{eq:cenns-theory:diff-cross-section-2}
\end{align}
where $m_a$ denotes the nuclear mass of the target material. It is easy to verify that the maximum induced recoil energy scales with the square of the incoming neutrino energy, i.e. $E_r^\text{max}=\nicefrac{2E_\nu^2}{m_a}$. As such it might seem beneficial to choose a neutrino source with the highest possible energy.\par

However, Eq.\ (\ref{eq:cenns-theory:diff-cross-section-1}) only holds in the limit of vanishing momentum transfer $q\rightarrow 0$ and as such $E_r\rightarrow 0$. The effective \xs/ actually decreases with increasing $q$, as the effective de Broglie wavelength $\nicefrac{h}{q}$ approaches the size of the nucleus. At this point the target does no longer recoil coherently, but rather scatters as a collection of individual nucleons. This loss of coherence becomes important for momentum transfers which satisfy $qR_A\gtrsim 1$, where $R_A$ is the radius of the nucleus \cite{drukier-01} and which is approximately given by $R_A=A^{\nicefrac{1}{3}}\cdot\SI{1.2}{\femto\meter}$, where $A$ is the mass number of the target element \cite{klein-01}. For \isotope{Cs}{133} and \isotope{I}{127} this corresponds to a momentum transfer on the order of $q\approx\SI{30}{\MeV}$. Following Lewin and Smith \cite{lewin-01}, this effect can be incorporated by introducing a form factor $F(q)$ that only depends on the momentum transfer and is independent of the nature of the interaction. It is the Fourier transform of the ground state mass density profile of the nucleus, normalized such that $F^2(q=0)=1$. The effective differential \xs/ can then be written as \cite{engel-01}
\begin{align}
\frac{\text{d}\sigma}{\text{d}E_r} = \frac{\text{d}\sigma_0}{\text{d}E_r} F^2\left(q\right)\quad\text{with}\quad q^2 = 2m_aE_r.\label{eq:cenns-theory:form-factor-inclusion}
\end{align}
A form factor $F(q)$ is adopted as proposed by Klein and Nystrand \cite{klein-01} that is based on the approximation of a Woods-Saxon nuclear density profile by a hard sphere of radius $R_A$ convoluted with a Yukawa potential of range $a$. The corresponding form factor can then simply be written as the product of the Fourier transformation of each individual density profile, i.e.
\begin{align}
F(q) &= \frac{4\pi\rho_0R_A^2}{Aq}j_1(qR_A)\frac{1}{1+a^2q^2}\label{eq:cenns-theory:form-factor}\\
     &= \frac{4\pi\rho_0}{Aq^3}\left(\sin\left(qR_A\right)-qR_A\cos\left(qR_A\right)\right)\frac{1}{1+a^2q^2}\\
\text{with}\quad\rho_0 &= \frac{3A}{4\pi R_A^3},\quad a = \SI{0.7}{\femto\meter},\quad R_A = A^{\nicefrac{1}{3}}\cdot\SI{1.2}{\femto\meter}
\end{align}
where $j_1$ denotes the first order spherical Bessel function of the first kind.\\

\begin{figure}[tb]
\begin{center}
\includegraphics[scale=1]{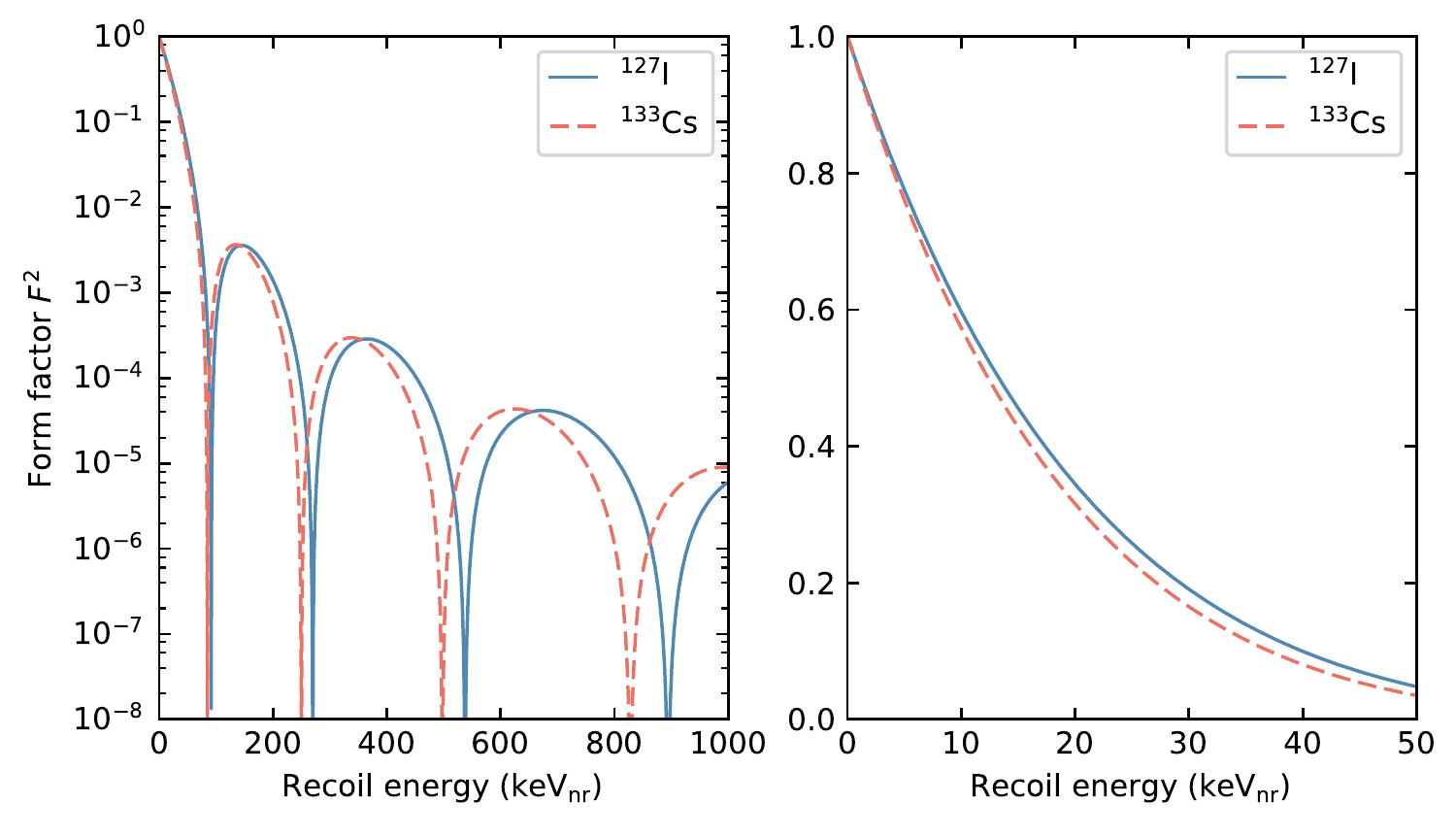}
\end{center}
\caption[Form factor $F^2$ versus recoil energy $E_r$ for \isotope{I}{127} (blue) and \isotope{Cs}{133} (red).]{Form factor $F^2$ versus recoil energy $E_r$ for \isotope{I}{127} (blue) and \isotope{Cs}{133} (red). The left panel shows the form factor over a wide range of recoil energies. The right panel focuses on the recoil energies of interest to a \acs*{cenns} search at the \acs*{sns}. A steep drop in $F^2$ for increasing recoil energies leads to a suppression of high energy recoils in the resulting \ac{cenns} spectrum.}
\label{fig:cenns-theory:form-factor}
\end{figure}

\begin{figure}[tb]
\begin{center}
\includegraphics[scale=1]{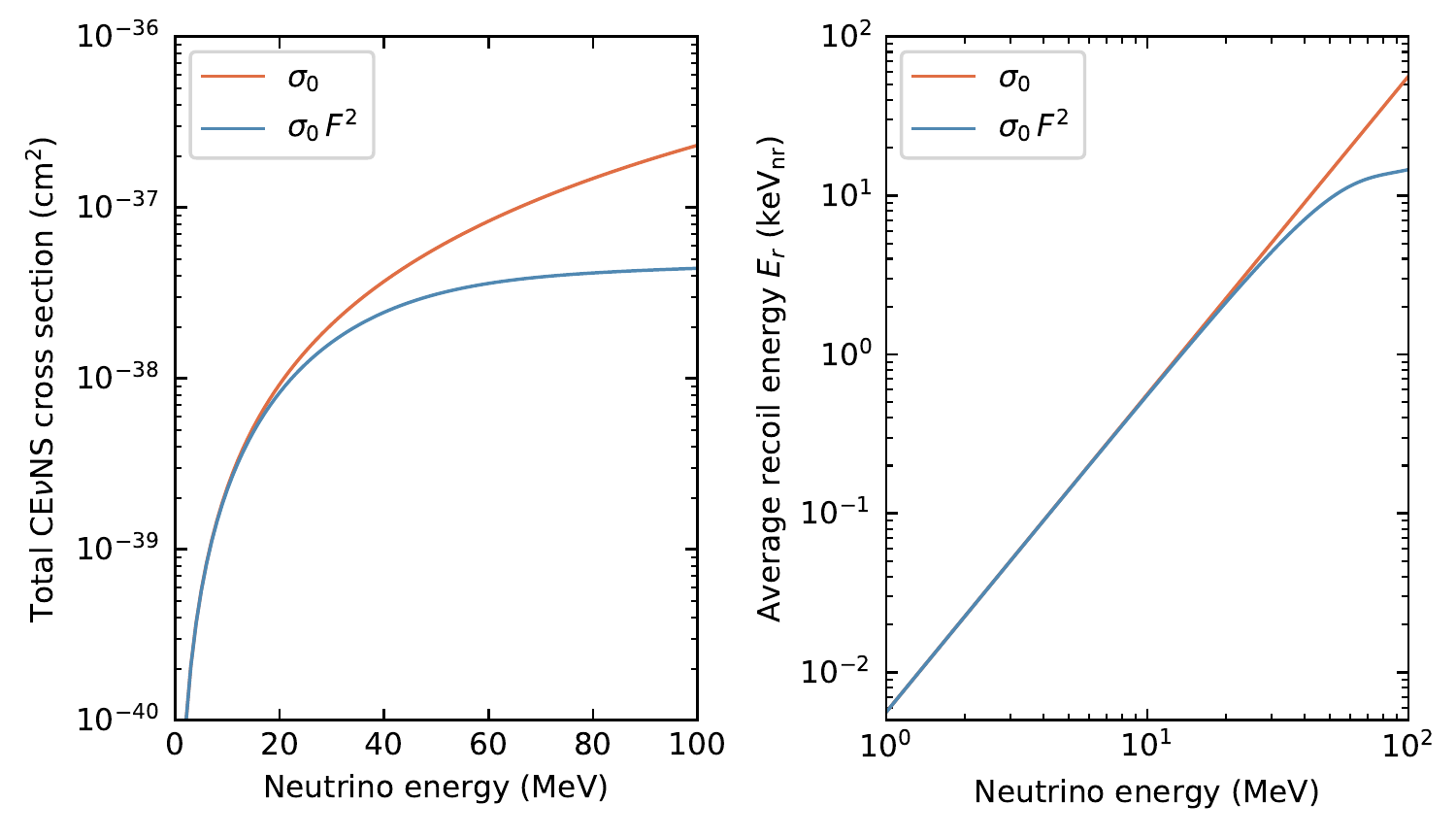}
\end{center}
\caption[Total \acs*{cenns} cross-section and average recoil energy]{Left: Total \acs*{cenns} cross section for \isotope{I}{127} versus incoming neutrino energy. Right: Average recoil energy of \isotope{I}{127} versus incoming neutrino energy. Both panels include calculations with and without the inclusion of the form factor.}
\label{fig:cenns-theory:total-cross-section}
\end{figure}

Fig.~\ref{fig:cenns-theory:form-factor} shows the form factor for cesium and iodine as calculated using Eq.~(\ref{eq:cenns-theory:form-factor}) where the momentum transfer was converted into the experimentally more accessible recoil energy of the target nucleus. The left panel shows the behavior over a wide range of recoil energies whereas the right panel focuses on the ones expected at the \ac{sns}. There is a steep drop in $F^2$ for increasing recoil energies. According to Eq.~(\ref{eq:cenns-theory:form-factor-inclusion}), this directly leads to a heavy suppression of the differential \xs/ for large recoil energies. The experimental consequence of this can be seen in Fig.~\ref{fig:cenns-theory:total-cross-section}. The left panel shows a comparison of the total \ac{cenns} \xs/ for \isotope{I}{127}, calculated with and without the inclusion of the form factor. For incoming neutrino energies of $E_\nu\gtrsim\SI{30}{\MeV}$ the total \xs/ starts to flatten out if $F^2$ is included. With increasing $E_\nu$ the available phase-space  to integrate the differential \xs/ increases, as $E_r^\text{max}\propto E_\nu^2$. However, large recoil energies are heavily suppressed by the form factor and as a result their contribution to the total \xs/ is negligible. The right panel of Fig.~\ref{fig:cenns-theory:total-cross-section} shows the average recoil energy induced in a hypothetical detector, which increases linearly with the incoming neutrino energy for $E_\nu\lesssim\SI{30}{\MeV}$ but then again tapers off. The exact rate and shape of this tapering is solely determined by the form factor and as such the choice of $F(q)$ can significantly influence the overall recoil spectrum. It is therefore important to note that there exist several different form factor models which are based on different representations of the nucleon density. A more detailed discussion on different form factor models can be found in \cite{lewin-01} and \cite{papoulias-01}. The latter reference also provides a more extensive and rigorous calculation of the \ac{cenns} \xs/.\\

More detailed \ac{cenns} \xs/ calculations were carried out within the \coherent/ collaboration including several different corrections to provide a precise \ac{sm} prediction \cite{barbeau-03, barbeau-04}. These include corrections such as the inclusion of axial vector currents and radiative corrections, which lead to an increase in the total \xs/ on the order of 5-\SI{10}{\percent} depending on the exact isotope. Different form factor models were tested ($\pm\mathcal{O}(\SI{5}{\percent})$), as well as the impact of differences in the proton and neutron form factor ($<\SI{1}{\percent}$). The inclusion of the strange quark radius was found to be negligible. The impact of strange quark contribution to the nuclear spin was found to be $\mathcal{O}(\SI{1}{\percent})$ for low $N$, and is negligible otherwise. A small contribution of weak magnetism to the \xs/ was also found to be negligible. The most important correction is the inclusion of different effective neutrino charge radii for different flavors \cite{sehgal-01,barbeau-03}. At low $q\approx0$ this can be incorporated by replacing $\sin^2\Theta_\text{W}$ with
\begin{align}
\sin^2\Theta_\text{eff} = \sin^2\Theta_\text{W} + \frac{\alpha}{6\pi}\ln\left(\frac{m_i^2}{m_\text{W}^2}\right)
\end{align}
where $m_i$ is the mass of the charged lepton associated with the neutrino $\nu_i$. This immediately leads to different \ac{cenns} \xs/s for different neutrino flavors, where the ratio between \xs/ can approximately be written as
\begin{align}
\left.\frac{\sigma(\nu_i)}{\sigma(\nu_j)}\right|_{q^2=0} \approx 1 + \frac{\alpha}{3\pi\sin^2\Theta_\text{W}}\ln\left(\frac{m_i^2}{m_j^2}\right).
\end{align} 
The \xs/ of muon neutrinos over electron neutrinos is therefore increased by a factor of $\sim\SI{3.5}{\percent}$. This effect grants the possibility to distinguishing different neutrino flavors via neutral current interactions once precision measurements of the \ac{cenns} \xs/ are achieved.

%% file: coherent-at-sns.tex
%
%
\chapter{COHERENT at the Spallation Neutron Source}
\label{chapter:coherent-at-the-sns}
\section{The Spallation Neutron Source}
\label{section:coherent-at-the-sns:sns}
The \acf{sns}, located at \acl{ornl}, provides a a large group of interdisciplinary researchers with the most intense, pulsed neutron beams in the world. A proton beam narrow in time with a timing \ac{fwhm} of approximately $\Delta t_\text{Beam}\sim\SI{380}{\ns}$ impinges on a liquid mercury target with a repetition rate of \SI{60}{\Hz}. The \ac{sns} is theoretically capable of delivering a total beam power on target of up to \SI{1.4}{\MW}. Combined with a typical proton energy of $E_p\approx\SI{1}{\GeV}$ this amounts to a total proton rate of $\Gamma_p\approx\SI{e16}{\per\second}$. Upon impact on a target nucleus, protons only interact with an individual nucleon instead of forming a compound, as their de Broglie wavelength is only $\sim\SI{0.1}{\femto\meter}$, and as such much smaller than the nucleus itself. Kinetic energy is transmitted from a proton to the nucleon via elastic collisions after which an intra-nuclear cascade ensues \cite{cugnon-01,krasa-01}. During this nucleon-cascade neutrons are spallated from the target nucleus, and also pions $\pi^\pm$ are produced. The pions are stopped within the target, where the $\pi^-$ are mostly recaptured by the mercury. The $\pi^+$, in contrast, decay at rest into a positive muon and a muon neutrino. The muon subsequently decays in-flight into a positron, an electron neutrino, and an anti-muon neutrino.\par

After the initial nucleon-cascade, which lasts about \SI{e-22}{\second}, the nucleus is left in a highly excited state. It then loses its remaining energy over $\sim\SI{e-16}{\second}$ mainly via neutron evaporation. Over the course of the intra-nuclear cascade and the following evaporation an average of 20-30 neutrons per incoming proton are spallated from the target~\cite{weiren-01}. Being produced in the intra-nuclear cascade the timing of this neutron emission is only associated directly with the beam. For the reminder of this thesis these beam-related neutrons are labeled as prompt. Additionally a total of $\sim0.08$ ($\pm\SI{10}{\percent}$) neutrinos per flavor per proton are produced during the full process \cite{collar-04} where all neutrinos are emitted isotropically from the source.\par

\begin{figure}[tb]
\begin{minipage}{3.25in}
\begin{center}
\includegraphics[width=3.25in]{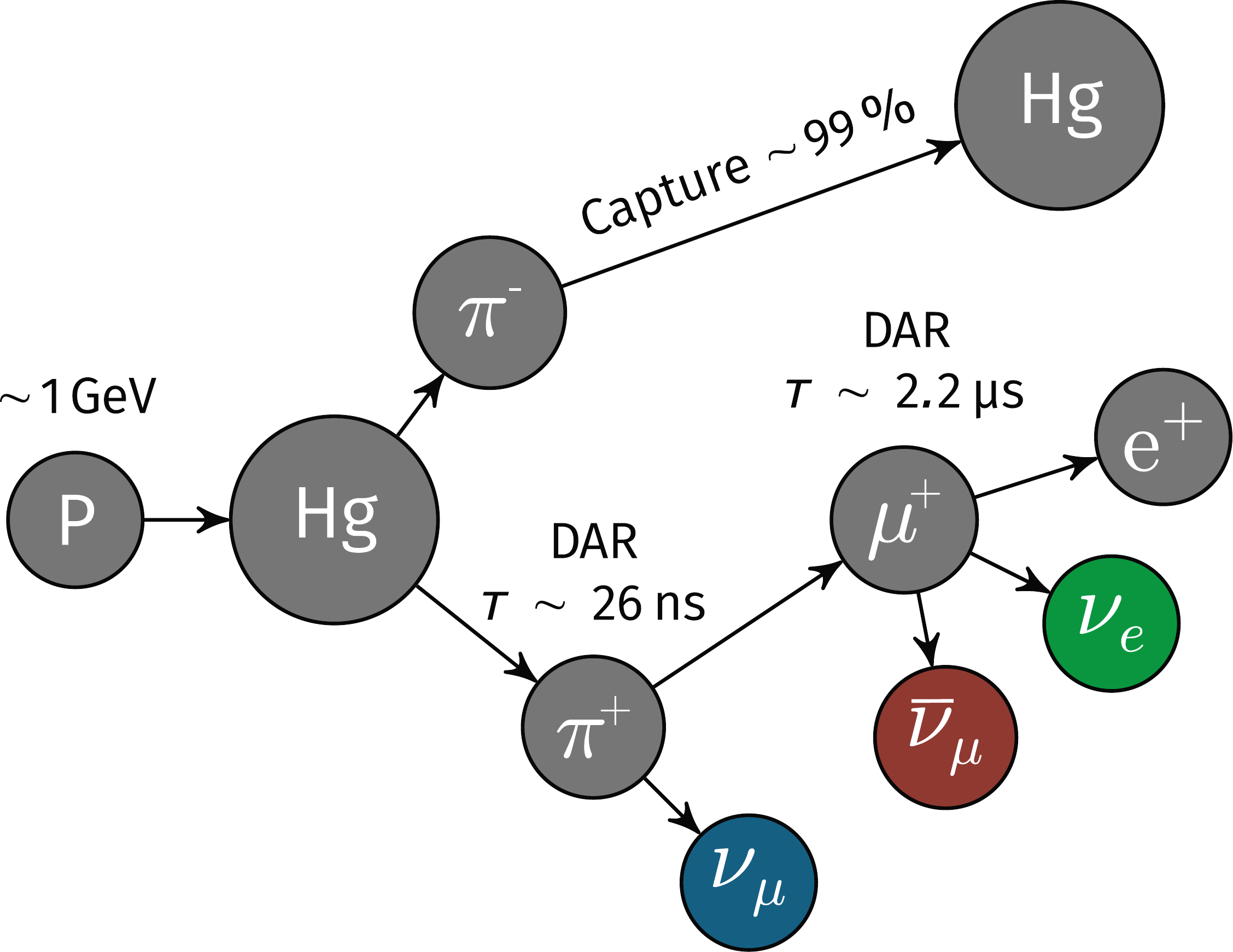}
\end{center}
\end{minipage}
\begin{minipage}{3.20in}
\begin{center}
\includegraphics[scale=1]{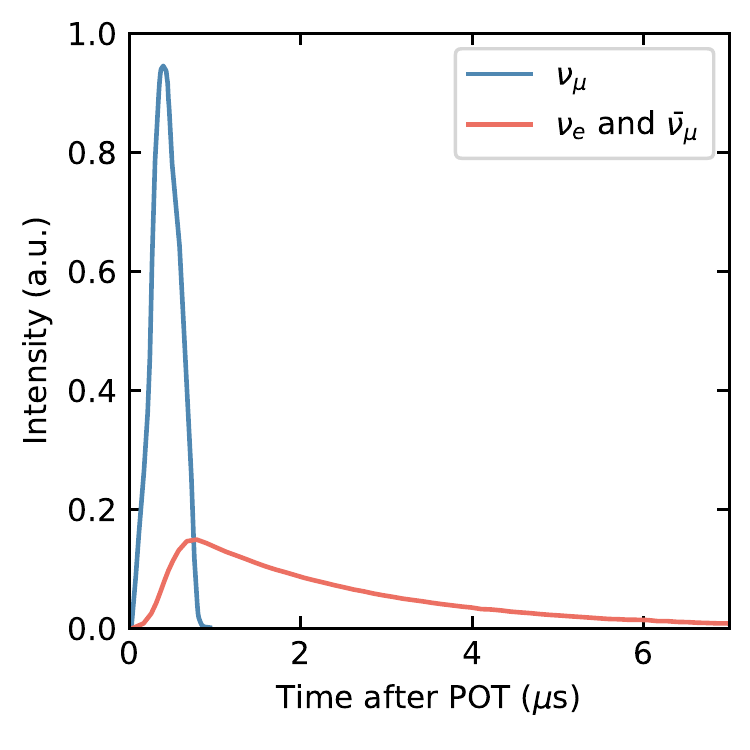}
\end{center}
\end{minipage}
\caption[Neutrino production mechanism at the \acs*{sns} and timing profile of neutrino emission]{\textbf{Left}: Neutrino production mechanism at the \acs*{sns}. A proton beam impinges on a mercury target producing pions. Three different types of neutrinos are produced within the subsequent decay chain. Secondary neutrino emission cascades are not shown as they only contribute a negligible amount of contamination. \textbf{Right}: Timing profile of neutrino emission following \acf*{pot} as determined by the Florida group of the \coherent/ collaboration using \geant/ simulations. A distinct difference in the arrival time is apparent between $\nu_\mu$, originating from pion decay, and the $\nu_e\;+\;\bar{\nu}_\mu$ emitted in the subsequent muon decay. The latter pair of neutrinos closely follows the distinct \SI{2.2}{\micro\second} muon decay profile.}
\label{fig:cenns:sns-neutrino-production-mechanism}
\end{figure}

A schematic of the neutrino production mechanism is illustrated in the left panel of Fig. \ref{fig:cenns:sns-neutrino-production-mechanism}. The right panel shows the timing profile of the neutrino emission and further illustrates the timing profile of each neutrino species. The $\nu_\mu$ are directly produced within the nucleon-cascade. As such their time profile closely follows the proton beam profile. The $\nu_e$ and $\bar{\nu}_\mu$ emission follows a convolution of the beam profile with the $\sim$\SI{2.2}{\micro\second} muon decay. The neutrino emission can therefore be categorized as prompt ($\nu_\mu$) and delayed ($\nu_e$, $\bar{\nu}_\mu$).\par

The pulsed neutrino emission profile is highly beneficial for background suppression. Potential \ac{cenns} signals can only arise within a couple of microseconds directly following the \ac{pot} trigger. As such it is possible to reduce the steady-state contribution from environmental and cosmic-ray induced radiation by approximately three to four orders of magnitude. However, the \ac{sns} also produces on the order of $10^{17}$ neutrons per second. Some of these can escape the shielding monolith surrounding the mercury target, leading to a potential beam-related background. This neutron background was assessed prior to the main detector deployment. It is described in chapter~\ref{chapter:background-studies} and found to be negligible.\par

\begin{figure}[tb]	
\begin{center}
\includegraphics[scale=1]{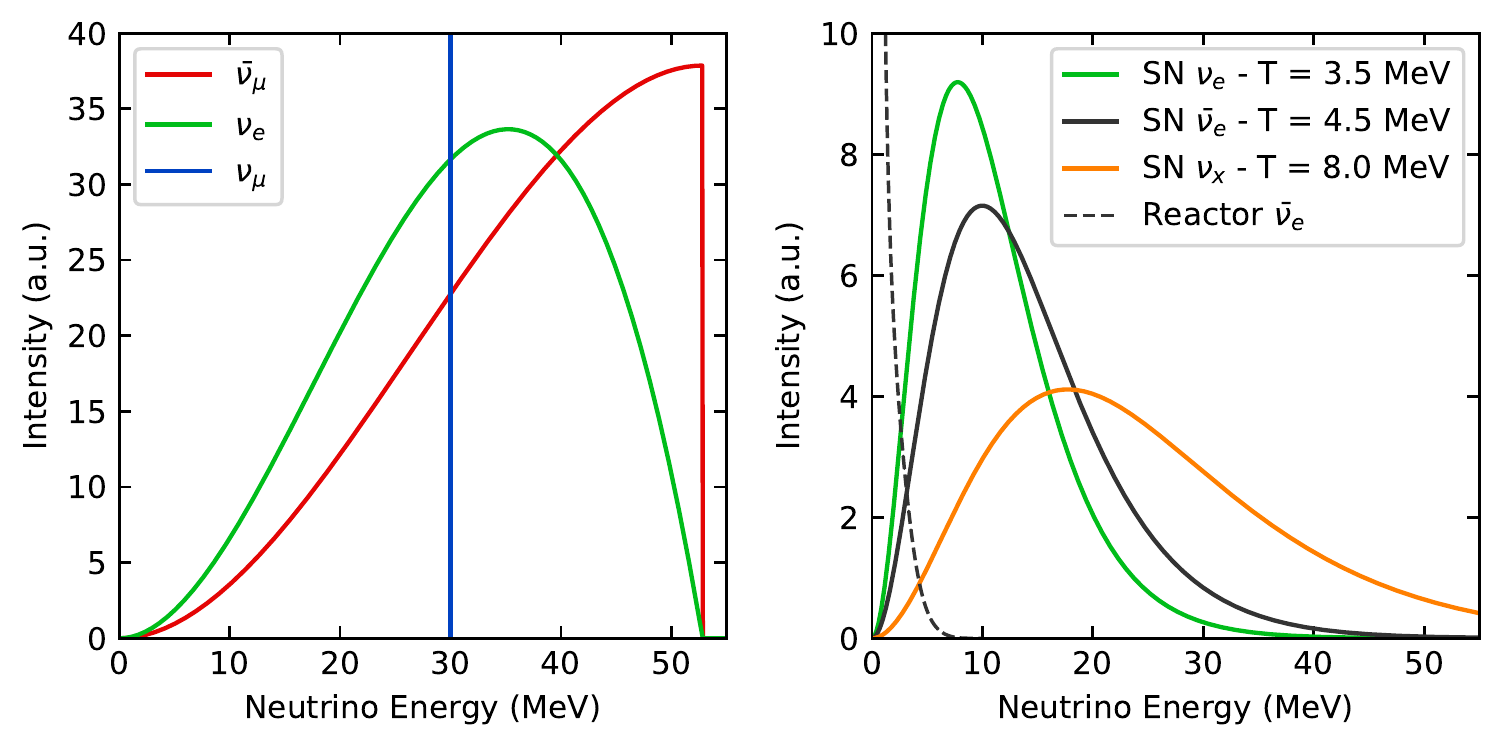}
\end{center}
\caption[\acs*{sns} and supernova neutrino emission spectrum]{\textbf{Left}: Neutrino spectra emitted from a stopped pion source, e.g. the \acs*{sns}. \textbf{Right}: Typical neutrino spectra emitted during stellar core collapse (solid) \cite{avignone-01} and emitted from a nuclear reactor (dashed) \cite{mueller-01}. Here $\nu_x$ denotes all other neutrino and anti-neutrino flavors. A large overlap between neutrino energies probed at the \acs*{sns} and the emission from supernovae is apparent.}
\label{fig:sns:sns-and-supernova-neutrino-spectrum}
\end{figure}

The exact energy emission spectra for all neutrino flavors can be analytically calculated as all processes involved in the neutrino production are well understood. The $\pi^+ \rightarrow  \mu^+ + \nu_\mu$ two body decay-at-rest results in a monochromatic neutrino energy of 
\begin{align}
\text{P}(E_\nu|\nu_\mu) = \delta\left(E_\nu - \frac{m_\pi^2 - m_\mu^2}{2m_\pi}\right) = \delta\left(E_\nu - \SI{29.79}{\MeV}\right)\label{eq:sns:nmu-emission}
\end{align}

The neutrino emission spectra for the $\mu^+\rightarrow e^+ + \nu_e + \bar{\nu}_\mu$ decay can be written as \cite{avignone-01}
\begin{align}
\text{P}(E_\nu|\nu_e) &= \frac{12}{W^4}E_\nu^2\left(W-E_\nu\right)\label{eq:sns:ne-emission}\\
\text{P}(E_\nu|\bar{\nu}_\mu) &= \frac{6}{W^4}E_\nu^2\left(W-\frac{2}{3}E_\nu\right),\label{eq:sns:anmu-emission}
\end{align}
where $W=\SI{52.83}{\MeV}$. The energy spectrum for each flavor is shown in Fig.~\ref{fig:sns:sns-and-supernova-neutrino-spectrum}. The emitted neutrinos meet the coherence criterion ($E_\nu\leq\SI{100}{\MeV}$) introduced in chapter~\ref{chapter:cenns-theory}. The right panel further shows the supernova neutrino emission spectra estimated using a Fermi-Dirac distribution with characteristic neutrino temperatures of $T(\nu_e)=\SI{3.5}{\MeV}$, $T(\bar{\nu}_e)=\SI{4.5}{\MeV}$ and $T(\nu_x)=\SI{8}{\MeV}$ \cite{avignone-01,keil-01}, i.e.

\begin{align}
f_\text{FD}(E_\nu,T) = \frac{E_\nu^2}{2T^3}\frac{1}{\exp^{\nicefrac{E_\nu}{T}} + 1}.
\end{align}

The overlap in neutrino energies probed at the \ac{sns} and in supernovae are apparent. The \ac{cenns} search at the \ac{sns} dsecribed in this thesis can therefore directly validate coherent scattering for neutrinos carrying energies similar to those produced in a stellar core collapse (chapter \ref{chapter:introduction}). The reactor neutrino emission spectrum as provided in \cite{mueller-01} is also shown. The emitted neutrino spectrum has a much lower energy than what is produced at the \ac{sns}. As such, there are some advantages to detect \ac{cenns}-induced nuclear recoils at the \ac{sns}, even if the total recoil rate is significantly lower.\par

\geant/~\cite{geant4} simulations were performed by the Florida group of the \coherent/ collaboration regarding the neutrino yield, neutrino spectra, and timing profiles as provided by the \ac{sns}. A negligible energy contamination of $<\SI{1}{\percent}$ from decay-in-flight and muon capture was found above the endpoint of the Michel spectrum~\cite{collar-04}.

\section{The COHERENT experiment at the SNS}
The work described in this thesis was done in the framework of the international \coherent/ collaboration. \coherent/ consists of approximately 80 researchers with strong backgrounds in rare event searches such as dark matter or $0\nu\beta\beta$-decay experiments. Yet, from the beginning it was clear that only a multi-target approach will be able to fully utilize the power of \ac{cenns} to probe and constrain neutrino physics beyond the \ac{sm}. The first observation of \ac{cenns} described in this thesis is a remarkable achievement and already provided improved constraints on non-standard interactions between neutrinos and quarks~\cite{collar-04}. However, due to the low event statistics and the large uncertainties involved, many of the more interesting physics questions beyond a first observation cannot be addressed yet.\par

\begin{figure}[!t]	
\begin{center}
\includegraphics[width=6.5in]{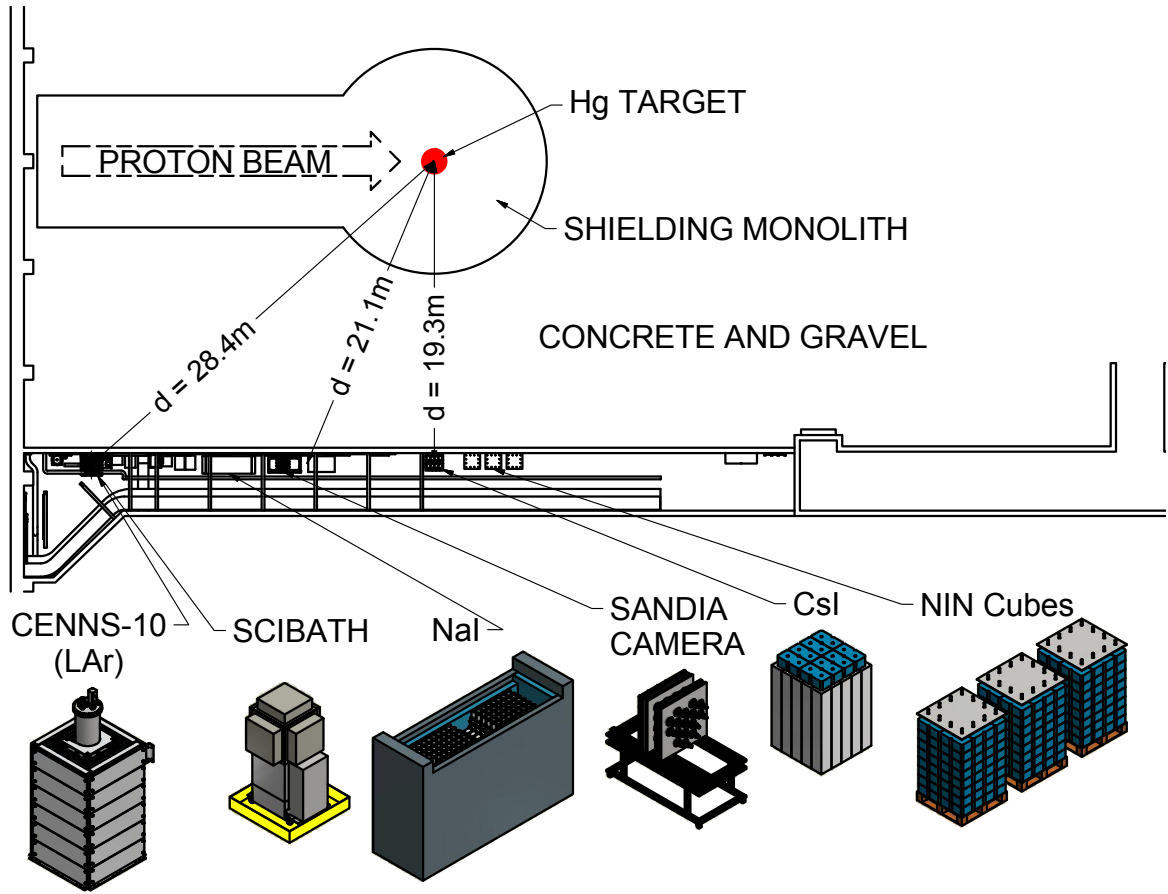}
\end{center}
\caption[Detectors deployed along the neutrino alley in the framework of the \coherent/ collaboration]{Past, current, and future \coherent/ detectors as positioned in a basement corridor at the \acs*{sns}. The corridor is now dubbed 'neutrino alley' and provides at least \SI{19}{\m} of shielding against beam-related backgrounds. It further offers a total of \SI{8}{\mwe} of overburden, helping to reduce cosmic-ray induced backgrounds. The CENNS-10, NaI and CsI detectors are designed to search for \ac{cenns} signals. The Sandia Camera and Scibath on the other hand measured the prompt neutron rate at multiple positions along the alley. The NIN cubes are currently measuring the \xs/ for neutrino-induced neutrons in Pb and Fe. The detector locations and their distance to the mercury target were determined using precision survey measurements performed by \ac{ornl}. These measurements are accurate to $\pm\SI{1}{\cm}$. Figure adapted from \cite{collar-04}.}
\label{fig:sns:neutrino-alley}
\end{figure}

Using multiple target materials the \coherent/ experiment will further probe the $N^2$ dependance of the scattering \xs/ as predicted by Eq.~(\ref{eq:cenns-theory:diff-cross-section-2}). Current operational \ac{cenns} detectors include \SI{14.57}{\kg} of \csi/ as described in this thesis, a \SI{28}{\kg} single-phase \ac{lar} detector and \SI{185}{\kg} of \nai/. An additional \ac{cenns} search using up to \SI{20}{\kg} of p-type point contact germanium detectors \cite{barbeau-01} is planned for deployment in late 2017. Recent advancements enable a threshold of a few hundred \SI{}{\keVee} in these detectors~\cite{aalseth-04}. All of these setups are located in a basement corridor at the \ac{sns}, which is now dubbed 'neutrino alley' (Fig. \ref{fig:sns:neutrino-alley}). This prime location provides at least \SI{19}{\m} of shielding against beam-associated backgrounds, such as prompt neutrons. It also offers an overburden of approximately \SI{8}{\mwe} (meters of water equivalent) helping to further reduce cosmic-ray induced backgrounds in the detectors.\par

In addition to the aforementioned detectors, \coherent/ also deployed several different neutron monitors to verify the low prompt neutron background expected from the source. The neutron flux was measured at several positions within the neutrino alley (Fig. \ref{fig:sns:neutrino-alley}). In the near future, \coherent/ will deploy another neutron monitor close to the \csi/ detector further substantiate the current neutron flux measurements.\par

Another potential background was identified in the framework of the \coherent/ collaboration. This background originates from \acp{nin} in the lead shield surrounding the detectors. Accordingly, another detector array was deployed that is dedicated to measuring the \xs/ of this process. A detailed description of these background measurements is presented in chapter~\ref{chapter:background-studies}.

%% file: background-studies.tex
\chapter{Background Studies}
\label{chapter:background-studies}

As discussed in the previous chapter several different neutron detectors were deployed along the 'neutrino alley' prior to the \csi/ experimentation. These measurements confirmed a negligible background level. This ensures a high signal-to-background level for the \ac{cenns} search described in this thesis. These measurements are described below.\par

Two early measurements of the prompt neutron flux used the Scibath~\cite{brice-01} and Sandia Camera~\cite{mascarenhas-01} neutron detectors, which were positioned at several locations along the basement corridor as well as in the \ac{sns} instrument bay. A neutron flux of approximately $1.5\times 10^{-7}\,$neutrons/cm$^2\,$s$(1-\SI{100}{\MeV})$ was measured in the vicinity of the location later used in the \csi/ deployment. However, these measurements carried a large uncertainty. To further constrain the prompt neutron flux and a potential \ac{nin} background, a neutron detector was positioned at the exact location later occupied by the \csi/ detector.\par

This neutron detector consisted of two \SI{1.5}{\liter} liquid scintillator cells filled with EJ-301, which were housed in a shielding described in \cite{fields-01,collar-02}. The innermost layer of ultra-low background lead was removed to accommodate the detector cells. The shielding was further surrounded by an additional \SI{3.5}{\inch} of neutron moderator, i.e., aluminum tanks filled with water. The scintillator cells were read out by ET9390 \acp{pmt}.\par

\begin{figure}[tbp]
\begin{center}
\includegraphics[width=5in]{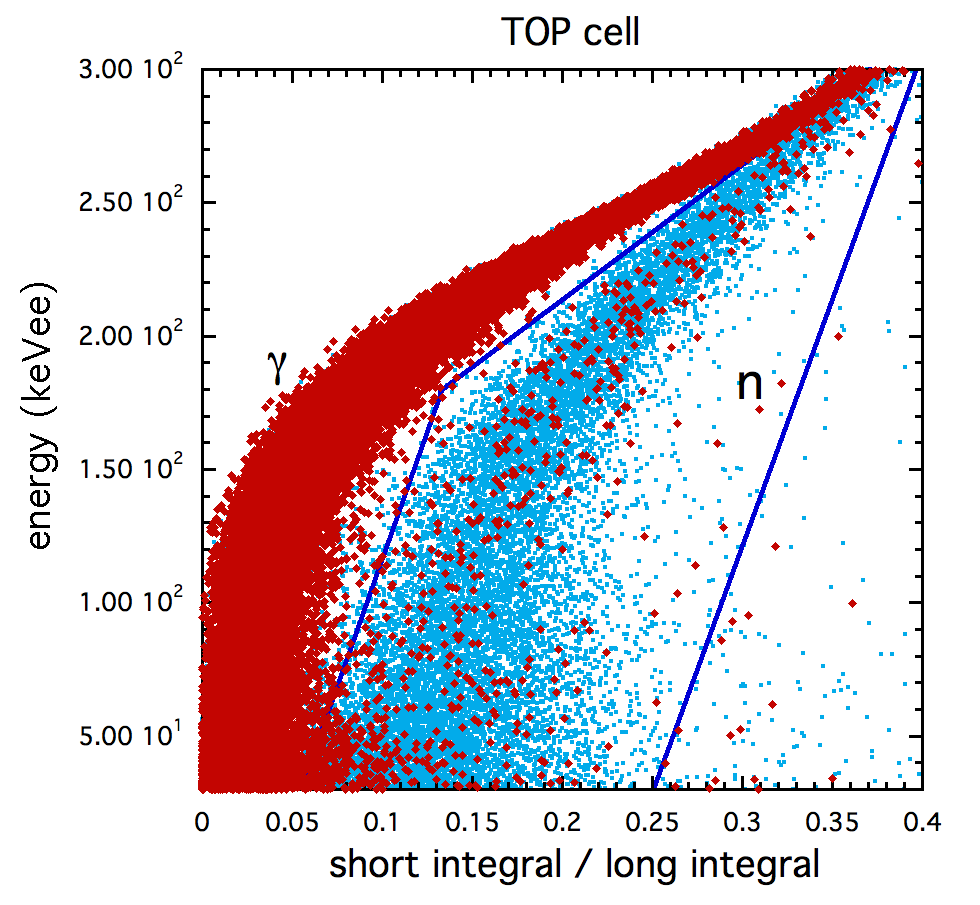}
\end{center}
\caption[\acs*{psd} calibration measurement for the EJ-301 scintillator cells]{Shown is the \acs*{psd} calibration measurement performed for one of the two EJ-301 detectors. The scintillation decay times for nuclear and electronic recoils are different for EJ-301. The nature of events can be discriminated using a pulse shape discrimination parameter. The ratio between the scintillation light present in the beginning of an event to its total light output was chosen for this analysis. Shown above is the energy visible in the detector on the y-axis and the pulse shape discrimination parameter on the x-axis. The red band represents events taken in the presence of a $\gamma$-source and therefore represents electronic recoils. The blue band represents nuclear recoils in the presence of a neutron source. A clear separation between nuclear and electronic recoils is apparent for intermediate energies. However, at low energies the $n$-$\gamma$ discrimination capability is limited as only a few photoelectrons are detected per event. For high energies both bands merge because of non-linearities caused by \ac{pmt} saturation. Events within the blue contour are accepted as neutron-like. Figure courtesy of Juan Collar.}
\label{fig:background-studies:psd}
\end{figure}

\begin{figure}[htb]
\begin{center}
\includegraphics[width=5in]{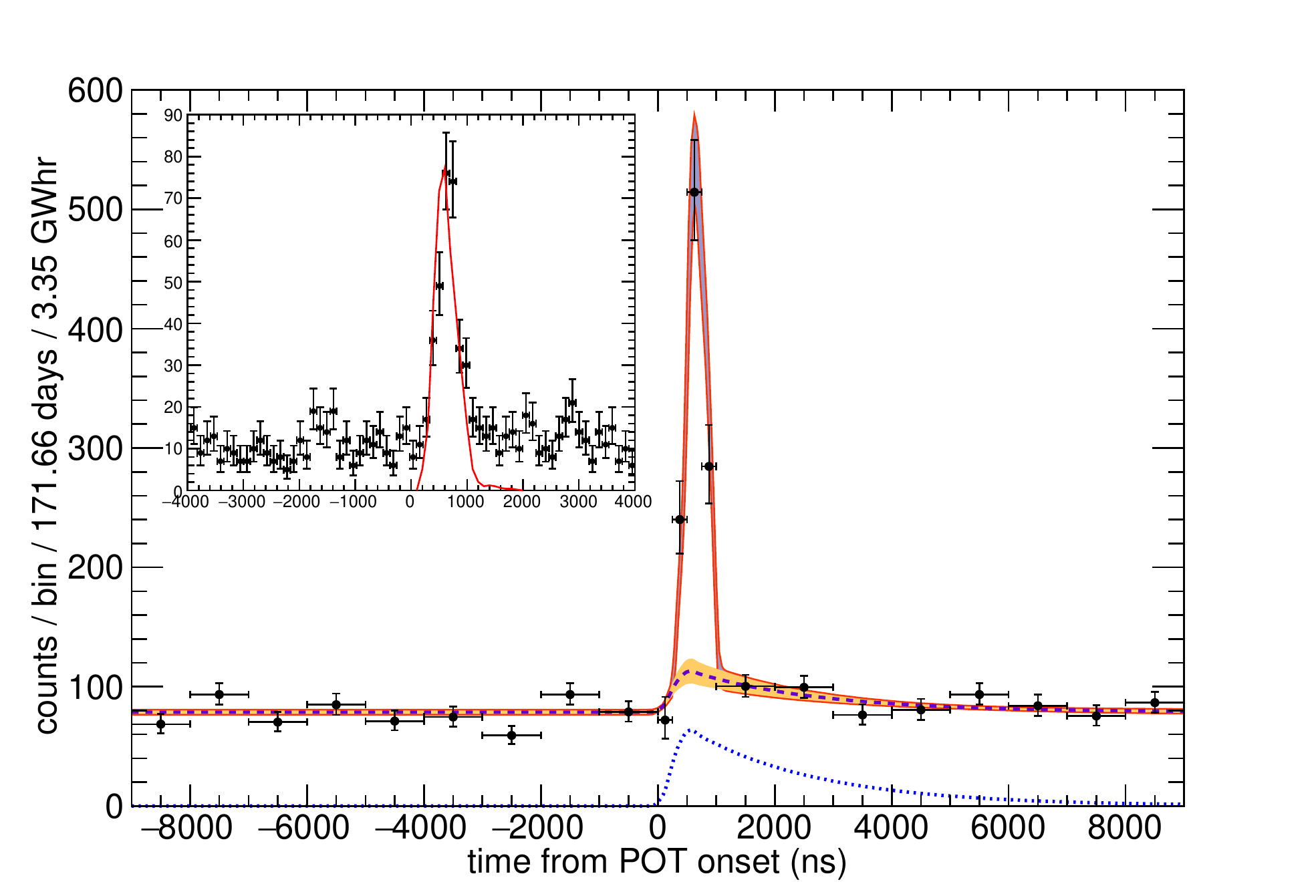}
\end{center}
\caption[Arrival time distribution of neutron-like events in the EJ-301 data set]{Arrival time distribution of neutron-like events in the EJ-301 data set \cite{collar-04}. Red lines represent the $1\sigma$ confidence interval of the three component model fit as described in the text. The dashed line represents the \ac{nin} component of the fit model. The dotted line shows the predicted \ac{nin} magnitude using rate predictions derived from \cite{lazauskas-01,lazauskas-02}. The inset shows the same data with a bin size of \SI{100}{\ns}. The red curve represents the normalized prompt neutron PDF as predicted from \geant/ simulations. Plot from \cite{collar-04}.}
\label{fig:background-studies:eljen-spectra}
\end{figure}

Both scintillator and \ac{pmt} used in this setup were selected primarily for their excellent neutron-gamma discrimination capability down to low energies \cite{luo-01,ronchi-01}. Using standard \ac{psd} techniques neutron-like events were selected with arrival times ranging from \SI{10}{\micro\second} before the \ac{pot} trigger to \SI{10}{\micro\second} after. Fig.~\ref{fig:background-studies:psd} illustrates the \ac{psd} used in this selection process. The energy range of these neutron-like events is limited from 30 to \SI{300}{\keVee}, where the lower bound is due to limitations in the $n$-$\gamma$ discrimination capability and the upper bound is due to non-linearities caused by \ac{pmt} saturation. The distribution of event arrival times with respect to the \ac{pot} trigger for neutron like events in the EJ-301 is shown in Fig.~\ref{fig:background-studies:eljen-spectra}. During the 171.7 days of experimentation the \ac{sns} provided a total integrated beam power on the mercury target of \SI{3.35}{\GWh}. In the following sections two beam-related backgrounds identified by the \coherent/ collaboration are discuss: First, prompt neutrons originating from the mercury target (section~\ref{section:prompt-neutrons}), and second, \acfp{nin} produced in the lead surrounding the \csi/ detector (section~\ref{section:backgrounds-studies:nins}).

\section{Prompt neutrons}
\label{section:prompt-neutrons}
A clear excess directly following the \ac{pot} trigger is visible in Fig.~\ref{fig:background-studies:eljen-spectra}. A prompt analysis window was defined from 200 to \SI{1100}{\ns} after the \ac{pot} trigger. The event rate in this window was found to be approximately 0.7 events per \ac{sns} live-day.\par

\begin{figure}[htbp]
\begin{center}
\includegraphics[width=5in]{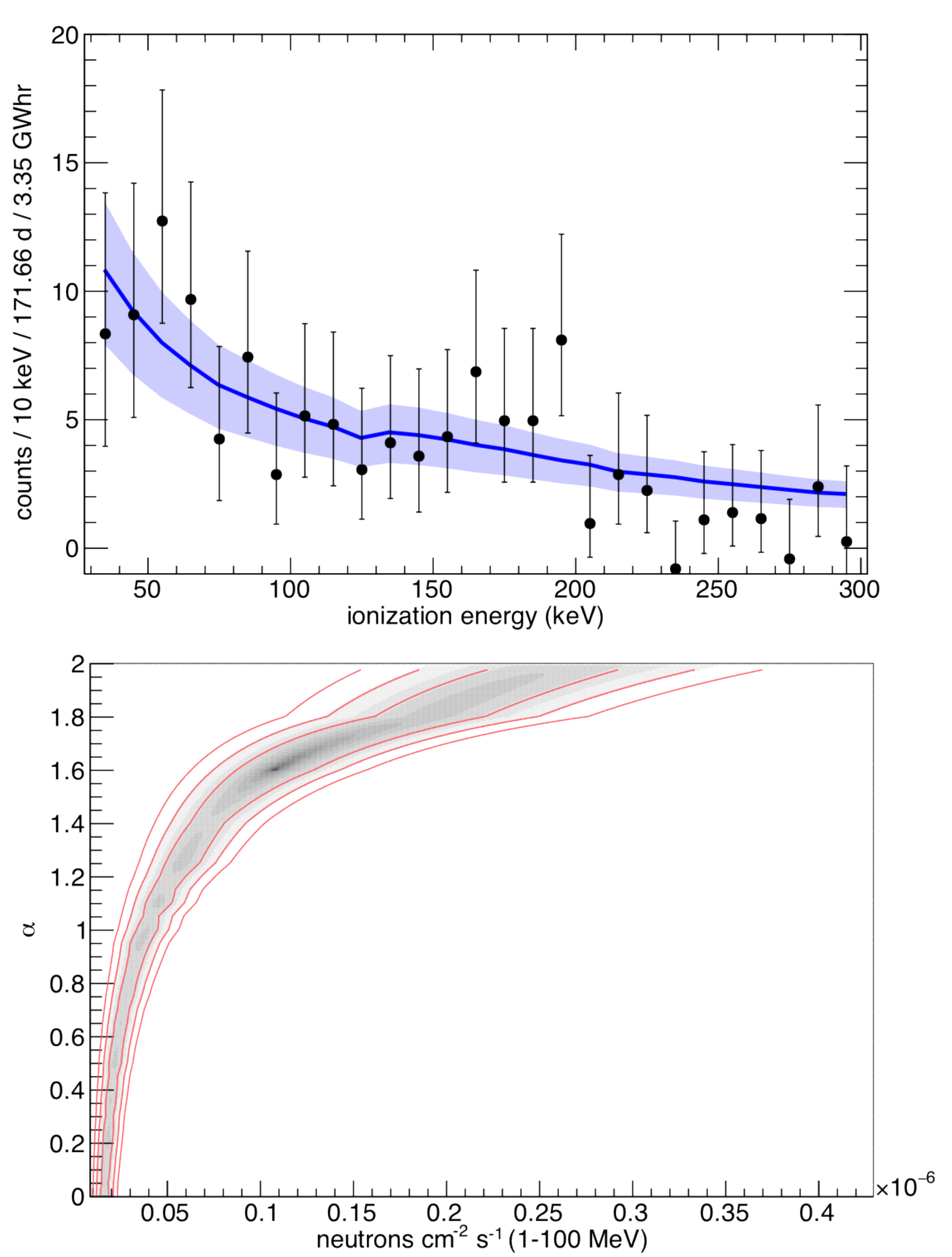}
\end{center}
\caption[Spectrum of neutron-like energy depositions within the EJ-301 with arrival times consistent with prompt neutrons]{\textbf{Top}: Spectrum of neutron like energy depositions within the EJ-301 with arrival times consistent with prompt neutrons, i.e. 200-\SI{1100}{\ns} after the \ac{pot} trigger. Contributions from environmental neutrons were removed from the spectrum using neutron-like events in anti-coincidence with the \ac{pot} trigger, i.e., arrival times of less than zero in Fig.~\ref{fig:background-studies:eljen-spectra}. The blue line represents the simulated EJ-301 response to the best-fit neutron spectral model tested. The blue shaded band shows the energy range corresponding to all parameter combinations within the $1\sigma$ band of the bottom panel. \textbf{Bottom}: Goodness-of-fit for all neutron models tested. Darker areas represent better fits. The red lines show the 1-$3\,\sigma$ levels of the fit. Plot from \cite{collar-04}.}
\label{fig:background-studies:prompt-neutrons}
\end{figure}

The energy spectrum between 30-\SI{300}{\keVee} for all neutron-like events within this prompt window is shown in the upper panel of Fig.~\ref{fig:background-studies:prompt-neutrons}. Multiple comprehensive \mcnp/ simulations were conducted to determine the EJ-301 response to incoming neutron spectra of differing spectral hardness. In these simulations the full detector and shielding geometry were unidirectionally exposed to neutrons coming from the \ac{sns} target. A power law, i.e., $P(E_n) \propto E_n[\text{MeV}]^{-\alpha}$, was chosen to model the spectral hardness of the incoming neutrons. This choice was informed by the previous measurements from the Scibath and Sandia Camera neutron detectors. The neutron energies simulated ranged from 1 to \SI{100}{\MeV}. Neutrons with energies below \SI{1}{\MeV} lose too much energy within the neutron moderators surrounding the EJ-301 cells and do not contribute significantly to the 30-\SI{300}{\keVee} energy range. The flux of neutrons with energies above \SI{100}{\MeV} was found to be negligible by the Scibath and Sandia Camera neutron detectors.

The \mcnp/ output was post-processed to incorporate the response of EJ-301 to nuclear recoils. The recoil energies were converted into an electron equivalent energy using the known quenching factors for hydrogen and carbon \cite{ficenec-01,verbinski-01,yoshida-01,hong-01}. The calculation of the total scintillation light produced in an event also included statistical fluctuations in the light generation.\par

The simulated energy spectrum were fitted to the experimental data by varying the neutron flux $\phi_n$ and spectral hardness $\alpha$. The goodness-of-fit between the simulated detector response and the experimental energy spectrum is shown in Fig.~\ref{fig:background-studies:prompt-neutrons}. The spectral hardness $\alpha$ is shown on the y-axis, whereas the corresponding neutron flux $\phi_n$ is shown on the x-axis. A darker area represents a better agreement between the simulated and measured energy spectra. Red lines represent the corresponding 1-3$\sigma$ confidence intervals. The best fit is given by
\begin{align}
\alpha = 1.6 \qquad\text{and}\qquad \phi_n = 1.09 \times 10^{-7} \frac{n}{\text{cm}^2\,\text{s}}.\label{eq:background-studies:prompt-neutron-fits}
\end{align}
The energy spectrum calculated using this parameter set is shown as a blue line in the upper panel of Fig.~\ref{fig:background-studies:prompt-neutrons}. The blue band consists of all parameter choices covered within the $1\sigma$ confidence interval. The calculated neutron flux is compatible with previous flux measurements using the Scibath and Sandia Camera neutron detectors, which measured a prompt neutron flux of $1.5\times 10^{-7}\,$neutrons/cm$^2\,$s$(1-\SI{100}{\MeV})$ close to the position occupied by the EJ-301 detector.\par

A second \mcnp/ simulation was used to estimate the beam-related background rate caused by prompt neutrons in the \csi/ detector during this \ac{cenns} search. A comprehensive model of the \csi/ setup was uniformly bathed in neutrons coming from the \ac{sns} target. The neutron spectral hardness and flux were fixed to the values given in Eq.~(\ref{eq:background-studies:prompt-neutron-fits}). During the post-processing of the simulation output the proper \csi/ response to nuclear recoils was used, i.e. light-yield and quenching factor as calculated later in this thesis (chapters~\ref{chapter:am-calibration} and \ref{chapter:quenching-calibration}). Poisson fluctuation were added to the number of \ac{pe} generated during an event. An energy spectrum was calculated from the simulated event energies and the signal acceptance model corresponding to the optimized choice of cut parameters used in the \ac{cenns} search (chapter~\ref{chapter:sns-analysis}) was applied to the spectrum.
The uncertainties of light yield (chapter~\ref{chapter:am-calibration}), quenching factor (chapter~\ref{chapter:quenching-calibration}), and signal acceptance (chapter~\ref{chapter:sns-analysis}) were properly propagated through this analysis exercise. The resulting background rate from prompt neutrons was found to be
\begin{align}
\Gamma_\text{prompt} = 0.92\,\pm\,0.23\,\frac{\text{events}}{\SI{}{GWh}}.\label{eq:background-studies:promp-rate}
\end{align}
The arrival time of the background events caused by prompt neutrons is highly concentrated at 200-\SI{1100}{\ns} after the \ac{pot} trigger (Fig.~\ref{fig:background-studies:eljen-spectra}). In chapter~\ref{chapter:sns-analysis} it is shown that this event rate is approximately $25\,$times smaller than the expected \ac{cenns} signal rate.

\section{Neutrino-induced neutrons (NINs)}
\label{section:backgrounds-studies:nins}
As discussed earlier neutrons produced by neutrinos interacting with lead \cite{lazauskas-01,lazauskas-02} constitute another relevant background. These \acp{nin} are produced by the charged-current reaction
\begin{center}$
\begin{aligned}
\nu_e\quad +\quad \isotope{Pb}{208}\quad\rightarrow &\quad \isotope{Bi}{208}^*\quad +\quad e^-\nonumber\\
																	    &\quad \downarrow\nonumber\\
																	    &\quad \isotope{Bi}{208-y}\quad +\quad x\times\gamma\quad +\quad y\times n\nonumber
\end{aligned}$
\end{center}
and the neutral-current interaction
\begin{center}$
\begin{aligned}
\nu_x\quad +\quad \isotope{Pb}{208}\quad\rightarrow &\quad \isotope{Pb}{208}^*\quad +\quad \nu_x'\nonumber\\
																	    &\quad \downarrow\nonumber\\
																	    &\quad \isotope{Pb}{208-y}\quad +\quad x\times\gamma\quad +\quad y\times n.\nonumber
\end{aligned}$
\end{center}
Here $x$ and $y$ denote the $\gamma$ and neutron multiplicity respectively. The \ac{nin} background is predominantly produced by the delayed $\nu_e$, as the charged-current interaction has the largest \xs/. An unbinned fit \cite{verkerke-01} to the EJ-301 arrival time data (Fig.~\ref{fig:background-studies:eljen-spectra}) was used to constrain this \ac{nin} background. The model used in this fit incorporated the following three neutron components; a random arrival time from environmental neutrons, prompt neutrons, and a \ac{nin} excess following the $\nu_e$ time-profile. The number of \acp{nin} found in this fit was converted into a \ac{nin} production rate using an \mcnp/ simulation. A homogeneous and isotropic \ac{nin} emission within the lead shield surrounding the EJ-301 was simulated. An energy spectrum corresponding to the highest stellar-collapse neutrino energies ($T=\SI{8}{\MeV}$) described in \cite{kolbe-01} is adopted for the \ac{nin} emission. The energy spectrum of these stellar-collapse neutrinos is slightly softer than the energy spectrum of $\nu_e$ at the \ac{sns} (Fig.~\ref{fig:sns:sns-and-supernova-neutrino-spectrum}). However, the change in \ac{nin} spectral hardness is negligible for different neutrino energies~\cite{kolbe-01}. The energy spectrum measured by the EJ-301 was calculated using an identical approach to the one described in section~\ref{section:prompt-neutrons}. The total \ac{nin} production rate was found to be
\begin{align}
\Gamma^\star_\text{nin} = 0.97\,\pm\,0.33 \frac{\text{neutrons}}{\SI{}{\GWh}\,\,\text{kg of Pb}}.
\end{align}
An additional simulation was performed using homogeneous and isotropic \ac{nin} emission in the lead surrounding the \csi/ used in the \ac{cenns} search and analyzed as described in section~\ref{section:prompt-neutrons}. The uncertainties of light yield (chapter~\ref{chapter:am-calibration}), quenching factor (chapter~\ref{chapter:quenching-calibration}), and signal acceptance (chapter~\ref{chapter:sns-analysis}) were properly propagated through this exercise. Scaling the simulated \ac{nin} event rate with the expected \ac{nin} production rate resulted in a final background rate of
\begin{align}
\Gamma_\text{nin}\,=0.54\,\pm\,0.18\,\frac{\text{events}}{\SI{}{GWh}}.\label{eq:background-studies:nin-rate}
\end{align}
In chapter~\ref{chapter:sns-analysis} it is shown that this event rate is approximately $43\,$times smaller than the expected \ac{cenns} signal rate.\par

These calculations indicate that both prompt neutrons, and \acp{nin} only contribute a negligible background to the \ac{cenns} search. The validity of the neutron transport simulations are further substantiated in section \ref{section:sns-analysis:cf-calibration}.

%% file: csi-setup.tex
%
%
\chapter{The \csi/ CE$\nu$NS search detector at the SNS}
\label{chapter:csi-setup}
\section{Overview and wiring of the detector setup}
\label{section:csi-setup:wiring}
\begin{figure}[tb]
\begin{center}
\reflectbox{\includegraphics[scale=0.9]{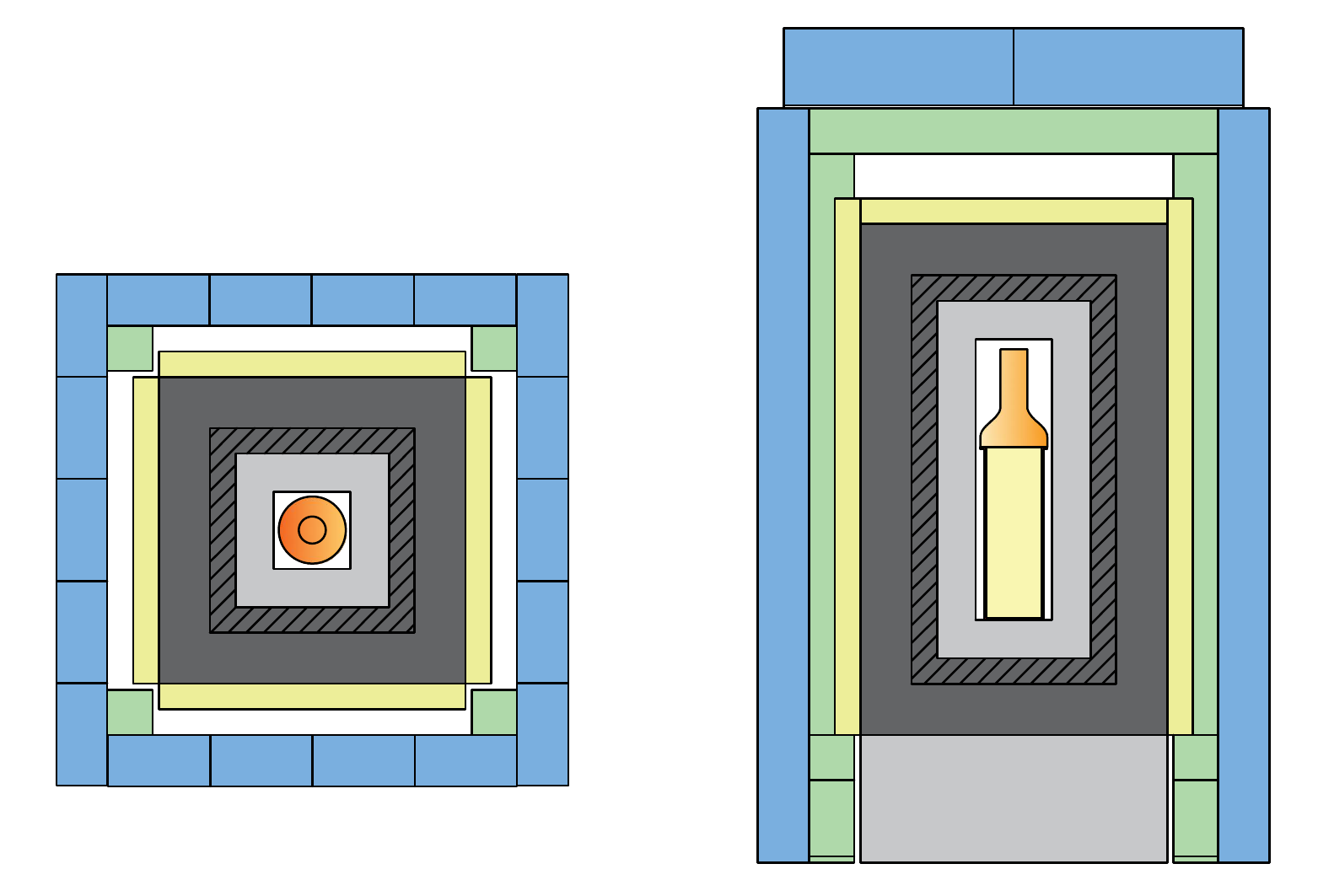}}
\end{center}
\caption[Schematic of the  \csi/ shielding at the \acs*{sns}]{Schematic of the \csi/ shielding at the \acs*{sns}. The shielding components from the inside out are as follows. (1) light gray: \SI{3}{\inch} of low-background \acs*{hdpe}. (2) hatched dark gray: \SI{2}{\inch} of low background lead. (3) dark gray: \SI{4}{\inch} of contemporary lead. (4) yellow: \SI{2}{\inch}-thick muon veto panels. (5) green: Aluminum Bosch-Rexroth extrusions. (6) blue: Aluminum tanks filled with water, on the sides, WaterBricks, on the top.}
\label{fig:csi-setup:csi-shielding-sketch}
\end{figure}

\begin{figure}[tb]
\begin{center}
\begin{minipage}{0.32\linewidth}
\includegraphics[width=\linewidth]{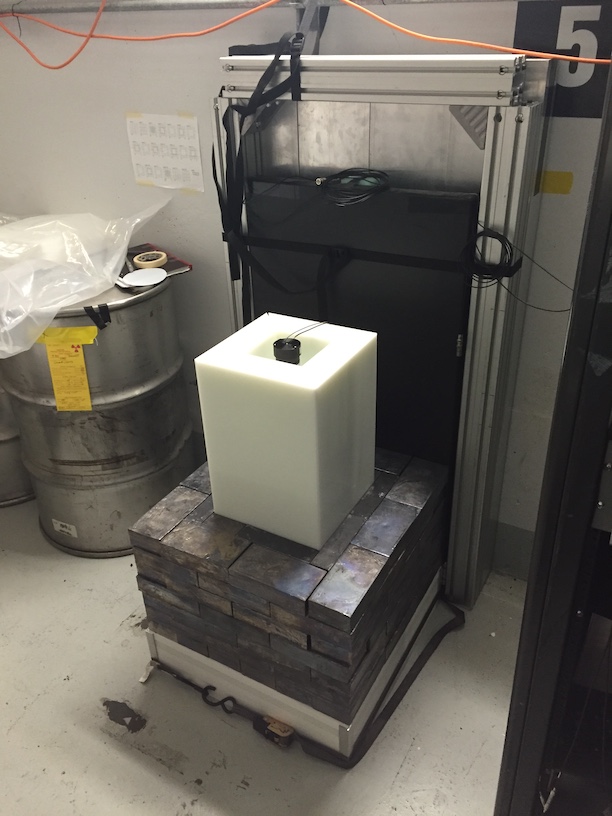}
\end{minipage}
\begin{minipage}{0.32\linewidth}
\includegraphics[width=\linewidth]{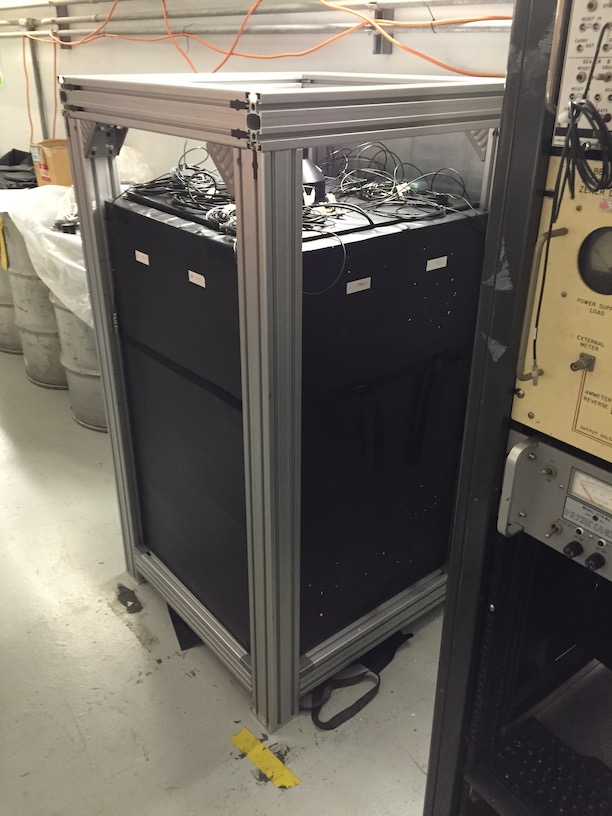}
\end{minipage}
\begin{minipage}{0.32\linewidth}
\includegraphics[width=\linewidth]{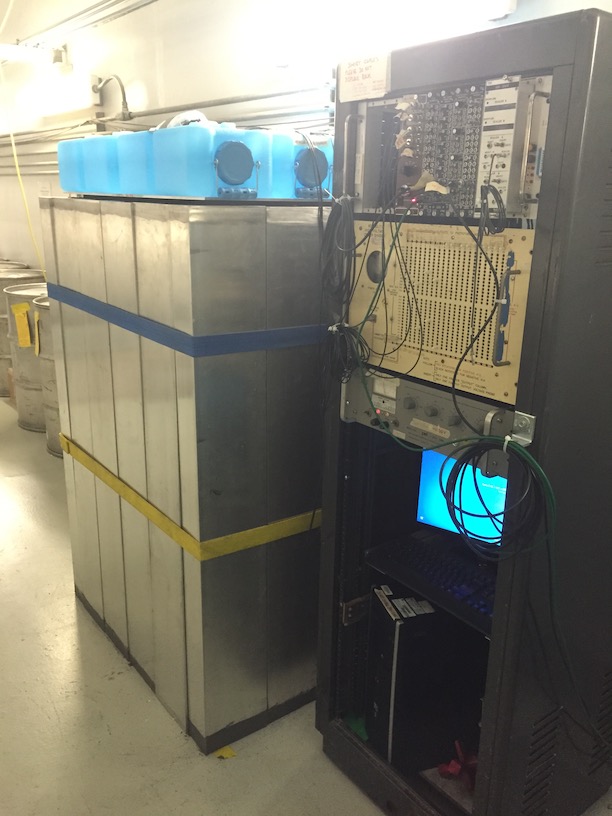}
\end{minipage}
\end{center}
\caption[Deployment of the CE$\nu$NS search detector in June 2015 at the SNS.]{Pictures of the CE$\nu$NS search detector deployment in June 2015 at the SNS. The left most picture highlights the inner-most layers of \acs*{hdpe} used to reduce the background originating from \ac{nin} production in the surrounding lead (chapter~\ref{chapter:background-studies}). The middle picture shows the setup once the aluminum frame was assembled. The right picture shows the full installation, including the NIM rack. Images courtesy of Juan Collar.}
\label{fig:csi-setup:deployment}
\end{figure}

A schematic of the \csi/ detector setup deployed at the \ac{sns} is shown in Fig.~\ref{fig:csi-setup:csi-shielding-sketch}. The assembly is located \SI{19.3}{\m} from the \ac{sns} mercury target in a basement corridor (Fig.~\ref{fig:sns:neutrino-alley}). The central detector is a \SI{34}{\cm} long sodium doped cesium iodine scintillator read out by a super-bialkali R877-100 \ac{pmt} by Hamamatsu. The detector is surrounded by \SI{3}{\inch} of \ac{hdpe}. The purpose of this innermost layer of \ac{hdpe} was to reduce the background caused by \ac{nin} production in the surrounding lead (chapter~\ref{chapter:background-studies}). This layer of \ac{hdpe} is followed by \SI{2}{\inch} of low-background lead with an approximate \isotope{Pb}{210} contamination of $\sim\SI{10}{\becquerel\per\kg}$. The outer $\gamma$-shield is made of an additional \SI{4}{\inch} of contemporary lead ($\sim\SI{100}{\becquerel\per\kg}$ of \isotope{Pb}{210}). This results in a minimum of \SI{6}{\inch} of lead surrounding the \csi/ in a any direction with a total mass of approximately 3.5 tons. The lead shield is encased by a \SI{2}{\inch}-thick high-efficiency muon veto on all vertical sides and the top. The setup rests upon a platform made of an additional \SI{6}{\inch} of \ac{hdpe} selected for its low-background properties. A heavy-duty frame made of aluminum Bosch extrusions is placed around the setup providing a secure anchor point for the muon veto panels. It provided a firm surface against which an additional layer of aluminum tanks was pushed, without disturbing the inner shielding configuration. These tanks were filled with tap water, providing at least an additional \SI{3.5}{\inch} of neutron moderator. Fig.~\ref{fig:csi-setup:deployment} shows the setup at three different stages during the installation process in June 2015.\par

\begin{figure}[tb]
\begin{center}
\includegraphics[width=6in]{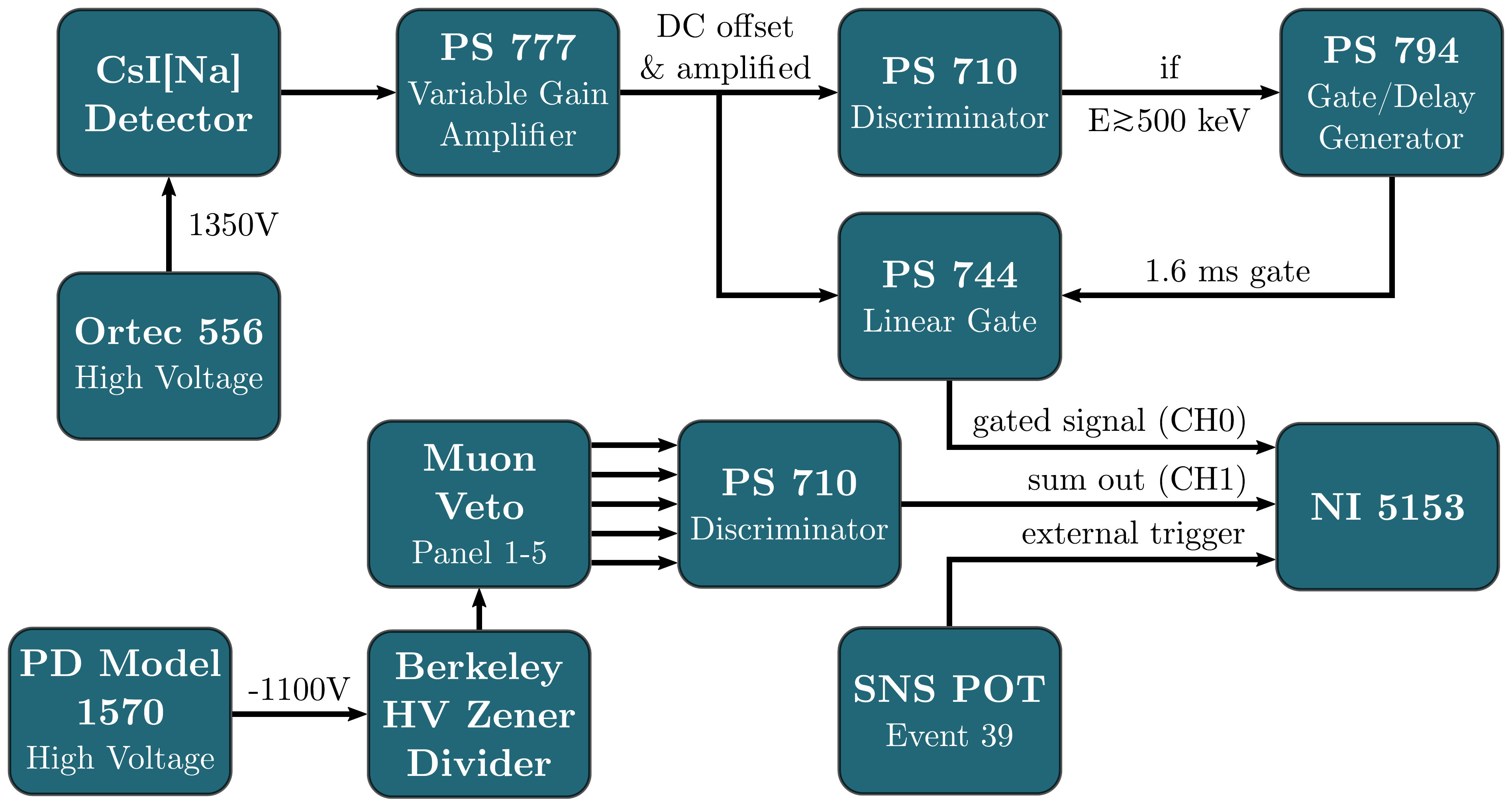}
\end{center}
\caption[Schematic of the data acquisition system used during the \acs*{cenns} search]{Schematic of the data acquisition system used during the \ac{cenns} search}
\label{fig:csi-setup:wiring-diagram-sns}
\end{figure}

A schematic of the data acquisition system is shown in Fig.~\ref{fig:csi-setup:wiring-diagram-sns}. The high-voltage of +\SI{1350}{\volt} for the central \csi/ detector is provided by an Ortec 556 power supply. The \ac{pmt} output signal is fed into a \ac{ps} 777 variable gain amplifier. The DC level of the signal is shifted by approximately +\SI{90}{\milli\volt} and amplified using the \ac{ps} 777. The DC shift was necessary to make use of the full data acquisition range. The additional amplification increased the \ac{spe} amplitude without increasing the baseline noise level, avoiding the need to operate the \ac{pmt} at an excessive high voltage. The \ac{ps} 777 output is split, one output is fed into a \ac{ps} 710 discriminator and the other one into a \ac{ps} 744 linear gate. The \ac{ps} 710 provides a logical signal to a \ac{ps} 794 gate/delay generator whenever the \csi/ signal containes an energy deposition of $\gtrsim\SI{500}{\keVee}$. The \ac{ps} 794 provides a $\sim\SI{1.6}{\milli\second}$ long logical signal which closes the \ac{ps} 744 linear gate. The output of the \ac{ps} 744 is fed into channel 0 of the \ac{ni} 5153 fast digitizer.\par

The linear-gate logic, consisting of the \ac{ps} 710, 794 and 744, was necessary as any ungated high-energy deposition in the \csi/, caused for example by a traversing muon, would reset the data acquisition card, rendering the data acquisition system unresponsive for approximately \SI{3}{\second}. The linear-gate logic prevents this reset and enables a continuous data acquisition. A gate length of \SI{1.6}{\milli\second} was chosen in order to cut most of the afterglow of any high-energy event while only introducing a minimal dead time of only $\mathcal{O}(\SI{1}{\percent})$.\par

The high-voltage for all muon veto panels is provided by a Power Designs Model 1570, which is set to -\SI{1100}{\volt}. The 1570 output is fed into a Berkeley HV Zener Divider, which in return provides nine distinct high-voltage output levels, one for each \ac{pmt} used within the muon veto. By operating each \ac{pmt} at an individual high voltage it was possible to match the gain of all \acp{pmt} (Fig.~\ref{fig:csi-setup:muon-veto-pmt-gain-curves}). The output of the five muon veto panels, i.e., four side panels and one top panel, is fed into the \ac{ps} 710 discriminator. The summed, discriminated output of all panels is fed into channel 1 of the \ac{ni} 5153. The external trigger for the data acquisition system is the \emph{event 39} signal provided by the \ac{sns}. \emph{Event 39} is generated when there is a potential extraction of protons from the storage ring, even if there are no protons stored in the ring, which approximately happens once in every 600 triggers. As such \emph{event 39} provides a stable \SI{60}{\hertz} trigger signal with an amplitude of $\sim\SI{2.5}{\volt}$ for all times no matter whether the \ac{sns} actually provided \ac{pot} or not. During the rest of this thesis this triggering signal is referred to as \ac{pot} trigger.\par

A \labview/ based data acquisition program was written, which is capable of achieving a \SI{100}{\percent} data throughput at a trigger rate of \SI{60}{\hertz}, as is the case at the \ac{sns}. For each trigger \SI{70}{\micro\second} long traces are recorded of both channels, i.e., the \csi/ and muon veto. The waveforms are sampled at \SI{500}{\mega\sample\per\second} with a digitizer depth of 8-bit and saved as raw binary files. The data acquisition software automatically compresses the binary data files into zip-archives. Given normal operations this amounts to approximately \SI{60}{\mega\byte} per minute or a manageable total of $\sim\SI{90}{\giga\byte}$ per day. An in depth look at the data structure is provided in section~\ref{sec:csi-setup:data-structure}. Once a run is completed the compressed data is pushed onto the HCDATA cluster of the \ac{ornl} Physics Division. The cluster keeps a spinning-disk copy of all data for fast access during the actual data analysis, as well as a long-term archive on tape.\par

Once the \csi/ detector was deployed an unexpected, minor complication was found in the data acquisition software. In order to minimize storage space the data is saved as I8 variables. However the data acquired from the \ac{ni} 5153 within \labview/ is actually provided as double precision floating point numbers in \SI{}{mV}. These floating point values are converted back to their \SI{8}{\bit} depth by applying
\begin{align}
V_\text{i8} = 2^8\times\frac{V_\text{dbl}}{V_\text{Range}},
\end{align}
where $V_\text{dbl}$ is the digitizer value in mV, $V_\text{Range}$ the current digitizer range and $V_\text{i8}$ the digitizer value in \SI{}{\adc}. Once a first batch of data had been analyzed it became apparent that $V_\text{Range}$ slightly differs from the setting provided, i.e., a digitizer range set to \SI{0.2}{\Vpp} actually shows a slightly larger range. This issue was confirmed by an \ac{ni} technician. The projection of $V_\text{dbl}$ onto $V_\text{i8}$ therefore resulted in some I8 values that never appear as no initial $V_\text{dbl}$ is mapped onto them. A correction needs to be applied during the analysis that maps the I8 values onto a continuous signal amplitude. This does not have any ill effect on the results presented here. Once this issue had been identified the impact of this on the data analysis was investigated. Some I8 values are simply omitted and as such the effective digitizer range is reduced by approximately \SI{20}{\adc}. But as all further analysis presented in this thesis was based on \SI{}{\adc} and not an absolute value in \SI{}{\mV} the analysis remained unaffected by this issue.\par
The exact transformation depends on $V_\text{Range}$. For the two digitizer ranges used in this thesis, the transformations that map the recorded $V_\text{i8}$ data onto a continuous set of signal amplitude values $V$ are given by
\begin{align}
V(V_\text{Range} = \SI{200}{\milli\Vpp}) &= \text{int}\left(V_\text{i8}\right) - \text{int}\left(\text{floor}\left(\frac{\text{double}\left(V_\text{i8}\right) + 5.0}{11.0}\right)\right)\label{eq:csi-setup:csi-transformation}\\
V(V_\text{Range} = \SI{500}{\milli\Vpp}) &= \text{int}\left(V_\text{i8}\right) + \text{sign}\left(V_\text{i8}\right)\text{int}\left(\text{floor}\left(\frac{4.0 - \text{double}\left(V_\text{i8}\right)}{11.0}\right)\right)\label{eq:csi-setup:muon-veto-transformation}
\end{align}
Here double() and int() represent data type casts into double precision floating point numbers and integers, respectively. The floor() function returns the the largest integer less than or equal to its argument.

\section{The central \csi/ detector}
\begin{figure}[tb]
\begin{center}
\includegraphics[width=3.25in]{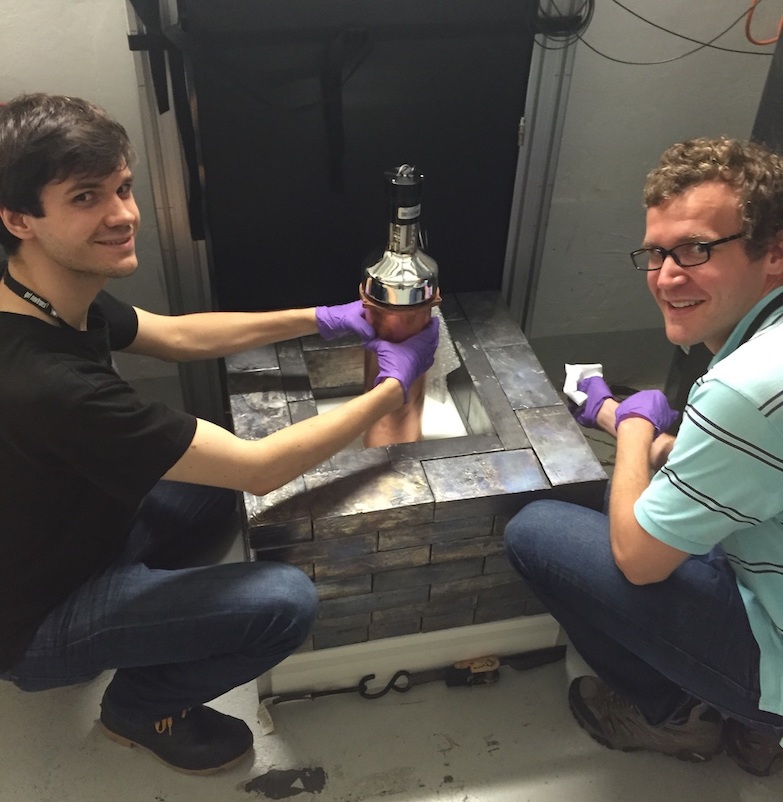}
\end{center}
\caption[Image of the central \csi/ detector during installation]{Image of the central \csi/ detector during installation. From left to right: Author of this thesis, Grayson Rich (University of North Carolina at Chapel Hill). Image courtesy of Juan Collar.}
\label{fig:csi-setup:qe-and-temp-stability}
\end{figure}

The inorganic scintillator \csi/ was chosen as the target material due to its many beneficial properties simplifying a \ac{cenns} detection and making it possible in the first place \cite{collar-02}. The crystal was grown by \amcrys/ in Ukrain, and is approximately \SI{34}{\cm} long with a total mass of \SI{14.57}{\kg}. It is wrapped in an expanded PTFE membrane reflector and encapsulated in a electroformed OFHC copper can, which was grown at \ac{pnnl}. Both \isotope{Cs}{133} and \isotope{I}{127} nuclei have a large number of neutrons, 78 and 74 respectively, which leads to a large coherent enhancement of the scattering \xs/, as shown in Eq.~(\ref{eq:cenns-theory:diff-cross-section-2}). The large size of the nucleus also allows for a larger momentum transfer between the incoming neutrino and the target nucleus before a loss of coherence sets in as governed by the form factor (Eq.~(\ref{eq:cenns-theory:form-factor})). Another benefit of this detector material is the very similar nuclear mass of cesium and iodine. As such, the \ac{cenns} induced recoil spectra are nearly indistinguishable from one another, leading to a much simpler data analysis. Yet, given the modest neutrino energies and the relatively large mass of the recoiling nucleus it is still necessary to achieve a low enough energy threshold to actually benefit from this enhancement.\par

\begin{figure}[tb]
\begin{center}
\includegraphics[width=6.5in]{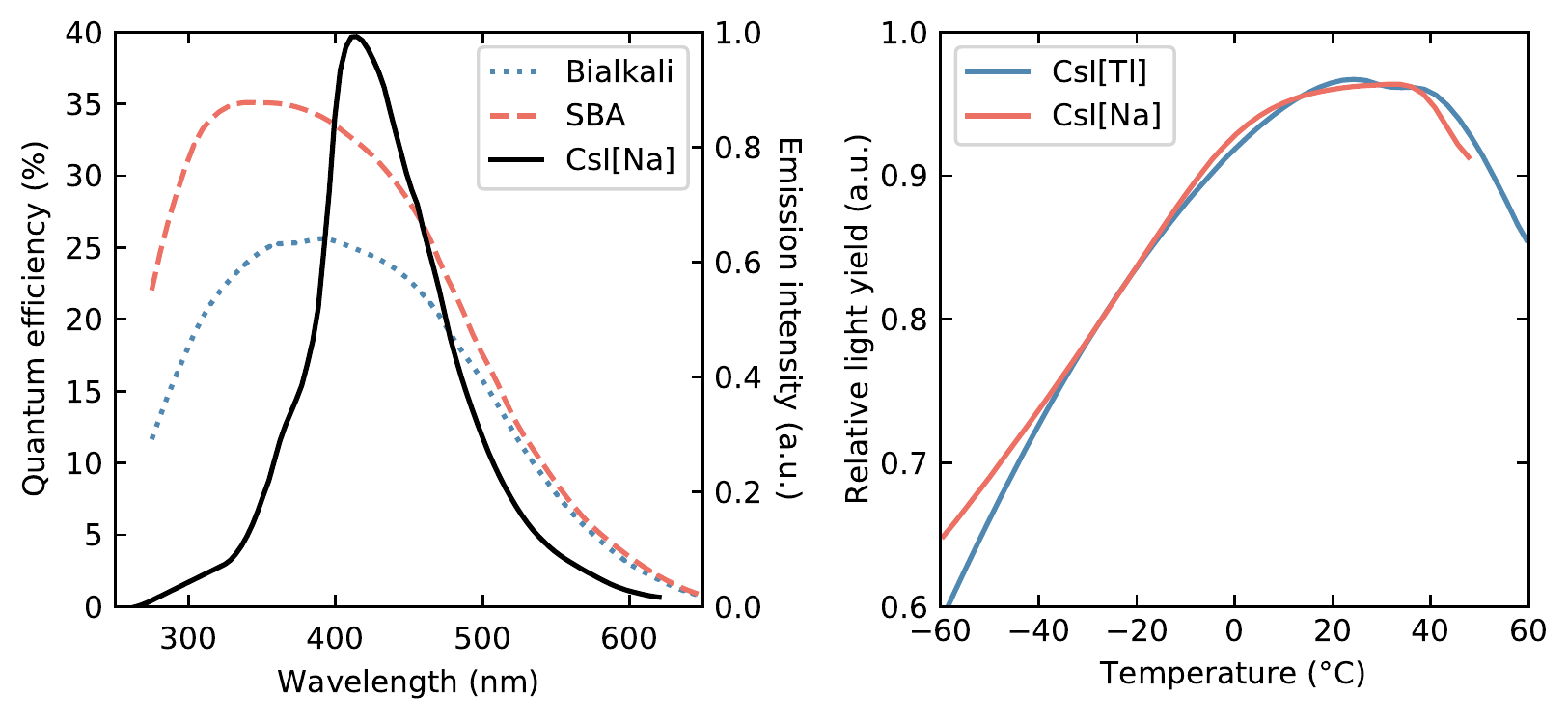}
\end{center}
\caption[Temperature stability of \csi/ and CsI(Tl)]{\textbf{Left:} Quantum efficiency of bialkali and super-bialkali (SBA) photocathodes are shown in red and blue respectively \cite{hamamatsu-01}. The emission profile of sodium doped CsI is shown in black \cite{haas-01}. The match between the quantum efficiency of the photocathodes and the scintillator emission profile is evident. \textbf{Right:} Temperature stability of \csi/ and CsI(Tl). The minimal change in light yield between 10-\SI{30}{\celsius} is readily visible. Adopted from \cite{csi-data-sheet}.}
\label{fig:csi-setup:qe-and-temp-stability}
\end{figure}

\begin{figure}[tb]
\begin{center}
\includegraphics[width=4.25in]{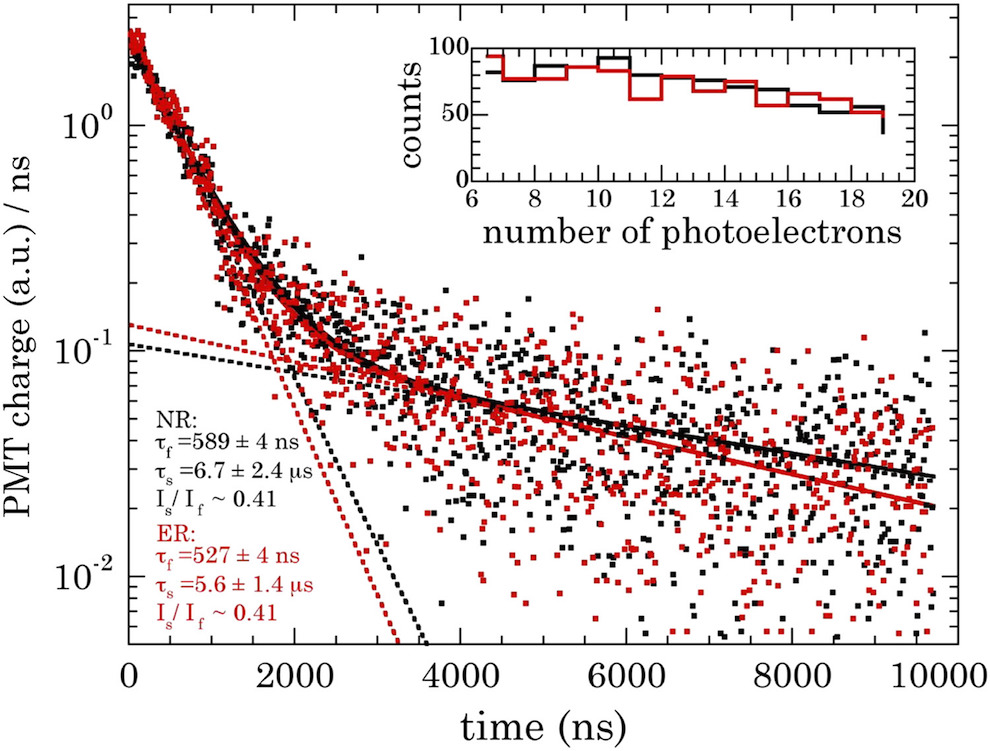}
\end{center}
\caption[Scintillation decay times of low-energy events in \csi/ caused by electronic or nuclear recoils]{Electronic and nuclear decay times of low energy events in \csi/~\cite{collar-02}. The decay profile was decomposited into a fast and slow component \cite{hrehuss-01}. \csi/ shows almost the same decay profile for both recoil types, which greatly simplifies the detector calibration (chapter~\ref{chapter:ba-calibration}). The \SI{60}{\ns} difference in the fast decay component might grant us the ability to statistically differentiate between nuclear and electronic recoils in the future. Figure courtesy of Juan Collar.}
\label{fig:csi-setup:csi-decay-times}
\end{figure}

\begin{figure}[tb]
\begin{center}
\includegraphics[width=3.75in]{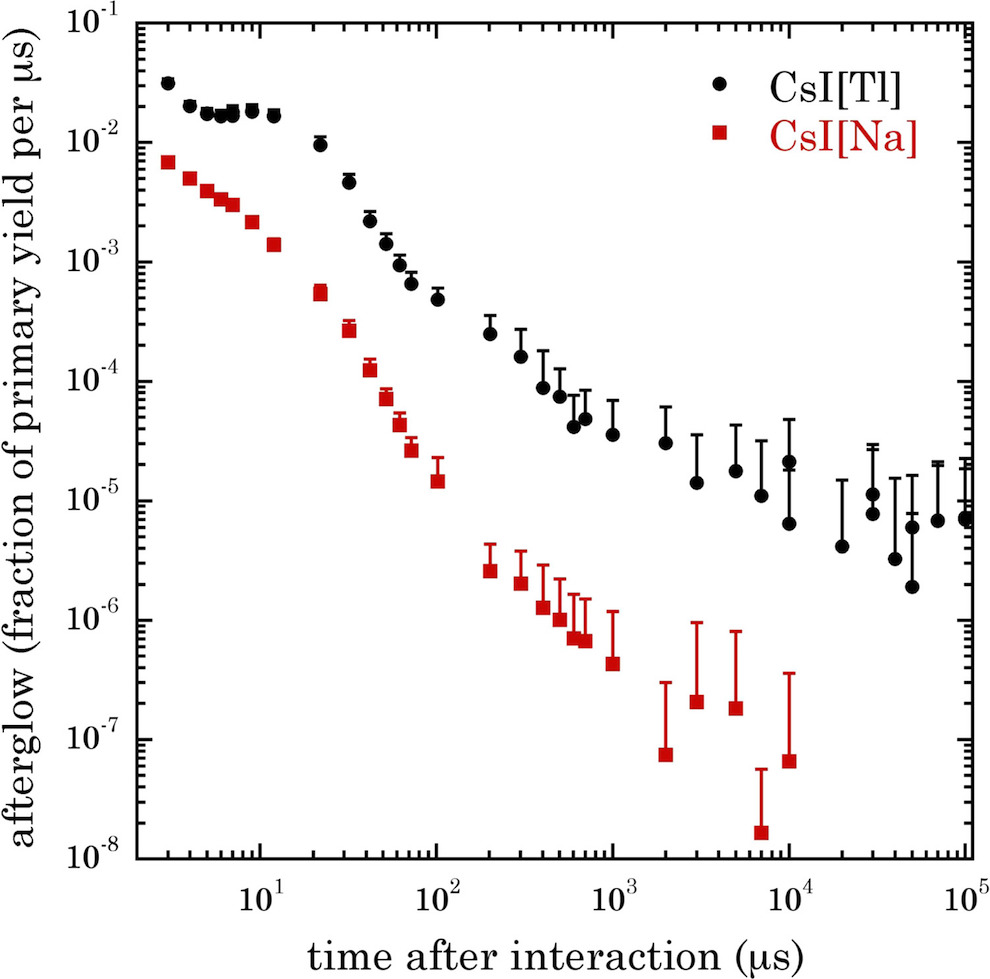}
\end{center}
\caption[Afterglow (phosphorescence) of \csi/ and CsI{[Tl]}.]{Afterglow (phosphorescence) of \csi/ and CsI{[Tl]} as a fraction of primary scintillation yield~\cite{collar-02}. The afterglow experienced by \csi/ is not as excessive as the afterglow in {CsI[Tl]}. Given the modest overburden of \SI{8}{\mwe} available at the \ac{cenns} search, this was crucial as otherwise most triggers would contain a significant number of \acp{spe} from phosphorescence following high energy, cosmic-ray induced events. Figure courtesy of Juan Collar.}
\label{fig:csi-setup:afterglow}
\end{figure}

In order to achieve such a low energy threshold, an R877-100 \SI{5}{\inch} flat-faced \ac{pmt} from Hamamatsu was chosen to read out the scintillation light. This \ac{pmt} uses a \ac{sba} photocathode which provides an increased \ac{qe} of $\sim\SI{34}{\percent}$ over the $\sim\SI{24}{\percent}$ \ac{qe} of common bialkali photo-cathodes~\cite{hamamatsu-01}. The \ac{sba} photocathode is further matched to the scintillation light of \csi/ as is shown in the left panel of Fig.~\ref{fig:csi-setup:qe-and-temp-stability}. The high \ac{qe}, paired with a light yield of \csi/ of $\sim$45\,photons per \SI{}{\keVee} \cite{haas-01} made it possible to achieve an energy threshold of approximately \SI{4.5}{\keVnr}. \csi/ has the additional advantage of short scintillation decay times (Fig.~\ref{fig:csi-setup:csi-decay-times}), as well as a manageable afterglow (phosphorescence), especially when compared to thallium-doped CsI (Fig.~\ref{fig:csi-setup:afterglow}). As such the detector can be operated at a modest overburden without contaminating a significant amount of \ac{sns} proton-on-target triggers with \acp{spe} from previous cosmic-ray induced events.\par

The right panel of Fig.~\ref{fig:csi-setup:qe-and-temp-stability} highlights yet another advantage of \csi/, namely the excellent light yield stability around room temperature. A change of $\pm\SI{10}{\celsius}$ only results in a negligible change of its light yield on the order of less than \SI{1}{\percent}. This has been highly beneficial as over the course of the last two years several other detectors were deployed within the \emph{neutrino alley}. As a result, the heat load was always changing which lead to fluctuations in the environmental temperature on the order of $\pm\SI{2}{\celsius}$. Yet the excellent temperature stability guaranteed a constant light emission throughout the full data run. It was also shown that \csi/ crystals show very low levels of internal radioactivity, i.e. \isotope{U}{238}, \isotope{Th}{232}, \isotope{K}{40}, \isotope{I}{126} and $^{134,\,137}$Cs, when grown from appropriate salts \cite{kim-01}. In her PhD thesis~\cite{fields-01}, Nicole Fields screened a boule slice of \csi/ from \amcrys/ using low-background $\gamma$ spectroscopy. She measured an activity per isotope on the order of $\mathcal{O}(\SI{10}{\milli\becquerel\per\kg})$. An in depth analysis of the salts used by \amcrys/ to grow the \csi/ used in this thesis, and most other materials used during the construction of this detector setup is provided in Chapter 4 of \cite{fields-01}.

\section{The muon veto}
\label{sec:csi-setup:muon-veto}
\begin{figure}[htbp]
\begin{center}
\includegraphics[scale=1]{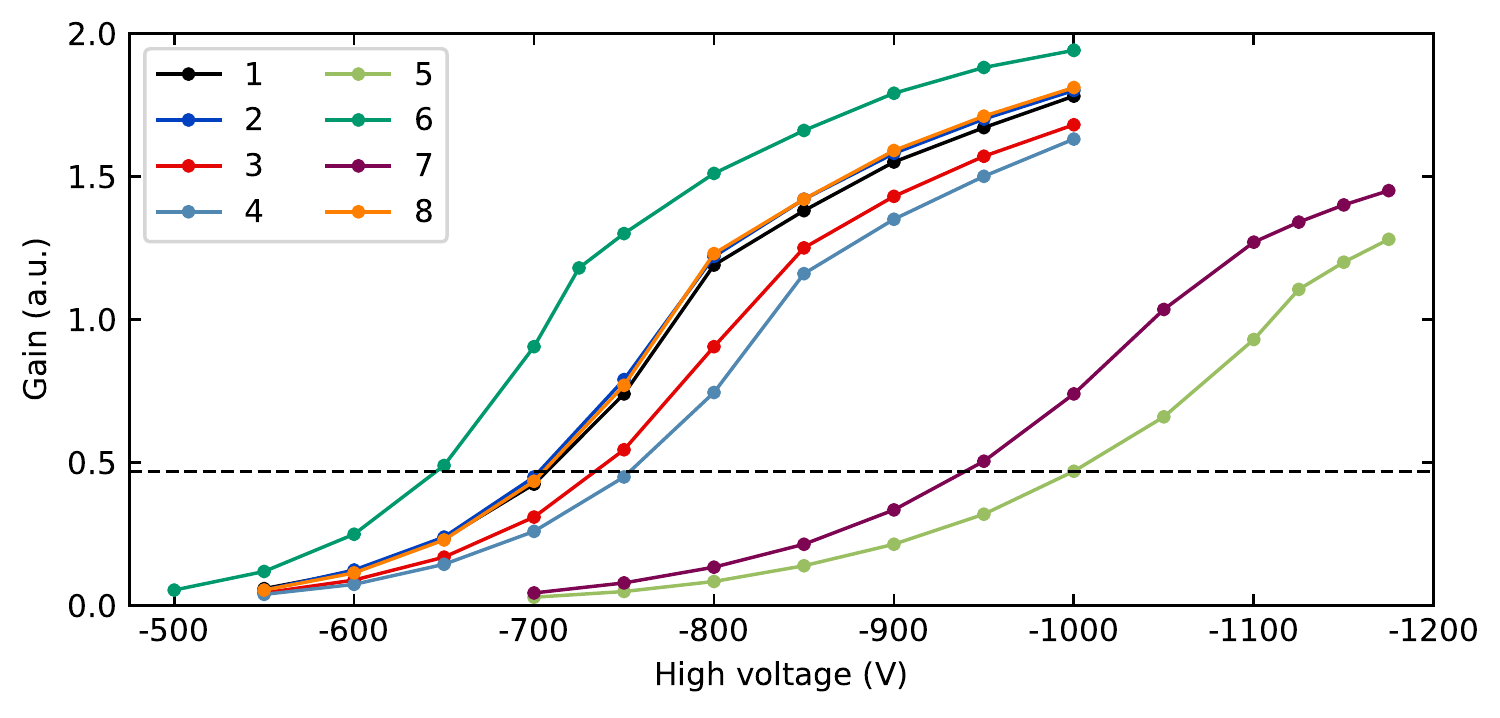}
\end{center}
\caption[Gain curves for each \acs*{pmt} used in the \csi/ muon veto]{Gain curves for each \acs*{pmt} used in the \csi/ muon veto. The gain was determined using a \isotope{Co}{57} source and a Pocket MCA. The dashed black line represents the gain to which all \acp{pmt} were matched. The high voltage chosen for each \acs*{pmt} is given in Table~\ref{tab:csi-setup:pmt-high-voltage-muon-veto}.}
\label{fig:csi-setup:muon-veto-pmt-gain-curves}
\end{figure}

The muon veto panel consists of four side panels and one top panel, each made out of EJ-200B. The dimension for each side panel is $42\times24\times2$ inch, the top panel is $24\times24\times2$ inch. Each side panel is read out by two DC-chained, gain matched ET 9102SB \acp{pmt} with a diameter of \SI{1.5}{\inch}. The top panel is read out by a single \SI{5}{\inch} flat faced \ac{pmt}.\par

Before assembly the gain curve of each ET 9102SB was mapped by attaching each \ac{pmt} to the same crystal and irradiating the setup with a \isotope{Co}{57} source, which provides $\gamma$s with an energy of \SI{122}{\keV}. An Amptek MCA 8000A (Pocket MCA) was used to record the energy spectrum for each \ac{pmt} operated at several different high voltages. The \isotope{Co}{57} peak position was recorded for each of these spectra. The resulting gain curves are shown in Fig.~\ref{fig:csi-setup:muon-veto-pmt-gain-curves}. Once each muon veto panel was wrapped with reflecting and light-blocking material the separation between the environmental $\gamma$ background and the high-energetic muon interactions was tested for different gain levels. The final operational high voltage for each \ac{pmt} is listed in Table~\ref{tab:csi-setup:pmt-high-voltage-muon-veto}.\par

\begin{figure}[htb]
\begin{center}
\includegraphics[width=3.5in]{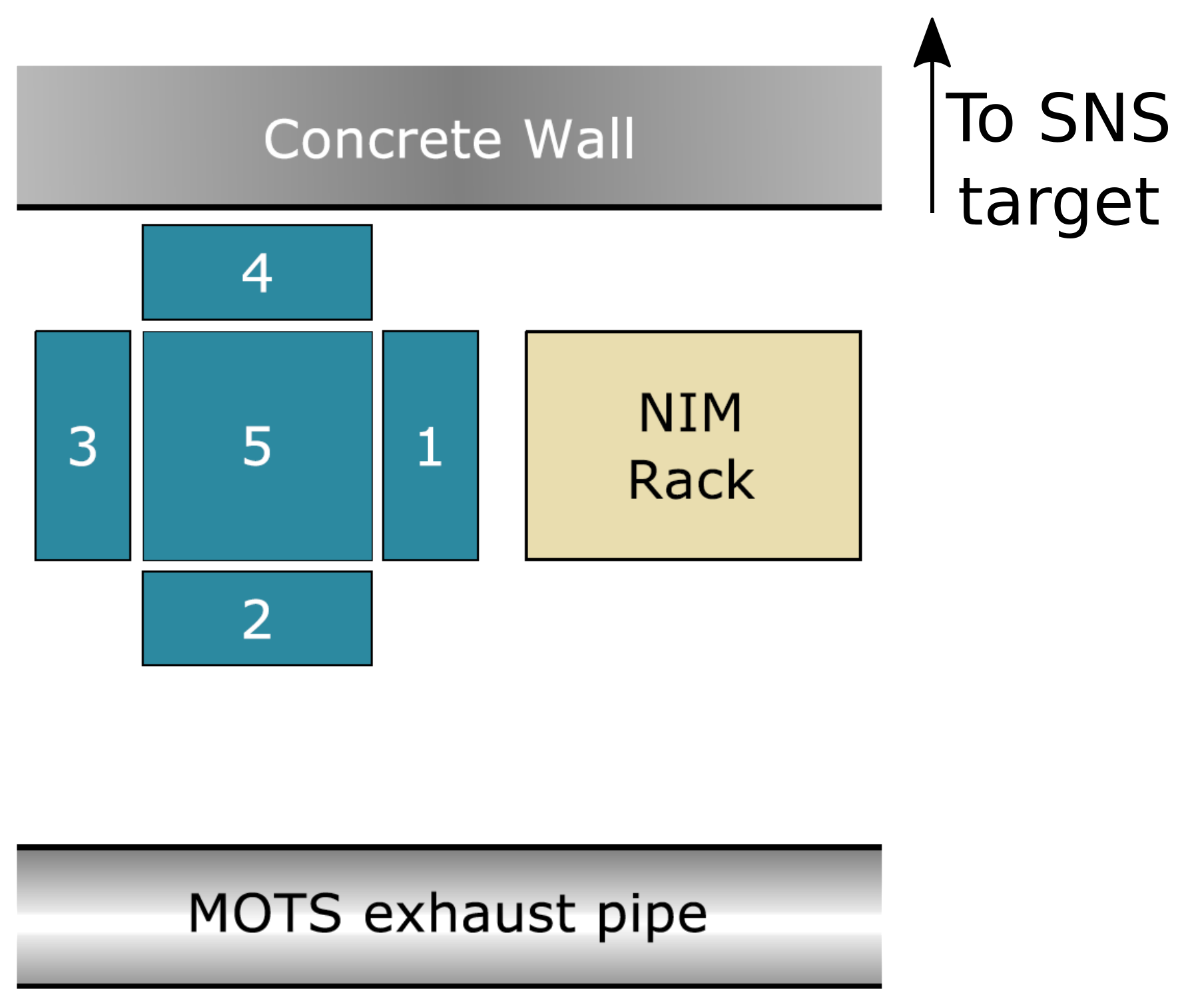}
\end{center}
\caption[Schematic of muon veto panel positioning]{Schematic of muon veto panel positioning. Drawing is not to scale.}
\label{fig:csi-setup:muon-veto-panel-labeling}
\end{figure}

The \ac{ps} 710 discriminator level for each panel is listed in Table~\ref{tab:csi-setup:pmt-high-voltage-muon-veto}. The levels for all side panels were set to achieve a maximum muon veto rejection while minimizing the triggering on the environmental $\gamma$ background. All side panels showed a similar triggering rate of \SI{20}{\hertz}. The top panel showed a triggering rate of approximately \SI{50}{\hertz}. All rates are in agreement with rate expectations based on the muon flux at sea level given the veto surface area. The individual panel locations with respect to the central \csi/ detector are shown in Fig.~\ref{fig:csi-setup:muon-veto-panel-labeling}, which also includes the \ac{mots} exhaust pipe~\cite{mots-01} running along the ceiling of the corridor. This pipe is a major source of $\gamma$-rays, mainly carrying an energy of \SI{511}{\keV}. As such, the overall triggering rate of all panels was found to be slightly elevated whenever the \ac{sns} was operational. The increase in rate depends on the exact panel positioning. Panel five experienced the largest increase with an overall triggering rate of $\sim\SI{100}{\hertz}$, whereas the side panels only experienced an increase in triggering rate on the order of 2-\SI{8}{\hertz}, depending on their positioning. 

\begin{table}[tb]
\begin{center}
\begin{tabular}{ccccc}
\toprule
PMT No. & Panel & High voltage (V) & Disc. level (mV) & Serial No.\\
\midrule
1 & 1 & -700 & \multirow{2}{*}{14.0} &107027 \\
2 & 1 & -700 & & 107010 \\
\cmidrule(lr){1-5}
3 & 2 & -720 & \multirow{2}{*}{14.5} & 106958 \\
4 & 2 & -740 & &64454 \\
\cmidrule(lr){1-5}
5 & 3 & -1000 & \multirow{2}{*}{13.7} & 64514 \\
6 & 3 & -660 & &64425 \\
\cmidrule(lr){1-5}
7 & 4 & -940 & \multirow{2}{*}{13.2} & 64463 \\
8 & 4 & -700 & &64467 \\
\cmidrule(lr){1-5}
9 & 5 & -1100 & 25.2 & N/A \\ 
\bottomrule
\end{tabular}
\end{center}
\caption[High voltages for each \acs*{pmt} used in the \csi/ muon veto]{High voltages for all \acp*{pmt} used in the \csi/ muon veto. Also shown is the panel number each \acs*{pmt} is attached to, as well as the discriminator level used for each panel.}
\label{tab:csi-setup:pmt-high-voltage-muon-veto}
\end{table}

\section{Data structure}
\label{sec:csi-setup:data-structure}
\begin{figure}[tb]
\begin{center}
\includegraphics[width=6in]{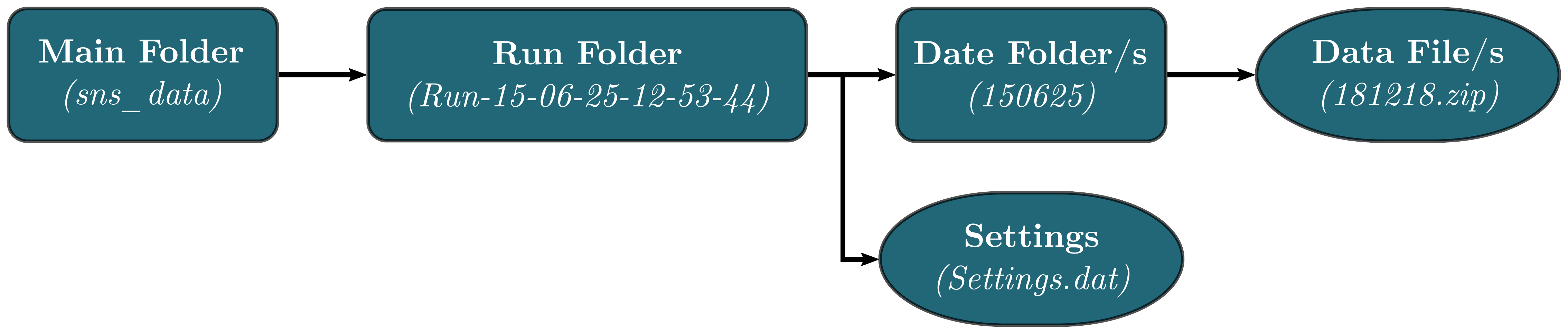}
\end{center}
\caption[Folder structure created by the data acquisition system used in the CE$\nu$NS search]{The main folder distinguishes between different detector setups. The run folder contains the time when the run was actually started. A new date folder is created every midnight. A new data file is created every minute per default.}
\label{fig:csi-setup:folder-structure}
\end{figure}

The folder structure created by the \labview/ DAQ system is shown in Fig.~\ref{fig:csi-setup:folder-structure}. The main folder is used to distinguish between different setups, e.g. calibrations or \ac{cenns} search runs. The run folders contain the year, month, day, 24-hour, minute and second in which this specific run was started. The date folders consist of six digits, representing year, month and day. During data acquisition a new date folder is created every midnight. A new data file is being created every minute per default.\par

During development both hardware and software were optimized to allow for a simultaneously data acquisition and data compression, without introducing any throughput limitations. The compression of a one minute binary data file with an initial size of \SI{240}{\mega\byte} into a zip-archive takes approximately \SI{45}{\second}. As such no data pile up is being introduced by compressing the data. This is highly beneficial for data storage as the compressed data only requires approximately a quarter of the disk space of the initial binary files. It also reduces the time needed to backup the data on the HCDATA cluster. Given a data transfer rate of 5-\SI{10}{\mega\byte\per\second} raw binary files could be transferred without introducing any data pile up. However, these transfer rates fall below \SI{5}{\mega\byte\per\second} for prolonged times on a regular basis. As such new data would be created faster than it can be moved onto the cluster. As such the data acquisition would have to be stopped once the local hard-drive is at full capacity. Being able to compress the data in parallel to the data acquisition fully eliminates this issue.\par

The settings file contains all DAQ settings used during the acquisition. For a full description of each value saved is presented in Table~\ref{tab:csi-setup:settings-file}. Some parameters are saved in the settings file even though they have no effect on the \ac{ni} 5153 digitizer, e.g. the vertical offset of either channel. These parameters are technically available in the sub-VIs provided by \ac{ni} to properly initialize the hardware, yet internally the \ac{ni} 5153 digitizer is not capable of adjusting these values. Fields marked as \emph{-reserved-} do not contain any meaningful parameters. They might contain non-zero entries as they were used during the development of the DAQ software. Yet, at this stage the values provided do no longer carry any meaningful information and can be disregarded.\par

\begin{figure}[tb]
\begin{center}
\includegraphics[width=6.5in]{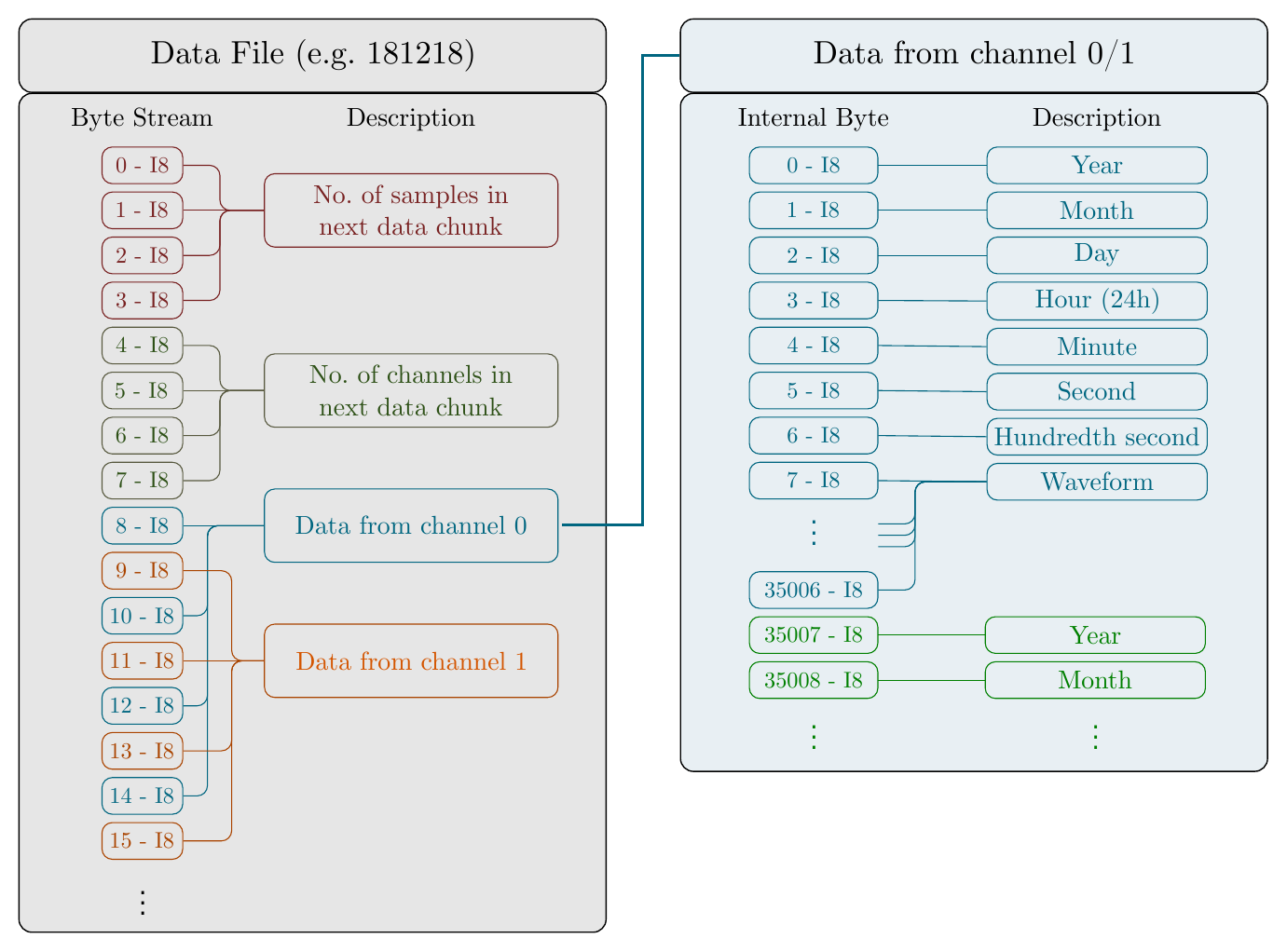}
\end{center}
\caption[Structure of the binary data files acquired during the CE$\nu$NS search]{Shown is the internal structure of the binary files saved by the LabView DAQ. Each binary file (left) consists of multiple, successive chunks of data. The first four bytes (I8, 0-3, red in the figure) of each chunk have to be merged into a single U32 which is the number of samples $N_S$ in each channel that belong to the current chunk. For the measurements presented in this thesis (\isotope{Am}{241},\isotope{Ba}{133} and SNS) $N_S$ has to be a multiple of \num{35007}. The second quartet of bytes (I8, 4-7, green in the figure) also has to be merged into a single U32 which is the number of channels $N_C$ in the current chunk. $N_C$ will always be 2 for this DAQ. Next, a total of $N_S \times N_C$ bytes follow which have to be decimated into two arrays. One array contains all data belonging to CH0 within this chunk and the other one contains all data belonging to CH1. One of these arrays is shown on the right. Each array consists of $N = \nicefrac{N_S}{35007}$ subsets of \num{35007} bytes. The first 7 bytes of each subset contain all the timing information needed for an event. The following \num{35000} bytes represent the waveform. The next LabView \emph{header} is encountered at byte index $i\,=\,N_S \times N_C + 8$ of the binary data file, where the procedure described above has to be repeated in order to extract the individual waveforms from the chunk.}
\label{fig:csi-setup:data-structure}
\end{figure}

As already stated, the acquired data is saved as binary files containing an array of I8 variables. Fig.~\ref{fig:csi-setup:data-structure} shows the internal structure of an example binary data file. Using \labview/ for the data analysis would provide built-in functions to properly read the chunks of data and decimate the corresponding arrays. However, due to the sheer size of the data set acquired the full analysis had to be performed on the HCDATA cluster, fully utilizing its potential of parallel data analysis. The analysis code used in this thesis is available at \cite{scholz-01}.

\begin{table}[hbtp]
	\begin{center}
		\begin{tabular}{clcl}
		\toprule
		\textbf{Index} & \textbf{Description} & \textbf{Index} & \textbf{Description}\\
		\midrule
		0 & CH0 Coupling [AC=0, DC=1]        & 13 & Trigger Source [CH0, CH1, Ext=3]  \\
		1 & CH0 Probe Attenuation            & 14 & Trigger Level [V]                 \\
		2 & CH0 Vertical Range [V]           & 15 & Hysteresis [V]                    \\
		3 & CH0 Vertical Offset [V]          & 16 & Holdoff [s]                       \\
		4 & CH1 Coupling [AC=0, DC=1]        & 17 & Trigger Delay [s]                 \\
		5 & CH1 Probe Attenuation	           & 18 & Trigger Type [Edge=0,Hyst=1]      \\
		6 & CH1 Vertical Range [V]           & 19 & - reserved -                      \\
		7 & CH1 Vertical Offset [V]          & 20 & - reserved -                      \\
		8 & Minimum Sample Rate [S/s]        & 21 & - reserved -                      \\
		9 & Minimum Record length [S] 			 & 22 & Start time stamp$^\star$          \\
		10 & Trigger Reference Position [\%] & 23 & End time stamp$^\star$            \\
		11 & Trigger Coupling [AC=0, DC=1]   & 24 & - reserved -                      \\
		12 & Triggering Slope [neg=0,pos=1]  & 25 & - reserved -                      \\
		\bottomrule
		\end{tabular}\\
		\footnotesize{$^\star$LabView not Unix, i.e. whole seconds after the Epoch 01/01/1904 00:00:00.00 UTC}
	\end{center}
	\vspace{-0.3cm}
	\caption[Description for all data values available in the settings file of each data run]{Description for all data values available in the settings file of each data run.}
	\label{tab:csi-setup:settings-file}
\end{table}

%% file: am-calibration.tex
%
%
\chapter{Light yield and light collection uniformity}
\label{chapter:am-calibration}

\begin{figure}[tb]
\begin{center}
\includegraphics[scale=1]{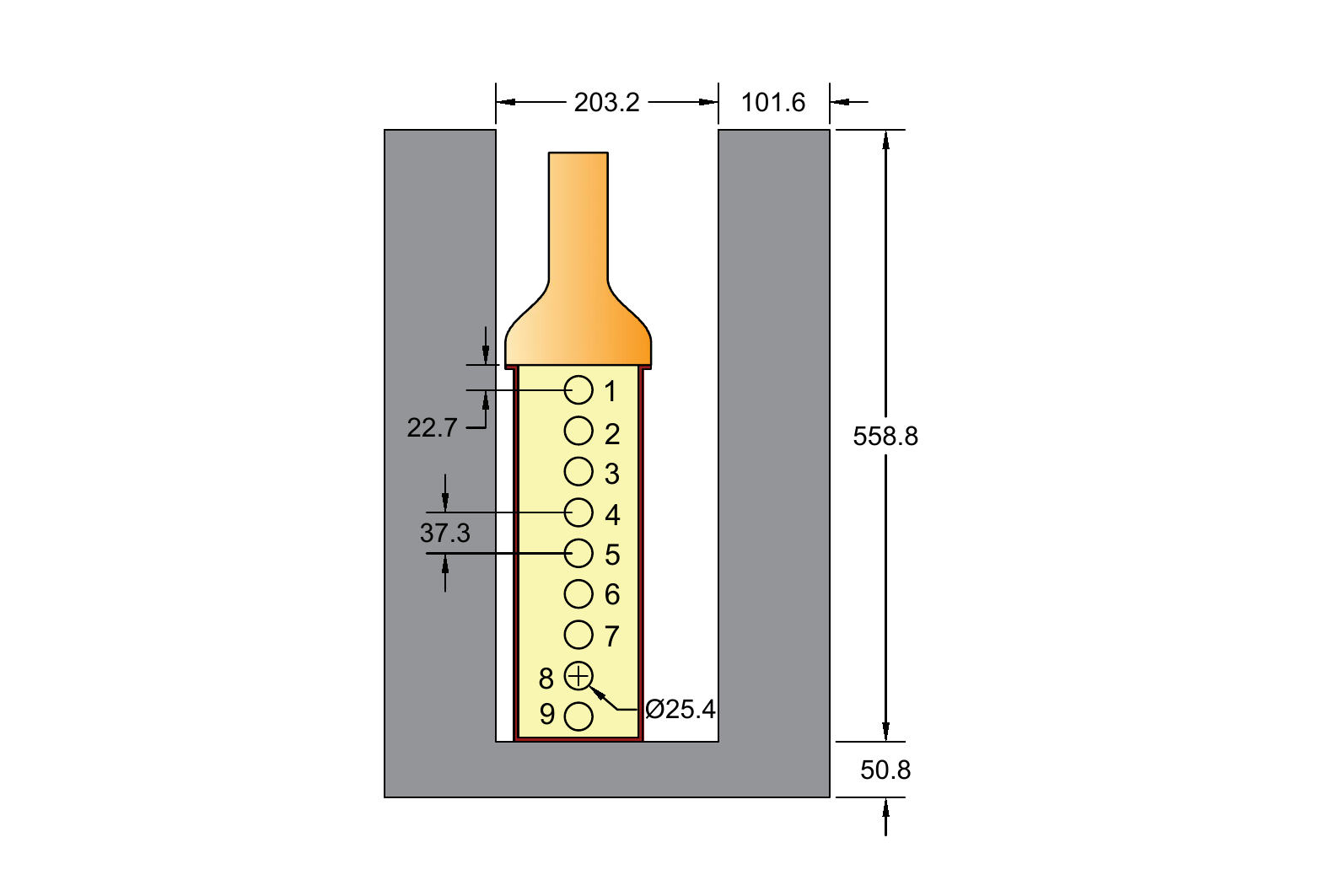}
\end{center}
\caption[Source placement during light yield and light collection uniformity calibrations]{Source placement and position numbering during the light yield and light collection uniformity calibrations using an \isotope{Am}{241} source. The lead shield surrounding the detector is shown in gray. All measurements are in mm.}
\label{fig:am-calibration:setup}
\end{figure}

Given the large size of the \csi/ detector, a crucial calibration was to probe the light collection uniformity along the crystal. An \isotope{Am}{241} source was placed on the outside of the copper can at nine equally-spaced positions along the length of the \csi/ detector (Fig.~\ref{fig:am-calibration:setup}) to measure the light collection uniformity in the \csi/ crystal. The main $\gamma$-ray emission of this isotope only carries an energy of $E_\gamma=\SI{59.54}{\keV}$. This low energy ensures that interactions with the crystal occur in close vicinity to the source location. As a result, only a small volume of the crystal is irradiated for each source location. Comparing the total light yield, i.e., the number of \ac{spe} produced per unit energy, found at each position provides a measure for the non-uniformity in the detector response. 

\begin{figure}[tb]
\begin{center}
\includegraphics[scale=1]{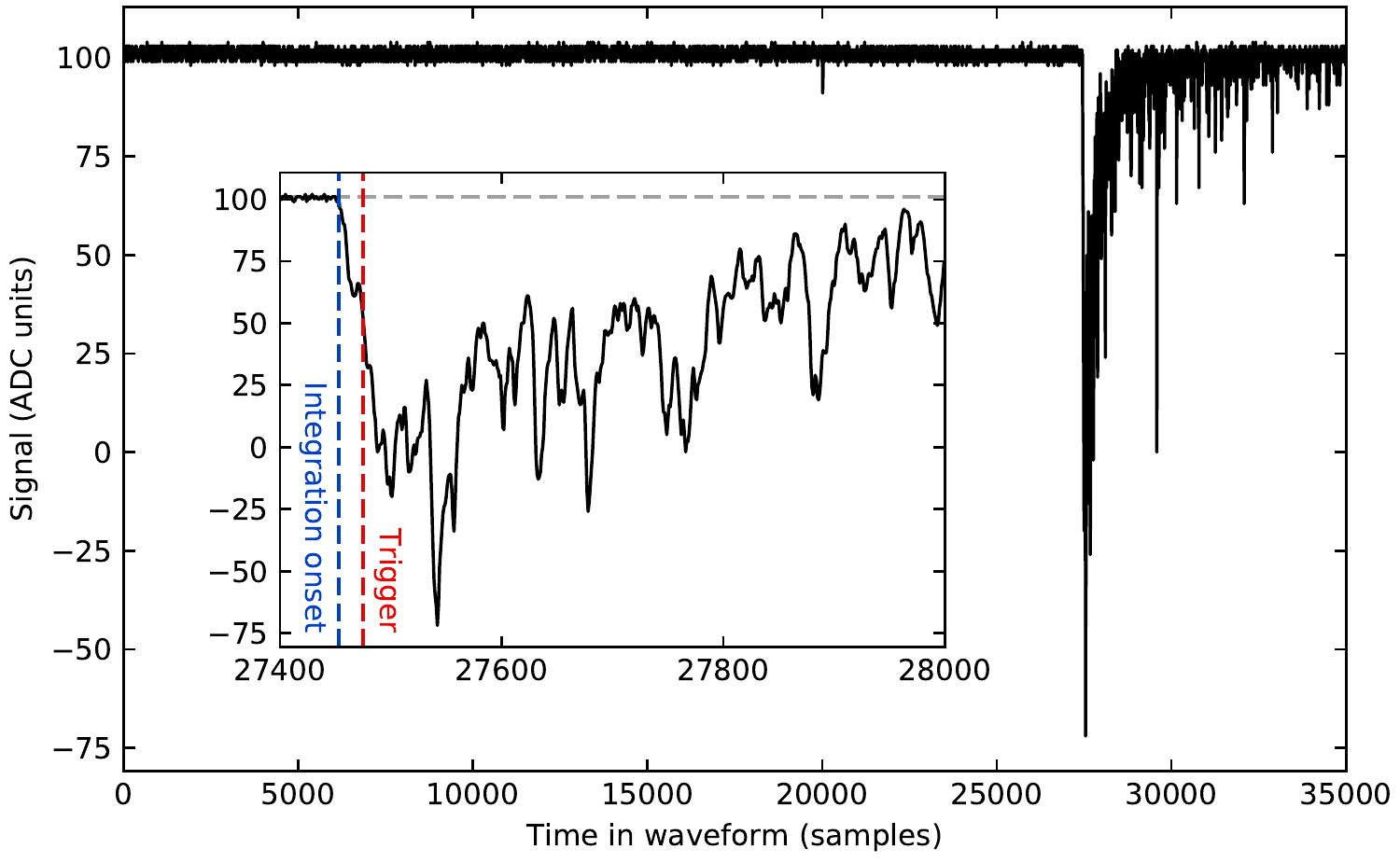}
\end{center}
\caption[Example waveform showing a typical \isotope{Am}{241} signal before data processing]{Example waveform showing a typical \isotope{Am}{241} signal before data processing. The inset provides a zoom into the triggering region. Sample \num{27400} marks the onset of the \acs{roi} which preceeds the hardware trigger by 75 samples. Highlighted is the hardware trigger (dashed red) and the onset for the charge integration window (dashed blue).}
\label{fig:am-calibration:example-waveform-Am-pulse}
\end{figure}

\section{Detector setup}
The calibrations were performed at a sub-basement laboratory at the University of Chicago, providing roughly \SI{6}{\mwe} of overburden as shielding from cosmic-rays. The central detector was placed within a well of contemporary lead, providing a total of \SI{2}{\inch} of $\gamma$-shielding on the bottom and \SI{4}{\inch} on the sides.\\

The detector was wired as shown in Fig.~\ref{fig:csi-setup:wiring-diagram-sns} with two exceptions. First, no muon veto was present during this calibration measurement. Second, the data was acquired triggering on the \csi/ signal itself instead of using an external trigger. The threshold of a standard edge trigger was set much lower than the typical amplitude of an \isotope{Am}{241} event. The threshold was further adjusted to achieve a data acquisition rate of $\mathcal{O}(\SI{60}{\hertz})$ to guarantee a \SI{100}{\percent} data throughput. The trigger position was set to \SI{78.5}{\percent} of the \SI{70}{\micro\second} long traces,i.e., at sample \num{27475}. The \csi/ signal was sampled at \SI{500}{\mega\sample\per\second} resulting in a total of \num{35000} samples per waveform. The digitizer range was set to \SI{200}{\milli\Vpp} with a digitizer depth of \SI{8}{\bit}. An example waveform is shown in Fig.~\ref{fig:am-calibration:example-waveform-Am-pulse}. The \csi/ signal $V$ was transformed using Eq.~(\ref{eq:csi-setup:csi-transformation}). For each source position data was acquired for approximately \SI{5}{\minute}.

\section{Waveform analysis}
\begin{figure}[htbp]
\begin{center}
\includegraphics[scale=1]{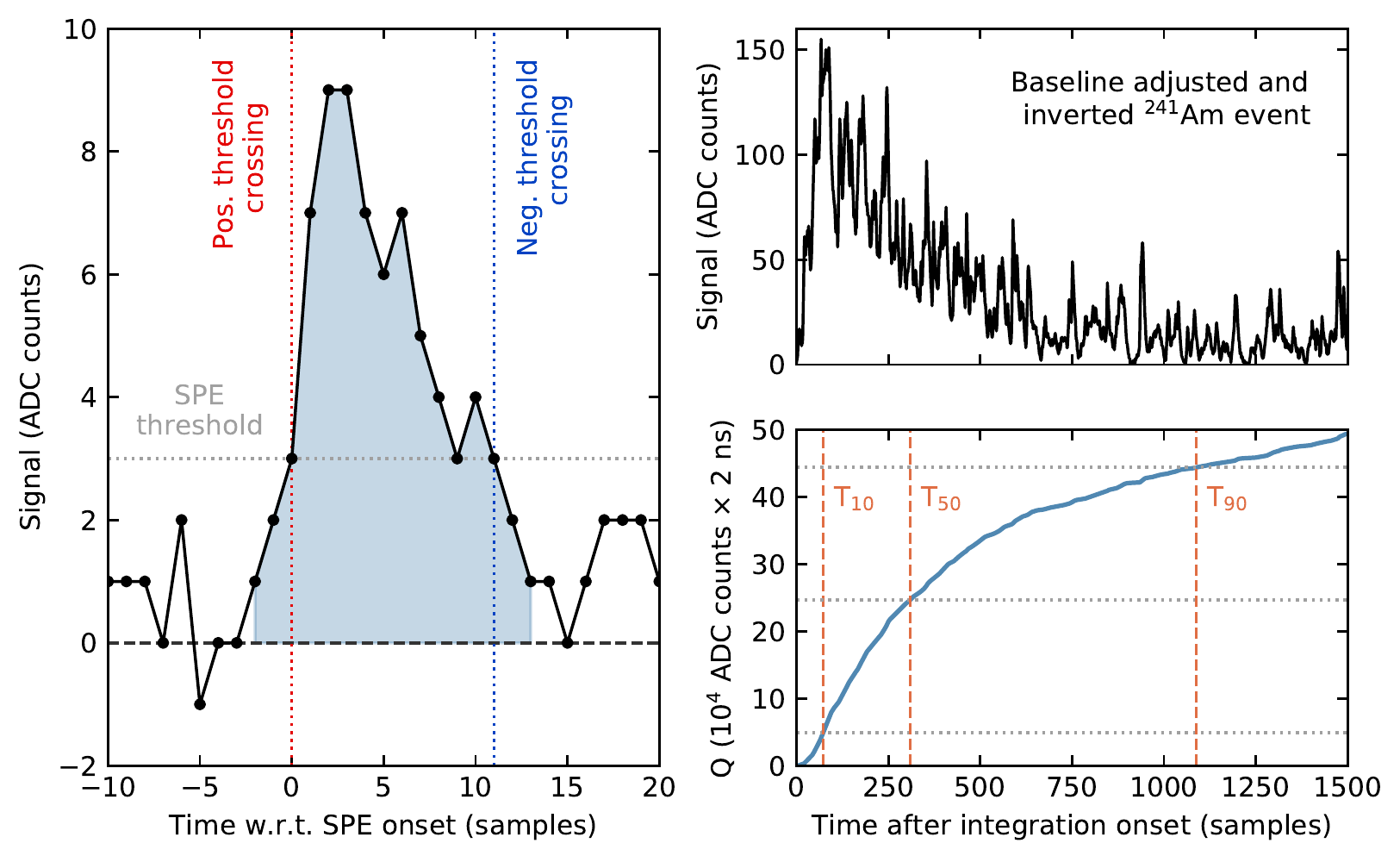}
\end{center}
\caption[Process used to determine the integration window of a \acs*{spe} and event rise-time definitions]{\textbf{Left}: Process used to determine the integration window of a \ac{spe}. The signal was already baseline-adjusted and inverted. The dashed gray line represents the detection threshold. A peak is detected when at least four consecutive samples show at least a value equal to the detection threshold. The positive (red) and negative (blue) threshold crossings are recorded. The integration window ranges from two samples before the positive crossing to two samples after the negative crossing (shaded blue area). \textbf{Right}: The top panel shows an example of a baseline adjusted and inverted \isotope{Am}{241} event. The onset was calculated using a standard threshold finding algorithm. Shown is the full \num{1500} sample long integration window. The bottom panel shows the corresponding integrated charge in blue. The dotted gray lines represent the 10, 50 and \SI{90}{\percent} levels of the total charge integrated for this event. The corresponding timings for each of these threshold crossings, i.e. T$_{10}$, T$_{50}$ and T$_{90}$, are shown in orange.}
\label{fig:am-calibration:spe-and-integration}
\end{figure}

The analysis pipeline for each of the nine data sets was identical and is described in the following: Each acquired \csi/ waveform was divided into two distinct regions. The first is the \acf{pt}, which spans sample 0 - \num{27399}. The second region is the \acf{roi}, spanning sample \num{27400} - \num{34999}. The overall baseline was estimated using the median $V_\text{median}$ of the first \num{20000} samples ($=\SI{40}{\micro\second}$) of the waveform. The \csi/ signal was shifted and inverted using
\begin{align}
\hat{V}_i = V_\text{median} - V_i\quad\text{for}\quad i\in[0,\num{35000}).
\end{align}
All peaks within the waveform were detected using a standard threshold crossing algorithm \cite{ni-01}. A peak was defined as at least four consecutive samples showing a minimum deviation of at least \SI{3}{\adc} from the baseline. The time of both positive and negative threshold crossings were recorded for each peak.

\subsection{Mean \acs*{spe} charge calibration}
\label{section:am-calibration:spe-charge}
The \ac{spe} charge spectrum can be calculated using the peaks found in the \acf{pt}. The charge of each peak was determined by integrating the signal $\hat{V}$ within an integration window, which ranges from two samples before the positive threshold crossing of a peak to two samples after the negative threshold crossing.\\

The left panel of Fig.~\ref{fig:am-calibration:spe-and-integration} illustrates the peak finding algorithm and the definition of the integration window. A new baseline $V_\text{median}^j$ was calculated for each peak $j$ using the median of the \num{250} samples directly preceding and following the integration window. The total integrated charge for peak $j$ was calculated using
\begin{align}
Q^j = \sum_{i=t_s}^{t_f} \left(\hat{V}_i - V_\text{median}^j\right)\label{eq:am-calibration:spe-charge-calculation},
\end{align}
where $t_s$ denotes the index of the beginning of the integration window and $t_f$ its end. The total charge $Q^j$ is given in unconventional units of $\SI{}{\adc}\times\SI{2}{\ns}$. These units can be convert into conventional charge units using
\begin{align}
Q\left(\text{pC}\right)=\frac{Q\left(\SI{}{\adc}\times\SI{2}{\ns}\right)}{32}
\end{align}
\begin{align}
\text{as}\quad\SI{1}{\adc}\times\SI{2}{\ns}\mathrel{\widehat{=}}\frac{\SI{0.2}{\mV}}{2^8\times\SI{50}{\ohm}}\times{\SI{2}{\ns}}=\SI{3.125e-14}{\coulomb},\nonumber
\end{align}
However, no additional information was gained by its conversion and it was therefore omitted throughout the rest of this thesis.\\

\begin{figure}[tbp]
\begin{center}
\includegraphics[scale=1]{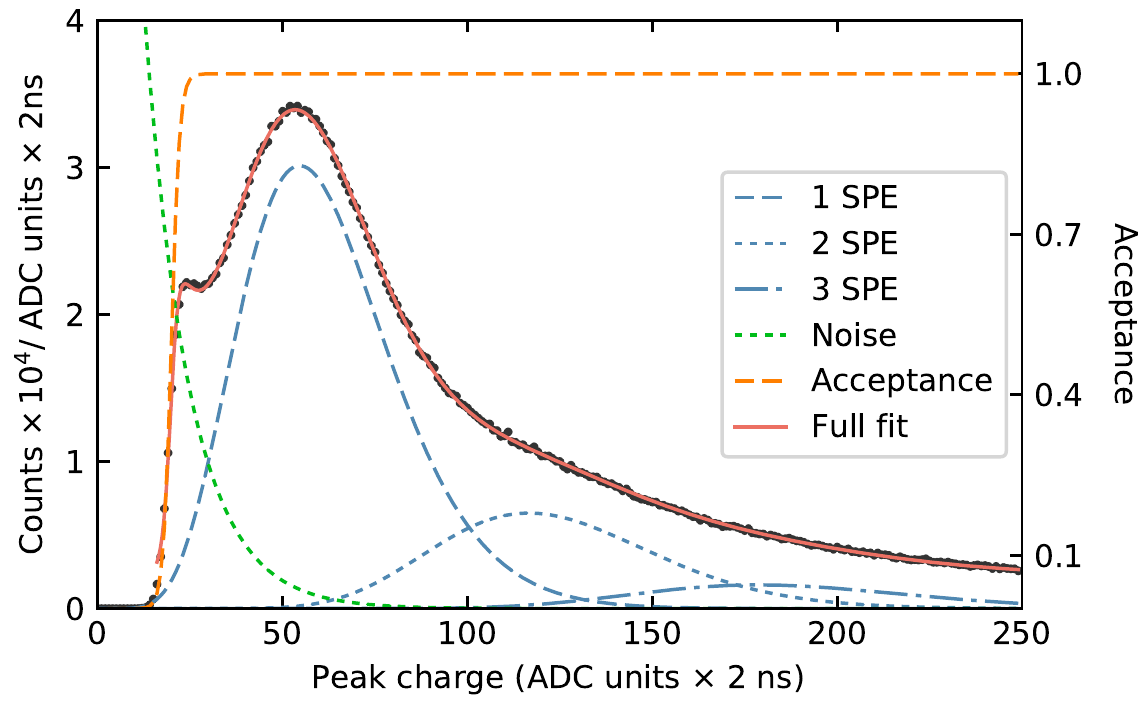}
\end{center}
\caption[\acs*{spe} charge distribution for \isotope{Am}{241} source position two of the light yield calibration]{\acs*{spe} charge distribution as determined for \isotope{Am}{241} source position two (Fig.~\ref{fig:am-calibration:setup}). Shown in red is the best fit of Eq.~(\ref{eq:am-calibration:spe-polya}) to the data. All individual fit components are also shown. Even though a Polya distribution was used as the \ac{spe} charge model, only a small deviation from a normal distribution is visible for the blue curves. The \ac{spe} charge distribution for all other \isotope{Am}{241} source positions showed similar fit qualities.}
\label{fig:am-calibration:example-spe-spectrum}
\end{figure}

Fig.~\ref{fig:am-calibration:example-spe-spectrum} shows the \ac{spe} charge spectrum calculated for data taken with the \isotope{Am}{241} source located at position two (Fig.~\ref{fig:am-calibration:setup}). The distribution consists of all integrated peaks found in the \ac{pt} region for all waveforms of this data set. The mean \ac{spe} charge \qspe/ was calculated by fitting a \ac{spe} charge distribution model to the data. Several different models were proposed in the past \cite{prescott-01,bellamy-01} and the average \qspe/ obtained by fitting these models to the data can slightly differ. In the following paragraph two competing models for the \ac{spe} charge distribution are introduced. The first uses a Gaussian distribution to describe the \ac{spe} charge spectrum, whereas the second uses a Polya distribution.\\

Most commonly the shape of the \ac{spe} charge spectrum is approximated by a Gaussian distribution \cite{bellamy-01}. The corresponding fit function is given by
\begin{align}
f_\text{gauss}\left(q,a_n,\sigma_n,\vec{a},\qspe/,\sigma,k,q_0\right)&=\;\left[a_n \text{e}^{-\nicefrac{q}{\sigma_n}} + \sum_{i=1}^{3}a_i\,g_i(q,\qspe/,\sigma)\right]\left(1.0 + \text{e}^{-k\left(q - q_0\right)}\right)^{-1}\label{eq:am-calibration:spe-gauss}\\[1ex]
\text{where}\quad g_i(q,\qspe/,\sigma) &= \Exp{\frac{(q-i\,\qspe/)^2}{2i\sigma^2}}.\nonumber
\end{align}
Equation~(\ref{eq:am-calibration:spe-gauss}) includes the following components: The exponential $\text{e}^{-\nicefrac{q}{\sigma_n}}$ represents events caused by  random baseline fluctuations that exceed the \ac{spe} search criteria. $g_i$ describes the $i$-th convolution of a Gaussian distribution with a mean of $\qspe/$ and a standard deviation of $\sigma$ with itself. The sum over $g_i$ therefore includes the charge distributions for one ($i=1$), two ($i=2$) and three ($i=3$) \ac{spe}. The multi-\ac{spe} peak amplitudes $a_{2,3}$ are allowed to float freely. The amplitudes $a_{1,2,3}$ are not restricted to follow a Poisson distribution, since the measurements of the \ac{spe} mean charge discussed throughout this thesis were not performed with a dedicated setup using an LED or laser pulse as trigger. The last factor in Eq.~(\ref{eq:am-calibration:spe-gauss}) describes a sigmoid-shaped acceptance curve, which represents the efficiency of the peak finding algorithm to detect a peak of charge $q$. In this acceptance model $k$ describes the steepness of the sigmoid and $q_0$ its midpoint.\\

\begin{figure}[tb]
\begin{center}
\includegraphics[scale=1]{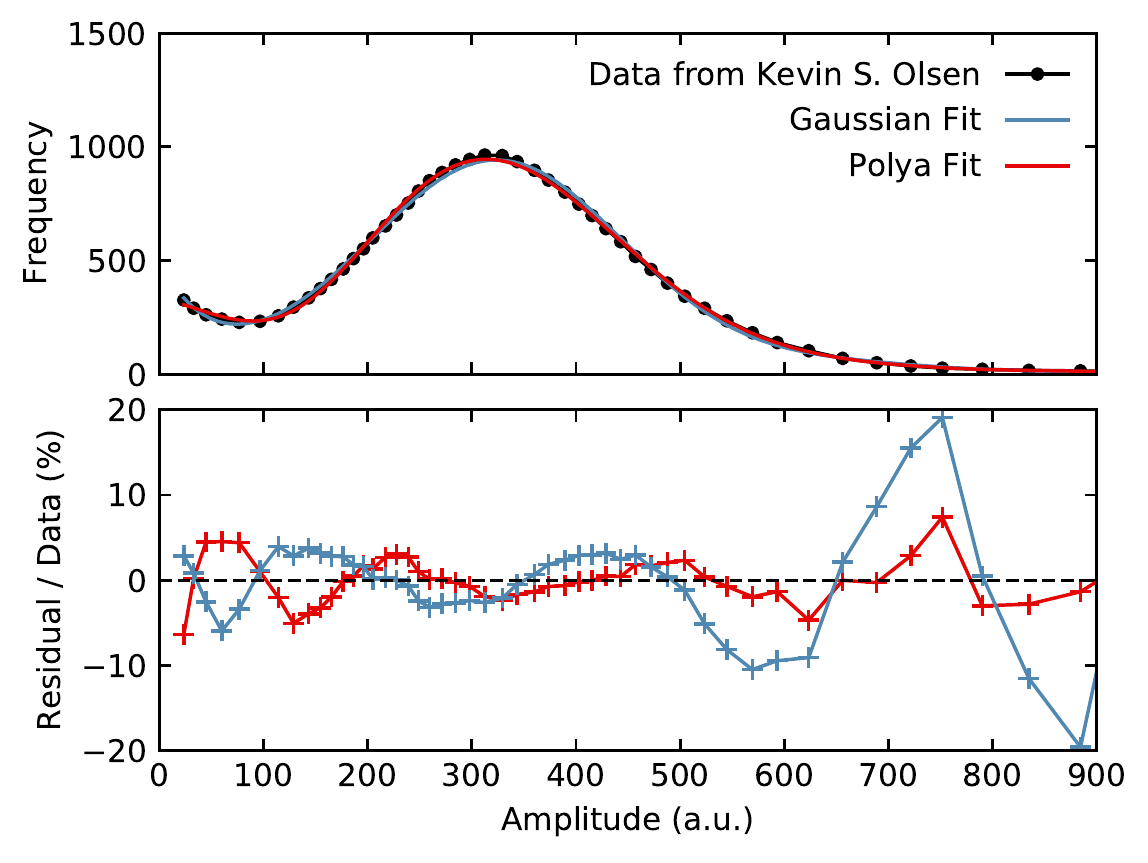}
\end{center}
\caption[Comparison of Polya and Gaussian shaped \acs*{spe} charge distribution models for the R877-100 \acs*{pmt}]{\textbf{Top}: The data shown represents a dedicated \ac{spe} charge measurement for a R877-100 \acs*{pmt} using a LED triggered setup, which was taken from \cite{olsen-01}. The \ac{spe} charge distribution was fitted using the models described in Eq.~(\ref{eq:am-calibration:spe-gauss}) (Gaussian, shown in blue) and (\ref{eq:am-calibration:spe-polya}) (Polya, shown in red). \textbf{Bottom}: Deviation between experimental data and the \ac{spe} charge model. Both models describe the peak region equally well. However, the overall deviation for the model based on the Polya distribution is smaller for the same degrees of freedom.}
\label{fig:am-calibration:polya-vs-gauss}
\end{figure}

For some \acp{pmt} the complex secondary electron emission at different dynode stages can lead to inconsistencies between the recorded experimental charge spectrum and a fitted model based on a Gaussian distribution~\cite{barbeau-04}. Prescott~\cite{prescott-01} instead proposed to use a Polya distribution to describe the \ac{spe} charge spectrum. Modifying Eq.~(\ref{eq:am-calibration:spe-gauss}) yields 
\begin{align}
f_\text{polya}\left(q,a_n,\sigma_n,\vec{a},\qspe/,\sigma,k,q_0\right)&=\;\left[a_n \text{e}^{-\nicefrac{q}{\sigma_n}} + \sum_{i=1}^{3}a_i\,p_i(q,\qspe/,\sigma)\right]\left(1.0 + \text{e}^{-k\left(q - q_0\right)}\right)^{-1}\label{eq:am-calibration:spe-polya}\\[1ex]
\text{where}\quad p_i(q,\qspe/,\sigma) &= \Exp{-\left(\sigma+1\right)\frac{q}{\qspe/}}\left(\frac{q}{\qspe/}\right)^{i\,\left(\sigma+1\right)-1},\nonumber
\end{align}
where all Gaussian distributions $g_i$ were replaced with Polya distributions $p_i$. Similar to the Gaussian case the subscript $i$ denotes the $i$-th convolution of a Polya distribution with itself. The noise contribution as well as the sigmoid shaped acceptance curve remain unchanged.\par

Both Eq.~(\ref{eq:am-calibration:spe-gauss}) and Eq.~(\ref{eq:am-calibration:spe-polya}) were fitted to a \ac{spe} charge spectrum that was taken for an identical R877-100 \ac{pmt} using an LED triggered setup, which was adapted from \cite{olsen-01}. The experimental charge spectrum including both fits is shown in the top panel of Fig.~\ref{fig:am-calibration:polya-vs-gauss}. The bottom panel shows the deviation of each model from the data. Both distributions describe the peak region equally well. However, the overall deviation for the model based on the Polya distribution is smaller. For the remainder of this thesis a Polya based \ac{spe} charge distribution model is used whenever a \ac{spe} charge distribution is fitted for the R877-100 \ac{pmt}.\par

The mean \ac{spe} charge \qspe/ was determined for each \isotope{Am}{241} source location by fitting Equation~(\ref{eq:am-calibration:spe-polya}) to the respective \ac{spe} charge spectra. Fig.~\ref{fig:am-calibration:example-spe-spectrum} shows the \ac{spe} charge spectrum calculated for the data acquired for source position two (Fig.~\ref{fig:am-calibration:setup}). The best fit of Equation~(\ref{eq:am-calibration:spe-polya}) is shown in red. The individual fit components are shown in green (noise), blue (\ac{spe} charges) and orange (acceptance). The orange curve indicates that only a tiny fraction of low-charge \acp{spe} are actually missed by the peak finding algorithm. Comparing the counts obtained from integrating the fitted \ac{spe} charge model without the inclusion of the acceptance sigmoid to the counts obtained when including it shows that only $\mathcal{O}(\SI{0.5}{\percent})$ of the \acp{spe} were missed by the peak finding algorithm. These \acp{spe} further carry only a small charge and as a result only a negligible amount of total charge ($\ll\SI{0.1}{\percent}$) remained unidentified.\\

\begin{figure}[tbp]
\begin{center}
\includegraphics[scale=1]{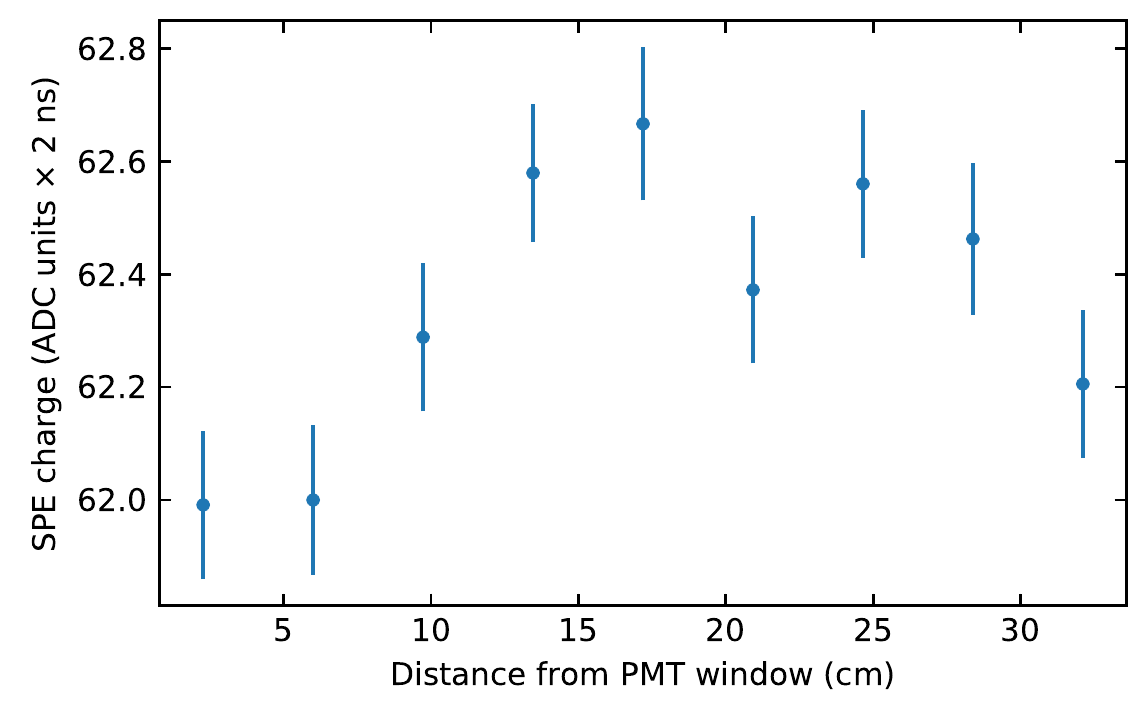}
\end{center}
\caption[Mean \acs*{spe} charge for all \isotope{Am}{241} source positions]{Mean \ac{spe} charge calculated for all \isotope{Am}{241} source positions. Only a \SI{1}{\percent} variation in \ac{spe} charge is visible among the different runs.}
\label{fig:am-calibration:spe-charge-vs-position}
\end{figure}

The mean \ac{spe} charges \qspe/ for all \isotope{Am}{241} source positions are shown in Fig.~\ref{fig:am-calibration:spe-charge-vs-position}. A slight increase in \qspe/ is apparent for \isotope{Am}{241} source distances of $\geq\SI{10}{\cm}$. This suggests that the \ac{pmt} was not completely warmed up during the data-taking for the closest positions, resulting in a slightly smaller \qspe/. After several minutes the amplification of the \ac{pmt} stabilized and resulted in a charge variation of less than \SI{1}{\percent}. This does not affect the light yield measurements in any significant way. The trigger threshold was chosen well below the amplitude expected from a \SI{59.54}{\keV} energy deposition. As a result all \isotope{Am}{241} events trigger the DAQ as a variation in their amplitude of $\sim\SI{1}{\percent}$ is too small to introduce any bias.\par

\begin{figure}[tb]
\begin{center}
\includegraphics[scale=1]{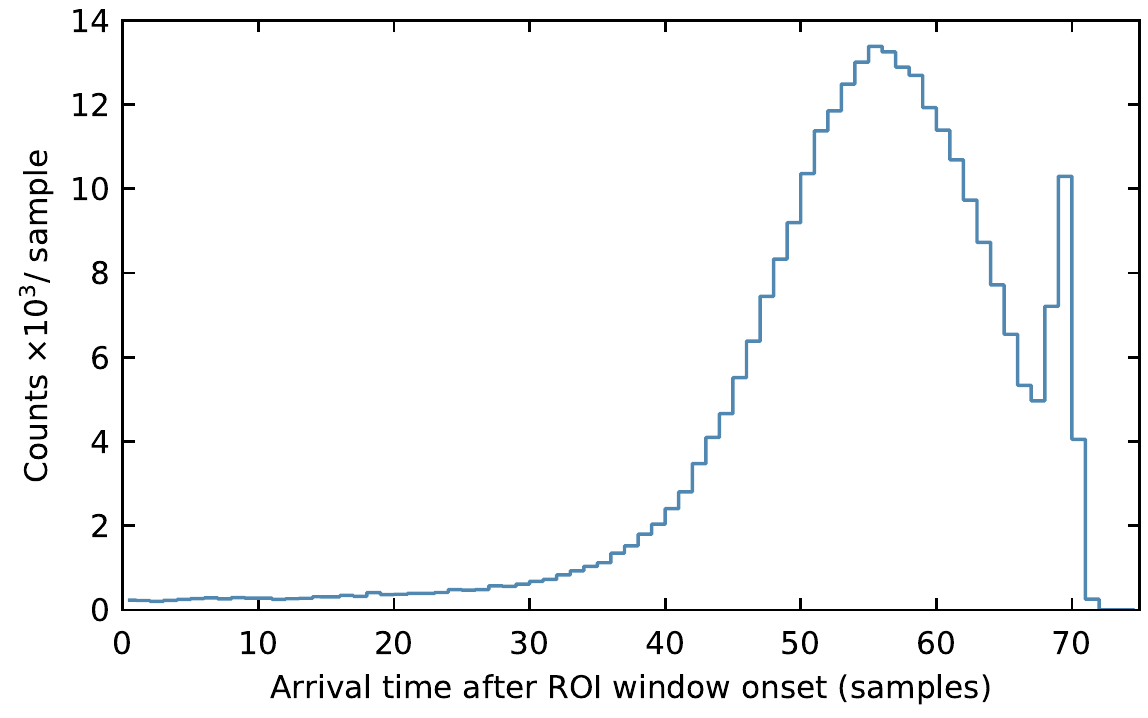}
\end{center}
\caption[Arrival time distribution for events in the light yield calibration]{Arrival time distribution of events triggering the data acquisition system in the \isotope{Am}{241} calibrations. A clear peak is visible at approximately 55 samples, arising from \isotope{Am}{241} events. As a result, it takes one of these scintillation light curves approximately 20 samples to rise from the baseline up to the trigger threshold. The sharp rise at 69 samples is caused by Cherenkov pulses, which are described in the text. These events show a much sharper rising edge.}
\label{fig:am-calibration:arrival-time-distribution}
\end{figure}

\subsection{Arrival time distribution}
Once the mean \ac{spe} charge was calculated for every \isotope{Am}{241} source position, the corresponding \isotope{Am}{241} charge spectra were calculated. For each waveform the onset of a potential \isotope{Am}{241} event was determined by applying a standard peak finding algorithm to the \acf{roi}. Peaks are defined by at least ten consecutive samples with an amplitude of at least \SI{3}{\adc} above the baseline. The arrival time $T_\text{arr}$, i.e., the positive threshold crossing of the first peak found in the \ac{roi} is recorded. Fig.~\ref{fig:am-calibration:arrival-time-distribution} shows the distribution of all recorded arrival times for all data sets. The beginning of \ac{roi} precedes the hardware trigger by 75 sample (inset of Fig.~\ref{fig:am-calibration:example-waveform-Am-pulse}), which limits potential arrival times to $T_\text{arr}\in[0,75]$ sample. Events caused by the \isotope{Am}{241} $\gamma$ emission are visible as a broad peak around 55 samples after the \ac{roi} window onset.\par

\begin{figure}[tb]
\begin{center}
\includegraphics[scale=1]{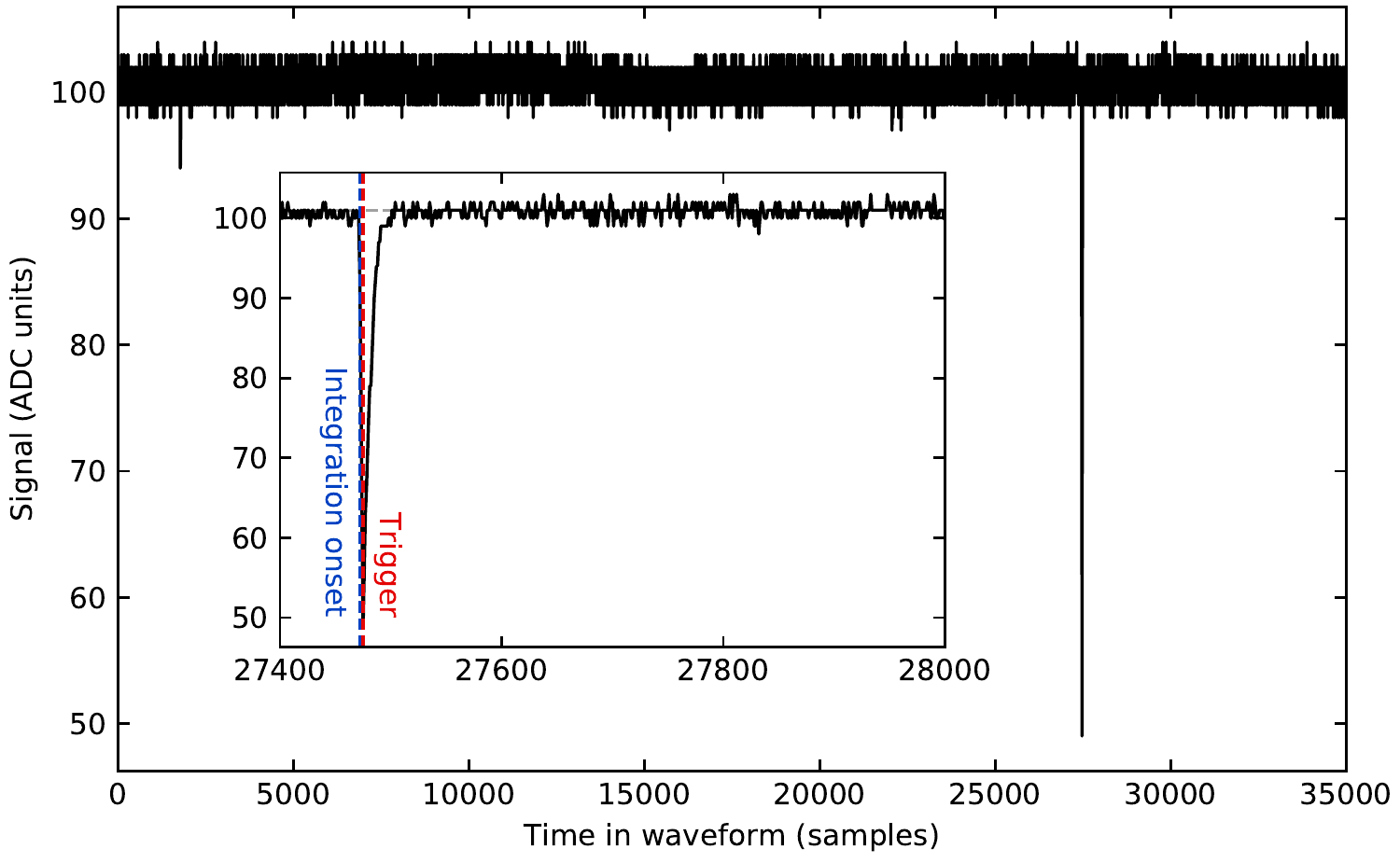}
\end{center}
\caption[Example waveform showing an event triggering on a Cherenkov signal]{Example waveform showing an event triggering on a Cherenkov signal arising from Cherenkov light emission in the \acs*{pmt} window. The large spike is typical of such an event and contains a charge equivalent to multiple \acs*{spe}. The inset shows the hardware trigger (dashed red) and the onset for the integration window as determined by the peak finding algorithm (dashed blue). It is apparent that for this type of background the hardware trigger and the peak onset determined by the peak finding algorithm almost coincide, leading to an arrival time $T_\text{arr}$ of $\sim70$ samples.}
\label{fig:am-calibration:example-waveform-Cherenkov-pulse}
\end{figure}

In contrast, the sharp rise at 69 samples after the \ac{roi} window onset is caused by afterpulses \cite{ma-01} or Cherenkov light emission in the \ac{pmt} window \cite{roodbergen-01}. The first of the two processes is caused by residual ionization in the gases caused by the accelerated photoelectrons inside the \ac{pmt}. The ions produced in this ionization are accelerated towards the photocathode. Upon impact on the photocathode an afterpulse is generated~\cite{ma-01}. The second is caused by small trace amounts of \isotope{U}{238}, \isotope{Th}{232} or \isotope{K}{40} within the \ac{pmt} window. The $\beta-$decays of such isotopes can produce Cherenkov light emission in the \ac{pmt} window or the electrons released in such a decay directly interact with the photocathode. The corresponding signal is a tight pulse large enough to surpass the trigger threshold (Fig.~\ref{fig:am-calibration:example-waveform-Cherenkov-pulse}). These Cherenkov events usually carry charges equivalent to two to fifteen \ac{spe} \cite{roodbergen-01,etenterprises-01}.\par

\subsection{Rise-time distributions}
To calculate the total charge for an event in \ac{roi} a \SI{3}{\micro\second} long integration window was defined. The onset of this window is given by two samples before $T_\text{arr}$ and spans a total of 1500 samples, i.e., \SI{3}{\micro\second}. A new baseline $V_\text{baseline}$ was estimated for this integration window using the median of the 500 samples (\SI{1}{\micro\second}) directly preceding it. The integrated scintillation curve $Q(t)$ is given by
\begin{align}
Q(t) = \sum\limits_{i = T_\text{arr}}^{T_\text{arr}\,+\,t} \left[\hat{V}_i - V_\text{baseline}\right]
\end{align}
For each event the total integrated charge $Q_\text{total}=Q(\SI{1499}{\sample})$ was recorded. In addition two different rise-times~\cite{ronchi-01,luo-01} were calculated using $Q(t)$. Rise-times are defined as the time it takes the integrated scintillation curve $Q(t)$ to rise from one predefined percentage of its maximum to another. The right panels of Fig.~\ref{fig:am-calibration:spe-and-integration} provide an illustration of their calculation.\par

The first rise-time was $T_{0-50}=T_{50}$, i.e. the time it took the integrated charge to rise from 0 to \SI{50}{\percent} of the total charge integrated in the window. The second was $T_{10-90} = T_{90} - T_{10}$, i.e. the time between passing the 10 and \SI{90}{\percent} charge threshold. Due to the length of the integration window both rise-times were limited to
\begin{align}
T_{0-50}\in[0,1500)\,\text{samples} = [0,3)\SI{}{\micro\second}\\
T_{10-90}\in[0,1500)\,\text{samples} = [0,3)\SI{}{\micro\second}.
\end{align}

\begin{figure}[htbp]
\begin{center}
    \includegraphics[width=6in]{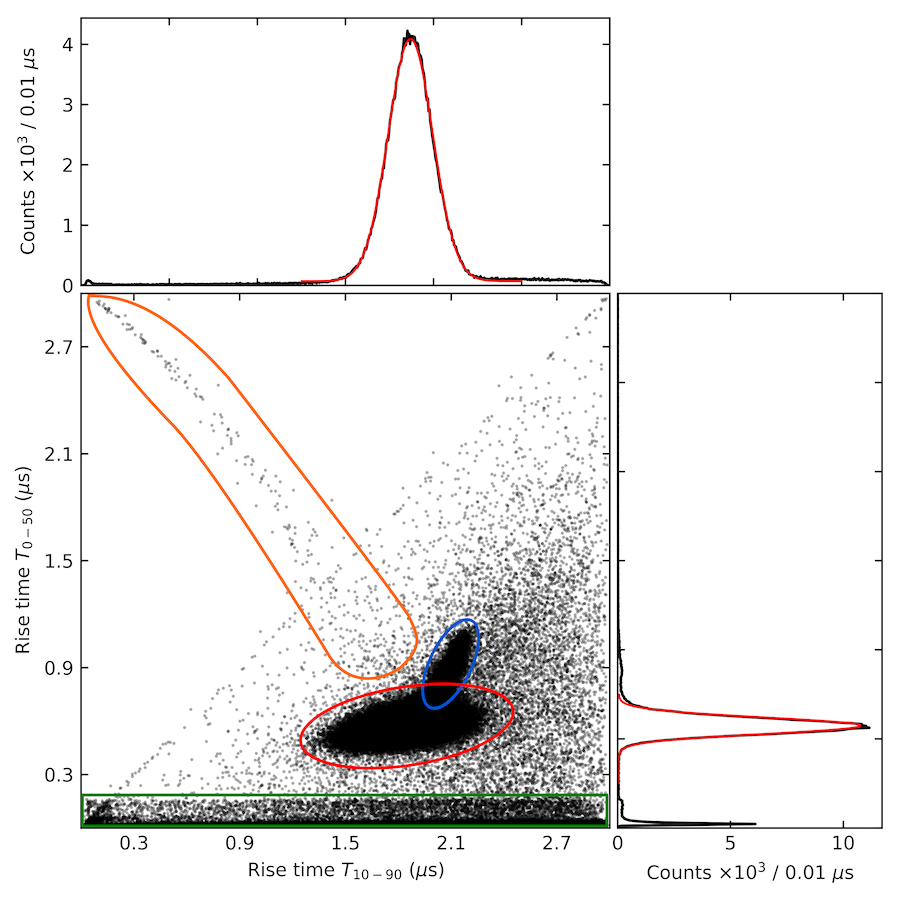}
\end{center}
\caption[Two-dimensional rise-time distribution for the light yield calibration]{Two-dimensional Rise-time distribution for all events recorded during the light yield calibration. The integration window was \SI{3}{\micro\second} long. Four distinct features are apparent. \textbf{Red}: Proper radiation induced events, mainly from $\gamma$s with an energy of $E=\SI{59.54}{\keV}$. \textbf{Blue}: Events exhibiting a digitizer range overflow. For these events the first couple of nanoseconds were clipped by the digitizer, which lead to longer rise-times. \textbf{Orange}: The pulse onset was misidentified due to a spurious \acs{spe} preceding the pulse. \textbf{Green}: Triggers on Cherenkov pulses with and without a leading \acs*{spe}.}
\label{fig:am-calibration:rt-distribution}
\end{figure}

The rise-time distributions are shown in Fig.~\ref{fig:am-calibration:rt-distribution}. Several different features are apparent in the two-dimensional rise-time distribution of all events recorded (bottom left panel). Four different phase-space regions are highlighted. Circled in red are correctly identified radiation induced events, mainly from \isotope{Am}{241} $\gamma$ emission (Fig.~\ref{fig:am-calibration:example-waveform-Am-pulse}). These events are centered at $T_{10-90}=\SI{1.87}{\micro\second}$ and $T_{0-50}=\SI{0.57}{\micro\second}$. It can be observed from the marginalized rise-time distributions on the top and right panel of Fig.~\ref{fig:am-calibration:rt-distribution}, that most events recorded fell within this category. A Gaussian was fitted to both marginalized rise-time distributions. Even though the ellipse covering these events appears tilted in the two-dimensional distribution an excellent agreement between data and fit was found. The fit results are given in Table~\ref{tab:am-calibration:rise-time-fit-values}.\par

\begin{table}[ht]
	\begin{center}
		\begin{tabular}{ccccc}
		\toprule
		\multicolumn{2}{c}{$T_{0-50}$} & &  \multicolumn{2}{c}{$T_{10-90}$}\\
		\midrule
		$\mu_{0-50}$    &  $0.57430 \pm 0.00048$\SI{}{\micro\second} & & $\mu_{10-90}$    & $1.86958 \pm 0.00059$\SI{}{\micro\second}\\
		$\sigma_{0-50}$ &  $0.04618 \pm 0.00037$\SI{}{\micro\second} & & $\sigma_{10-90}$ & $0.12726 \pm 0.00050$\SI{}{\micro\second}\\
		\bottomrule
		\end{tabular}
	\end{center}
	\caption[Mean and standard deviation of a Gaussian fitted to the marginalized rise-time distributions of $T_{0-50}$ and $T_{10-90}$]{Mean and standard deviation of a Gaussian fitted to the marginalized rise-time distributions of $T_{0-50}$ and $T_{10-90}$. The fits are shown in Fig.~\ref{fig:am-calibration:rt-distribution}.}
	\label{tab:am-calibration:rise-time-fit-values}
\end{table}

Circled in blue are events that exceeded the digitizer range. Both rise-times were biased towards higher values as the respective event signal was truncated in the early part of the integration window. The arrival time $T_\text{arr}$ of events circled in orange was misidentified due to a preceding spurious peak in the \ac{roi}, leading to a misaligned integration window. Last, a fourth population consisting of triggers on a Cherenkov pulse (Fig.~\ref{fig:am-calibration:example-waveform-Cherenkov-pulse}) are circled in green. Their spread is caused by spurious peaks before and after the triggering Cherenkov spike.\par

\section{Determining the light yield and light collection uniformity}
The information provided above was used to reject background events. First all events were rejected which showed at least one sample exceeding the digitizer range. Next, the following cuts on rise-time were implemented to remove Cherenkov triggers and events with a misidentified onset:
\begin{align}
T_{0-50} &> \SI{0.25}{\micro\second} = 125\,\text{samples}\\
T_{10-90} &> T_{0-50}
\end{align}

\begin{figure}[tb]
\begin{center}
\includegraphics[scale=1]{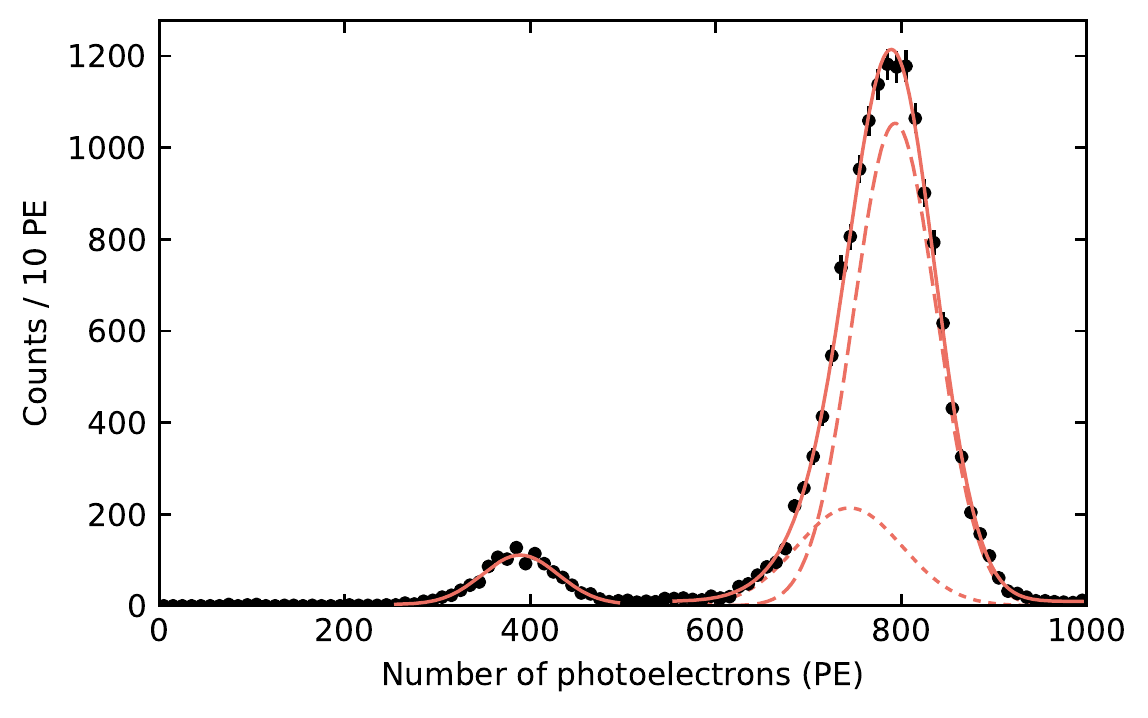}
\end{center}
\caption[Energy spectrum recorded during the light yield calibration for the \isotope{Am}{241} source location two]{Energy spectrum recorded for the \isotope{Am}{241} source position two. The integrated charges were converted to a corresponding number of \acs*{pe} using \qspe/. The main \isotope{Am}{241} emission peak is apparent (dashed), which also contains contributions from the L-shell escape peak of cesium and iodine (dotted). A smaller peak, which is visible at lower energies, represents the corresponding K-shell escape peak.}
\label{fig:am-calibration:example-energy-spectrum}
\end{figure}

These rise-time cuts were restrictive enough to remove most background events without rejecting any \isotope{Am}{241} events. In addition, events with more than $\agc/=3$ peaks in the \ac{pt} were rejected to minimize any bias from afterglow from a preceding event. The charge of all remaining events was converted into an equivalent number of \ac{spe} $N_\text{pe}$ using the corresponding mean \ac{spe} charge \qspe/. It is
\begin{align}
N_\text{pe}\,=\,\frac{Q_\text{total}}{\qspe/}
\end{align}

The energy spectrum for \isotope{Am}{241} source position two (Fig.~\ref{fig:am-calibration:setup}) is shown in Fig.~\ref{fig:am-calibration:example-energy-spectrum}. The energy spectra for all other source locations were comparable. The main \isotope{Am}{241} emission peak is apparent at $\sim\SI{800}{\pe}$. The shoulder towards lower energies is caused by the L-shell escape peaks of cesium and iodine. A secondary feature at $\sim\SI{400}{\pe}$ represents the K-shell escape peaks from cesium and iodine. Due to their low energy, $\gamma$-rays from \isotope{Am}{241} interact with the \csi/ crystal close to the surface making such escapes likely.\par

For each source position the average number of photoelectrons $N_\text{pe}^\text{am}$ produced in an \isotope{Am}{241} event was determined from the respective energy spectra. Using $N_\text{pe}^\text{am}$ the local light yield $\mathcal{L}$ for an incoming energy of \SI{59.54}{\keV} was calculated by
\begin{align}
\mathcal{L} = \frac{N_\text{pe}^\text{am}}{\SI{59.54}{\keV}}
\end{align}

\begin{figure}[htbp]
\begin{center}
\includegraphics[scale=1]{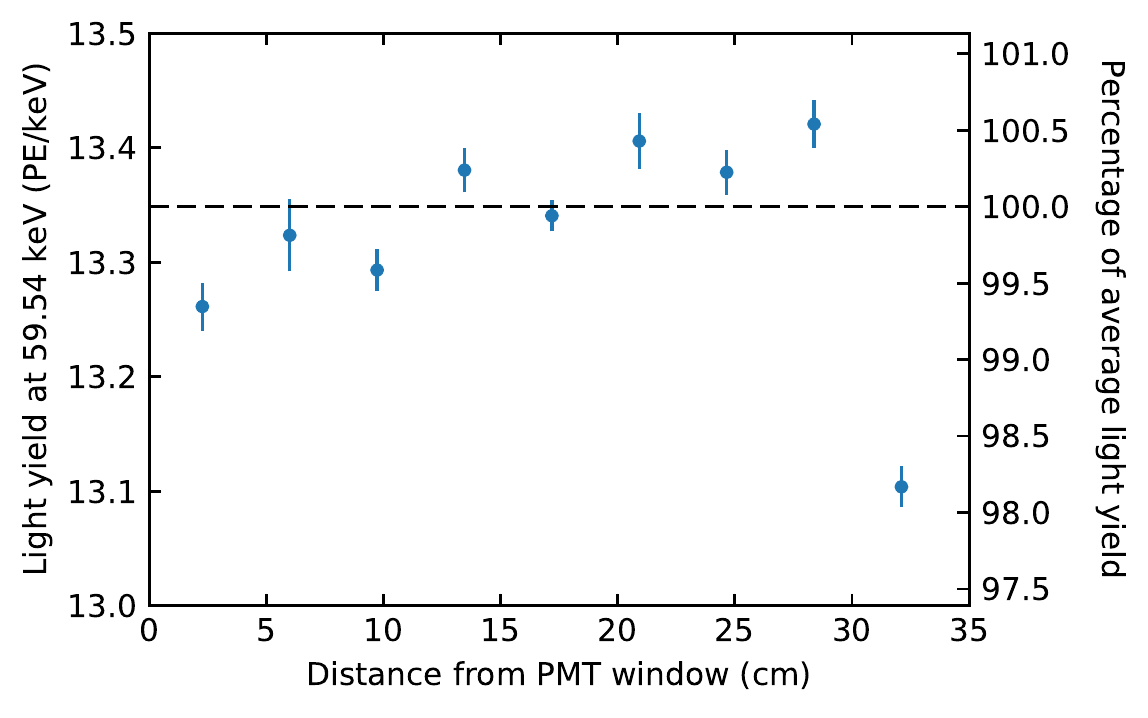}
\end{center}
\caption[Light yield and light collection uniformity of the \csi/ crystal]{Light yield for every source position. The eight positions closest to the \acs*{pmt} show a variation in light yield of $\mathcal{O}(\SI{0.5}{\percent})$. The largest deviation was found for the source position farthest from the \acs*{pmt} and closest to the backing reflector. A dotted line shows the weighted average of the closest eight positions at $\mathcal{L}_\text{CsI} = 13.348\,\nicefrac{\text{PE}}{\SI{}{\keVee}}$.}
\label{fig:am-calibration:light-yield-vs-position}
\end{figure}

Fig.~\ref{fig:am-calibration:light-yield-vs-position} shows the light yield calculated for every source position. Only a small variation in $\mathcal{L}$ is apparent for all source locations, which was expected from discussions with the manufacturer. The largest deviation was found at the position farthest away from the \ac{pmt} window and closest to the back reflector. The uncertainty weighted average of the closest eight positions is given by 
\begin{align}
\mathcal{L}_\text{CsI} = 13.348 \pm 0.019\,\frac{\text{PE}}{\SI{}{\keVee}}.
\end{align}
The overall variation with respect to $\mathcal{L}_\text{CsI}$ within the first eight locations is $\mathcal{O}(\SI{0.5}{\percent})$. The position farthest away shows a deviation of \SI{1.8}{\percent}. The light yield values and their respective deviation from $\mathcal{L}_\text{CsI}$ for each position are given in Table~\ref{tab:am-calibration:light-yield-vs-position}. This measurement confirmed an almost perfect light collection uniformity throughout the crystal. For the rest of this thesis a constant light yield given by $\mathcal{L}_\text{CsI}$ was assumed

\begin{table}[htbp]
	\begin{center}
		\begin{tabular}{cccc}
		\toprule
		Distance from \acs*{pmt} (\SI{}{\cm}) & Light yield (PE/\SI{}{\keVee}) & Variation from $\mathcal{L}_\text{CsI}$ (\SI{}{\percent})\\
		\midrule
			2.3  & $13.261 \pm 0.021$ & $-0.65$ \\
			6.0  & $13.324 \pm 0.032$ & $-0.19$ \\
			9.7  & $13.293 \pm 0.018$ & $-0.41$ \\
			13.5 & $13.380 \pm 0.019$ & $0.24$ \\
			17.2 & $13.341 \pm 0.014$ & $-0.059$ \\
			20.9 & $13.406 \pm 0.024$ & $0.43$ \\
			24.7 & $13.379 \pm 0.020$ & $0.23$ \\
			28.4 & $13.421 \pm 0.020$ & $0.54$ \\
			32.1 & $13.104 \pm 0.018$ & $-1.8$ \\
		\bottomrule
		\end{tabular}
	\end{center}
	\caption{Variation of the light yield at \SI{59.54}{\keV} along the crystal axis. $\mathcal{L}_\text{CsI}$ was found to be $13.348 \pm 0.019\,$(PE/\SI{}{\keVee}).}
	\label{tab:am-calibration:light-yield-vs-position}
\end{table}

%% file: ba-calibration.tex
%
%
\chapter{Barium calibration of the CE$\nu$NS detector}
\label{chapter:ba-calibration}
After establishing that the variation of light collection efficiency along the \csi/ crystal axis was negligible, an additional low energy calibration was performed. The goal of this measurement was to build a library of low energy events taking place within the \csi/ crystal that can be use to define and quantify the acceptance of several different cuts used later in the analysis of the \ac{cenns} search data. Given the minimal difference in the decay times for nuclear and electronic recoils in \csi/ at low energies (Fig.~\ref{fig:csi-setup:csi-decay-times}), this calibration was performed using electronic recoils induced by $\gamma$ interactions instead of relying on a neutron source to induce nuclear recoils. The \csi/ detector used for these calibrations was later used for the \ac{cenns} search (chapter~\ref{chapter:sns-analysis}).\par

The total data acquisition for this calibration took place over the course of three months, i.e., between March 27, 2015 and June 15, 2015. This measurement provided an ideal test environment for the long term stability of the detector setup. No issues were identified and as a result the detector was deployed at the \ac{sns} shortly after the calibration. The detector setup was located at the same sub-basement laboratory where the light yield calibrations described in chapter~\ref{chapter:am-calibration} were performed. In this location the \csi/ detector was shielded against activation from cosmic rays for a prolonged time by $\sim\SI{6}{\mwe}$, reducing the backgrounds present in the \ac{cenns} search.

\section{Detector setup}
\begin{figure}[!h]
	\begin{center}
		\includegraphics[scale=1]{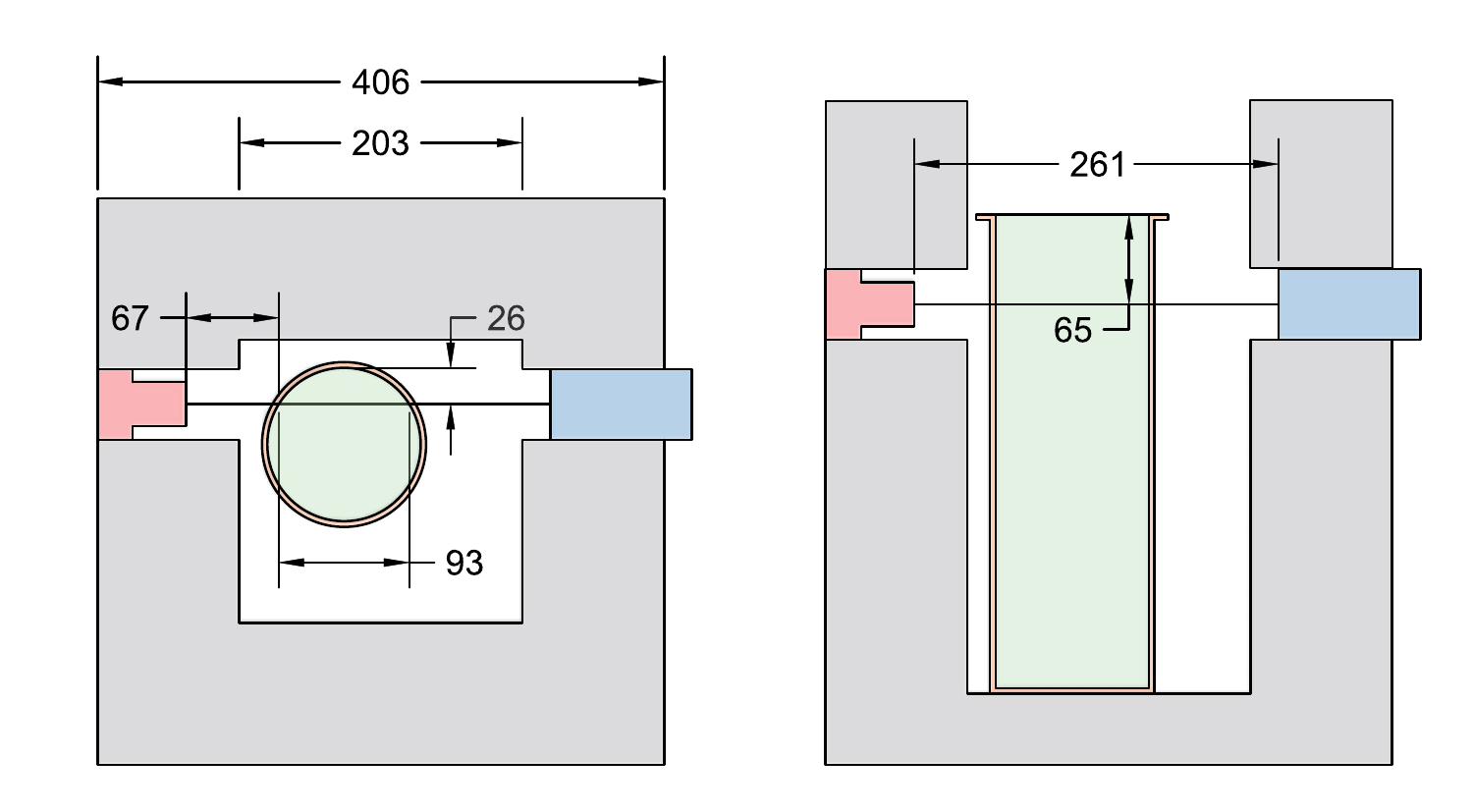}
	\end{center}
	\caption[Schematic of the detectors and shielding used in the \isotope{Ba}{133} calibration measurement]{Schematic of the detectors and shielding used in the \isotope{Ba}{133} calibration measurement. Measurements are in \SI{}{\mm}. A top-down view is shown on the left, and a side view is provided on the right. The \ac{pmt} and the lead shield on top were omitted for clarity. The red object represents the lead collimator used to constrain the isotropic emission of the \isotope{Ba}{133} button source to a pencil beam. The blue object represents the \brillance/ backing detector. The pencil beam only traverses the edge of the crystal at a given height. However, as established in chapter~\ref{chapter:am-calibration}, the variation in the light collection efficiency along the crystal is negligible. As a result, this calibration is representative of events happening anywhere in the \csi/ crystal.}
	\label{fig:ba-calibration:experimental-setup-sketch}
\end{figure}
\begin{figure}[t]
	\begin{minipage}{0.49\linewidth}
	\begin{center}
		\includegraphics[width=2.5in]{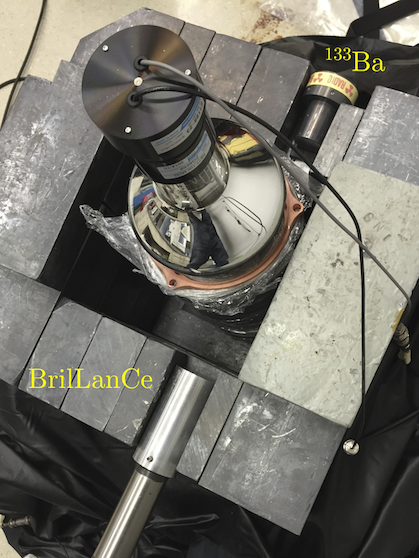}
	\end{center}
	\end{minipage}
	\begin{minipage}{0.49\linewidth}
	\begin{center}
		\includegraphics[width=2.5in]{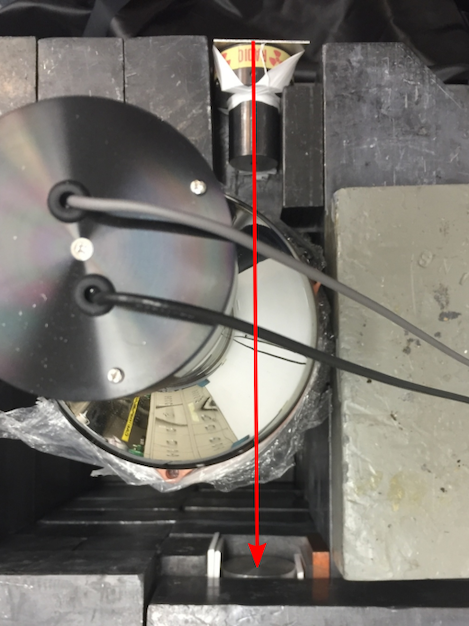}
	\end{center}
	\end{minipage}
	\caption[Pictures of the \isotope{Ba}{133} source alignment]{Pictures of the \isotope{Ba}{133} calibration measurement. The left picture highlights the \isotope{Ba}{133} source within its lead collimator on the top and the \brillance/ backing detector on the bottom. The right picture shows the alignment of source and backing detector by highlighting the center line from one to the other.}
	\label{fig:ba-calibration:experimental-setup}
\end{figure}

In order to create an event library containing interactions with only a few to a few tens of \ac{spe} within the crystal, a highly collimated pencil beam of \isotope{Ba}{133} $\gamma$-rays was sent through the \csi/. This pencil beam was produced by placing a \isotope{Ba}{133} button source in a lead collimator. The lead collimator provided a pinhole with a diameter of $\diameter=\SI{1.2}{\mm}$ and a total depth of $d=\SI{44.5}{\mm}$. This constrained the maximum aperture of the collimated \isotope{Ba}{133} $\gamma$-beam to $\theta_c\sim\SI{1.9}{\degree}$. The setup triggered on forward Compton-scattered gammas detected using a \brillance/ backing detector. A schematic of the experimental setup is shown in Fig.~\ref{fig:ba-calibration:experimental-setup-sketch}. Pictures of the setup are shown in Fig.~\ref{fig:ba-calibration:experimental-setup}. The maximum single forward scattering angle of a $\gamma$-ray being detected in the backing detector is given by $\theta_\text{max}\sim\SI{12}{\degree}$, leading to a maximum energy deposition of $E_\text{max}\sim\SI{6}{\keV}$.\par

\begin{figure}[t]
	\begin{center}
		\includegraphics[width=6in]{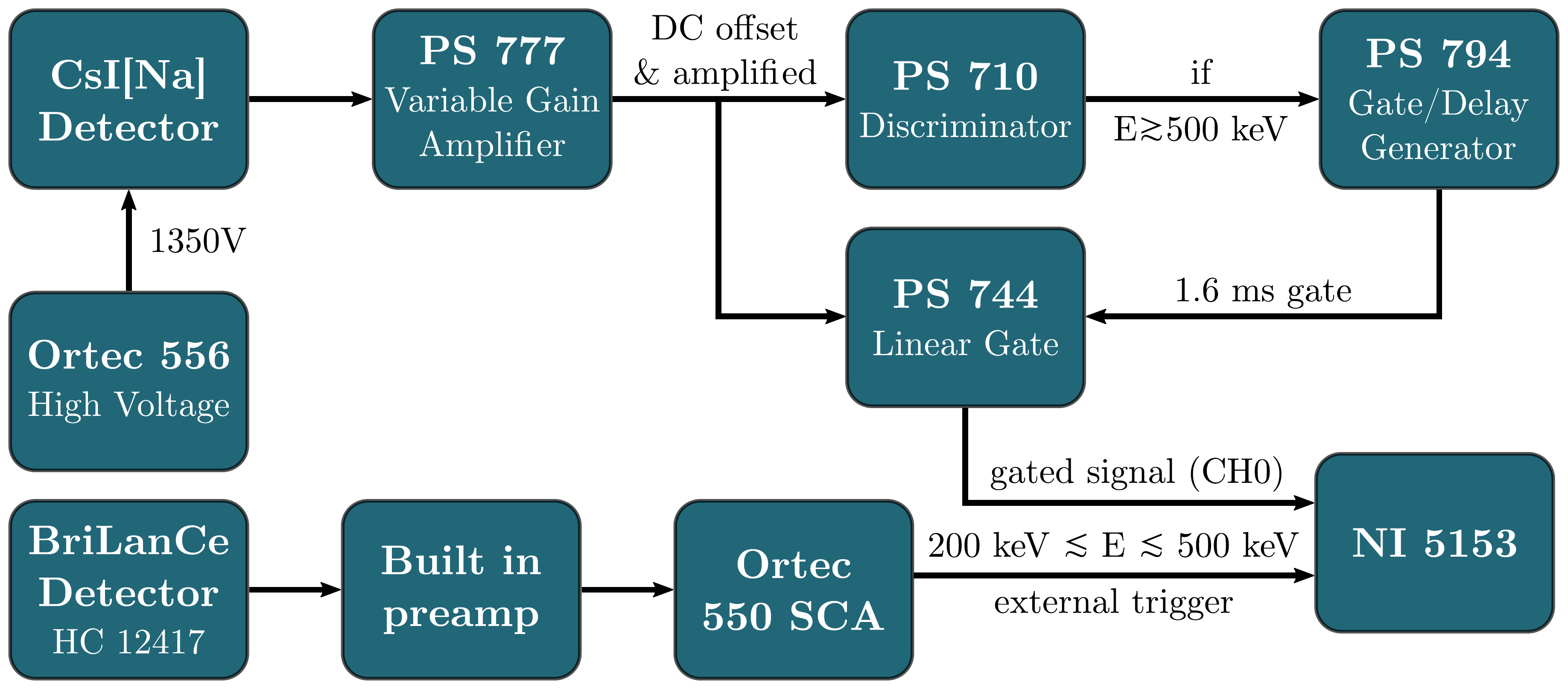}
	\end{center}
	\caption[Data acquisition system used in the \isotope{Ba}{133} calibration measurement]{Data acquisition system used in the \isotope{Ba}{133} calibration measurement.}
	\label{fig:ba-calibration:wiring-diagram}
\end{figure}

\begin{figure}[t]
	\begin{center}
		\includegraphics[width=6in]{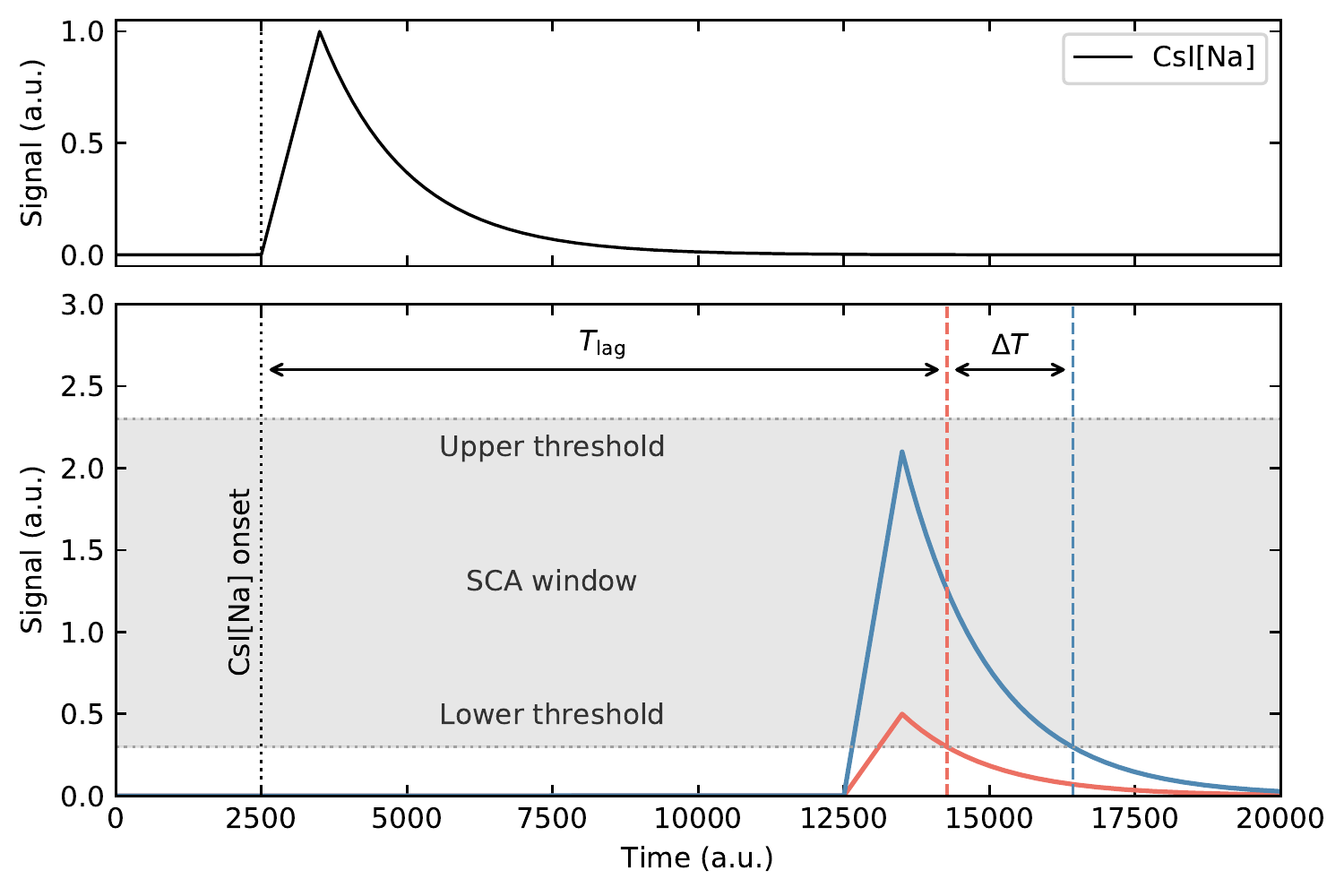}
	\end{center}
	\caption[Lag time between the \csi/ signal and trigger from the \brillance/ detector]{Lag time between the \csi/ signal and trigger from the \brillance/ backing detector. The \csi/ signal is shown on the top, whereas the corresponding \brillance/ signal is shown on the bottom. The red (blue) curve represents an event depositing an energy close to $\sim$\SI{200}{\keVee} ($\sim$\SI{500}{\keVee}) in the backing detector. The difference in the actual trigger position of both events, i.e when a signal actually leaves the \acs*{sca} window, is readily visible.}
	\label{fig:ba-calibration:lag-time}
\end{figure}

The schematic of the data acquisition system used in this measurement is shown in Fig.~\ref{fig:ba-calibration:wiring-diagram}. The linear-gate logic rejecting high-energy events in the \csi/ signal is identical to the one described in chapter \ref{chapter:csi-setup}. However, a different triggering signal was used to acquire data. The pre-amp output of the \brillance/ backing detector was fed into an Ortec 550 \ac{sca}. This modul provided a logical signal, acting as external trigger, whenever an event with an equivalent energy of 200-\SI{500}{\keVee} was detected within the \brillance/ backing detector. This ensured that a trigger was active for all small angle Compton scattered gammas of all major emission lines of \isotope{Ba}{133}. For each trigger waveforms with a total length of \num{35000} samples at a sampling rate of \SI{500}{\mega\sample\per\second}, i.e., \SI{70}{\micro\second}, were recorded. In contrast to the previous calibration, the trigger location was set to \SI{100}{\percent} of the trace, i.e., at \SI{70}{\micro\second} into the trace. This is necessary to account for the jitter introduced by using the \ac{sca}. The \ac{sca} produces a trigger whenever the analyzed signal exits the analysis window (Fig.~\ref{fig:ba-calibration:lag-time}). This behavior causes the timing of the trigger produced by the \ac{sca} to be directly dependent on the energy deposited in the backing detector. Consequently, events with higher energy show larger lag times between the \csi/ signal and the trigger from the backing detector (Fig.~\ref{fig:ba-calibration:lag-time}).\par

While adjusting the \ac{sca} analysis levels, the \csi/ output was closely monitored to ensure that the maximum delay, i.e., $T_\text{lag} + \Delta T$, between a trigger from the \ac{sca} and a scattering event in the \csi/ was less than \SI{20}{\micro\second}. Therefore, all Compton-scattered events occur no sooner than \SI{50}{\micro\second} into the waveform. However, not all triggers by the backing detector are due to small angle Compton-scattered events. Even though the \brillance/ detector was surrounded by some lead, some triggers were caused by environmental radiation that deposited the right energy within the backing detector. As the triggering particle did not interact with the \csi/ prior to hitting the backing detector, the corresponding waveforms of these events only consist of random coincidences between the \csi/ and the backing detector.

\section{Waveform analysis}
\label{section:ba-calibration:waveform-analysis}
As the low energy events of interest consist of only a few \ac{pe}, it is challenging to determine whether a singular event was acquired due to a true Compton-scattered trigger or due to environmental background radiation. However, true and random coincidences can be discriminated on a statistical basis by using all \num{21275438} waveforms acquired over the course of the three months.\par

\begin{figure}[htb]
	\begin{center}
		\includegraphics[width=6in]{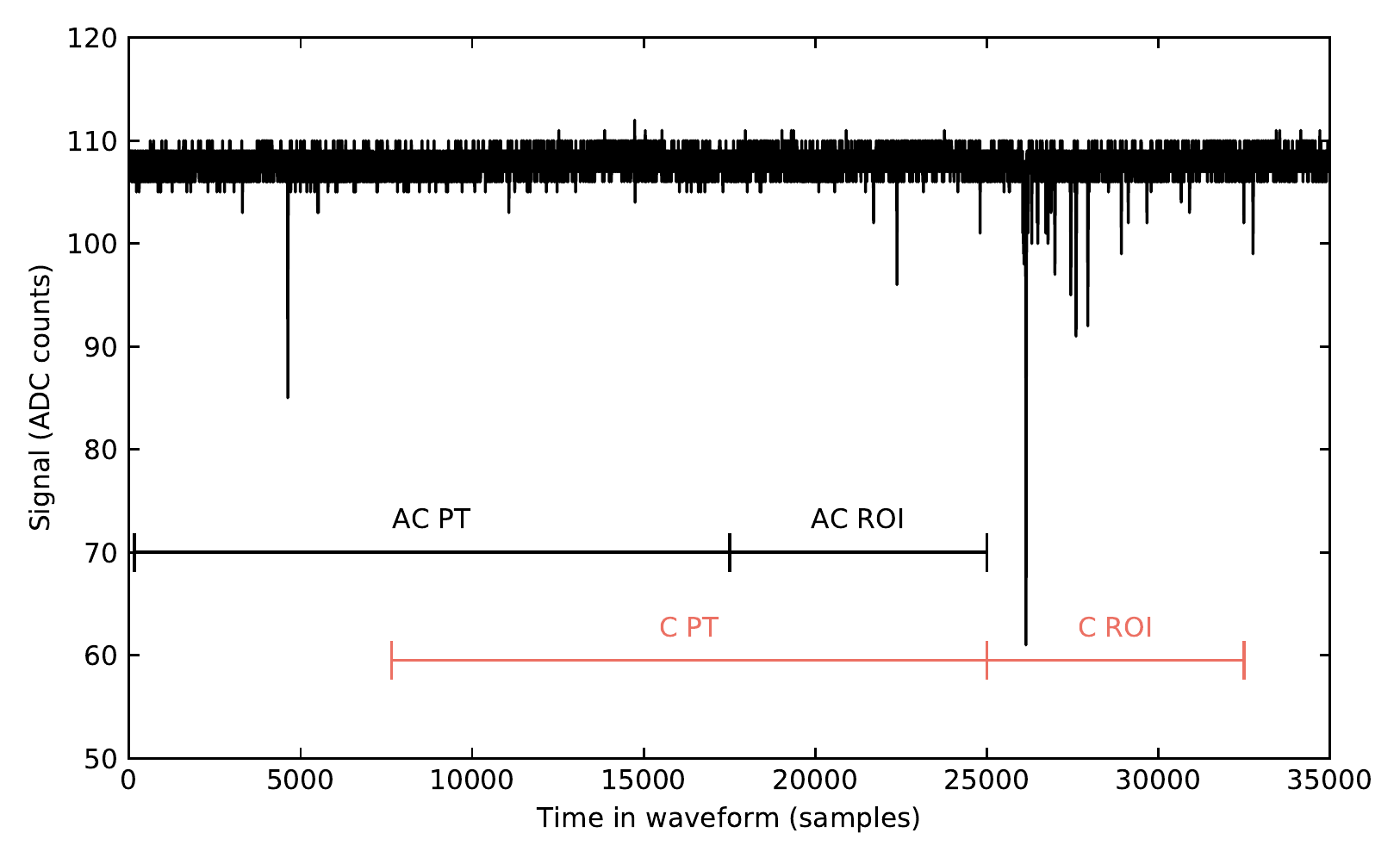}
	\end{center}
	\caption[Example waveform for the \isotope{Ba}{133} calibration highlighting different analysis regions]{Example waveform for the \isotope{Ba}{133} calibration highlighting different analysis regions. The trigger from the Ortec 550 \acs*{sca} was set to sample \num{35000}. No true coincidence event can occur prior to sample \num{25000}. The two main analysis regions are also shown. The anti-coincidence region is depicted in black and the coincidence region in red. Their respective sub-regions are also included, i.e., \acf{pt} and \acf{roi}.}
	\label{fig:ba-calibration:example-waveform}
\end{figure}

To achieve this statistical discrimination, two slightly overlapping regions are defined in each waveform (Fig.~\ref{fig:ba-calibration:example-waveform}). The first interval is the anti-coincidence (AC) region that extends from sample \num{150} to \num{25000}. As discussed earlier it is impossible for a true coincidence event to occur before sample \num{25000} (= \SI{50}{\micro\second} into the waveform). This immediately implies that this region can only contain random coincidences between events in the \csi/ and the backing detector. The second interval denotes the coincidence (C) region, extending from sample \num{7650} to \num{32500}. This region contains both, random coincidences from environmental triggers and true coincidence events from small angle Compton-scattering. Both regions are analyzed using an identical analysis pipeline. Through this process two independent n-tuples are created for each waveform, one for the C and one for the AC region. These n-tuples contain parameters characterizing each event, e.g., the average basline or the number of peaks present in an analysis region (Table~\ref{tab:ba-calibration:parameters-extracted-per-wf}). These two distinct sets of n-tuples are denoted as \cDS/ and \acDS/ data sets throughout the remainder of this chapter.\par

To extract a certain feature exhibited by low-energy events, the statistics of the feature can be extracted by subtracting its distribution derived from \acDS/ from the distribution derived from \cDS/. The residual \rDS/=\cDS/-\acDS/ only contains a statistical contribution from low-energy events only, as long as both regions were treated identically throughout the full analysis. The analysis pipeline for each waveform is detailed below.\par

\begin{table}[htbp]
	\begin{center}
		\begin{tabular}{ccccc}
		\toprule
		& \multicolumn{2}{c}{AC region} & \multicolumn{2}{c}{C region}\\
		 \cmidrule(lr){2-3}\cmidrule(lr){4-5}
		& Start & End & Start & End \\
		\midrule
		PT  &   \num{150} & \num{17500} &  \num{7650} & \num{25000} \\
		ROI & \num{17500} & \num{25000} & \num{25000} & \num{32500} \\
		\bottomrule
		\end{tabular}
	\end{center}
	\caption[Start and end samples of each analysis region used in the \isotope{Ba}{133} calibration]{Start and end sample of each analysis region. The starting sample is included in the respective region whereas the ending sample is excluded. Here PT denotes the \acl{pt} and ROI the \acl{roi}.}
	\label{tab:ba-calibration:pt-roi-ranges}
\end{table}

 For each waveform it is determined whether the signal is fully contained within the digitizer range. A digitizer overflow is determined as any sample that shows as +127 or \SI{-128}{\adc}, i.e., the upper and lower limits of an I8 variable. If such an overflow is detected, the corresponding flag $f_\text{o}\,=\,1$ is set. Next, the waveform is checked for the presence of a linear-gate. Such an event will appear as a baseline of \SI{0}{\adc}, as the \csi/ signal is blocked for \SI{1.6}{\milli\second} whenever a linear gate is initiated due to a high energy deposition within the \csi/ (chapter~\ref{chapter:csi-setup}). In contrast, the normal baseline for the \csi/ detector is set to $\sim\SI{100}{\adc}$. It is possible to to implement a computationally-inexpensive way to check for these linear gates by comparing the number of times a waveform crosses a threshold of \SI{18}{\adc} in a falling ($n_f$) and rising ($n_r$) manner. A linear-gate is present if $n_f\neq n_r$, which is indicated in the data by the corresponding linear-gate flag $f_\text{g}\,=\,1$.\par

To determine a global baseline $V_\text{median}$, the median of the first \num{20000} samples of the current waveform is used. The signal $V$ is adjusted and inverted, i.e.,
\begin{align}
\hat{V}_i = V_\text{median} - V_i\quad\text{for}\quad i\in[0,\num{35000}).
\end{align}
The location of any potential \ac{spe} can be identified using a peak-finding algorithm as described in the light yield calibration (section~\ref{section:am-calibration:spe-charge}). As stated, a peak was detected when at least four consecutive samples had an amplitude of at least \SI{3}{\adc}. Both positive and negative threshold crossings were recorded for each peak. Similar to the procedure in section~\ref{section:am-calibration:spe-charge}, the charge of each peak is calculated using Eq.~(\ref{eq:am-calibration:spe-charge-calculation}).\par

Instead of calculating a single, overall \ac{spe} charge distribution for the full data set, the data is subdivided into \SI{5}{\minute} intervals to monitor the stability of the mean \ac{spe} charge \qspe/. For each interval an independent charge distribution is created. The distributions are fitted using Eq.~(\ref{eq:am-calibration:spe-polya}) to extract the mean \ac{spe} charge \qspe/ for the corresponding data period. In the further data analysis the appropriate \qspe/ is used to establish a proper energy scale.\par

\begin{figure}[htbp]
	\begin{center}
		\includegraphics[scale=1]{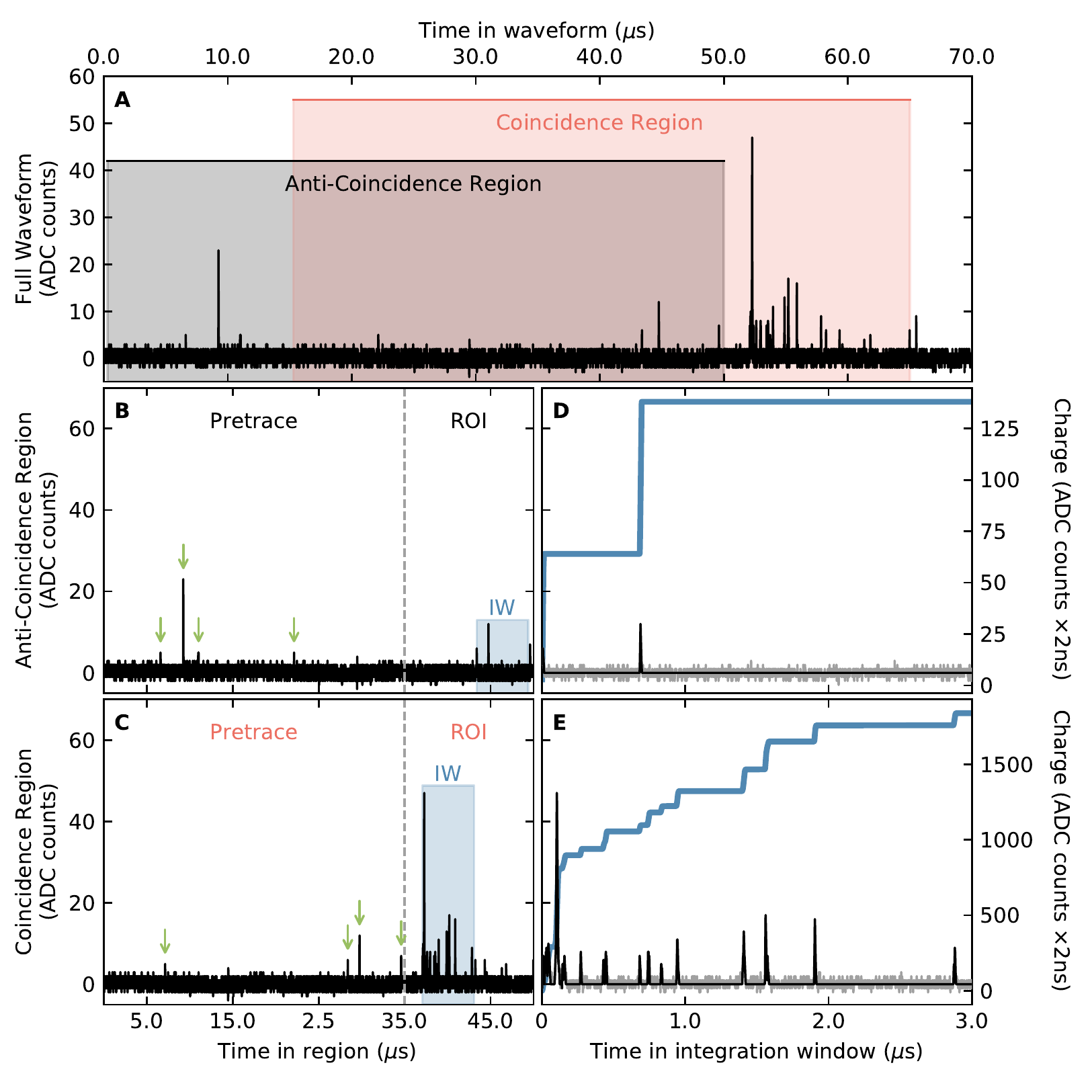}
	\end{center}
	\caption[Analysis of an example waveform for the \isotope{Ba}{133} calibration]{Analysis of an example waveform for the \isotope{Ba}{133} calibration. The full waveform is shown in panel \textbf{A} with the full AC and C regions highlighted. Panel \textbf{B} and \textbf{C} provide a zoom into AC and C, respectively, with the \ac{pt} and \ac{roi} sub-regions annotated. Green arrows mark all peaks in the respective \ac{pt} regions as identified by the threshold finding algorithm. The shaded blue region highlights the \SI{3}{\micro\second} long integration window, starting with the first peak in the \ac{roi}. Panel \textbf{D} and \textbf{E} provide a further zoom into these integration windows. The gray curve shows the baseline-adjusted signal. The black one in contrast represents the signal after zero-suppressing all samples that were not identified as part of a peak. The blue curve represents the charge integration $Q(t)$ of the zero-suppressed signal that is later used to determine the total charge as well as the individual rise-times.}
	\label{fig:ba-calibration:example-waveform-analysis}
\end{figure}

In what follows the procedure applied to the analysis regions, i.e., C and AC, is described. The analysis is identical for both regions and illustrated in Fig.~\ref{fig:ba-calibration:example-waveform-analysis}. A full example waveform is shown in panel \textbf{A}. The C (AC) region is shown in shaded red (gray). A zoom into both regions is shown in panel \textbf{B} (AC) and \textbf{C} (C), respectively. The non-overlapping \ac{pt} and \ac{roi} sub-regions are highlighted. The \acf{pt} is fairly long, spanning \num{17350} samples. Its position was chosen such that it could not possibly contain a signal from a coincident Compton-scatter event for neither analysis region. Its main purpose is to provide a veto against contamination from afterglow. As discussed in chapter~\ref{chapter:csi-setup}, and shown in Fig.~\ref{fig:csi-setup:afterglow}, \csi/ can phosphorescence for up to $\mathcal{O}(\SI{10}{\milli\second})$ after an actual event. A large energy deposition, e.g., from a muon traversing the crystal, could potentially add several \acp{spe} from phosphorescence in a subsequent trigger and introduce a non-negligible bias in the analysis. To remove these potentially contaminated events from the data set, events were rejected based on the total number of peaks detected in the corresponding \ac{pt}. The higher the number of peaks, the likelier it is that these were caused by phosphorescence.\par
The \acf{roi} is much shorter, spanning \num{7500} samples. The AC \ac{roi} is aligned such that it can not physically contain a coincident Compton scatter event, whereas the C \ac{roi} could contain these events.\par

Peaks within both regions, i.e., \ac{pt} and \ac{roi}, are identified using the results from the initial peak search. The total number of peaks $N_\text{pt}$ in the \ac{pt} is recorded. For the example waveform shown in Fig.~\ref{fig:ba-calibration:example-waveform-analysis} it is $N^C_\text{pt} = N^{AC}_\text{pt} = 4$ as indicated by the green arrows. The location of the first peak within the \ac{roi} defines the onset of the \num{1500} sample (=\SI{3}{\micro\second}) long integration window. These windows are highlighted in shaded blue in panels \textbf{B} and \textbf{C}. A new baseline for each integration window is determined using the median of the \SI{1}{\micro\second} immediately preceding the window. A zoom into both integration windows after adjustment with the new baseline is shown in gray in panels \textbf{D} and \textbf{E}. Any sample not belonging to a peak is zero-suppressed. The resulting signal $V^\star$ is shown in black. It can be observed that all peaks are preserved whereas the baseline noise is completely suppressed. The integrated scintillation curve $Q(t)$ (shown in blue) for both integration windows is
\begin{align}
Q(t) = \sum\limits_{i = T_\text{arr}}^{T_\text{arr}\,+\,t} V_i^\star\label{eq:ba-calibration:integrated-scintillation-curve}
\end{align}
Besides the total integrated charge $Q_\text{total} = Q(\SI{1499}{\sample})$ for each event, the rise-times $T_{10}$, $T_{50}$ and $T_{90}$ are also recorded, as described in chapter~\ref{chapter:am-calibration} and shown in Fig.~\ref{fig:am-calibration:spe-and-integration}.\par
The parameters extracted for each event are given in Table ~\ref{tab:ba-calibration:parameters-extracted-per-wf}, where a separate and independent n-tuple is created for both the AC and C region containing their respective parameters where applicable. The procedure described above is repeated for each waveform and the individual n-tuples are gathered in the \acDS/ and \cDS/ data sets.\par 

\begin{table}[tb]
	\begin{center}
		\begin{tabular}{cl}
		\toprule
		Parameter & Description \\
		\midrule
		$t$   & Timestamp of current event in seconds since epoch \\
		$f_\text{o}$ & The waveform contains samples exceeding the digitizer range \\
		$f_\text{g}$ & The waveform contains a linear gate event \\
		$V_\text{median}$ & Median baseline of the first \num{20000} samples \\
		$V_\text{avg}$ & Average baseline of the first \num{20000} samples (excluding peak regions) \\
		$\sigma_v$ & Standard deviation of the first \num{20000} samples (excluding peak regions) \\
		$N_\text{pt}$ & Peaks found in the \ac{pt} \\
		$N_\text{roi}$ & Peaks found in the \ac{roi} \\
		$N_\text{iw}$ & Peaks found in the integration window \\
		$T_\text{arr}$ & Arrival time, i.e., time of first peak in \ac{roi} with respect to \ac{roi} onset \\
		$Q_\text{total}$ & Total integrated charge in integration window \\
		$T_{10}$, $T_{50}$,$T_{90}$ & Rise-times of integrated scintillation curve \\
		\bottomrule
		\end{tabular}
	\end{center}
	\caption[Overview of the parameters recorded for each waveform in the \isotope{Am}{241} calibration, the \isotope{Ba}{133} calibration and the \acs*{cenns} search]{Overview of the parameters recorded for each waveform in the \isotope{Am}{241} calibration, the \isotope{Ba}{133} calibration and the \ac{cenns} search. Two independent data sets are created, one for the C and one for the AC region.}
	\label{tab:ba-calibration:parameters-extracted-per-wf}
\end{table}

\begin{center}
\begin{figure}[htbp]
	\begin{center}
		\includegraphics[scale=1]{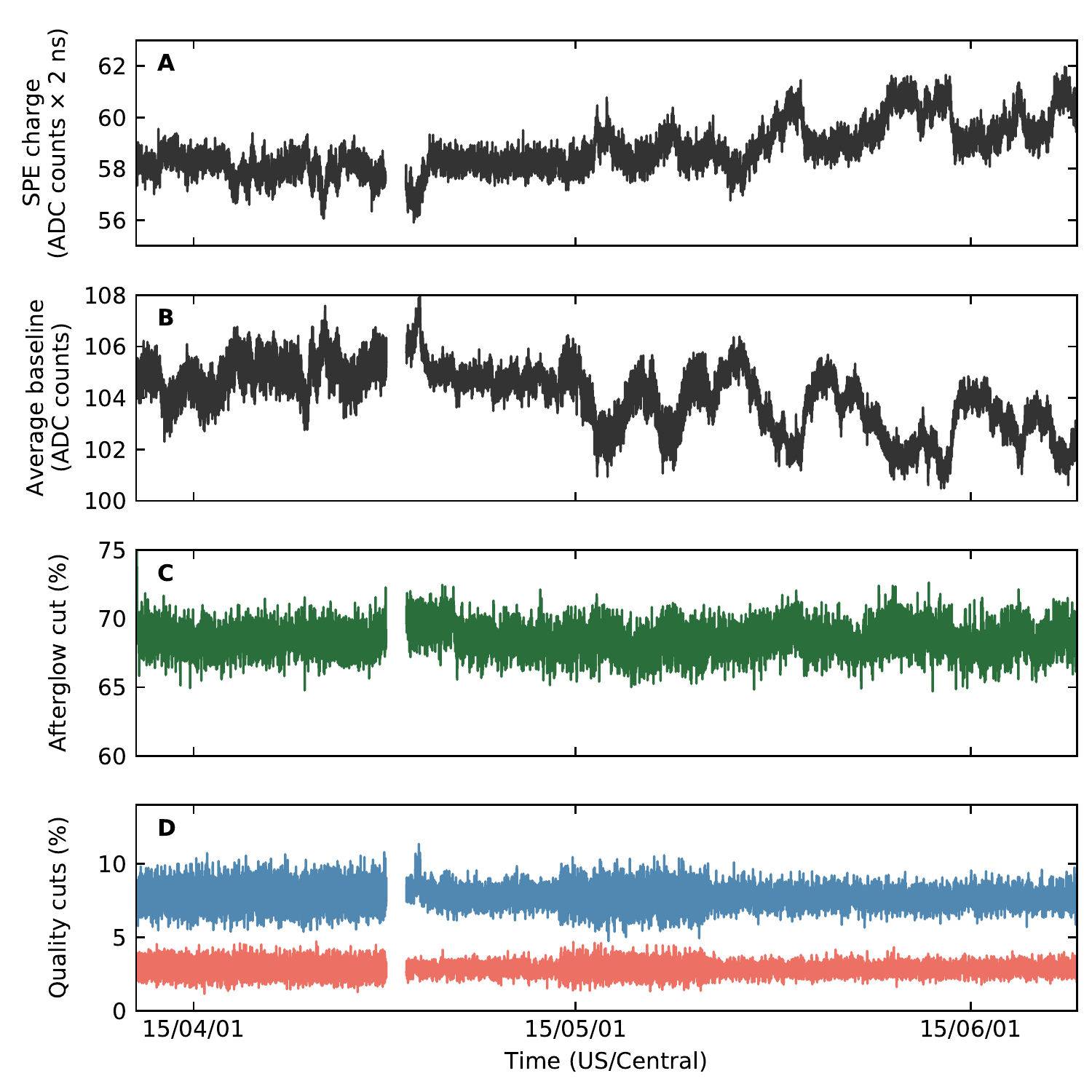}
	\end{center}
	\caption[Stability tests performed for the \isotope{Ba}{133} calibration]{Stability tests performed for the \isotope{Ba}{133} calibration. Every data point represents the average of a five minute interval. Panel \textbf{A} shows the evolution of the mean \ac{spe} charge. Panel \textbf{B} shows the average baseline of the first \SI{40}{\micro\second} of each waveform. An anti-correlation between both parameters is evident. A change of \SI{1}{\adc} in the average baseline corresponds to $\sim$\SI{1}{\adcq}, which could point towards a bias in the \ac{spe} charge calculation. However, the integration window of an \ac{spe} typically spans $\sim$17 samples. A significant bias in the baseline calculation of the \ac{spe} charge integration window would therefore result in a change in the \ac{spe} charge on the order of $\mathcal{O}(17\si{\adcq})$. A change of $\sim$\SI{1}{\adcq} is therefore deemed acceptable. In addition no increase or decrease in the average number of \ac{pe} detected per five minute interval was observed.}
\label{fig:ba-calibration:stability-tests}
\end{figure}
\end{center}

\begin{center}
\begin{figure}[t]
\ContinuedFloat
\caption[]{(Continued) Panel \textbf{C} shows the percentage of waveforms with more than ten peaks in the anti-coincidence pretrace. The high percentage of waveforms with such a large number of peaks in the pretrace can be attributed to the high level of phosphorescence in the crystal due to the presence of the \isotope{Ba}{133} source. No significant variation over time was seen in the percentage of waveforms rejected. No correlation with the mean \ac{spe} charge (panel \textbf{A}) or the average baseline (panel \textbf{B}) is apparent. The percentage of waveforms showing more than ten peaks in the coincidence pretrace shows the same level of phosphoresence and is therefore not depicted. The last panel \textbf{D} shows the percentage of waveforms showing either a linear gate (red) or digitizer overflow (blue), which are again stable throughout the whole data taking period.}
\end{figure}
\end{center}

\section{Detector stability performance over the data-taking period}
After all recorded waveforms were analyzed, the detector stability over the course of the data-taking period was tested. There are five main parameters of interest, which are all shown in Fig.~\ref{fig:ba-calibration:stability-tests}. Each data point represents the average over a five minute long time interval. The evolution of \qspe/ over time is shown in panel \textbf{A}. Panel \textbf{B} shows the average baseline derived from the first \num{20000} samples for the same five minute intervals. A slight variation in the mean \ac{spe} charge over the course of the measurement that appears to be anti-correlated to the fluctuations in the average baseline can be seen. Yet no significant increase or decrease in the average number of \ac{pe} detected per five minute interval was observed. Therefore the performance of the \ac{spe} finding algorithm was deemed sufficient. Using the appropriate mean \ac{spe} charge for every time period eliminates any potential bias introduced by these fluctuations.\par

\begin{figure}[htb]
	\begin{center}
		\includegraphics[scale=1]{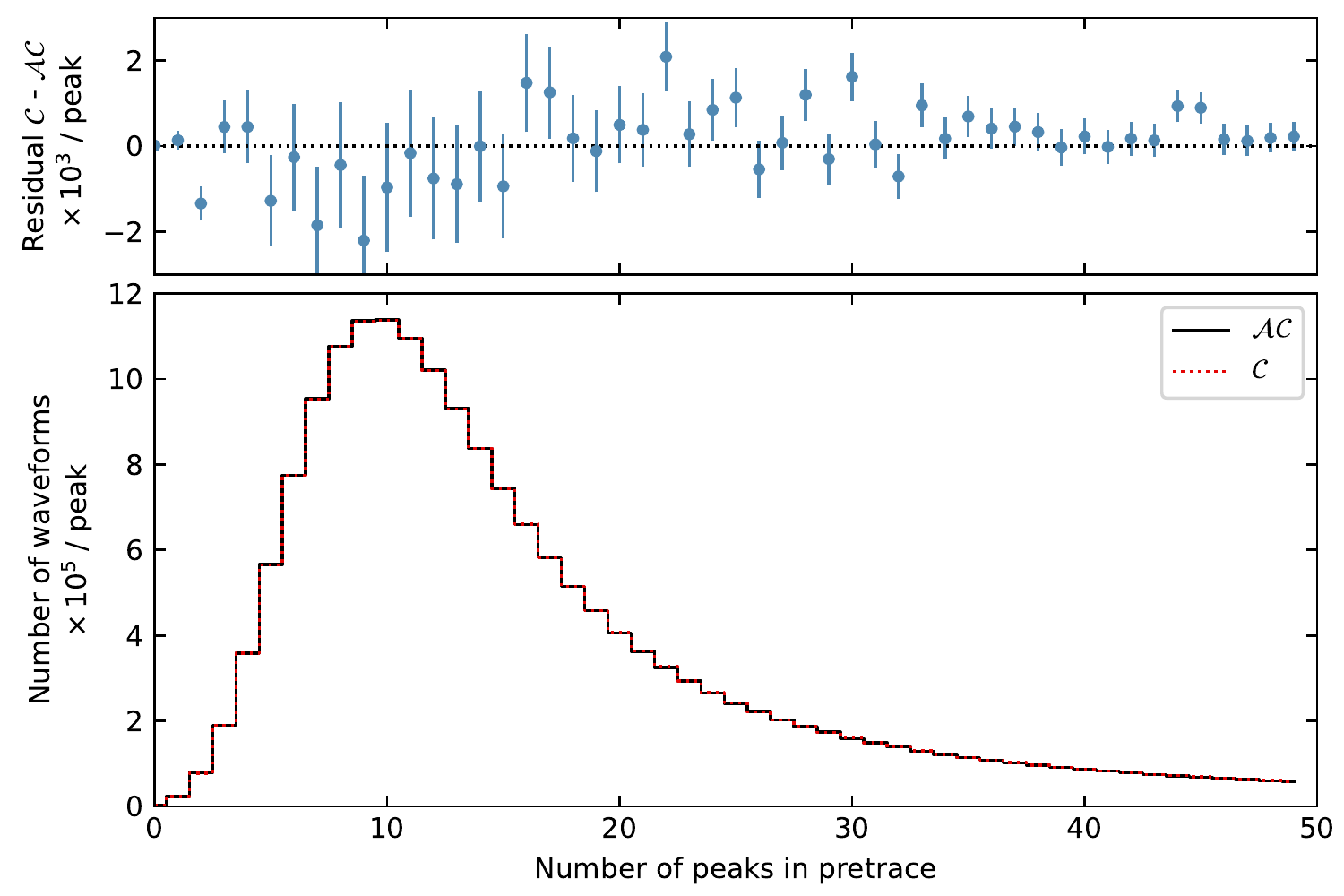}
	\end{center}
	\caption[Number of peaks in the pretrace for both AC and C regions in the \isotope{Ba}{133} calibration]{Number of peaks in the pretrace for the \isotope{Ba}{133} calibration. The bottom panel shows the raw distributions for both \acDS/ (black) and \cDS/ (red), whereas the top panel shows the residual \rDS/ of both data sets. A slight deviation from zero is visible for low $N_\text{pt}$. However, the average difference between both distributions is only \SI{0.24}{\percent}, which is deemed negligible.}
	\label{fig:ba-calibration:npt-distribution}
\end{figure}

Panel \textbf{C} shows the number of waveforms with more than ten peaks in the AC \ac{pt}. Almost \SI{70}{\percent} of all waveforms recorded showed a significant amount of scintillation in the \ac{pt}, which is caused by afterglow due to the presence of the \isotope{Ba}{133} source. However, the percentage of waveforms showing such a high level of afterglow within their waveforms remained constant throughout the data taking period. There was also no correlation visible between the performance of the afterglow cut and the average baseline or the mean \ac{spe} charge. The distribution of peaks in the C and AC \ac{pt} marginalized over the full data set is shown in Fig.~\ref{fig:ba-calibration:npt-distribution}. No significant difference between \acDS/ and \cDS/ was found. As a result, the afterglow cut does not introduce any bias in the event selection and the afterglow cut for \cDS/ is not shown as it would not add any further information.\par

Panel \textbf{D} of Fig.~\ref{fig:ba-calibration:stability-tests} shows the percentages of events with either a linear gate event (red) or a digitizer range overflow (blue). Both showed no significant variation over the course of the data taking and are independent of any variation in other parameters. It is therefore assumed that the data quality cuts do not introduce any bias in the event selection.\par

In summary, the detector performed satisfactory over the full three months of data-taking. No time periods from the data analysis need to be excluded and the full data set containing $\sim21$ million triggers could be used to quantify the cut acceptances.

\section{Definition and quantification of data cuts}
After establishing that the \csi/ detector performed as expected throughout the full data-taking period of this \isotope{Ba}{133} calibration, the statistical data analysis was performed. As discussed above, separate n-tuples were created for each trigger and each analysis region. All n-tuples belonging to the AC region are denoted as \acDS/ data set and all n-tuples belonging to the C region are denoted as \cDS/ data set. For both data sets an additional entry was added to each n-tuple by converting the total charge integrated in the integration window into an equivalent number of \ac{pe} $N_\text{pe}$. The conversion used the corresponding mean \ac{spe} charge $Q_\text{spe}$ determined for the five minute interval that contains the event, i.e.,
\begin{align}
N_\text{pe}=\frac{Q_\text{total}}{Q_\text{spe}}
\end{align}

First a series of data quality cuts were applied to both \acDS/ and \cDS/. All events containing a linear gate $f_g$ and/or overflow flag $f_o$ were removed from analysis. As these flags are raised for the full waveform they are shared between both sets and the events removed were identical for \acDS/ and \cDS/. Second, all events that showed more than \agc/ peaks in their respective \ac{pt} were removed from each data set. The purpose of this data cut is to minimize any bias introduced by excessive afterglow from previous energy depositions. Given the slight difference in the onset of the AC and C \ac{pt}, an individual event can be deemed good in one data set, whereas it was cut in the other. However, the overall number of events passing this cut is neither biased towards \acDS/ nor \cDS/ (Fig.~\ref{fig:ba-calibration:npt-distribution}). The total number of events cut is also dependent on the exact choice of \agc/ and is further examined later on. This concludes the discussion of quality cuts for the barium calibration. In the following the acceptance calculations for the data cuts used in the \ac{cenns} search are presented.

\subsection{The Cherenkov cut}
\label{section:ba-calibration:cherenkov-cut}
The first data cut that was applied is the so-called \emph{Cherenkov cut}. As discussed in earlier, a single Cherenkov pulse usually carries a charge equivalent to two to fifteen \ac{spe} (chapter~\ref{chapter:am-calibration}), and as such would pose an unwanted background to the \ac{cenns} search. However, such an event typically only produces a single, sharp spike in the digitizer trace, whereas a \ac{cenns} event appears as multiple individual \ac{spe} peaks following a scintillation decay time as shown in Fig.~\ref{fig:csi-setup:csi-decay-times}. The cut is therefore set to allow only events where the total number of peaks found in the integration window is at least \chc/. This cut was applied to events with a total energy of $N_\text{pe} \leq 40$ to avoid high energy events from being rejected by the Cherenkov cut. With increasing $N_\text{pe}$, i.e., increasing energy, individual \ac{spe} will merge into larger peaks. As the peak-finding algorithm only tags peaks by their threshold crossings this would lead to a decreasing number of peaks found. Applying the Cherenkov cut only to data with $N_\text{pe} \leq 40$ guarantees that no high energy event was mistakenly rejected by this cut.

\begin{figure}[tb]
	\begin{center}
		\includegraphics[scale=1]{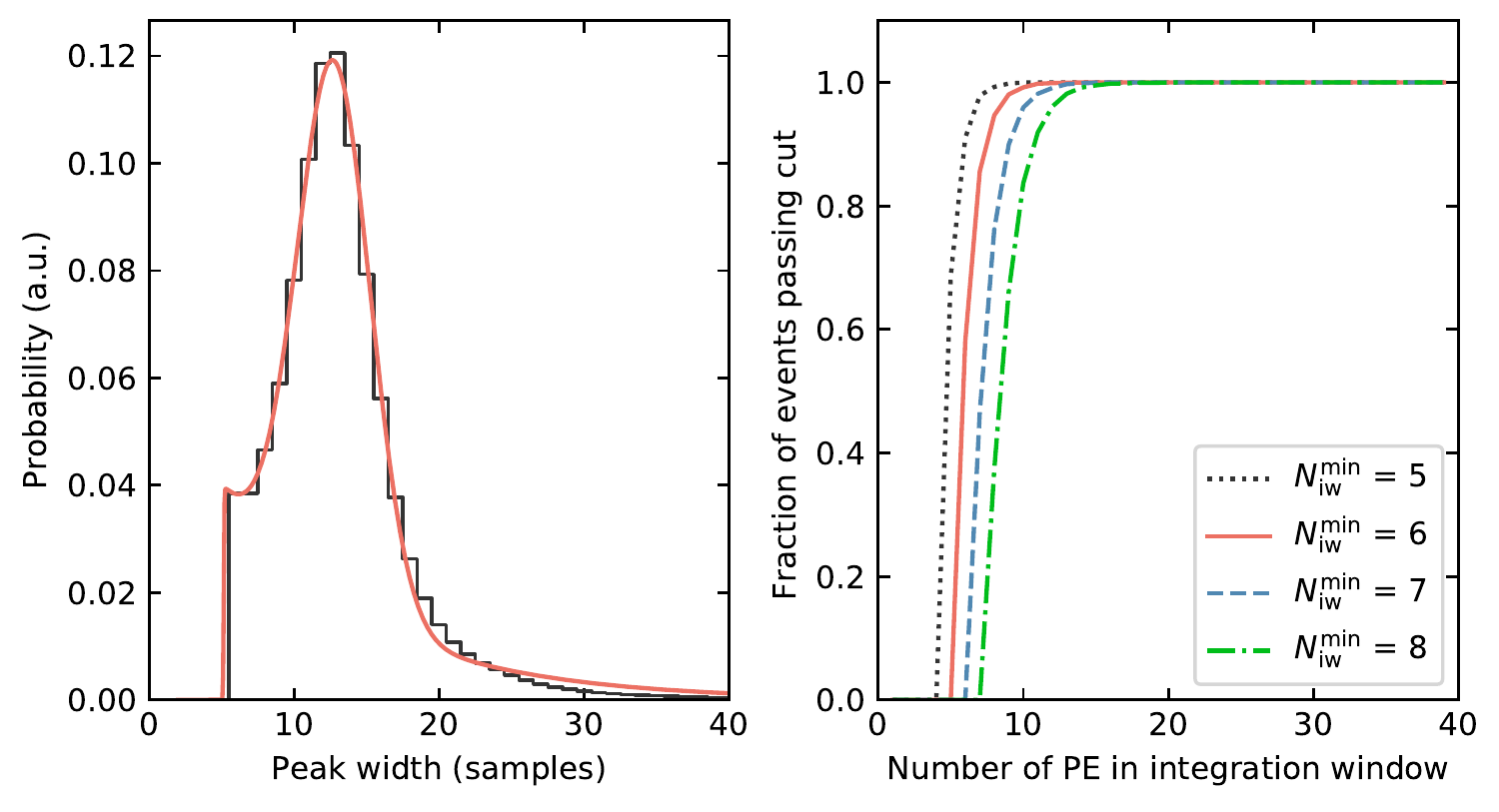}
	\end{center}
	\caption[SPE peak-width distribution and Monte Carlo simulation of Cherenkov cut survival rate for radiation induced events in the \isotope{Ba}{133} calibration]{\textbf{Left}: SPE peak-width distribution as measured during the barium calibration. \textbf{Right}: \acs*{mc} simulation of the Cherenkov cut survival rate for radiation induced events. A fixed peak width of \SI{13} samples and a fixed \ac{spe} charge for each peak are assumed. Including a proper distribution for both parameters leads to a broader rise of the survival fraction, but does not diminish the \SI{100}{\percent} survival rate for larger $N_\text{pe}$ }
	\label{fig:ba-calibration:mc-cherenkov-cut}
\end{figure}

Fig.~\ref{fig:ba-calibration:npt-distribution} shows that the average number of peaks found in each \ac{pt}, i.e. AC and C region of triggers in the \isotope{Ba}{133} calibration, is on the order of ten. Assuming that this distribution is also reflective of the rest of the waveform, the estimated average number of random peaks in the \SI{3}{\micro\second} integration window is approximately given by 0.75. The Cherenkov cut should therefore provide an excellent rejection of events containing Cherenkov spikes as the probability for additional, spurious \ac{spe} within the integration window is low. In order to ensure that the cut is not overly aggressive, which would also remove possible \ac{cenns} events, the performance of the cut was tested with an additional \ac{mc} study. First the average width of each peak was calculated using information gathered while determining the mean \ac{spe} charge \qspe/. The left panel of Fig.~\ref{fig:ba-calibration:mc-cherenkov-cut} shows the width distribution of each peak found in each waveform in black. The red fit function is given by
\begin{align}
f\left(q,a_n,\sigma_n,a,\mu,\sigma,k,q_0\right)&=\;\left[a_n \text{e}^{-\nicefrac{q}{\sigma_n}} + a\,g(q,\mu,\sigma)\right]\left(1.0 + \text{e}^{-k\left(q - q_0\right)}\right)^{-1}\label{eq:ba-calibration:spe-height}\\[1ex]
\text{where}\quad g(q,\mu,\sigma) &= \Exp{\frac{(q-\mu)^2}{2\sigma^2}}.\nonumber
\end{align}
which consists of an exponential representing the noise and a single Gaussian representing the width of an individual \ac{spe}. The last term reflects the efficiency in detecting a peak with the peak finding algorithm. The average \ac{spe} width is $\Delta t_\text{spe} = \SI{12.7}{\sample}$. A \SI{3}{\micro\second}=\num{1500} sample long array is populated with $N_\text{pe}$ \ac{pe}, where a decay profile following Fig.~\ref{fig:csi-setup:csi-decay-times} is assumed. Each of these \ac{spe} is set to have a fixed width of \SI{13}{\sample}. The total number of separate peaks is recorded and it was checked if current waveform still passed a cut of \chc/. The procedure was repeated \num{10000} times for varying number of photoelectrons $N_\text{pe}$ and different choices of \chc/ and the percentage of simulated waveforms passing the cut was scored. The results are presented in the right panel of Fig.~\ref{fig:ba-calibration:mc-cherenkov-cut}. The survival rate of the Cherenkov cut quickly rises towards unity and stays at that level for the full cut window, i.e $N_\text{pe} \leq 40$. While the Cherenkov cut could be extended to apply to events with higher energy, given the charges to be expected from a Cherenkov spike, the $0\leq N_\text{pe}\leq 40$ cut range is deemed sufficient.

\subsection{Rise-time cuts}
\label{section:ba-calibration:rt-cuts}
The second type of data cuts are based on several possible rise-times of the integrated charge of an event. Conceptually these cuts use the known decay profile of a low energy radiation induced event to separate bona fide events from spurious collections of \ac{spe} by using rise-time characteristics.  Similar \ac{psd} approaches are discussed in~\cite{luo-01,ronchi-01}. Given a scintillation decay profile as shown in Fig.~\ref{fig:csi-setup:csi-decay-times}, the cumulative distribution function of the charge profile can be written as
\begin{align}
F_\text{Q}(t) & = \frac{1}{1+r} F(t,\tau_\text{fast}) + \frac{r}{1+r} F(t,\tau_\text{slow})\label{eq:ba-calibration:emission-curve}\\[1ex]
\text{with}\quad F(t,\tau) & = \frac{1 - \text{e}^{-\nicefrac{-t}{\tau}}}{1 - \text{e}^{-\nicefrac{-t_\text{max}}{\tau}}}.\nonumber
\end{align}
Here $\tau_\text{fast}=\SI{527}{\ns}$, $\tau_\text{slow}=\SI{5.6}{\micro\second}$ and $r=0.41$ as shown in \cite{collar-02}. The integration window used in this \isotope{Ba}{133} calibration limits $t_\text{max}=\SI{3}{\micro\second}$. As already introduced in chapter~\ref{chapter:am-calibration}, two distinct rise-times are of interest, i.e., $T_{0-50}$ and $T_{10-90}$. The theoretical threshold crossing times, i.e., $T_{10}$, $T_{50}$, and $T_{90}$, can be calculated by evaluating $F_\text{Q}(T_{10})=0.1$, $F_\text{Q}(T_{50})=0.5$ and $F_\text{Q}(T_{90})=0.9$. The resulting rise-times based on Eq.~\ref{eq:ba-calibration:emission-curve} are thus given by
\begin{equation}
\begin{aligned}
T_{0-50} &= T_{50} = \SI{506.4}{\ns}\label{eq:ba-calibration:electronic-rise-times}\\
T_{10-90} &= T_{90} - T_{10} = \SI{1876.3}{\ns},
\end{aligned}
\end{equation}
whereas for a flat, i.e completely random, distribution the expectation is $T_{0-50}=\SI{1.5}{\micro\second}$ and $T_{10-90}=\SI{2.4}{\micro\second}$. Therefore, the rise-times provide an excellent method to reject events consisting of spurious, random \ac{spe}.\par

\begin{figure}[htb]
	\begin{center}
		\includegraphics[scale=1]{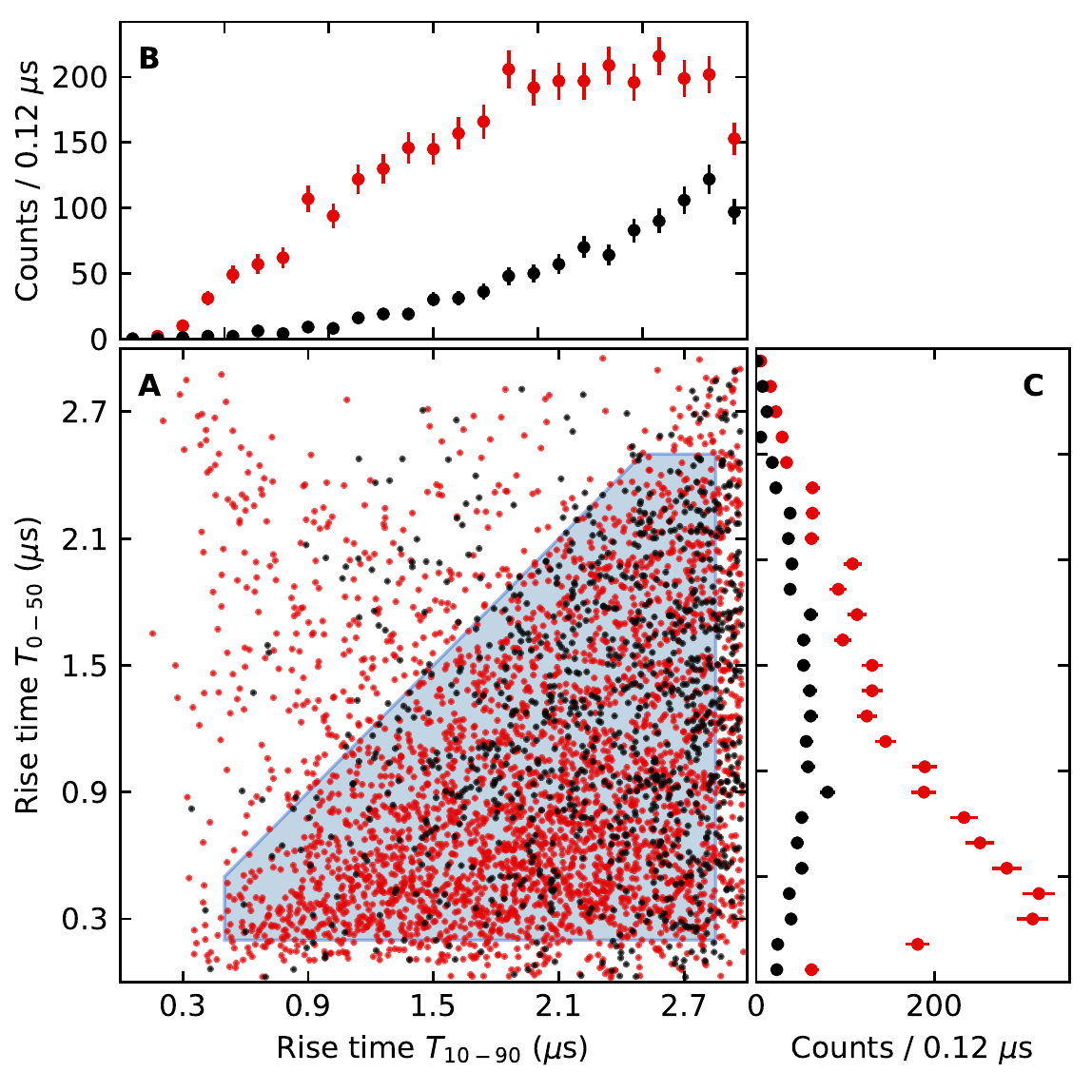}
	\end{center}
	\caption[Example rise-time distribution for the \isotope{Ba}{133} calibration with a Cherenkov cut of 5 for events with an energy of $5\,\leq\,N_\text{pe}\,\leq\,20$]{\textbf{A}: Two dimensional rise-time distributions for events in the \isotope{Ba}{133} calibration data set that are pass all quality cuts with an additional Cherenkov cut of $\chc/=5$. Only events with an energy of $5\,\leq\,N_\text{pe}\,\leq\,20$ are shown. Red (black) data points represent the \cDS/ (\acDS/) data set. The shaded blue region represents one of the proposed rise-time cuts. \textbf{B}: $T_{10-90}$ distribution marginalized over $T_{0-50}$. \textbf{C}: $T_{0-50}$ marginalized over $T_{10-90}$. An excess of coincidence events over anti-coincidences is readily visible in all panels.}
	\label{fig:ba-calibration:rt-distribution-01}
\end{figure}

An example of the different rise-time distributions as measured for this \isotope{Ba}{133} calibration data set is shown in Fig.~\ref{fig:ba-calibration:rt-distribution-01}. Panel \textbf{A} shows the two-dimensional distribution of rise-times for events with an energy of $5\,\leq\,N_\text{pe}\,\leq\,20$, that pass all quality cuts and an additional Cherenkov cut of $\chc/\,=\,5$. Red (black) data points represent the \cDS/ (\acDS/) data set. Panel \textbf{B} shows the $T_{10-90}$ distribution marginalized over $T_{0-50}$ and panel \textbf{C} shows $T_{0-50}$ marginalized over $T_{10-90}$. An excess of coincidence events over anti-coincidences is readily visible in all panels. Panel \textbf{B} suggests that the anti-coincidences (\acDS/, black) show a much longer $T_{10-90}$ rise-time than the events present in the coincidence region (\cDS/, red), whereas panel \textbf{C} shows that the corresponding \acDS/ $T_{0-50}$ is centered around \SI{1.5}{\micro\second}. Both of these observations are consistent with the assumption that these anti-coincidence events mainly arise from spurious events.\par

\begin{figure}[htb]
	\begin{center}
		\includegraphics[scale=1]{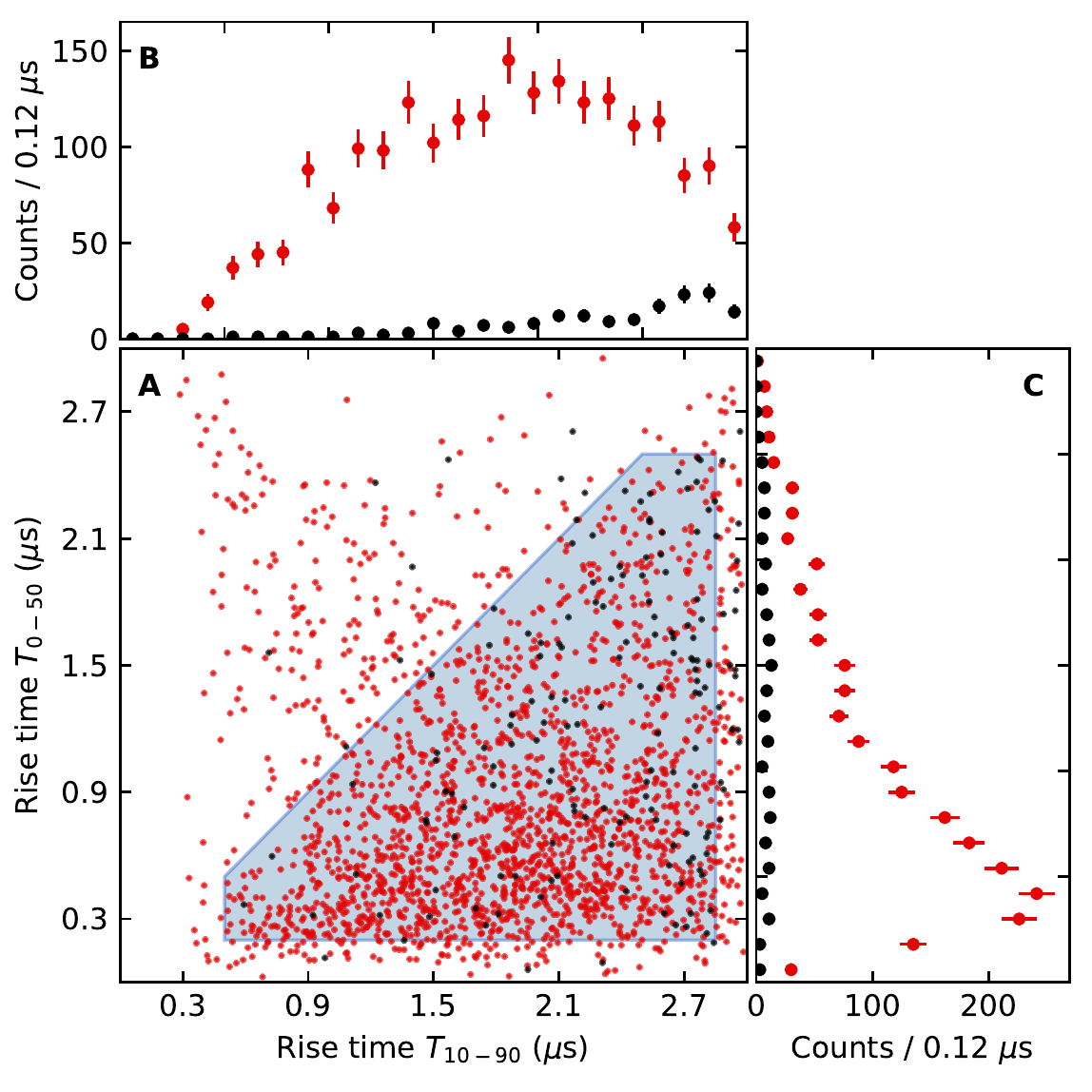}
	\end{center}
	\caption[Example rise-time distribution for the barium calibration with a Cherenkov cut of 7 for events with an energy of $5\,\leq\,N_\text{pe}\,\leq\,20$]{Labeling as in Fig.~\ref{fig:ba-calibration:rt-distribution-01}. Data quality cuts are identical, whereas the Cherenkov cut was set to $\chc/=7$. Most of the \acDS/ data is cut whereas the excess in \cDS/ remains almost untouched.}
	\label{fig:ba-calibration:rt-distribution-02}
\end{figure}

The shaded blue region in panel \textbf{A} represents one of the proposed rise-time cuts, i.e., all events within the polygon are accepted, whereas outside events are rejected. The shape of the area incorporates two main ideas. First, several events above the diagonal of $T_{0-50} = T_{10-90}$ can be seen. As discussed in chapter~\ref{chapter:am-calibration}, the onset of these events was misidentified due to a preceding \ac{spe} in the \ac{roi}. As a result, the integration window is misplaced and only covers part of the event, which distorts both the absolute charge and the individual rise-times. This cut is termed the diagonal rise-time cut, i.e., $T_{0-50} < T_{10-90}$. Second, additional rise-time cuts are defined that reject fringe cases, i.e., events showing rise-times close to 0 or \SI{3}{\micro\second}. These cuts form a rectangular acceptance window in the plane spanned by $T_{0-50}$ and $T_{10-90}$. Therefore, these cuts are referred to as orthogonal rise-time cuts.\par

To illustrate the effect of the Cherenkov cut on both \cDS/ and \acDS/ data sets, the rise-time distributions were plotted for different \chc/ values. Fig.~\ref{fig:ba-calibration:rt-distribution-02} shows a rise-time distribution plot similar to the one in Fig.~\ref{fig:ba-calibration:rt-distribution-01}, where with an increase in \chc/ from 5 to 7. It is apparent that the \acDS/ data were almost completely rejected by the data quality and the Cherenkov cuts alone. The low energy events in the \cDS/ data set were almost perfectly preserved.

\subsection{Calculating cut acceptances}
\label{section:ba-calibration:calculating-acceptances}
The signal acceptances for different combinations of the Cherenkov and rise-time cuts need to be quantified. As the procedure is identical for each cut parameter combination, this section describes the acceptance calculation using one particular set of cuts as an example, it is

\begin{equation}
\begin{aligned}
\text{Cherenkov}\qquad\,\,N^\text{min}_\text{iw} &=8\label{eq:ba-calibration:example-cut-parameters}\\
\text{Orthogonal rise-time cut}\qquad\,T_{0-50} &\in [0.2,2.5]\SI{}{\micro\second}\\
\text{Orthogonal rise-time cut}\qquad T_{10-90} &\in [0.5,2.85]\SI{}{\micro\second}\\
\text{Diagonal rise-time cut}\qquad\,\,T_{0-50} &< T_{10-90}.
\end{aligned}
\end{equation}

The signal acceptances calculated in the following reflect the acceptances corresponding to the orthogonal rise-time and the Cherenkov cuts and were calculated based on the \isotope{Ba}{133} data set. All remaining signal acceptances later used in the \ac{cenns} search (chapter \ref{chapter:sns-analysis}) were calculated using the \ac{cenns} search data recorded at the \ac{sns} itself, rather than the \isotope{Ba}{133} data set. The rational for this is as follows: First, since the \csi/ crystal was exposed to a $\gamma$ source for the \isotope{Ba}{133} calibration, the phosphorescence in the crystal was much higher compared to the \ac{cenns} search data set. Second, the onset of a forward scattered Compton event is not perfectly aligned with the beginning of the \acf{roi} window in the \isotope{Ba}{133} calibration measurement. The combination of these two effects creates a larger amount of radiation induced events in the \isotope{Ba}{133} measurement, that are preceded by spurious \acp{spe} from afterglow. The onset for these events is therefore misidentified. As events with a misidentified onset are the main types of events removed by the diagonal rise-time cut, the \isotope{Ba}{133} calibration can therefore not be used to predict the percentage of events cut in the \ac{cenns} data set. The diagonal rise-time cut is further discussed at a later point in this thesis.\par

In order to calculate the signal acceptance for the chosen cut parameters, an energy spectrum calculated from the \acDS/ and \cDS/ data sets with only the quality cuts and a diagonal rise-time cut applied was compared to a spectrum with all data cuts applied. The former is termed \emph{uncut} and the latter is termed \emph{cut} spectrum, making reference to the Cherenkov and orthogonal rise-time cuts only. The full list of data cuts applied to each set is shown in Table ~\ref{tab:ba-calibration:cuts-for-data-sets}.\par

\begin{table}[tb]
	\begin{center}
		\begin{tabular}{lcc}
		\toprule
		Cut type & Uncut data set & Cut data set \\
		\midrule
			Overflow             & $\checkmark$ & $\checkmark$ \\
			Linear gate          & $\checkmark$ & $\checkmark$ \\
			Afterglow            & $\checkmark$ & $\checkmark$ \\
			Diagonal rise-time   & $\checkmark$ & $\checkmark$ \\
			Cherenkov            & -            & $\checkmark$ \\
			Orthogonal rise-time & -            & $\checkmark$ \\
		\bottomrule
		\end{tabular}
	\end{center}
	\caption[Definition of \emph{uncut} and \emph{cut} data sets during the barium calibration]{Definition of \emph{uncut} and \emph{cut} data sets during the barium calibration. The naming convention only makes reference to the Cherenkov and orthogonal rise-time cuts.}
	\label{tab:ba-calibration:cuts-for-data-sets}
\end{table}

\begin{figure}[htbp]
	\begin{center}
		\includegraphics[scale=1]{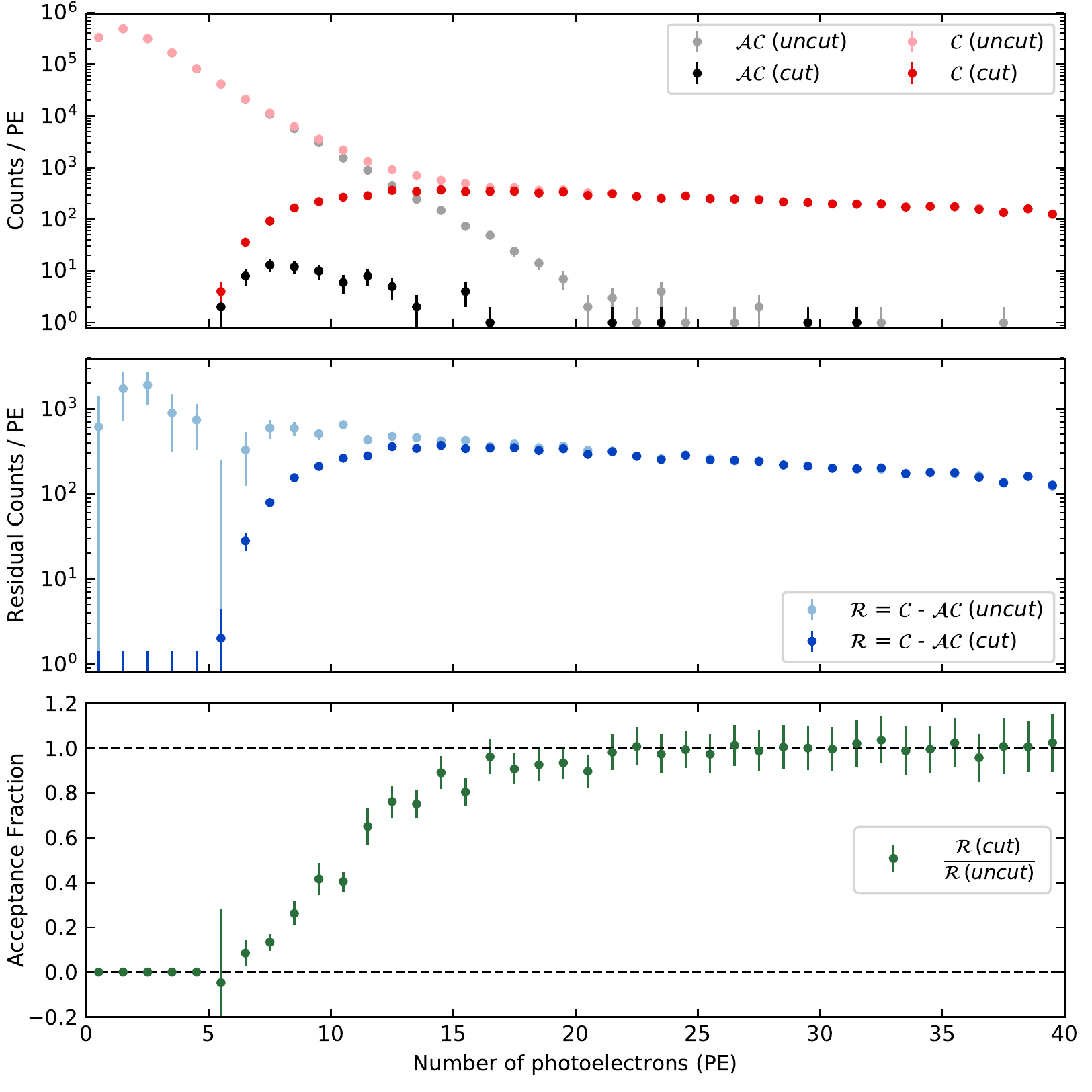}
	\end{center}
	\caption[Example of the acceptance calculation for a given set of cut parameters]{Example of the acceptance calculation for a given set of cut parameters. The cuts are as stated in Eq.~(\ref{eq:ba-calibration:example-cut-parameters}), as well as $N_\text{pt}=8$. \textbf{Top}: Energy spectra for both \cDS/ and \acDS/, for the \emph{uncut} and \emph{cut} data sets. The \emph{uncut} set is shown in desaturated colors. The roll off at low energies, i.e. $\npe/\rightarrow 0$ in the \emph{cut} spectra is due to the additional Cherenkov and orthogonal rise-time cuts. \textbf{Middle}: Residual spectra \rDS/=\cDS/-\acDS/ for both data sets. The \emph{uncut} residual is shown in desaturated colors. The residual spectrum of each set only includes trigger associated with coincident events as random coincidences were removed by the subtraction of \acDS/. \textbf{Bottom}: Acceptance fraction calculated using the ratio between the \emph{cut} and \emph{uncut} residual spectra. }
	\label{fig:ba-calibration:npe-spectrum}
\end{figure}

The resulting \acDS/ and \cDS/ energy spectra are shown in the top panel of Fig.~\ref{fig:ba-calibration:npe-spectrum}, where an afterglow cut of $\agc/=8$ was chosen. The \emph{uncut} data is shown in desaturated colors, whereas the \emph{cut} spectra are shown in saturated colors. Several interesting features stand out. First, there is a large excess above \SI{15}{\pe} in the \cDS/ region over the \acDS/ region for both \emph{uncut} and \emph{cut} sets. Second, no significant difference between the \cDS/ region of both \emph{uncut} and \emph{cut} sets above  \SI{15}{\pe} is visible. Third, the rise for both \cDS/ and \acDS/ in the \emph{uncut} spectrum at low energies, i.e., below \SI{15}{\pe}, is apparent. In contrast, the \emph{cut} spectra both decline towards zero for $\npe/\rightarrow 0$ due to the additional Cherenkov and rise-time cuts. Fourth, these additional cuts have a much larger impact on the \acDS/ data than the \cDS/ data, confirming that radiation-induced events are mostly preserved, whereas events containing spurious \acp{spe} are rejected.\par

Once the individual \cDS/ and \acDS/ energy spectra had been determined, the residual spectra were calculated as $\rDS/\,=\,\cDS/-\acDS/$ for both \emph{uncut} and \emph{cut} sets. The residuals are shown in the middle panel of Fig.~\ref{fig:ba-calibration:npe-spectrum}. As discussed earlier, the residual spectrum only contains small angle Compton-scattered events, as all random coincidences from environmental radiation were subtracted out. The \emph{cut} and \emph{uncut} residual spectra can be compared to find the percentage of events that survive the additional data cuts. The ratio between the \emph{cut} and \emph{uncut} residual spectrum is referred to as signal acceptance fraction, which is shown in the bottom panel of Fig.~\ref{fig:ba-calibration:npe-spectrum}.\par

\begin{figure}[tb]
	\begin{center}
		\includegraphics[scale=1]{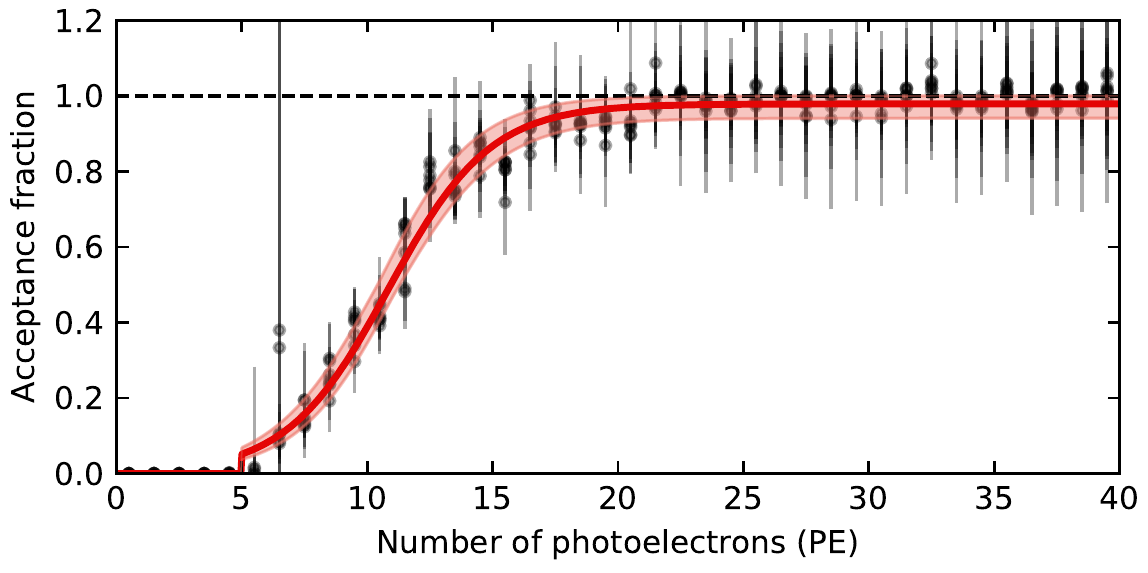}
	\end{center}
	\caption[Fit of the signal acceptance model $\eta(N_\text{pe},a,k,x_0)$ to the acceptance fractions calculated for the \isotope{Ba}{133} calibration]{The gray data points represent acceptance fractions calculated from the \isotope{Ba}{133} calibration data as shown in Fig.~\ref{fig:ba-calibration:npe-spectrum} for all $\agc/\in[3,10]$. The red curve represents the best fit of the signal acceptance model $\eta(N_\text{pe},a,k,x_0)$ to all data simultaneously. The red band represents the $1\sigma$ confidence interval derived using a percentile bootstrapping approach.}
	\label{fig:ba-calibration:acceptance-fit}
\end{figure}

As noted previously, the average background level of the \isotope{Ba}{133} calibration was much higher than of the \ac{cenns} data set due to the presence of the \isotope{Ba}{133} source. To confirm that this did not introduce any bias in the signal acceptances derived from this calibration, the acceptance calculation was repeated for all $\agc/\in[3,10]$. A comparison of the resulting signal acceptance fractions found no bias was introduced by the choice of \agc/, i.e., the calculated acceptance fractions were independent of \agc/. However, in order to incorporate the uncertainty based on the choice of \agc/, a sigmoid-shaped signal acceptance function $\eta_\text{ba}(N_\text{pe})$ was fit to all acceptance fractions (bottom panel of Fig.~\ref{fig:ba-calibration:npe-spectrum}) calculated for all $\agc/\in[3,10]$ simultaneously. The fit model is given by
\begin{align}
\eta_\text{ba}(N_\text{pe},a,k,x_0) =\frac{a}{1 + \text{e}^{-k(N_\text{pe} - x_0)}} \Theta_\text{H}\left(N_\text{pe} - 5\right),\label{eq:ba-calibration:acceptance-function}
\end{align}
where $a$ describes the maximum amplitude, $k$ the width and $x_0$ the location of the sigmoid. $\Theta_\text{H}$ denotes the Heaviside step function and represents an ultimate lower bound, below which the acceptance is always zero.\par

The acceptance fractions, i.e., the ratio between the \emph{cut} and the \emph{uncut} residual energy spectrum, are calculated for all $\agc/\in[3,10]$. The resulting acceptances are shown in gray in Fig.~\ref{fig:ba-calibration:acceptance-fit}. The best fit of Eq.~(\ref{eq:ba-calibration:acceptance-function}) to all of these acceptances simultaneously is shown in solid red.\par

\begin{figure}[tb]
	\begin{center}
		\includegraphics[scale=1]{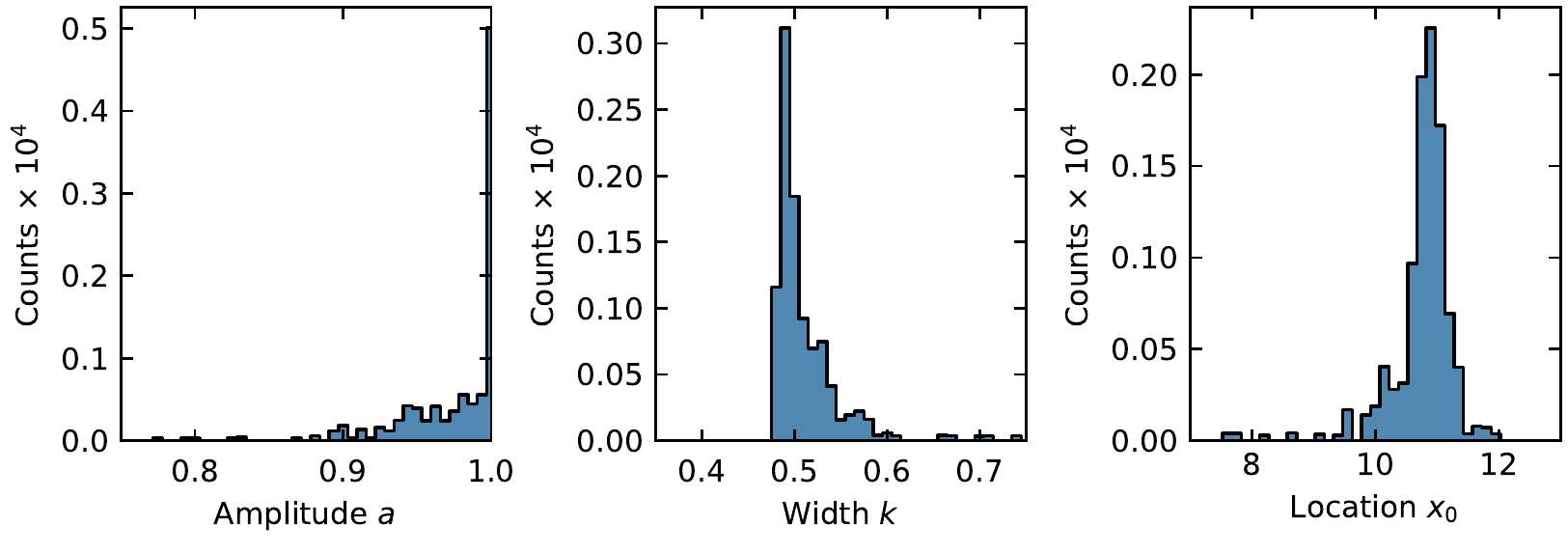}
	\end{center}
	\caption[Bootstrapped parameter distributions of $a$, $k$ and $x_0$ obtained by using a bootstrap approach of resampling residuals]{Bootstrapped parameter distributions of $a$, $k$ and $x_0$, i.e., the parameters used in the acceptance function $\eta_\text{ba}$ (Eq.~\ref{eq:ba-calibration:acceptance-function}). This acceptance function quantifies the acceptance of the Cherenkov and orthogonal rise-time cuts using the \isotope{Ba}{133} calibration data. The parameter distributions were obtained by using a bootstrap approach of resampling residuals which is further described in the text.}
	\label{fig:ba-calibration:bootstrapped-distributions}

\end{figure}

In order to estimate the uncertainty on the acceptance function, the underlying distributions of $a$, $k$ and $x_0$ were estimated using a bootstrap resampling procedure \cite{wehrens-01,wu-01}. First, the model $\eta(N_\text{pe})$ was fitted to the data using a least squares approach, and the residuals $\vec{r}$ between the best fit and each acceptance value (gray in Fig.~\ref{fig:ba-calibration:acceptance-fit}) were calculated. For each individual acceptance value a residual was randomly drawn from $\vec{r}$ and added to its original value. As a result a new synthetic set of acceptance fractions was created. The model $\eta(N_\text{pe})$ was then refit to the new synthetic set of acceptance fractions and the resulting fit parameters $a$, $k$ and $x_0$ were recorded. The last two steps, i.e., the creation of a new synthetic set of acceptance fractions and the fit of the acceptance model, were repeated $N_\text{B}=\num{10000}$ times to properly sample the distribution of $a$, $k$ and $x_0$. The result is shown in Fig.~\ref{fig:ba-calibration:bootstrapped-distributions}.\par

These bootstrapped parameter distributions can be used to calculate the $1\sigma$ confidence interval of each individual parameter $a$, $k$ and $x_0$, by determining the 15.865 and the 84.135 percentile of the respective bootstrapped distribution \cite{dixon-01}. The final $1\sigma$ confidence interval of the acceptance function $\eta_\text{ba}$ is shown in shaded red in Fig.~\ref{fig:ba-calibration:acceptance-fit}. The best fit parameters for the choice of cuts (Eq.~\ref{eq:ba-calibration:example-cut-parameters}) presented in this chapter are given by
\begin{equation}
\begin{aligned}
a   &= 0.979^{+0.021}_{-0.038}\label{eq:ba-calibration:best-fit-w-drt}\\
k   &= 0.494^{+0.034}_{-0.013}\\
x_0 &=  10.85^{+0.18}_{-0.40}
\end{aligned}
\end{equation}
This approach only covers the acceptances due to the Cherenkov and orthogonal rise-time cuts. Comparing the calculated signal acceptance to the one obtained with a simple \ac{mc} approach (shown in Fig.~\ref{fig:ba-calibration:mc-cherenkov-cut}) shows that the real acceptance rises much slower towards unity. However, the acceptance is finite for events with $N_\text{pe}<8$. This shows the superiority of the \isotope{Ba}{133} measurement over the \ac{mc} approach in quantifying the signal acceptance function. The following section details how the acceptance fraction of the diagonal rise-time can be obtained.\par

\begin{figure}[tb]
	\begin{center}
		\includegraphics[scale=1]{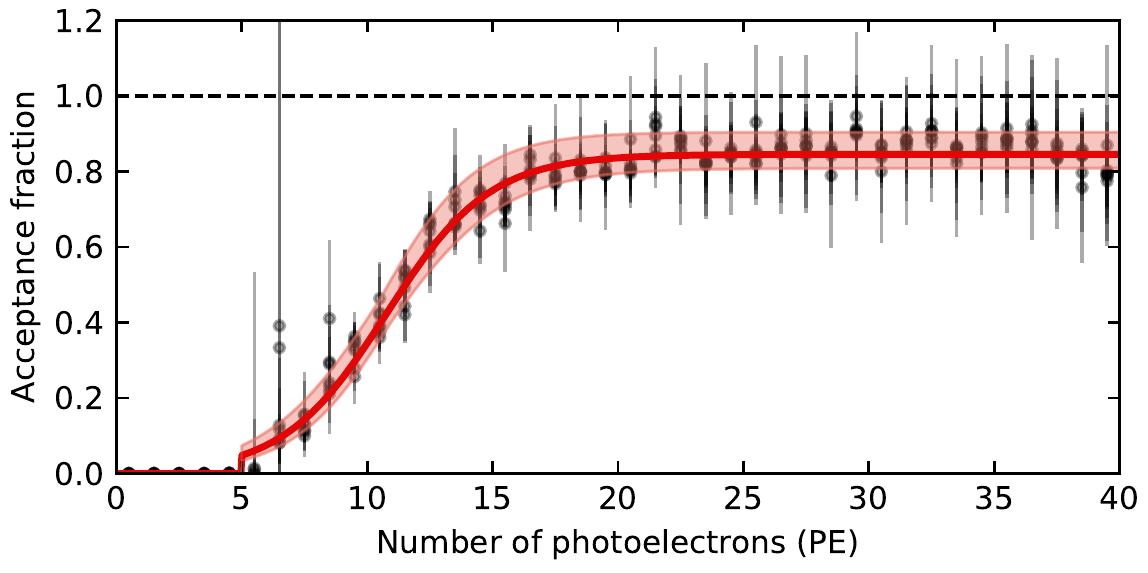}
	\end{center}
	\caption[Fit of the signal acceptance model $\eta(N_\text{pe},a,k,x_0)$ to acceptances that include a diagonal rise-time cut]{Fit of the signal acceptance model $\eta(N_\text{pe},a,k,x_0)$ to acceptances, which were calculated based on an \emph{uncut} residual energy spectrum that did not include a diagonal rise-time cut. As a result, the overall acceptance is smaller than what can be seen in Fig.~\ref{fig:ba-calibration:acceptance-fit}.}
	\label{fig:ba-calibration:acceptance-fit-WO-DRT}
\end{figure}

\begin{figure}[tb]
	\begin{center}
		\includegraphics[scale=1]{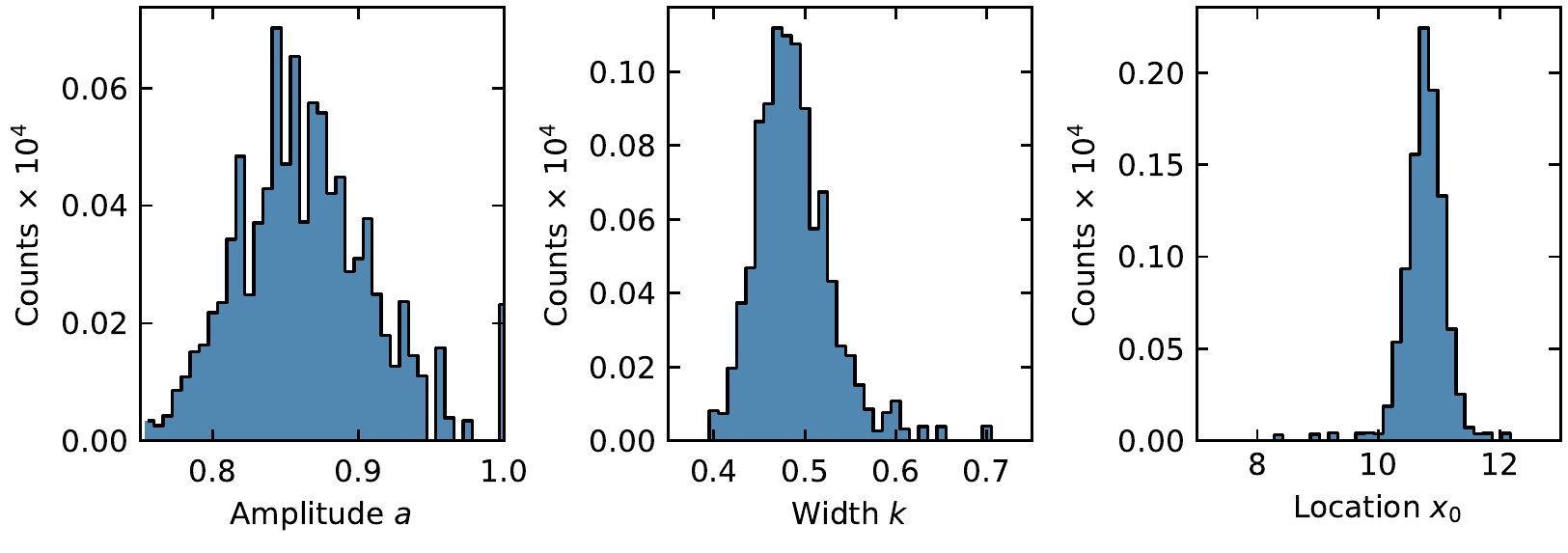}
	\end{center}
	\caption[Bootstrapped parameter distributions of $a^\star$,$k^\star$ and $x^\star_0$ obtained by using a bootstrap approach of resampling residuals]{Bootstrapped parameter distributions of $a^\star$,$k^\star$ and $x^\star_0$, i.e., the parameters used in the acceptance function $\eta_\text{ba}$ (Eq.~\ref{eq:ba-calibration:acceptance-function}). This acceptance function quantifies the acceptance of the Cherenkov and orthogonal rise-time cuts using the \isotope{Ba}{133} calibration data. The parameter distributions were obtained by using a bootstrap approach of resampling residuals which is further described in the text. The acceptance fractions fitted in this exercise were determined using an \emph{uncut} spectrum that did not include the diagonal rise-time cut.}
	\label{fig:ba-calibration:bootstrapped-distributions-WO-DRT}
\end{figure}

The signal acceptance was calculated as described above, but with an \emph{uncut} data set where the diagonal rise-time cut was not applied. In this case the resulting signal acceptance reflects all cuts, Cherenkov, orthogonal and diagonal rise-time cuts. First, the $\rDS/\,=\,\cDS/-\acDS/$ residuals for both \emph{uncut} and \emph{cut} data sets were calculated and the resulting energy spectra were compared. Due to the missing diagonal rise-time cut there are more events in the \emph{uncut} spectrum that are not present in the \emph{cut} spectrum. This results in an overall reduction of the signal acceptance (Fig.~\ref{fig:ba-calibration:acceptance-fit-WO-DRT}, gray points). As described above, the acceptances calculated for different afterglow cuts \agc/ were fitted simultaneously using Eq.~(\ref{eq:ba-calibration:acceptance-function}). The best fit is shown as solid red line in Fig.~\ref{fig:ba-calibration:acceptance-fit-WO-DRT}. The $1\sigma$ errors for each individual parameter ($a^\star$, $k^\star$, and $x^\star_0$) were determined with a percentile bootstrap resampling approach as described above. The bootstrapped parameter distributions are shown in Fig.~\ref{fig:ba-calibration:bootstrapped-distributions-WO-DRT}. The fit values for an \emph{uncut} data set without a diagonal rise-time cut is given by
\begin{equation}
\begin{aligned}
a^\star   &= 0.845^{+0.059}_{-0.036}\label{eq:ba-calibration:best-fit-wo-drt}\\
k^\star   &= 0.487^{+0.032}_{-0.042}\\
x^\star_0 &= 10.76^{+0.21}_{-0.34}
\end{aligned}
\end{equation}

It can be observed that both $k^\star $ and $x^\star_0$ remain almost unchanged with respect to the previous approach for which $k= 0.494^{+0.034}_{-0.013}$ and
$x_0 =  10.85^{+0.18}_{-0.40}$ were found. However, $a^\star$ is much smaller than $a = 0.979^{+0.021}_{-0.038}$.\par

\begin{figure}[htb]
	\begin{center}
		\includegraphics[scale=1]{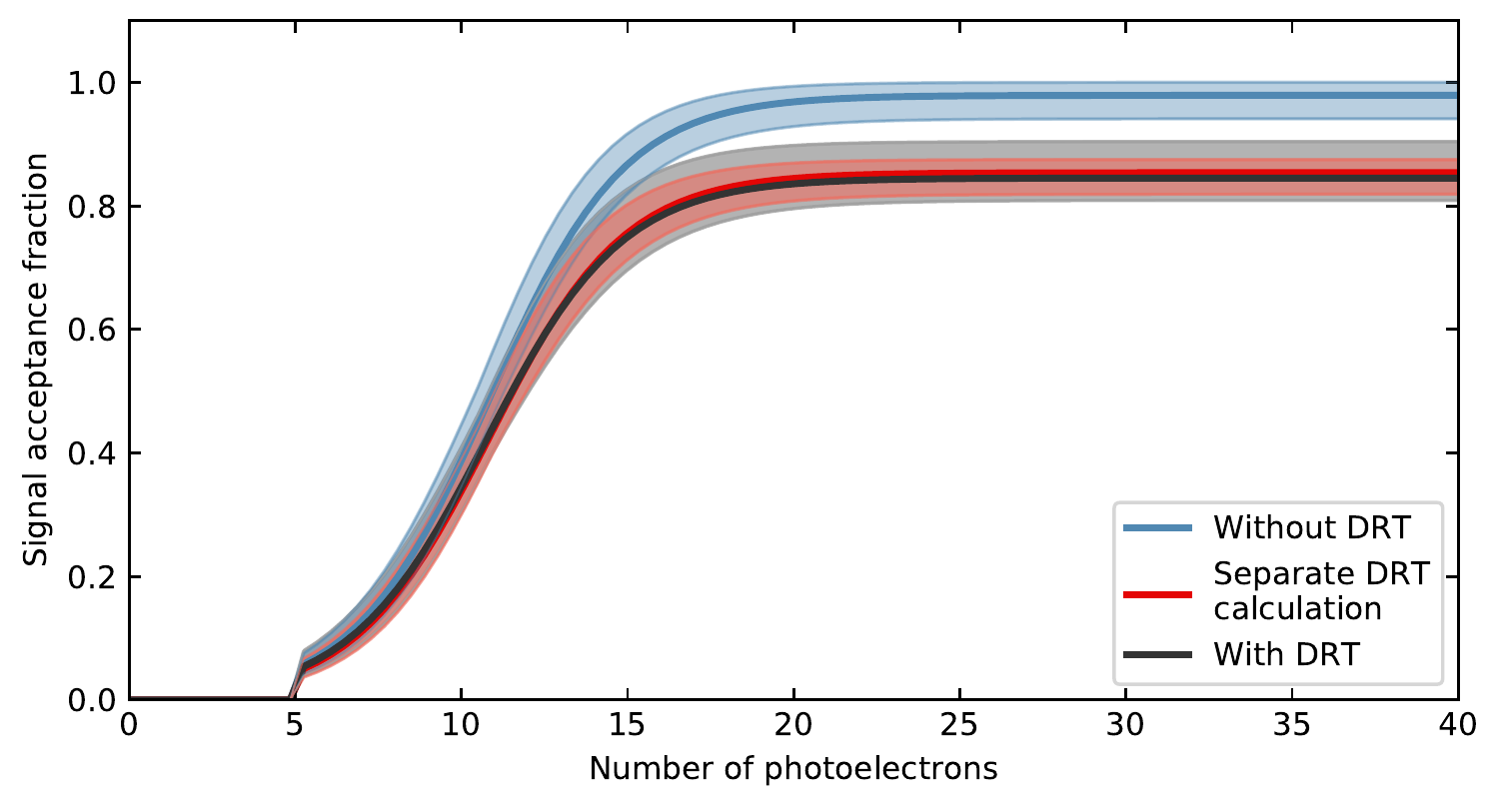}
	\end{center}
	\caption[Comparison of different approaches to incorporate the diagonal rise-time cut into the signal acceptance function]{Comparison of different approaches to incorporate the diagonal rise-time cut into the signal acceptance function. Shown in blue is the acceptance model fit using the \emph{uncut} data set that excludes all events with $T_{0-50}\geq T_{10-90}$. The corresponding fit values are given in Eq.~(\ref{eq:ba-calibration:best-fit-w-drt}). Black shows the signal acceptance fit obtained for an \emph{uncut} data set that did not exclude events with$T_{0-50}\geq T_{10-90}$. The corresponding fit values are given in Eq.~(\ref{eq:ba-calibration:best-fit-wo-drt}). The red line shows the blue model after a separate calculation of the diagonal rise-time cut acceptance was included. Once this additional scaling factor was included, the difference between the black and red model is negligible.}
	\label{fig:ba-calibration:drt-estimation-and-comparison}
\end{figure}

In order to scale $a$ to match $a^\star$, the acceptance of the diagonal rise-time cut was calculated separately. Since the percentage of events with a misidentified onset is independent of the energy deposited, high-energy depositions in the \csi/ crystal can be used to estimate the overall acceptance $\eta_\text{drt}$ of the diagonal rise-time cut. To this end, the fraction of events with $T_{0-50} > T_{10-90}$ in the energy region of 50-\SI{150}{\pe} was computed. The same fraction was computed for all $N_\text{pt}\in[3,10]$ and a constant was fitted to the data. The best fit is given by
\begin{align}
\eta_\text{drt} = 0.872 \pm 0.011
\end{align}
The diagonal rise-time acceptance fraction $\eta_\text{drt}$ was then incorporated into the previous acceptance model. It is
\begin{align}
\hat{\eta}(N_\text{pe},a,k,x_0) = \eta(N_\text{pe},a,k,x_0)\times\eta_\text{drt}.
\end{align}
Fig.~\ref{fig:ba-calibration:drt-estimation-and-comparison} shows a comparison between all signal acceptance models. The blue curve shows the signal acceptance as derived from an \emph{uncut} data set from which all events with $T_{0-50} \geq T_{10-90}$ were removed. The black curve shows the model for an \emph{uncut} data set that includes these events. The red curve represents the blue model after scaling with $\eta_\text{drt}$. As the diagonal rise-time cut only provides an overall scaling of the acceptance function, both amplitudes can be compared. It is
\begin{align}
a^\star & = 0.845^{+0.059}_{-0.036}\\
a\times\eta_\text{drt} & = 0.979^{+0.021}_{-0.038} \times 0.872 \pm 0.011 = 0.854^{+0.021}_{-0.035}
\end{align}
The agreement between both approaches is apparent. As a result, the scaling approach in which the signal acceptance $\eta_\text{drt}$ of the diagonal rise-time cut is calculated independently was later used to incorporate the acceptance from the diagonal rise-time cut in the \ac{cenns} search data.

%% file: qf-measurements.tex
\chapter{Measurement of the low-energy quenching factor in \csi/}
\label{chapter:quenching-calibration}
As discussed in chapter~\ref{chapter:cenns-theory}, the only visible signal from a \ac{cenns} interaction is a nuclear recoil within the detector. These recoils carry only a small amount of energy. Their detection is made even more difficult due to a process typically referred to as quenching: For a low energy nuclear recoil only a small amount of energy is converted into scintillation or ionization, and the rest is dissipated via secondary nuclear recoils and heat. The quenching factor can be defined by comparing the scintillation or ionization yield of a nuclear recoil of given energy to the scintillation or ionization yield of an ionizing particle of the same energy, i.e., a particle which predominantly loses its energy through electronic recoils in the detector. The quenching factor plays a crucial role in establishing an energy scale for nuclear recoils. It is needed to convert the predicted \ac{cenns} nuclear recoil energies from \si{\keVnr} to \si{\keVee}. Using the light yield calibration measured in chapter~\ref{chapter:am-calibration} using an \isotope{Am}{241} source, the electron equivalent energy can further be converted into an equivalent number of photoelectrons \npe/, i.e., a quantifiable detector response.

To add to previous measurements of the quenching factor of \csi/ two new and independent measurements of the quenching factor were performed in the framework of the \coherent/ collaboration at \ac{tunl}. The data acquisition system used in the measurement of the \csi/ quenching factor described in this chapter was different from the one used in the light yield (chapter~\ref{chapter:am-calibration}) and \isotope{Ba}{133} calibration (chapter~\ref{chapter:ba-calibration}).

\section{Experimental setup}
\begin{figure}[htbp]
\begin{center}
\includegraphics[width=0.495\linewidth]{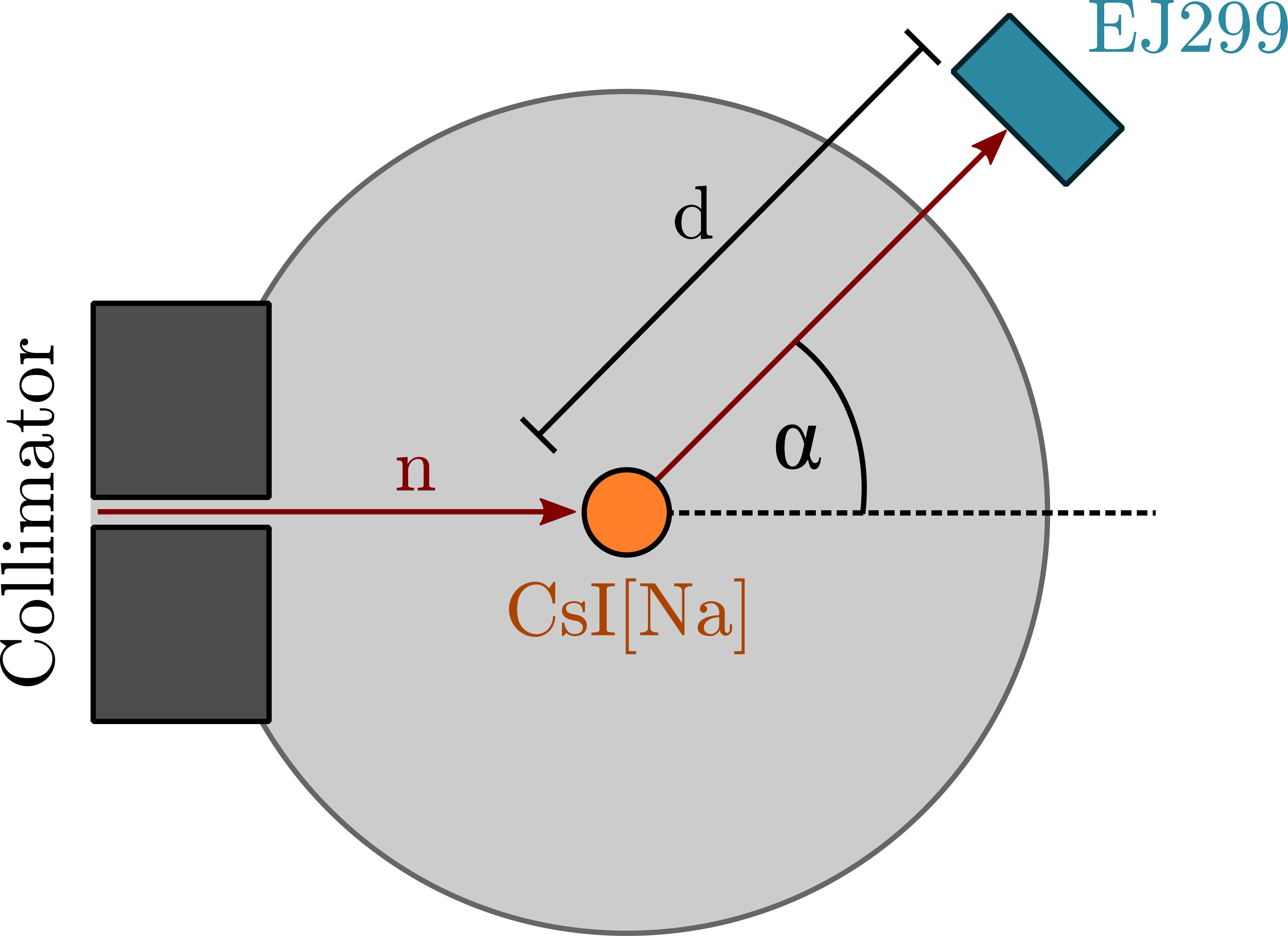}
\includegraphics[width=0.495\linewidth]{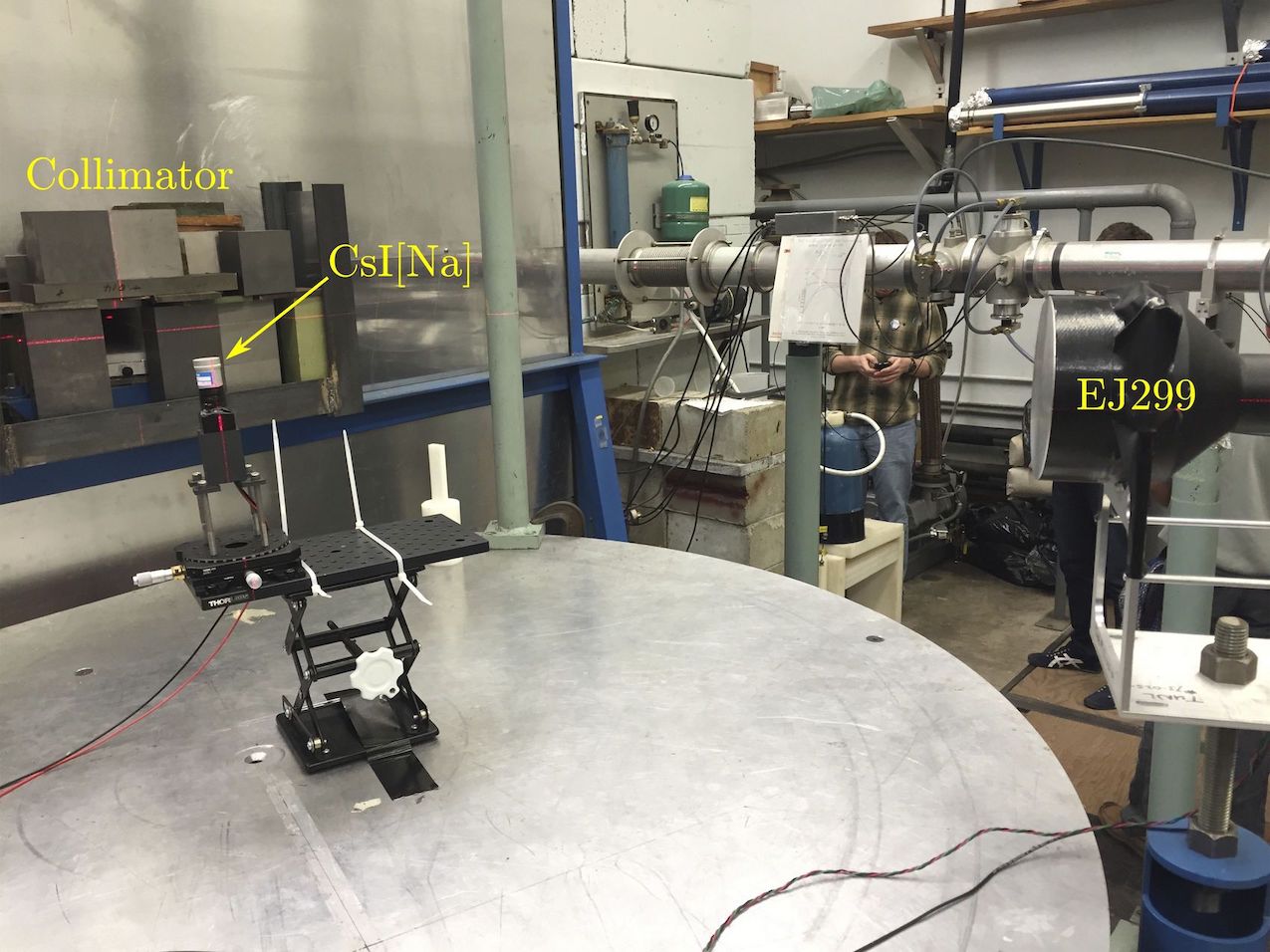}
\end{center}
\caption[Experimental setup of the \csi/ quenching factor measurements at \acs*{tunl}]{Experimental setup of the \csi/ quenching factor measurements at \acs*{tunl}. A highly collimated neutron beam (red) is directed towards a small \csi/ detector (orange). By knowing the angle $\alpha$ between the incoming neutron beam and the \ej299/ (blue), the energy deposited in the \csi/ by a triggering neutron can be kinematically inferred.}
\label{fig:quenching:setup}
\end{figure}

\begin{figure}[htbp]
\begin{center}
\includegraphics[width=1\linewidth]{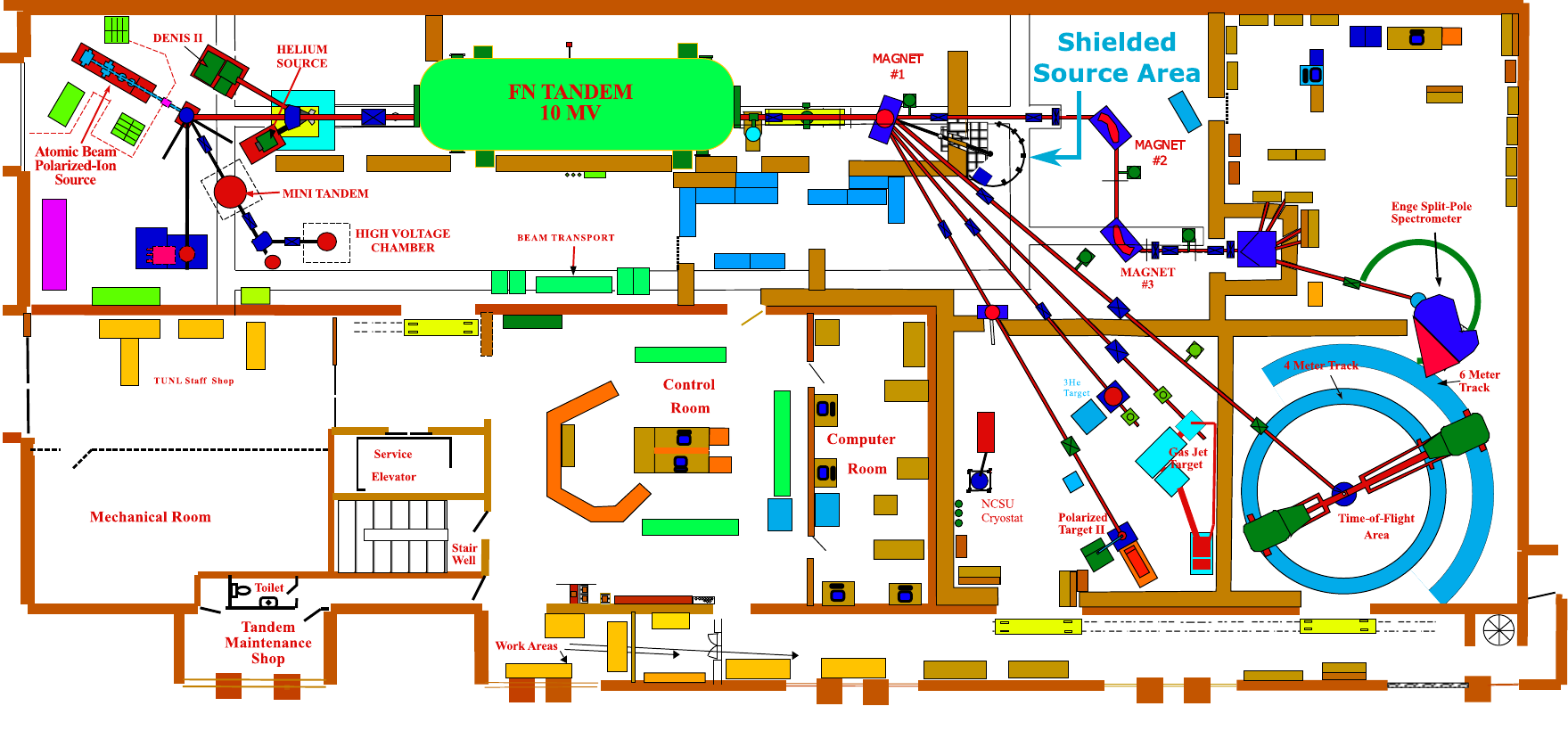}
\end{center}
\caption[Schematic of the \acs*{tunl} facilities]{Schematic of the \acs*{tunl} facilities. The quenching factor measurements were performed at the shielded source area, which provides a highly collimated neutron beam of known energy. Adapted from \cite{tunl-01}.}
\label{fig:quenching:tunl}
\end{figure}

The \csi/ quenching factor was measured at \acf{tunl} (Fig.~\ref{fig:quenching:tunl}). The experimental setup was located in the \ac{ssa} of the facility (Fig.~\ref{fig:quenching:setup}). Deuterium ions were accelerated by a FN tandem Van de Graaff accelerator and directed towards a deuterium gas chamber. The subsequent reaction of deuterium ions with the deuterated target produced a highly collimated neutron beam of known energy. A small \csi/ detector ($l=\SI{55}{\mm}$, $\diameter=\SI{22.4}{\mm}$), which was procured from the same manufacturer as the \ac{cenns} detector, using an identical growth method and sodium dopant concentration of 0.114 mole \si{\percent}, was positioned in the center of the neutron beam. The size of the detector was chosen such, that predominantly single nuclear recoils are produced within the crystal. The \csi/ crystal was read out by an ultra-bialkali PMT \cite{nakamura-01}, which made it possible to probe recoil energies down to $\sim\SI{3}{\keVnr}$. The high voltage for the ultra-bialkali PMT was provided by a Stanford Research Systems Inc. PS350 power supply and set to \SI{-935}{\volt}.\\

A \ej299/ plastic scintillator ($l=\SI{47.6}{\mm}$, $\diameter=\SI{114.3}{\mm}$) was used to detect neutrons that scattered off the \csi/ detector. This scintillator was read out by a 5-inch 9390B \ac{pmt} from ET Enterprises. The \ej299/ is capable of $n$-$\gamma$ discrimination using standard \acf{psd} techniques \cite{pozzi-02}. \ej299/ signals produced by nuclear recoils exhibit different scintillation decay times from those produced by electronic recoils. As a result, the percentage of the total scintillation light emitted in the first few nanoseconds of an event is different for nuclear and electronic recoils. This can be used for \ac{psd} and is further examined in section~\ref{section:quenching-calibration:ej299-calib}. The $n$-$\gamma$ discrimination was used in the quenching factor runs to reduce the background caused by triggers on environmental $\gamma$ radiation. The high voltage for the 9390B PMT was provided by an Agilent E3631A and was set to \SI{-750}{\volt}. The \ej299/ output was fed into an Ortec 934 \ac{cfd}. The 934 \ac{cfd} logical output provided the trigger for the U1071A Acqiris 8-bit fast digitizer. The raw output of the ultra-bialkali PMT reading out the \csi/ crystal was fed into channel 1 of this fast digitizer, whereas the raw output of the 9390B \ac{pmt} reading out the \ej299/ was fed through a \SI{6}{\dB} attenuator and recorded as channel 2. The full energy range of neutron induced events is contained in the digitizer range using the attenuator in the data acquisition system.\\

For each trigger \SI{6}{\micro\second}-long waveforms were recorded for both channels, i.e., the raw output of the \csi/ and the \ej299/ detectors. The sampling rate was set to \SI{500}{\mega\sample\per\second} with a trigger position set to \SI{2}{\micro\second} and a digitizer range of $\pm\SI{25}{\mV}$. The overall triggering rate for this setup in the environmental radiation field, which consists of mostly $\gamma$s, was $\mathcal{O}(\SI{250}{\hertz})$. Using a \isotope{Na}{22} source that emits back-to-back \SI{511}{\keV} annihilation-radiation $\gamma$-rays, and which was positioned in the midpoint of the distance between both detectors, an offset of \SI{20}{\ns} between the individual detector channels was found. This offset was corrected for in the analysis.\\

The simple geometry of this quenching factor measurement allowed for the selection of energy of recoiling cesium and iodine nuclei by varying the scattering angle $\alpha$. The recoil energy can be calculated using the simple kinematic relation \cite{minowa-01}
\begin{align}
	E_\text{R} = 2\left(1 + A\right)^{-2}\left(1+A-\cos^2\alpha-\cos\alpha\sqrt{A^2-1+\cos^2\alpha}\right)E_n,\label{eq:quenching:recoil-kinematics}
\end{align}
where $E_n$ is the energy of the incident neutron, $\alpha$ the scattering angle and $A$ the mass ratio between the target nucleus and the incoming neutron. Data was taken for a total of seven different runs with different angles $\alpha$ between the incident neutron beam and the \ej299/. The quenching factor measurement described in this chapter covered the full energy range of nuclear recoils that is of interest to the \ac{cenns} search at the \ac{sns} (Fig.~\ref{fig:quenching:quenching-factor}). The \csi/ detector position remained unchanged for every run. The exact \ej299/ detector locations as well as additional information regarding the expected recoil energy and individual run times are given in Table~\ref{tab:quenching:experiment-setup}. Besides decreasing $\alpha$, the distance $d$ between both detectors was increased in order to avoid triggering on beam-related neutrons that were not  scattered off the \csi/. The following sections will first discuss the calibrations performed for each detector followed by the evaluation of the quenching factor of \csi/.

\begin{table}[htbp]
\begin{center}
\begin{tabular}{cccc}
\toprule
Angle $\alpha$ ($\SI{}{\degree}$)  & Distance $d$ (cm) & Recoil Energy ($\SI{}{\keVnr}$) & Run Time (min)\\
\midrule
45 & 80 & 18.44 & 165 \\
39 & 90 & 14.04 & 182 \\
33 & 90 & 10.17 & 225 \\
27 & 100 & 6.86 & 207 \\
24 & 100 & 5.45 & 178 \\
21 & 100 & 4.19 & 203 \\
18 & 110 & 3.09 & 71 \\
\bottomrule
\end{tabular}
\end{center}
\caption[\ej299/ detector positions, recoil energies and run times d]{All \ej299/ detector positions used in the measurement of the quenching factor of \csi/. The corresponding recoil energies were calculated using Eq.~(\ref{eq:quenching:recoil-kinematics}). By increasing the distance for smaller angles the triggering of on beam-related neutrons that did not scatter off the \csi/ was avoided. The measurement for \SI{18}{\degree} was cut short due to time constrains imposed by the scheduling of other experiments at \acs*{tunl}.}
\label{tab:quenching:experiment-setup}
\end{table}

\section{Detector calibrations\label{sec:quenching:detector-calibrations}}
\subsection{\csi/ calibrations}
\label{section:quenching-calibration:csi-calib}
\begin{figure}[!t]
\begin{center}
\includegraphics[width=1\linewidth]{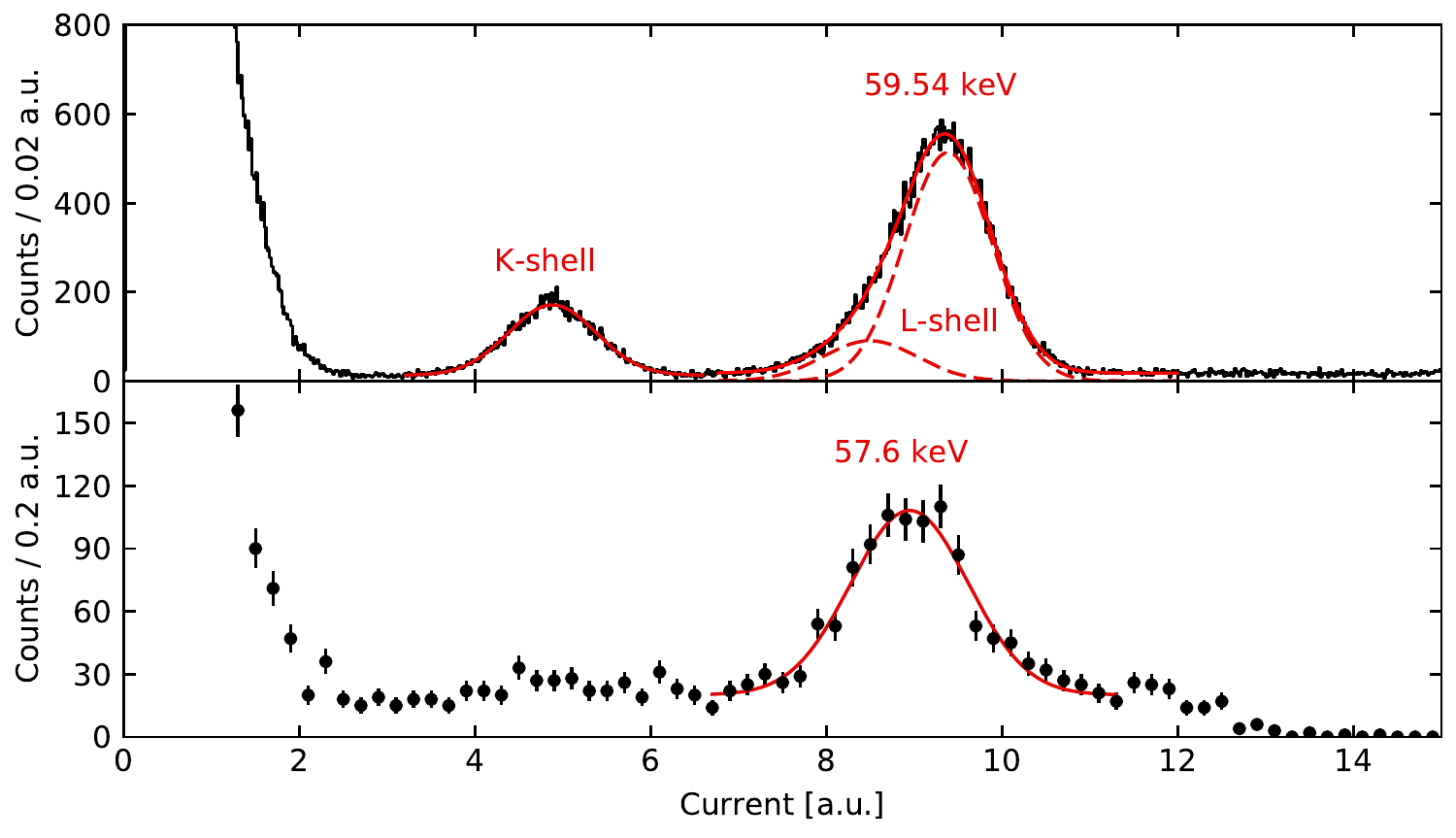}
\end{center}
\caption[\csi/ energy calibration using a \isotope{Am}{241} source and the in-situ $\isotope{I}{127}(n,n'\gamma)$ reaction]{Energy calibration of the \csi/ detector used in the quenching factor measurements at \ac{tunl}. \textbf{Top:} Dedicated \isotope{Am}{241} spectrum taken after the quenching factor measurements. The \isotope{Am}{241} main emission peak at \SI{59.54}{\keV} as well as the K- and L-shell escape peaks from Cs and I can be observed. \textbf{Bottom:} Energy calibration using the in-situ $\isotope{I}{127}(n,n'\gamma)$ reaction. The drop after $\sim 12\,$a.u. is due to overflows that were excluded from the analysis.}
\label{fig:quenching:csi-energy-calibration}
\end{figure}

To calibrate the light yield of the \csi/ detector a dedicated data set using an \isotope{Am}{241} source was taken. For this measurement waveforms were acquired using the \csi/ detector signal as a trigger, rather than the 934 \ac{cfd} output. The trigger position was again set to \SI{2}{\micro\second} into the digitized traces. For each trigger the DC baseline of the corresponding waveform was determined as the median of its first \SI{2}{\micro\second}. The exact pulse onset $t_\text{csi}^0$ was defined by a threshold crossing of \SI{0.6}{\mV}, i.e., three digitizer steps, followed by at least ten consecutive samples above threshold. The \csi/ signal was integrated starting at two samples before t$_\text{csi}^0$ for a total of \SI{3}{\micro\second} while ignoring samples below \SI{0.6}{\mV}. The resulting charge spectrum is shown at the top of Fig.~\ref{fig:quenching:csi-energy-calibration}. The main $\gamma$-emission peak at \SI{59.54}{\keV} is readily visible. Due to the low energy emission of this isotope and the small detector size most of the energy depositions occur in close proximity to the source, i.e., close to the surface. Therefore the K- and L-shell escape peaks can also be resolved. These peaks consist of two distinct energies, one from Cs and one from I, which are merged due to the limited energy resolution. In order to be able to test the two competing \ac{spe} charge distribution models described in what follows, the output was at this point not converted to \ac{spe}.\par

In addition to the external \isotope{Am}{241} calibration, a light yield calibration was performed using the in-situ $\isotope{I}{127}(n,n'\gamma)$ reaction. During de-excitation a $\gamma$-ray with an energy of \SI{57.6}{\keV} is produced. This light yield calibration uses the data acquired during actual quenching factor runs. As a result the triggering condition were different from the \isotope{Am}{241} light yield calibration described above. The DC baseline was determined using the first \SI{1}{\micro\second}. Only events with a signal onset between \SI{1}{\micro\second} and \SI{3}{\micro\second} into the waveform are recorded. This guaranteed that each signal could be integrated over the full \SI{3}{\micro\second} integration window following the onset. The corresponding charge spectrum is shown at the bottom of Fig.~\ref{fig:quenching:csi-energy-calibration}. Fitting a gaussian to the peak gave a total of 834 events under the $\isotope{I}{127}(n,n'\gamma)$ peak. The best fit results for all peaks are given in Table~\ref{tab:quenching:csi-energy-calibration}.\par

The calibration results were converted into the light yield of the detector, i.e., the number of \ac{pe}s produced per \SI{}{\keVee} for a given energy. For this purpose the mean \ac{spe} charge \qspe/ for this particular detector-\ac{pmt} assembly was determined. For each trigger all potential \ac{spe} peaks in the pretrace, i.e., the first \SI{1}{\micro\second}, were identified. A \ac{spe} was defined as at least four consecutive samples with an amplitude of at least \SI{0.6}{\mV}. The charge of each \ac{spe} was determined by integrating over a corresponding integration window which is defined by two samples before and after the threshold crossings in the sample.\par

\begin{table}[tbp]
\begin{center}
\begin{tabular}{ccc}
\toprule
Energy (keV) & Centroid $\mu$ (a.u.) & Sigma $\sigma$ (a.u.)\\
\midrule
$30.93\;\&\;28.57$ & $4.877\pm0.006$ & $0.498\pm0.007$\\
$55.60\;\&\;55.26$ & $8.495\pm0.181$ & $0.534\pm0.063$\\
$57.60$ & $8.946\pm0.033$ & $0.667\pm0.038$\\
$59.54$ & $9.382\pm0.029$ & $0.499\pm0.011$\\ 
\bottomrule
\end{tabular}
\end{center}
\caption[Best fit results for the energy calibration of the \csi/ detector]{Best fit results for the energy calibration of the small \csi/ detector used in the quenching factor measurements at \acs*{tunl}. The K- and L-shell escape peaks consist of two distinct lines which can not be resolved due to the limited energy resolution.}
\label{tab:quenching:csi-energy-calibration}
\end{table}

\begin{figure}[tbp]
\begin{center}
\includegraphics[scale=1]{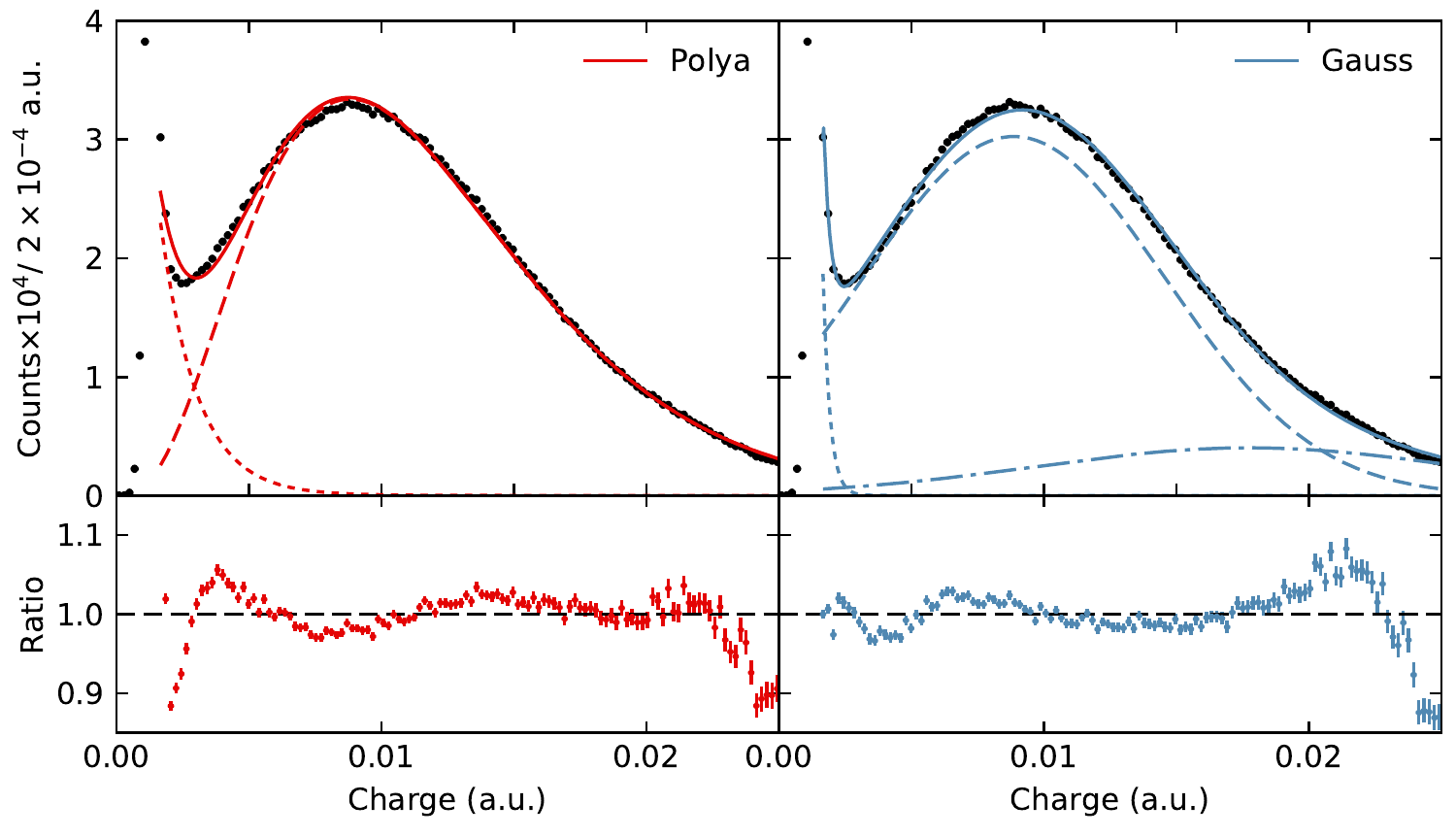}
\end{center}
\caption[\acs*{spe} calibration spectrum fitted by two different charge distribution models]{\acs*{spe} charge spectrum for the \isotope{Am}{241} calibration run. Two different models for the charge distribution were fitted to the data. \textbf{Left}: The Polya distribution model fails to properly fit the valley between the \acs*{spe} peak and the rise due to noise. The fit also suggests a negligible contribution from the two \acs*{spe} peak, which is unphysical given the presence of the \isotope{Am}{241} source.
\textbf{Right}: The Gaussian distribution model performs better at fitting the valley and also includes a significant contribution from the two \ac{spe} peak.}
\label{fig:quenching:spe-spectra}
\end{figure}

The resulting charge spectrum for the \isotope{Am}{241} run can be seen in Fig.~\ref{fig:quenching:spe-spectra}. As discussed in chapter~\ref{chapter:am-calibration}, the mean \ac{spe} charge can be extracted by fitting a model to the spectrum. In chapter~\ref{chapter:am-calibration} the Polya distribution provided a better charge model for the R877-100 \ac{pmt} used in the \ac{cenns} search. However, the \csi/ detector used in the quenching factor measurement described in this chapter, consists of a different crystal and used an ultra-bialkali \ac{pmt} instead of the R877-100 \ac{pmt}. To determine if the Polya distribution represents a better \ac{spe} charge model than a simple Gaussian, a comparative fit procedure is used.\par

The competing charge models described by Eq.~(\ref{eq:am-calibration:spe-gauss}) (Gauss) and Eq.~(\ref{eq:am-calibration:spe-polya}) (Polya) were fitted to the \ac{spe} charge spectrum. The results are shown in the top panels of Fig.~\ref{fig:quenching:spe-spectra}. The deviation of the model from the data is shown in the bottom panels. The overall deviation of both models is well within \SI{5}{\percent} for most parts of the fit, except for the low and high charge regions. The discrepancy in the low charge region could indicate under-amplified \ac{spe}. The deviation in the high charge region is due to the fact that only contributions from up to two \ac{spe} are included in the fit. The mean \ac{spe} charge for all other runs was determined in an analog fashion, which all showed the same fit quality.

The variance in the calculated mean \ac{spe} charges over all quenching factor runs was negligible. As a result, a single mean \ac{spe} charge value is adopted, which is used for all quenching factor runs. It is
\begin{align}
\qspe/(\text{polya}) = 0.0118 \pm 0.003\;\text{a.u.}\qquad \qspe/(\text{gauss}) = 0.0089 \pm 0.003\;\text{a.u.} \label{eq:quenching-factor:qspe}
\end{align}
The energy calibration can be expressed in terms of the light yield, i.e., \ac{pe}s per \si{\keVee}, shown in Table~\ref{tab:quenching:light-yield-calibration}. The slight increase in light yield around \SI{30}{\keV} compared to that at $\approx\SI{60}{\keV}$ is in agreement with the non-proportional $\gamma$-response of \csi/ that was measured in~\cite{mengesha-01,beck-01}. For the remainder of the analysis the light yield obtained from the \isotope{Am}{241} line at \SI{59.54}{\keV} was used. This choice was made as the \ac{cenns} detector was also calibrated using a \isotope{Am}{241} source (chapter~\ref{chapter:am-calibration}). The nominal light yield at \SI{59.54}{\keV} was linearly extrapolated to very low energies. This avoided any dependence on a particular light emission model and any uncertainty associated. Any energy deposition by nuclear recoils can still be converted directly into units of electron equivalent for any \csi/ detector using the quenching factor data, as long as the light yield at $\sim$\SI{60}{\keV} is known for a $\SI{3}{\micro\second}$ long integration window. Previous measurements by Park \textit{et al.} \cite{park-01} and Guo \textit{et al.} \cite{guo-01} used the same approach, which greatly simplifies the direct comparison of the results with their data.

\begin{table}[tbp]
\begin{center}
\begin{tabular}{ccccc}
\toprule
	& \multicolumn{2}{c}{Polya} & \multicolumn{2}{c}{Gauss} \\
\cmidrule(lr){2-3} \cmidrule(lr){4-5}
Energy & Centroid & Light Yield & Centroid & Light Yield \\
\scriptsize{(\si{\keVee)}}  & \scriptsize{(PE)} & \scriptsize{(PE/\si{\keVee})} & \scriptsize{(PE)} & \scriptsize{(PE/\si{\keVee})} \\
\midrule
$29.75$ & $413 \pm 11$ & $13.88 \pm 0.37$ & $548 \pm 18$  & $18.42 \pm 0.61$\\
$55.43$ & $720 \pm 24$ & $12.99 \pm 0.43$ & $954 \pm 38$  & $17.21 \pm 0.69$\\
$57.60$ & $758 \pm 19$ & $13.16 \pm 0.33$ & $1005 \pm 34$ & $17.45 \pm 0.59$ \\
$59.54$ & $795 \pm 20$ & $13.35 \pm 0.34$ & $1054 \pm 36$ & $17.70 \pm 0.60$\\ 
\bottomrule
\end{tabular}
\end{center}
\caption[Light yield calibration for the \csi/ detector used in the quenching factor measurements]{Light yield calibration for the \csi/ detector used in the quenching factor measurements at \acs*{tunl}. The energies for the Cs and I K- and L-shell escape peaks were combined to a single central energy assuming an equal contribution from both isotopes.}
\label{tab:quenching:light-yield-calibration}
\end{table}

\subsection{\ej299/ calibrations}
\label{section:quenching-calibration:ej299-calib}
The \ej299/ plastic scintillator was calibrated using several different gamma (\isotope{Na}{22}, \isotope{Cs}{137}) and neutron (\isotope{Cf}{252}) sources. During the calibrations the trigger was set to the \ej299/ output and the \csi/ channel was disregarded. The trigger position was set to \SI{2}{\micro\second} into the \SI{6}{\micro\second} long waveforms. For each trigger the pulse onset was determined by looking for at least ten consecutive samples above the \SI{0.6}{\mV} threshold. For each peak two charge integrals were computed as follows,
\begin{align}
	Q_\text{long} = \sum\limits_{t = \SI{0}{\nano\second}}^{\SI{420}{\nano\second}}V(t) \quad \text{and} \quad Q_\text{tail} = \sum\limits_{t = 60\text{ns}}^{420\text{ns}}V(t)\label{eq:quenching:ej-integrals}
\end{align}
where $V(t)$ represents the digitized voltage values of the \ej299/ output and $t=\SI{0}{\nano\second}$ denotes the pulse onset (Fig.~\ref{fig:quenching:example-waveform}). The long integral $Q_\text{long}$ is a measure of the total energy deposited in the detector. In contrast, the relative amount of charge contained in the tail integral $Q_\text{tail}$ depends on the type of the incident particle and is used for $n$-$\gamma$ \ac{psd}. The timings for both integrals were chosen following \cite{pozzi-02} to maximize the $n$-$\gamma$ discrimination capability for this particular scintillator.\par

\begin{figure}[htbp]
\begin{center}
\includegraphics[scale=1]{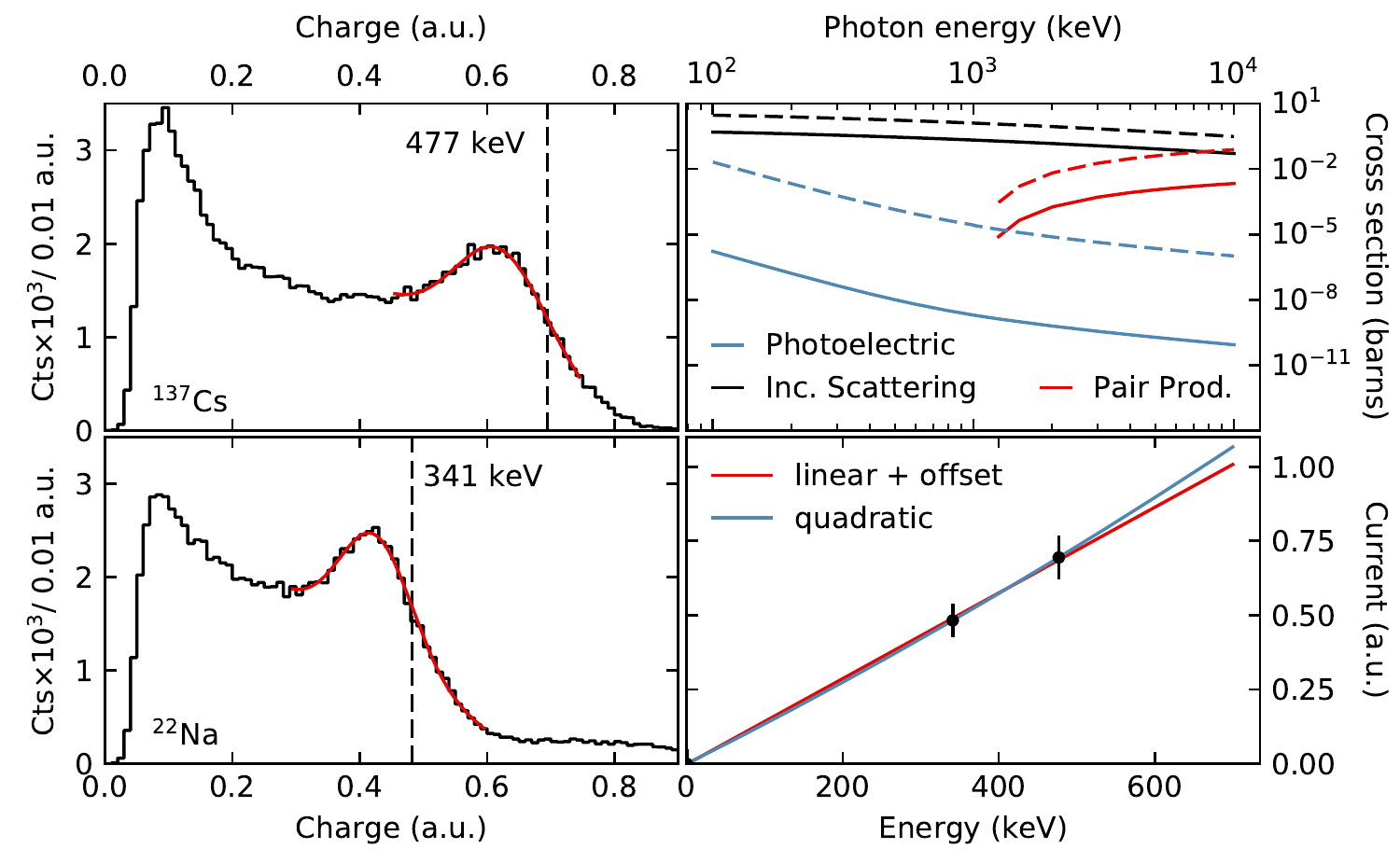}
\end{center}
\caption[\ej299/ energy calibration and photon cross-sections]{\textbf{Left top \& bottom}: Energy calibration spectra recorded with the \ej299/ detector for two different $\gamma$ sources. Due to the limited energy resolution the Compton edge is not perfectly well defined. The edge was defined as the $1\sigma$ deviation from the peak (dashed lines). \textbf{Top right}: Hydrogen (solid) and carbon (dashed) $\gamma$-ray cross sections. Data taken from \cite{xcom}. \textbf{Bottom right}: Linear and quadratic fit to the \ej299/ calibration data. There is only a minor ($\leq\SI{5}{\percent}$) difference between both calibration types for energies below \SI{500}{\keVee}.}
\label{fig:quenching:ej-calibration}
\end{figure}

First an energy calibration was performed using the \isotope{Na}{22} and \isotope{Cs}{137} gamma sources. These provide $\gamma$-lines at \SI{511}{\keV} and \SI{661.7}{\keV}, respectively. Due to the low $Z$ number of both hydrogen and carbon and the relatively low $\gamma$ energies for both sources, only the Compton edge and not a full energy peak is visible in the spectrum. The top right panel of Fig.~\ref{fig:quenching:ej-calibration} shows the individual cross sections for the photoelectric effect, incoherent scattering and pair production in hydrogen and carbon. It can be observed that the scattering cross section in hydrogen is almost seven orders of magnitude larger than the one for the photoelectric effect at the energies of interest. In addition, the $\gamma$ energies are still well below the energy required for pair production. The maximum energy transfer in a single scatter event is given by
\begin{align}
E_T = E_\gamma\left(1-\frac{1}{1+\frac{2E_\gamma}{m_ec^2}}\right).
\end{align}
The expected Compton edge energies are given by \SI{341}{\keV} and \SI{477}{\keV} for a \isotope{Na}{22} and \isotope{Cs}{137} source, respectively. Due to the poor energy resolution of \ej299/, both Compton edges appear as broad peaks. The Compton edge was therefore defined as the $1\sigma$ positive deviation from the maximum of the edge (vertical dashed lines in the left, top and bottom panels of Fig.~\ref{fig:quenching:ej-calibration}). A linear fit to the data points, shown in the right of Fig.~\ref{fig:quenching:ej-calibration}, provides a charge-energy conversion of
\begin{align}
	E[\text{keV}] = 693.95\,Q_\text{long}[\text{a.u.}] + 1.13.
\end{align}
Fitting a quadratic charge-energy relation to the data showed no significant deviation, i.e., $\leq\SI{5}{\percent}$, between \SI{30}{\keVee} and \SI{500}{\keVee}. This backs up the linearity of the \ej299/ detector and substantiates the choices made in the definition of the Compton edges.\par

\begin{figure}[htbp]
\begin{center}
\includegraphics[width=6in]{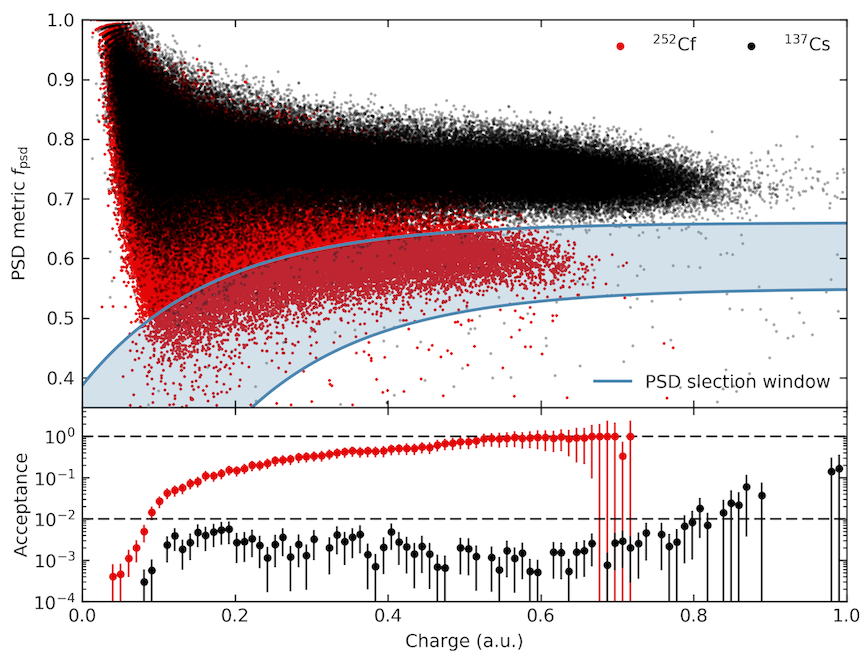}
\end{center}
\caption[Pulse shape discrimination metric for \isotope{Cf}{252} and \isotope{Cs}{137}]{\textbf{Top}: \ac{psd} metric $f_\text{psd}$ as defined in Eq.~(\ref{eq:quenching:psd-metric}) for the \ej299/ \acf{psd} calibrations using \isotope{Cf}{252} (red) and \isotope{Cs}{137} (black). The \isotope{Cs}{137} only emits $\gamma$s and as a result only electronic recoils were induced in the \ej299/ in the presence of this source. The electronic recoil band covered by the \isotope{Cs}{137} data is apparent. The \isotope{Cf}{252} emits both $\gamma$s and neutrons. As a result, events measured in the \ej299/ in the presence of the \isotope{Cf}{252} source are composed of nuclear and electronic recoils. The electronic recoil component is masked by the black data. The nuclear recoil component forms a separate band below the electronic band. The \isotope{Cf}{252} data was taken without the \SI{6}{\dB} attenuator. To directly compare it to the \isotope{Cs}{137} data set, its signal was divided by 1.995 prior to the analysis. The \isotope{Cf}{252} data only covers charges of up to $0.7\,$a.u. as events with higher charges experienced digitizer range overflows. The shaded blue region shows the \ac{psd} cut contour. Only events within this band were accepted as nuclear recoils. \textbf{Bottom}: Fraction of events accepted by the \ac{psd} cut for both data sets. For a wide range of energies over \SI{99}{\percent} of the $\gamma$-rays from the \isotope{Cs}{137} calibration are rejected.}
\label{fig:quenching:psd-feature}
\end{figure}

During the quenching factor measurements the \ej299/ triggered with a rate of approximately \SI{250}{\Hz}. Most of these triggers originated from the ambient radiation field, i.e., $\gamma$s not related to the beam. The environmental background can be reduced by exploiting the $n$-$\gamma$ \ac{psd} capabilities of the \ej299/ plastic scintillator. Due to the different decay times \cite{pozzi-02,ej299-data-sheet} for nuclear (13, 50, \SI{460}{\ns}) and electronic (13, 35, \SI{270}{\ns}) recoils, the percentage of charge visible in the tail region of an event is different for these different types of interactions. Following \cite{pozzi-02}, the \ac{psd} metric below was adopted:
\begin{align}
f_\text{psd} = \frac{Q_\text{long} - Q_\text{tail}}{Q_\text{long}}\label{eq:quenching:psd-metric},
\end{align}
where $Q_\text{long}$ and $Q_\text{tail}$ are defined as in Eq.~(\ref{eq:quenching:ej-integrals}). $f_\text{psd}$ ranges from 0 to 1. Given the aforementioned decay times, events produced by electronic recoils experience a higher $f_\text{psd}$.\par

Two \ac{psd} calibration measurements were obtained using a \isotope{Cs}{137} and \isotope{Cf}{252} source. The \isotope{Cs}{137} source only emits $\gamma$-rays. As a result all interactions measured in the \ej299/ in the presence of this source only consist of electronic recoils. When the \ac{psd} feature $f_\text{psd}$ of each event is plotted as a function of its energy $E$ measured in the \ej299/ these events form a band (black data points in Fig.~\ref{fig:quenching:psd-feature}). In contrast, the \isotope{Cf}{252} source emits both $\gamma$-rays and neutrons. As a result both electronic and nuclear recoils are induced in the \ej299/. Plotting $f_\text{psd}$ as a function of the measured energy for this data set results in the formation of two bands. One associated with nuclear recoils and one with electronic recoils (red data in Fig.~\ref{fig:quenching:psd-feature}). Due to the magnitude of the environmental $\gamma$ background, a \ac{psd} cut was implemented to provide $\geq\SI{99}{\percent}$ $\gamma$ rejection. The cut is shown in shaded blue in Fig.~\ref{fig:quenching:psd-feature}. The acceptance fraction is shown in the bottom panel of the same figure.

\section{Quenching factor data analysis}
\label{section:quenching-factor:qf-analysis}
\begin{figure}[htb]
\begin{center}
\includegraphics[scale=1]{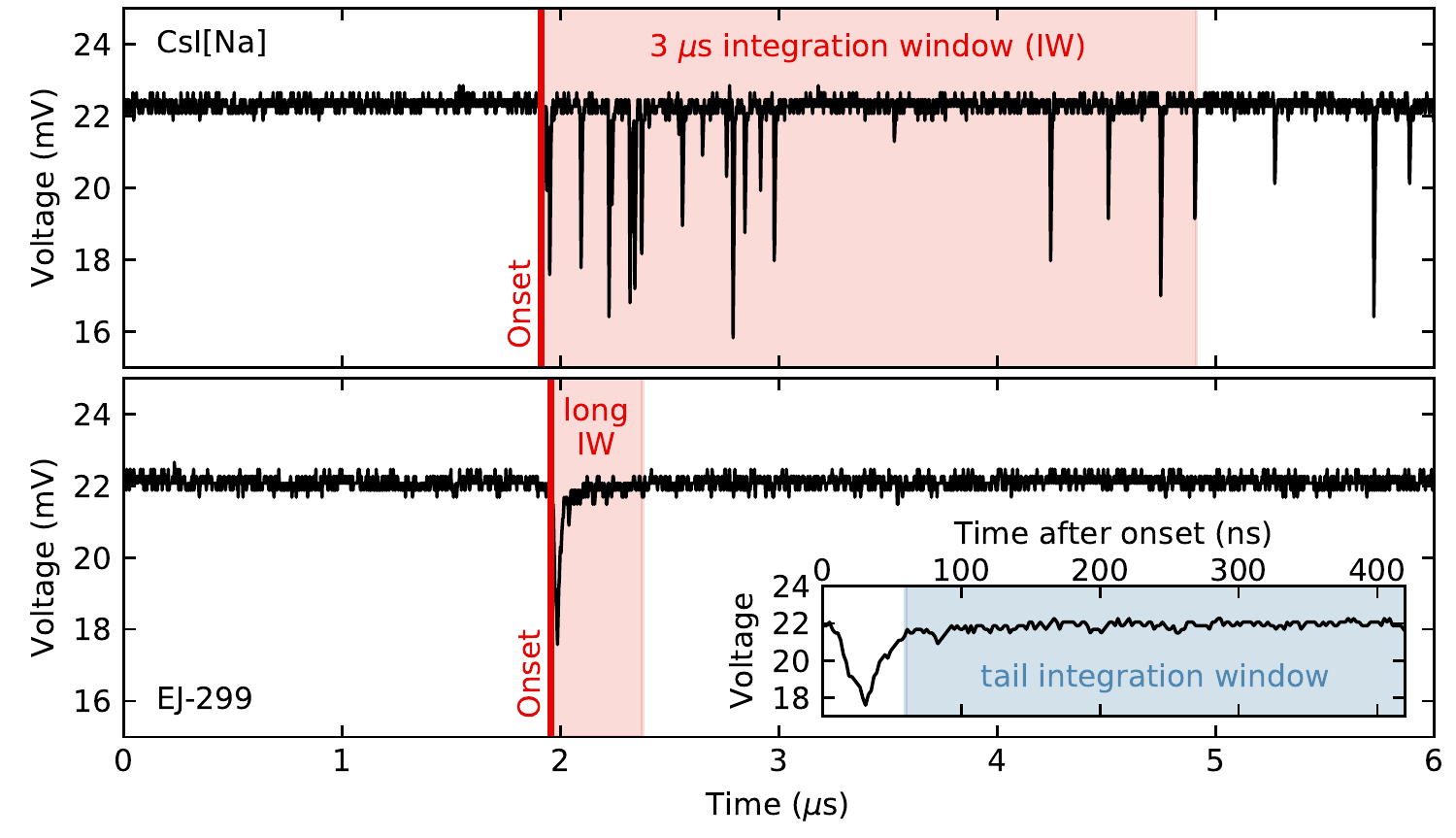}
\end{center}
\caption[Example waveform taken from the \SI{45}{\degree} data set of the quenching factor measurement]{Example waveform taken from the \SI{45}{\degree} data set of the quenching factor measurement. The \csi/ signal is shown on the top, the EJ299 signal on the bottom. The onset (solid red line) in the \csi/ crystal happened approximately \SI{34}{\nano\second} prior to the \ej299/ onset, compatible with the lag time expected for an event triggered by a scattered neutron. The \SI{3}{\micro\second} long integration window applied to the \csi/ signal is shown in shaded red as is the long integration window $Q_\text{long}$ for the \ej299/. The inset in the lower plot shows the \ej299/ signal for the full long integration window and highlights the tail integration window $Q_\text{tail}$ in shaded blue. It can be observed that the main emission is omitted in the tail integral. Due to the different decay times for nuclear and electronic recoils, the ratio between the $Q_\text{long}$ and $Q_\text{tail}$ can be used for particle identification as shown in Eq.~(\ref{eq:quenching:psd-metric})~\cite{pozzi-02}.}
\label{fig:quenching:example-waveform}
\end{figure}

Quenching factor data was taken using the Ortec 934 \ac{cfd} output as trigger signal for the data acquisition system used in the quenching factor measurements. The trigger position was set to \SI{2}{\micro\second} into the digitized waveforms (Fig.~\ref{fig:quenching:example-waveform}). First, a baseline for each trigger was determined for both channels as the median of the first \SI{1}{\micro\second}. Second, the onset $t_\text{ej}$ of the \ej299/ pulse was identified as described in section~\ref{sec:quenching:detector-calibrations}. The \csi/ trace was scanned for \ac{spe} between $t_\text{ej}-\SI{1}{\micro\second}$ and $t_\text{ej}+\SI{1}{\micro\second}$, where a \ac{spe} is defined as at least three consecutive samples above \SI{0.6}{\mV}. The first \ac{spe} marks the onset $t_\text{csi}$ of a potential \csi/ signal. The lag time is given by $\Delta t\,=\,t_\text{csi}-t_\text{ej}$, where a negative lag time corresponds to an event with a \csi/ onset prior to the trigger. Given a neutron beam energy of $\sim\SI{3.8}{\MeV}$ (section~\ref{section:quenching-calibration:simulations}), a lagtime of $\Delta t \approx \SI{-37}{\ns}$ for a measured distance of \SI{1}{\m} between the detectors is expected. After determining $t_\text{csi}$ the \csi/ signal is integrated for a total of \SI{3}{\micro\second}, where only samples associated with a \ac{spe} contribute to the integral. The total charge $Q_\text{total}$ is converted into a number of photoelectrons $N_\text{pe}$ using either \qspe/(polya) or \qspe/(gauss) (Eq.~\ref{eq:quenching-factor:qspe}).\par

\subsection{Determining the experimental residual spectrum}
\label{section:quenching-calibration:fast-scintillator}
\begin{figure}[tbp]
\begin{center}
\includegraphics[width=6in]{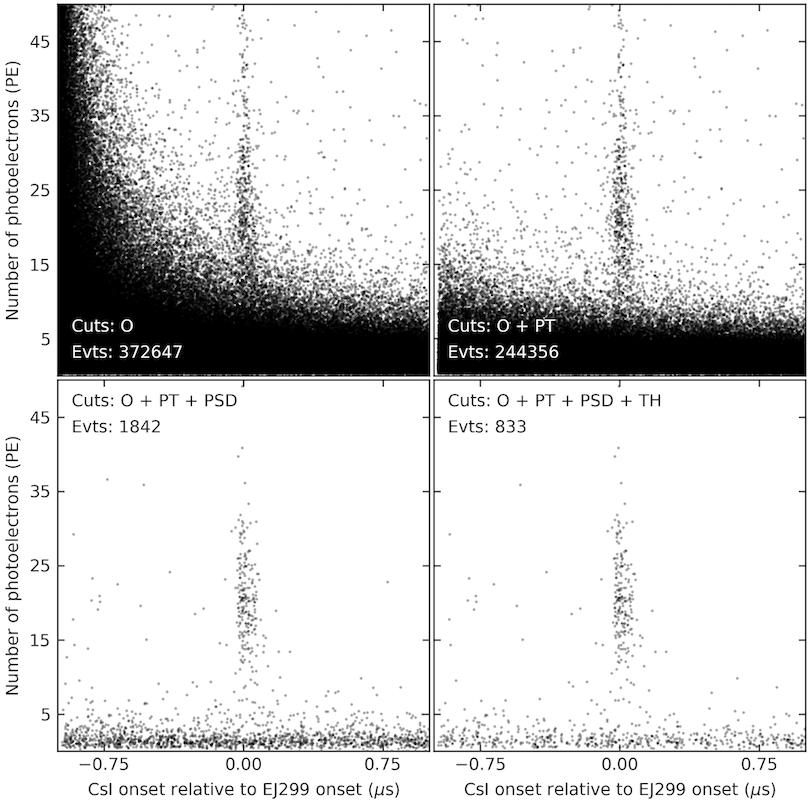}
\end{center}
\caption[Applying quality cuts on the data set acquired for a \ej299/ angle of \SI{45}{\degree} in the quenching factor measurements]{Applying quality cuts on the data set acquired for a \ej299/ angle of \SI{45}{\degree} in the quenching factor measurments. The \qspe/(gauss) model (section~\ref{sec:quenching:detector-calibrations}) was used for the charge to \ac{spe} conversion. Negative values on the x-axis indicate events for which the onset in the \csi/ occurred before the \ej299/ trigger. The cuts applied to each data set are shown in the respective plots as is the number of total events passing them. After all cuts are applied, a clear excess around \SI{0}{\micro\second} and \SI{20}{\pe} becomes evident. The cuts are abbreviated as follows. \textbf{O}: Events showing no digitizer overflow. \textbf{PT}: Events with at most one peak in the pretrace. \textbf{PSD}: Events showing a neutron-like energy deposition in the \ej299/. \textbf{TH}: Events with a minimum energy deposition of \SI{400}{\keVee} in the \ej299/.}
\label{fig:quenching:cuts}
\end{figure}

In order to optimize the signal-to-background ratio, a number of different cuts were applied to the data. First, all events with digitizer range overflow in either channel are excluded. Second, triggers showing more than one peak in the pretrace, i.e., in the first \SI{1}{\micro\second} are removed. Third, only events with a neutron-like \ac{psd} feature $f_\text{psd}$ are accepted. Last, a threshold cut on the energy deposited in the plastic scintillator is applied. All of these cuts are independent of the energy deposited in the \csi/ and therefore do not contribute to any acceptance bias. The impact of all of these cuts on the data set is exemplified in Fig.~\ref{fig:quenching:cuts} for the \SI{45}{\degree} data set. No clear excess is visible in the top left panel, whereas a clear signal arises after all cuts were applied in the lower right panel at $\Delta t \approx \SI{0}{\micro\second}$ and $N_\text{pe}\approx 20$.\par

To define a signal region in $\Delta t$, a relevant time limit for which an event in the \csi/ could have actually been caused by a scattered neutron needs to be defined. The \ac{tof} between the \csi/ and \ej299/ detectors can be calculated for non-relativistic neutrons using the energy-velocity relationship 
\begin{align}
E = \frac{1}{2}mv^2\quad\rightarrow\quad v\left(\frac{\text{cm}}{\text{s}}\right) = 1.3822\cdot 10^6 \sqrt{E\left(\text{eV}\right)}
\end{align}
For a neutron energy of \SI{3.8}{\MeV} (section~\ref{section:quenching-calibration:simulations}) and a measured distance between the detectors of one meter, the \ac{tof} is approximately $\SI{37}{\ns}$. However, the spread in neutron energy and the change in distance for different \ej299/ angles $\alpha$, the signal window onset was defined as $t_0 = \SI{-62}{\ns}$ with respect to the \ej299/ hardware trigger. Due to the low amount of energy deposited in the crystal and the low number of \ac{spe} created, the stochastic nature of the light emission of these \ac{pe} can lead to a non-negligible spread in the arrival time $T_\text{arr}$ of the first \ac{spe} after the interaction. Assuming a double exponential scintillation decay profile as measured in \cite{collar-02}, the arrival probability density function of an \ac{spe} can be written as
\begin{align}
P_\text{spe}(t)\;=\;\frac{1}{1+r}\;\frac{\Exp{-\frac{-t}{\tau_\text{fast}}}}{\tau_\text{fast}} + \frac{r}{1+r}\;\frac{\Exp{-\frac{-t}{\tau_\text{slow}}}}{\tau_\text{slow}},\label{eq:quenching:pe-arrival}
\end{align}
where for nuclear recoils $\tau_\text{fast}\,=\,\SI{589}{\nano\second}$,  $\tau_\text{slow}\,=\,\SI{6.7}{\micro\second}$, and  $r\,=\,0.41$ (Fig.~\ref{fig:csi-setup:csi-decay-times})~\cite{collar-02}. The probability of an event with $N_\text{pe}$ photoelectrons to show an arrival time of $T_\text{arr} < \tau_i$ can be calculated. It is
\begin{align}
P_\text{arr}\left(T_\text{arr} < \tau_i,N_\text{pe}\right)\,=\,1\,-\,\left(1\,-\int\limits_{t'=0}^{\tau_i}P_\text{spe}(t')\text{d}t'\right)^{N_\text{pe}}.
\end{align}

For each $N_\text{pe}\in[1,50]$, the time after interaction $\tau_\text{99}$ for which \SI{99}{\percent} of neutron induced signals already have shown their first \ac{pe}, i.e., $P_\text{arr}(T_\text{arr} < \tau_{99})=0.99$ was calculated.

\begin{figure}[tbp]
\begin{center}
\includegraphics[scale=1]{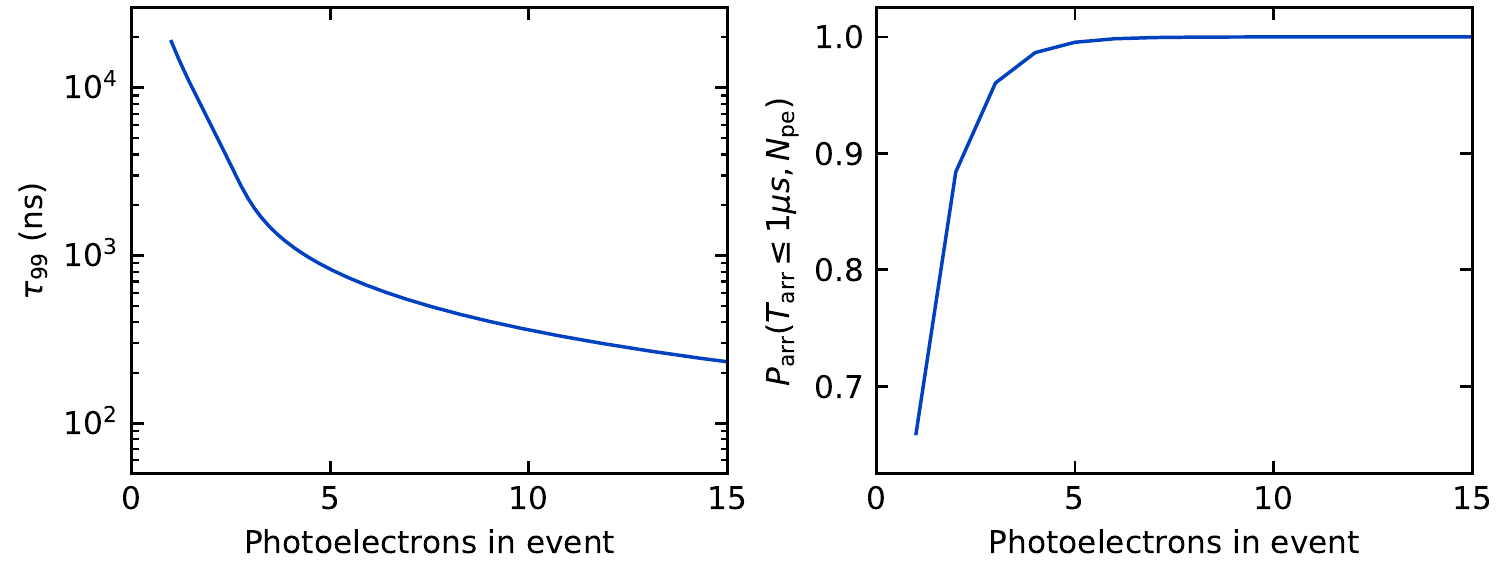}
\end{center}
\caption[Theoretical arrival times for events in the quenching factor measurements]{Theoretical arrival times for events in the quenching factor measurements. \textbf{Left}: \SI{99}{\percent} of all events in the quenching factor measurements with a given number of photoelectrons show an arrival time of equal or less than $\tau_{99}$. A rapid decay for $N_\text{pe}\leq 5$ can be observed. In the \ac{cenns} search (chapter~\ref{chapter:sns-analysis}) only events with at least eight individual peaks are accepted. This substantiates that the onset of events in the  data is well defined and as a result does not suffer from the spread experienced in this quenching factor measurement. \textbf{Right}: Percentage of events with an arrival time of one microsecond or less. Events with at least five $N_\text{pe}$ are fully contained in the analysis window.}
\label{fig:quenching:pe-arrival-times}
\end{figure}

The left panel of Fig.~\ref{fig:quenching:pe-arrival-times} shows the evolution of $\tau_{99}$ for an increasing number of \ac{pe} created in an event. A rapid decline in $\tau_{99}$ for $N_\text{pe} \leq 4$ and a more gradual decline for $N_\text{pe} > 4$ are visible. It also becomes evident that $\tau_{99}>\SI{1.037}{\micro\second}$ for $N_\text{pe}\leq4$, yet the analysis window is limited to arrival times of $T_\text{arr}\,\leq\,\SI{1.037}{\micro\second}$. The right panel of Fig.~\ref{fig:quenching:pe-arrival-times} shows the percentage of events exhibiting an arrival time of \SI{1.037}{\micro\second} or less. For events with 1, 2, 3 or 4 \ac{pe} this probability, i.e. $P_\text{arr}(T_\text{arr} < \SI{1.037}{\micro\second})$, is 65.9, 88.4, 96.0 and \SI{98.6}{\percent}, respectively. Events with more than four \ac{pe} exhibit an arrival time of less than \SI{1.037}{\micro\second} more than \SI{99}{\percent} of the time. An upper bound for the signal acceptance window can be defined as
\begin{align}
t_\text{max}\left(N_\text{pe}\right) =
  \begin{cases}
   t_0 + \tau_{99} + \SI{50}{\ns} & \text{for}\,\,N_\text{pe} \geq 5 \\
   t_0 + \SI{1.037}{\micro\second} + \SI{25}{\ns}  & \text{otherwise.}
  \end{cases}
\end{align}

\begin{figure}[htbp]
\begin{center}
\includegraphics[scale=1]{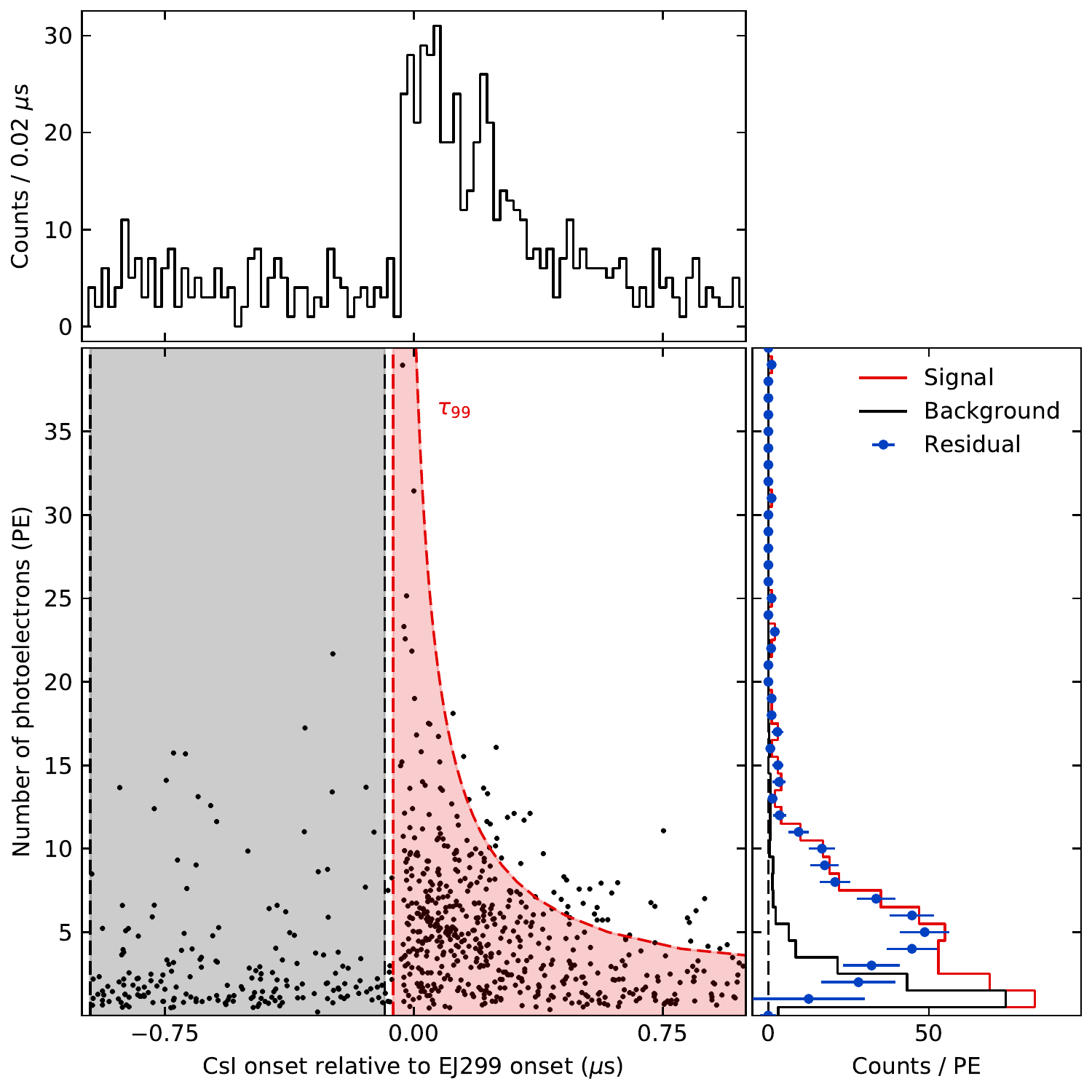}
\end{center}
\caption[Calculating the residual energy spectrum for the \SI{24}{\degree} data set of the quenching factor measurement]{Calculating the residual energy spectrum for the \SI{24}{\degree} data set of the quenching factor measurement. \textbf{Central panel}: Shown are events passing all cuts. The shaded red region marks lag times for possible neutron scatter events. Shaded black regions mark random backgrounds. \textbf{Top panel}: The excess caused by neutron recoils at $\Delta t\,\approx\,0$ as well as the spread in arrival times due to the low number of \ac{pe} involved can be observed. \textbf{Right panel}: Energy spectra for both signal (red) and background (black) regions. The background event spectrum was scaled to match the total time over which the signal region was marginalized. The acceptance was corrected for the resulting residual spectrum (blue).}
\label{fig:quenching:residual-calculation}
\end{figure}

The additional \SI{50}{\ns} in the first case offset the \SI{25}{\ns} buffer introduced in $t_0$ and provide another \SI{25}{\ns} buffer at the end of the window. However, the signal window for the second case already covers the full analysis window in $\Delta t$. \SI{25}{\ns} are added to offset the buffer in $t_0$. The resulting signal region in $\Delta t$ is shown in shaded red in Fig.~\ref{fig:quenching:residual-calculation}. By definition at least \SI{99}{\percent} of events for a given number of \ac{pe} exhibit arrival times within the signal region. The exception are events with 1-4 \ac{pe}. In addition, a background region is defined as $\Delta t\in[-975,-80]\SI{}{\ns}$, containing random coincidences only, shown in shaded black. Both regions are marginalized over $\Delta t$ and binned in $1\,$\ac{pe} wide bins. Each bin of the resulting background spectrum is scaled to match the exposure of the signal region, i.e.
\begin{align}
\eta\left(N_\text{pe}\right) = \frac{t_\text{max}\left(N_\text{pe}\right)-t_0}{\SI{895}{\ns}}.
\end{align}

The scaled background spectrum is subtracted from the signal spectrum, removing any contribution from random coincidences within the signal region. The resulting residual spectrum therefore only includes events caused by nuclear recoils induced by neutrons scattering off \csi/ nuclei.\par

For events with $N_\text{pe}\in[1,4]\,\SI{}{\pe}$ the residual needs to be scaled to match the efficiency of all other bins due to the limited analysis window available. A correction factor of $\nicefrac{99}{65.9}$, $\nicefrac{99}{88.4}$, $\nicefrac{99}{96.0}$ and $\nicefrac{99}{98.6}$ was applied for 1 to 4 $N_\text{pe}$, respectively. The right panel of Fig.~\ref{fig:quenching:residual-calculation} shows an example of all three spectra. The red (black) histogram shows the marginalized spectrum for the signal (background) region, where the background data was already scaled to match the exposure of the signal region. The blue data points show the acceptance corrected residual spectrum.\par

\subsection{Simulating the detector response using \mcnp/}
\label{section:quenching-calibration:simulations}
The effective quenching factor for a given recoil energy can be calculated by comparing the experimental residual spectrum with a simulated spectrum of energy depositions within the \csi/ detector. Consequently, a comprehensive \mcnp/ \cite{pozzi-01} simulation spanning $10^9$ neutron histories for each of the seven \ej299/ positions was run. The simulated geometry is shown in the left-hand side of Fig.~\ref{fig:quenching:setup}, where the collimator is comprised of a concrete wall and a paraffin shield. Both detector geometries were implemented including their specific encapsulation. The energy spectrum of the incident neutron beam was derived by analyzing neutron \ac{tof} data taken during the measurements \cite{grayson-01}, as well as by evaluating simulations of the deuterium-deuterium reaction in the gas cell~\cite{tunl-01}.\par

\begin{figure}[tbp]
\begin{center}
\includegraphics[scale=1]{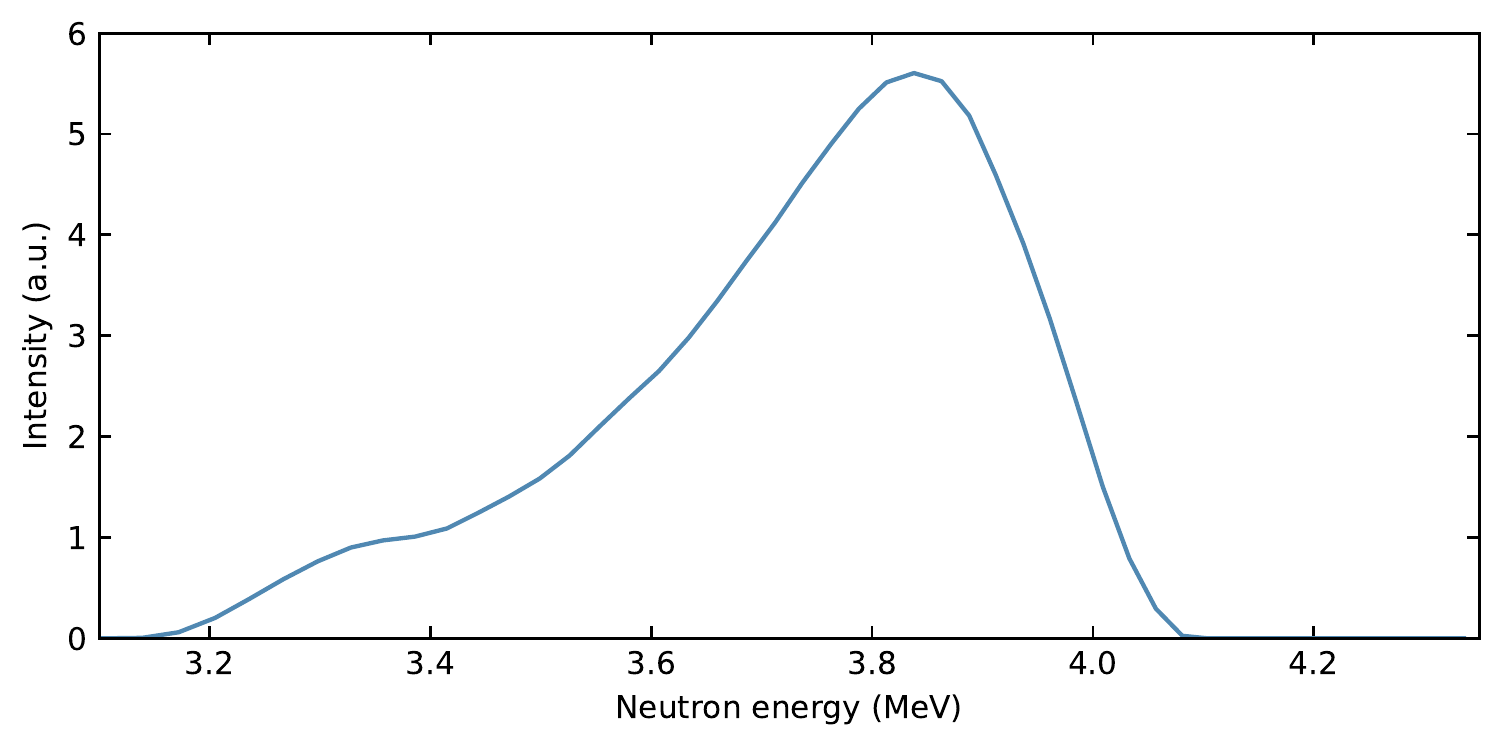}
\end{center}
\caption[Neutron beam energy spectrum used in the \mcnp/ simulations of the quenching factor measurments.]{Neutron beam energy spectrum as determined using both \ac{tof} measurements and simulations of the deuterium-deuterium reaction in the gas cell. Data provided by Grayson Rich (University of North Carolina at Chapel Hill) \cite{grayson-01}. This spectrum was used as the initial neutron energy distribution in the \mcnp/ simulations of the quenching factor measurments.}
\label{fig:quenching:neutron-beam-energy-spectrum}
\end{figure}

\begin{figure}[tb]
\begin{center}
\includegraphics[scale=1]{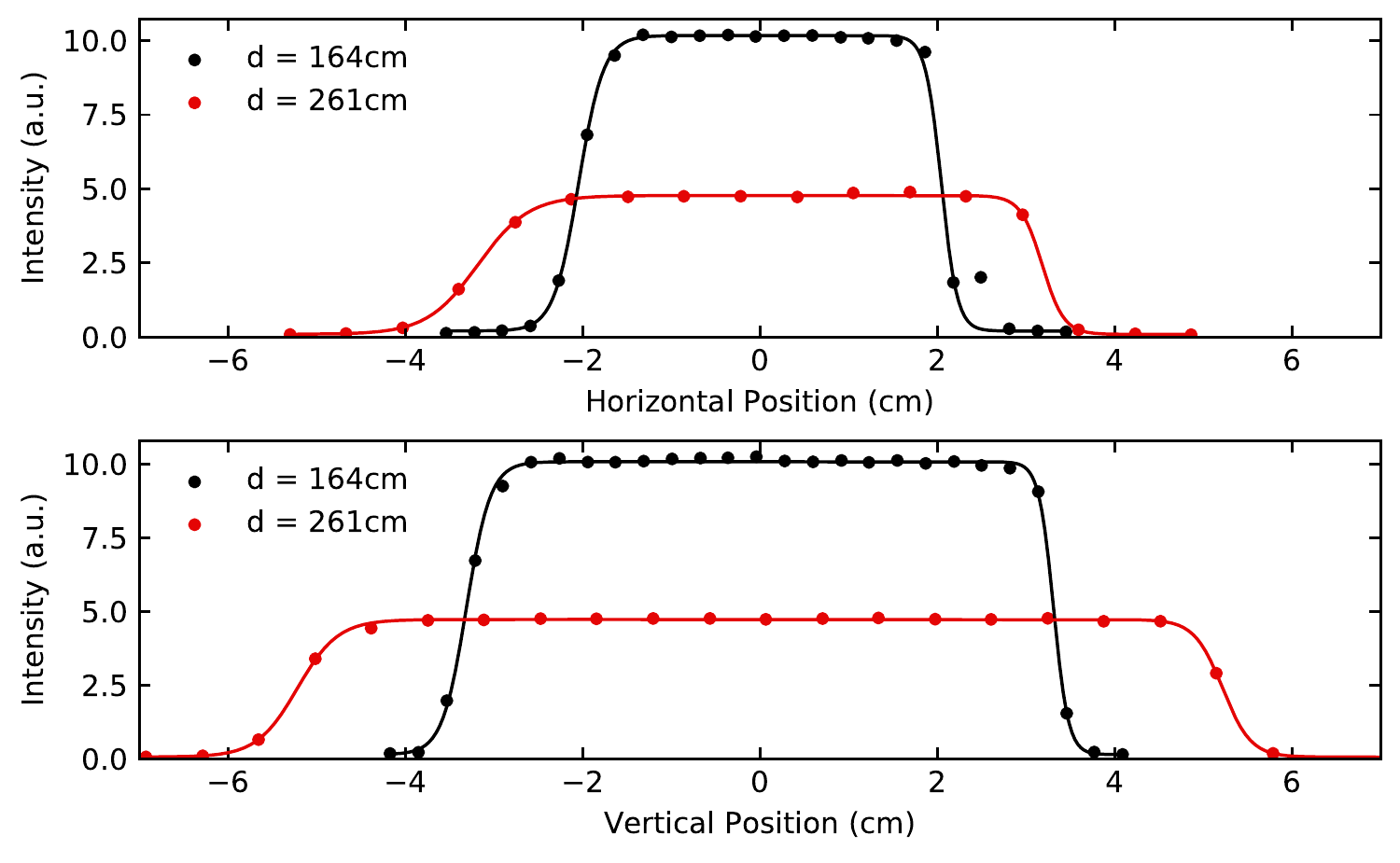}
\end{center}
\caption[Neutron beam transversal profile available in the \acs*{ssa} at \acs*{tunl}]{Transversal neutron beam profile available in the \acf*{ssa} at \acs*{tunl} as measured at a distance of \SI{164}{\cm} (black) and \SI{261}{\cm} (red) from the tip of the gas cell. The end of the plastic collimator (Fig.~\ref{fig:quenching:setup}) is located at \SI{145.7}{\cm}. The solid lines represent a fit of two sigmoids to the data in order to estimate the FWHM. The beam divergence according to these measurements was implemented in the \mcnp/ simulations.}
\label{fig:quenching:neutron-beam-profile}
\end{figure}

The resulting energy spectrum of the incident neutron beam, which was ultimately implemented in the simulations, is shown in Fig.~\ref{fig:quenching:neutron-beam-energy-spectrum}. The spectrum peaks at approximately \SI{3.8}{\MeV} and has a \ac{fwhm} of \SI{0.4}{\MeV}. The two-dimensional beam profile was measured at two different stand-off distances from the gas cell and is shown in Fig.~\ref{fig:quenching:neutron-beam-profile}. A beam profile and spread according to these measurements was included in the simulations.\par

\begin{figure}[tbp]
\begin{center}
\includegraphics[scale=1]{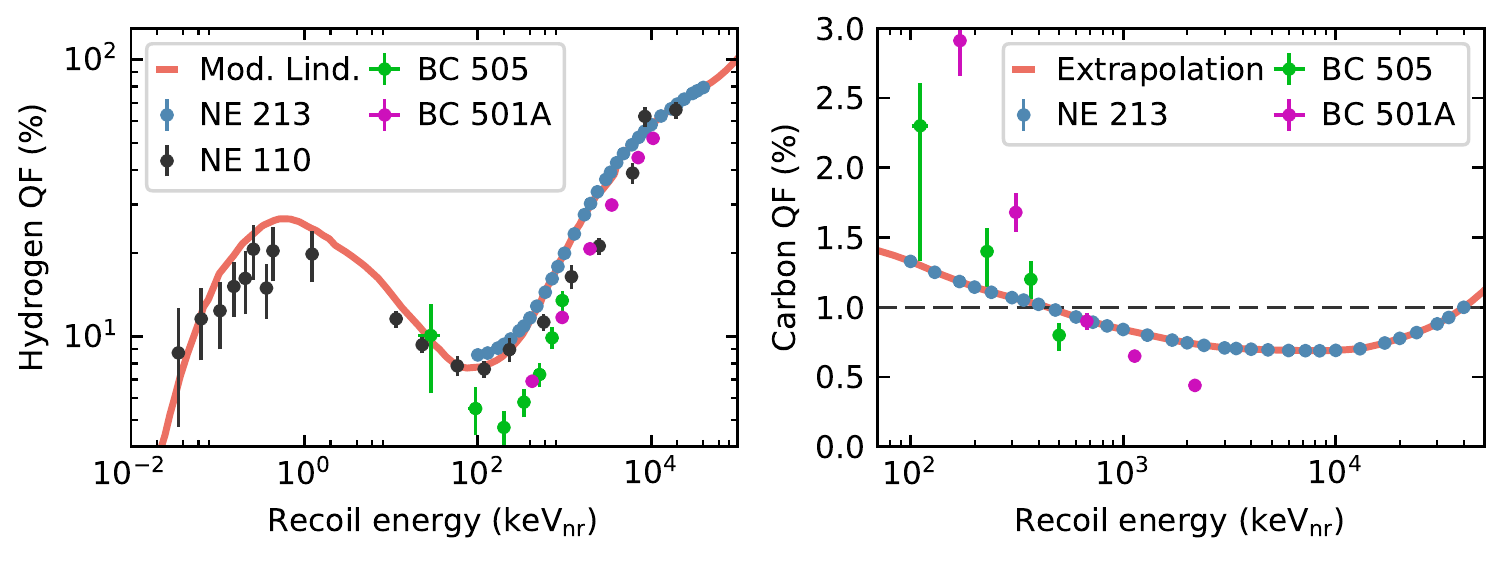}
\end{center}
\caption[Hydrogen and carbon quenching factors for different organic scintillators]{Hydrogen and carbon quenching factors for different organic scintillators. NE-213 is a liquid scintillator and is equivalent to EJ-301 and BC-501 \cite{eljen-01}. BC-505 is another liquid scintillator whereas NE-110 is a plastic scintillator \cite{eljen-02}. The density of hydrogen and carbon atoms for each of these scintillators is comparable and on the order of \SI{5e22}{\atom\per\cubic\cm}. Data shown for NE-110 is taken from \cite{ficenec-01}, NE-213 from \cite{verbinski-01}, BC-501A from \cite{yoshida-01} and BC-505 from \cite{hong-01}. The left panel uses a logarithmic scale, whereas the right panel uses a linear scale. The modified Lindhard model (shown as red curve in the left panel) was adopted for hydrogen recoils. A constant quenching factor of \SI{1}{\percent} was assumed for carbon recoils.}
\label{fig:quenching:h-c-quenching-factor}
\end{figure}

For each simulated neutron history, scattering at least once in both detectors, the energy deposited at each vertex for both detectors, as well as the type of recoiling nucleus was recorded. For scatter vertices happening within the \ej299/ detector, the energy deposited was converted from \SI{}{\keVnr} to \SI{}{\keVee} using a modified Lindhard model for hydrogen recoils \cite{ficenec-01,collar-03} and a constant quenching factor of \SI{1}{\percent} for carbon recoils (Fig.~\ref{fig:quenching:h-c-quenching-factor}), \cite{yoshida-01, hong-01, verbinski-01}.\par

No significant difference was observed for the quenching factor results if the carbon quenching factor is extrapolated as shown in red in Fig.~\ref{fig:quenching:h-c-quenching-factor}. This is unsurprising, as the energy transfer is dominated by proton recoils, with approximately \SI{68}{\percent} of all vertices scatter off of hydrogen and contributing approximately \SI{78}{\percent} of the total unquenched energy in the \ej299/. The sum of all individually quenched recoil energies represents the total energy visible in the plastic scintillator for that particular neutron history.\par

\begin{figure}[tbp]
\begin{center}
\includegraphics[width=6in]{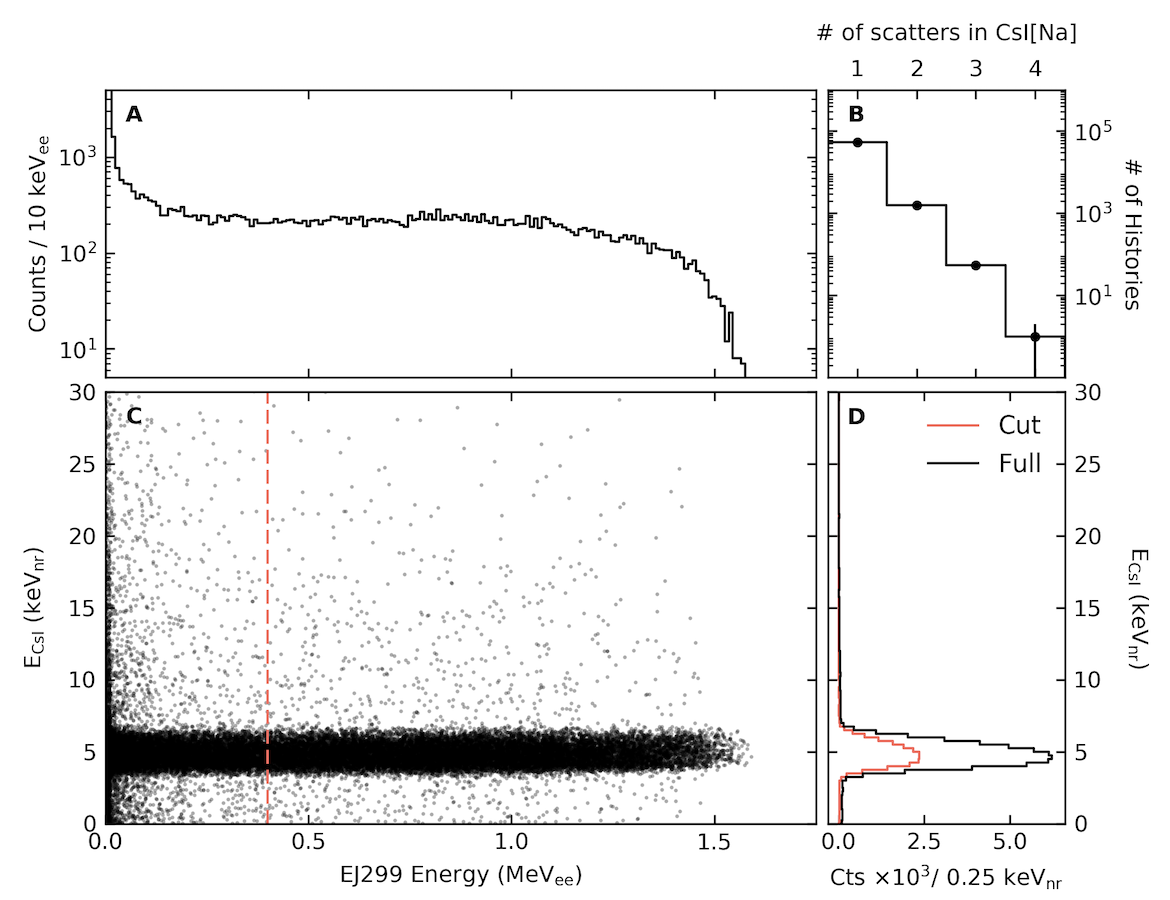}
\end{center}
\caption[Results of the \mcnp/ simulations for a \ej299/ position of \SI{24}{\degree}]{Results of the \mcnp/ simulations for a \ej299/ position of \SI{24}{\degree} in the quenching factor measurements. Panel \textbf{A} shows the energy depositions within the \ej299/ detector. The carbon and proton recoils were converted to ionization energies using the known response of a closely related organic scintillator EJ-301 \cite{verbinski-01} and a modified Lindhard model for low energies \cite{ficenec-01}. Panel \textbf{B} shows the number of times a triggering neutron scatters within the CsI[Na] crystal. Approximately \SI{93}{\percent} of all neutrons scatter only once. Panel \textbf{C} shows the two-dimensional distribution of the unquenched energy deposited in the CsI[Na] detector versus ionization energy deposited in the \ej299/. No correlation between the two energies was found. Panel \textbf{D} shows the unquenched energy deposited in the \csi/ detector. A distinct peak arises from the kinematic selection of scattered neutrons. No change in the peak shape and location is visible if events below \SI{400}{\keVee} (dashed red line in the bottom left plot) are cut.}
\label{fig:quenching:sim-results}
\end{figure}

For the \csi/ detector in contrast approximately $\SI{93.4}{\percent}$ of all histories only scatter once within the crystal due to the small size of the detector (Fig.~\ref{fig:quenching:sim-results} panel \textbf{B}). To simplify the analysis a constant quenching factor for each angle is assumed. This allows summing over the energy deposited at each vertex in the \csi/ crystal to extract the total energy deposited.\par

Fig.~\ref{fig:quenching:sim-results} shows the results of the simulation corresponding to the \SI{24}{\degree} data set. Panel \textbf{A} shows the visible energy in the \ej299/ from beam related neutrons after accounting for the quenching factor of hydrogen and carbon. Panel \textbf{D} shows the unquenched energy spectrum deposited in the \csi/ crystal. A distinct peak at an energy approximately predicted by Eq.~(\ref{eq:quenching:recoil-kinematics}), \SI{5.45}{\keV} in this case, can be observed. Panel \textbf{C} shows the energy deposited in both detectors for each history. The energy deposited in the \csi/ is uncorrelated to the energy deposited in the \ej299/ above $E_\text{ej299}\gtrsim\SI{50}{\keVee}$. A simple threshold cut is therefore truly energy independent and does not introduce any bias.\par

\subsection{Extracting the quenching factor}
To calculate the best estimate for the quenching factor $Q_f$ at a given energy, the simulated recoil spectrum was compared to the experimental residual spectrum.
First, the nuclear recoil energy of every simulated interaction in the \csi/ was converted into a an electron equivalent energy using a constant quenching factor $Q_f$. The electron equivalent energy was further converted into a number of photoelectrons using the the light yield calculated earlier (section~\ref{section:quenching-calibration:csi-calib}).For each neutron history $k$ scattering $m_k$ times within the \csi/ detector, the average number $\mu^k_\text{pe}$ of \ac{pe} produced is given by
\begin{align}
\mu^k_\text{pe} = \sum_{i=1}^{m_k} Q_f\,\mathcal{L}\,E^k_i
\end{align}
where $Q_f$ represents the quenching factor, $\mathcal{L}$ is the light yield and $E^k_i$ is the energy deposited at vertex $i$ for the neutron history $k$. A Poisson spreading was applied to every $\mu^k_\text{pe}$ to calculate the simulated energy spectrum $n$. For each bin $j$, where $j$ is the number of \ac{pe} produced, it is
\begin{align}
n^j = \sum_k P\left( j | \mu^k_\text{pe}\right) \quad\text{with}\quad P\left( x | \mu \right) = \frac{{e^{ - \mu } \mu ^x }}{{x!}}.
\end{align}
The simulated spectrum needs to be scaled to match the exposure acquired in the experiment, i.e., the number of neutrons incident on the \csi/ need to match in both cases. The final simulated energy spectrum $N_\text{sim}$ is therefore given by
\begin{align}
n^j_\text{sim} = n^j\frac{t_\alpha\phi_n}{10^9}, \label{eq:quenching-factor:neutron-flux}
\end{align}
where $t_\alpha$ is the time measured for angle $\alpha$ in seconds and $\phi_n$ the neutron flux provided at the neutron beam facility. The $\chi^2$ goodness of fit statistic for the chosen quenching factor and neutron scaling is
\begin{align}
\chi^2 = \sum_j\frac{\left(n^j_\text{res} - n^j_\text{sim}\right)^2}{n^j_\text{res}},
\end{align}
where $n^j_\text{res}$ and $n^j_\text{sim}$ represent the experimental residual and the simulated spectrum, respectively. The best fit is given by the minimum of the two dimensional $\chi^2$ distribution.\par

\begin{figure}[htb]
\begin{center}
\includegraphics[scale=1]{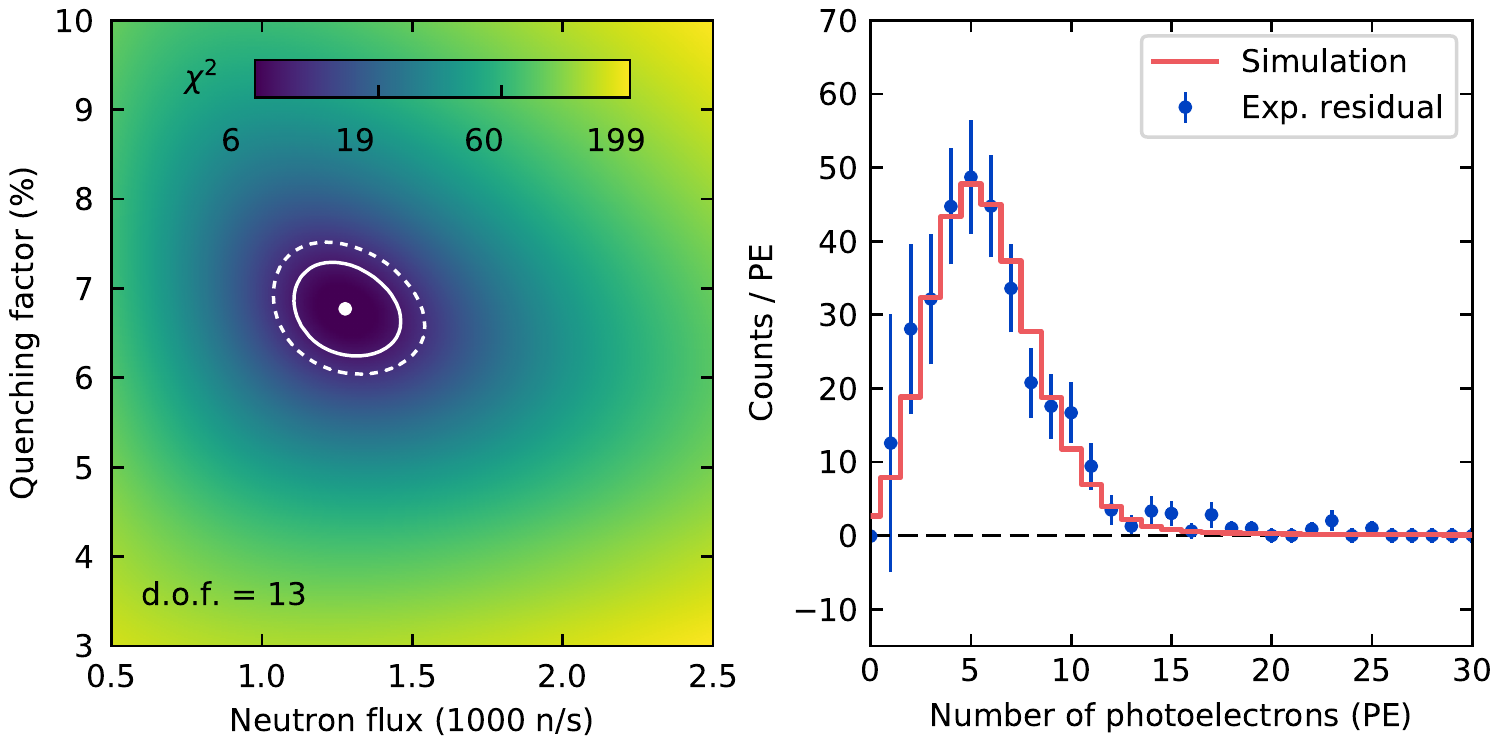}
\end{center}
\caption[Fitting the simulated recoil data to the experimental residual spectrum]{Fitting the simulated recoil data to the experimental residual spectrum for a \ej299/ position of \SI{24}{\degree} in the quenching factor measurements. \textbf{Left}: Two dimensional $\chi^2$ distribution for the quenching factor fit. The white dot marks the best fit. The solid and dashed white lines represent the 1- and 2-$\sigma$ contours respectively, i.e., $\Delta\chi^2=2.30$ and $\Delta\chi^2=4.61$ \cite{press-01}. \textbf{Right}: Best fit of the simulation data (red histogram) to the residuals (blue). The residual data was calculated as shown in Fig.~\ref{fig:quenching:residual-calculation}.}
\label{fig:quenching:fitting-example}
\end{figure}
Fig.~\ref{fig:quenching:fitting-example} shows this process for the \SI{24}{\degree} data set. The two-dimensional $\chi^2$ distribution is shown on the left. A white dot represents the minimum and 1- and 2-$\sigma$ contours are shown as solid and dashed white lines respectively, where the 1- and 2-$\sigma$ contours are defined by an increase of $\chi^2$ with respect to the minimum by $\Delta\chi^2=2.30$ and $\Delta\chi^2=4.61$, respectively \cite{press-01}. The right panel shows the best fit of the simulated recoil spectrum (red) to the experimental residual data (blue). The excellent agreement between simulation and experimental data is readily visible. The fit quality for all other data sets is similar to the one presented.\par

To properly propagate any uncertainties, the fitting procedure was repeated by varying the \ac{spe} mean charge and the light yield within their respective errors. To explore the uncertainty associated with the choice of $t_0$ and $t_\text{max}$ the signal window was varied by $\pm\SI{24}{\ns}$. Further the \ac{psd} cut contours were floated by $\pm\SI{2}{\percent}$ and different threshold cuts were applied on the minimum total energy deposited in the \ej299/, i.e., $E_\text{threshold} \in[250,300,350,400,450,500]$. The best fit values for $Q_f$ and $\phi_n$ as well as the minimum $\chi^2$ were recorded for each of these results.
The best estimate of the quenching factor $Q_f$ and the neutron flux $\phi_n$ is given by the best fit found for the initial choice of cut parameters. In contrast, the uncertainty represents the square root of the unweighted variance of all fit results added in quadrature to the 1-$\sigma$ uncertainty of the initial best fit. The final best fit values for each angle are given in Table~\ref{tab:quenching:quenching-factor-data}.\par

\begin{figure}[tb]
\begin{center}
\includegraphics[scale=1]{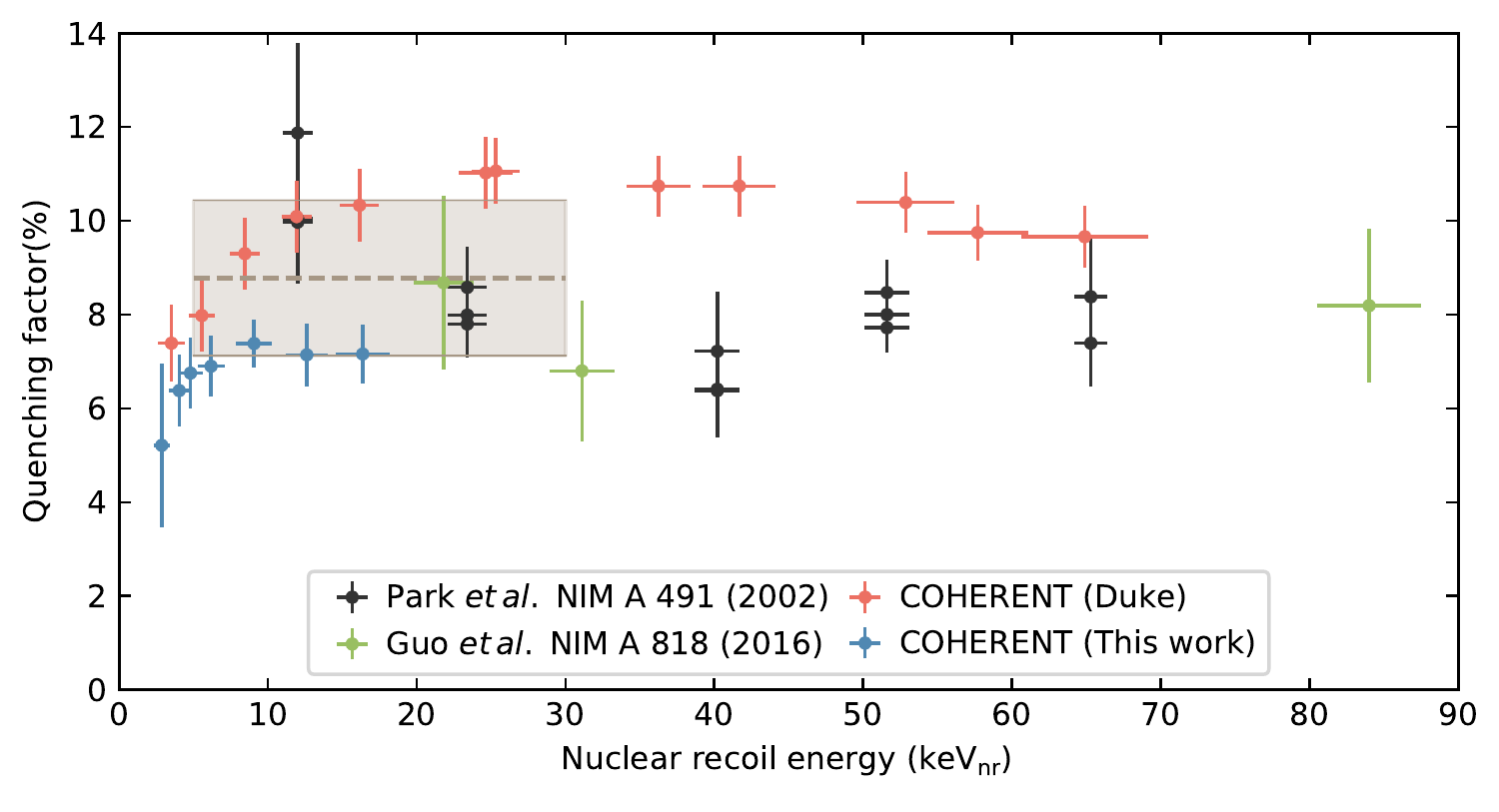}
\end{center}
\caption[Comparison of the \csi/ quenching factor measurements with previous experiments]{Comparison of the \csi/ quenching factor measurement with previous experiments. The results of this thesis are shown in blue. Previous measurements are shown in black \cite{park-01} and green \cite{guo-01}. A second measurement of the quenching factor performed by the \coherent/ collaboration is shown in red. The data for the second measurement was also acquired at \ac{tunl} with the same crystal-\ac{pmt} assembly and neutron source, but differed in data acquisition system and data analysis. The grayed region spans the energy region of interest for the \ac{cenns} search presented in chapter~\ref{chapter:sns-analysis}. The dotted line represents the choice of quenching factor used in the \ac{cenns} search. The shaded area shows the corresponding $1\sigma$ uncertainty band.}
\label{fig:quenching:quenching-factor}
\end{figure}

Fig.~\ref{fig:quenching:quenching-factor} shows the measurement presented in this thesis in blue. A decreasing quenching factor for low energy recoils can be seen, where previous measurements suggested a rise (black, green). The quenching factor measurement provided in \cite{collar-02} was not included as the quenching factor definition employed there slightly differed from the one used in the experiments shown here. A second measurement performed by the Duke group of the \coherent/ collaboration (red) found a slightly higher quenching factor, but also observed a decreasing scintillation efficiency at low energies. As semi-empirical quenching models fail to reproduce this behavior \cite{tretyak-01}, a constant quenching factor is assumed for the energy region of interest in the \ac{cenns} search discussed in this thesis. To define this constant quenching factor, all available data in the \ac{cenns} search region, i.e., $\sim$5-\SI{30}{\keVnr}, was included in the calculation. The lower bound of this energy range is set by the threshold achieved in the \ac{cenns} search (chapter~\ref{chapter:sns-analysis}). The upper bound represents the energy at which the \ac{cenns} recoil rate becomes negligible. The best fit value was determined by weighing each data point with its uncertainty \cite{collar-04}. To estimate the uncertainty on the best fit quenching factor the unweighted standard deviation of all data points was used. The final quenching factor is given by
\begin{align}
Q_f = 8.78 \pm \SI{1.66}{\percent}.
\end{align}
Fig.~\ref{fig:quenching:quenching-factor} shows the quenching factor adopted in the \ac{cenns} search as a dotted brown line. The uncertainty band is shown as a shaded brown region.\par

\begin{table}[htbp]
\begin{center}
\begin{tabular}{cccc}
\toprule
Angle$\,(\SI{}{\degree})$&Recoil energy$\,(\SI{}{\keVnr})$&Quenching factor$\,Q\,(\%)$ & Neutron flux$\,\phi_n\,(1000\,\nicefrac{n}{s})$\\
\midrule
18 & 2.87$\,\pm\,$0.56 & 5.21$\,\pm\,$1.74 & 1.20$\,\pm\,$0.43\\
21 & 4.04$\,\pm\,$0.72 & 6.38$\,\pm\,$0.76 & 1.15$\,\pm\,$0.33\\
24 & 4.80$\,\pm\,$0.80 & 6.75$\,\pm\,$0.75 & 1.31$\,\pm\,$0.25\\
27 & 6.18$\,\pm\,$0.90 & 6.90$\,\pm\,$0.65 & 1.34$\,\pm\,$0.34\\
33 & 9.07$\,\pm\,$1.21 & 7.38$\,\pm\,$0.51 & 1.77$\,\pm\,$0.39\\
39 & 12.61$\,\pm\,$1.43 & 7.14$\,\pm\,$0.67 & 1.24$\,\pm\,$0.41\\
45 & 16.37$\,\pm\,$1.81 & 7.16$\,\pm\,$0.63 & 1.50$\,\pm\,$0.44\\
\bottomrule
\end{tabular}
\end{center}
\caption[\csi/ quenching factor and neutron flux for all \ej299/ positions]{\csi/ quenching factor and neutron flux for all \ej299/ positions. The quenching factor data is also shown in Fig.~\ref{fig:quenching:quenching-factor} where previous measurements by others were included. The neutron flux calculated in this analysis is shown in Fig~\ref{fig:quenching:neutron-flux}, including the flux derived from an in-situ measurement of the inelastic \isotope{I}{127}$\left(n,n^\prime\gamma\right)$ reaction rate.}
\label{tab:quenching:quenching-factor-data}
\end{table}

\begin{figure}[tb]
\begin{center}
\includegraphics[scale=1]{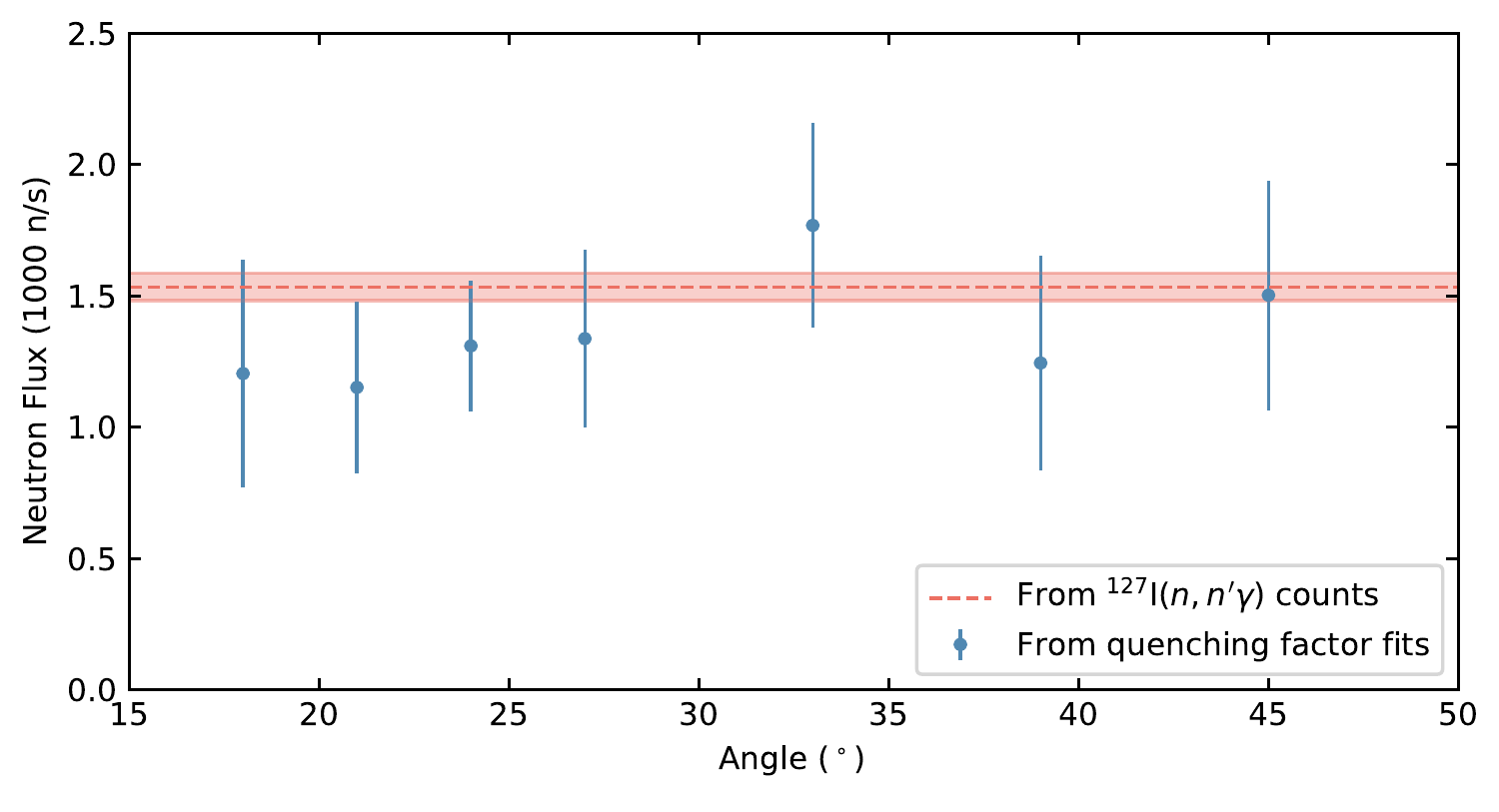}
\end{center}
\caption[Best fit neutron flux for each detector position]{Best fit neutron flux derived for each fit. The large uncertainties are due to the uncertainty in the cut efficiency of the \ac{psd}. The average neutron flux derived from the inelastic counts is $1534\pm53$ neutrons per second, compatible with all positions at the $1\sigma$ level.}
\label{fig:quenching:neutron-flux}
\end{figure}

Fig.~\ref{fig:quenching:neutron-flux} shows the best fit neutron flux derived from the scaling factor (Eq.~(\ref{eq:quenching-factor:neutron-flux})) for each angle. The large uncertainty of the scaling factor mainly arises from the uncertainty of the \ac{psd} cut efficiency. The efficiency shown in Fig.~\ref{fig:quenching:psd-feature} does not reflect the true efficiency for accepting a neutron-induced event as the \isotope{Cf}{252} calibration still contains a large portion of $\gamma$-like triggers. As the acceptance simply compares the number of events surviving the cut after marginalizing over $f_\text{psd}$, the true efficiency should be larger. This error increases for smaller threshold energies. Therefore, the best fit probably slightly underestimates the neutron flux. The constant neutron flux derived for every angular position of the \ej299/ detector substantiates that no threshold effects were introduced in this analysis~\cite{collar-05}.\par

The true neutron flux variation between runs was estimated to be less than \SI{10}{\percent}~\cite{barbeau-04}, being compatible with the variation shown for most angles, the exception being \SI{33}{\degree}. Due to the large uncertainties on the neutron flux, this deviation appears not to be concerning. In addition to the neutron flux estimates from the scaling factor of the quenching factor fits, an in-situ measurement of the inelastic scattering rate \isotope{I}{127}$\left(n,n^\prime\gamma\right)$ can be used to estimate the neutron flux incident on the detector. Due to the triggering condition employed (beginning of section~\ref{section:quenching-factor:qf-analysis}), the coincidence rate $\Gamma_c$ of an inelastic scatter event in the \csi/ detector being visible in the analysis window, i.e., $t_\text{ej}-\SI{1}{\micro\second}$ and $t_\text{ej}+\SI{1}{\micro\second}$ (beginning of section~\ref{section:quenching-factor:qf-analysis}), can be calculated using
\begin{align}
\Gamma_c = \Gamma_t\Gamma_i\Delta t
\end{align}
Here $\Gamma_t \approx 240-\SI{265}{\Hz}$ is the average total triggering rate for each angular \ej299/ position, $\Gamma_i$ the rate of inelastic scatter events in the \csi/ and $\Delta t = \SI{2}{\micro\second}$ is the width of the analysis window in each waveform. $\Gamma_i$ was calculated using a dedicated \mcnp/ simulation including $10^9$ neutron histories and is dependent on the exact neutron flux $\phi_n$ delivered by the accelerator. It is
\begin{align}
\phi_n = \frac{N_\text{c}\,N_n}{N_i\,N_\text{t}\,\Delta t},
\end{align}
where $N_\text{c} = 834$ is the number of inelastic scatter events found in the analysis that coincide with the trigger. $N_n = 10^9$ is the total number of neutrons simulated, $N_i = \num{14518617}$ the number of inelastic events produced in the simulation and $N_t = \num{18723150}$ the total number of triggers taken in all of the individual quenching factor runs. The average neutron flux throughout all measurements is given by $\phi_n\,=\,1534\pm53$ neutrons per second, which is shown in Fig.~\ref{fig:quenching:neutron-flux} as a red band.\par

This secondary measurement again suggests that the true neutron flux was slightly underestimated in the quenching factor fits. However, as the neutron flux only provides an overall scaling factor this slight bias is considered negligible.

%% file: sns-analysis.tex
%
%
\chapter{CE$\nu$NS search at the SNS}
\label{chapter:sns-analysis}
The previous chapters discussed all calibration and background measurements performed prior to the \ac{cenns} search at the \ac{sns}, which is described in this chapter. The \csi/ detector was described in chapter~\ref{chapter:csi-setup}, which also included a schematic of the data acquisition system and the data format. The \ac{sns} beam facility was discussed in detail in section~\ref{section:coherent-at-the-sns:sns}. Between June 25th, 2015 and May 26th, 2017 data was almost continuously acquired at the \ac{sns} with a \SI{60}{\hertz} triggering rate. The total number of triggers acquired in this time period was \num{2825705648}.\\

Two \SI{70}{\micro\second} long waveforms were acquired for each \acf{pot} trigger. The first channel contains the \csi/ signal, the second contains the discriminated output from the muon veto panels (Fig.~\ref{fig:csi-setup:wiring-diagram-sns}). Both traces were sampled at \SI{500}{\mega\sample\per\second}, resulting in \num{35000} sample-long waveforms. The trigger position was set to \SI{78.5}{\percent} of the total waveform, i.e., at sample \num{27475}. As discussed in section~\ref{section:csi-setup:wiring}, the \ac{pot} trigger, i.e., \emph{event 39}, provided by the \ac{sns} was used as external trigger. The \ac{pot} trigger occurs at a constant rate of \SI{60}{\hertz} regardless of the operational status of the \ac{sns}.\\

For about two thirds of the recorded data, the \ac{sns} was operational, i.e., protons were impinging on the mercury target and neutrinos were produced. For the remaining third the \ac{sns} underwent planned maintenance and no neutrinos were emitted. For the remainder of this thesis, the sub-set of data for which neutrinos were produced is referred to as \onDS/ data set. Likewise, periods for which no power on target was provided are referred to as \offDS/ data set.\par


\section{Waveform analysis}
A large fraction of the \ac{cenns} search data analysis mimicked the analysis presented in chapter~\ref{chapter:ba-calibration}. This enabled the \isotope{Ba}{133} calibration (chapter~\ref{chapter:ba-calibration} to be used to quantify the acceptances of Cherenkov and rise-time cuts employed in the \ac{cenns} search. Two overlapping sub-regions were defined for each individual \csi/ waveform, i.e., the coincidence (C) and anti-coincidence (AC) region. The analysis pipeline used for both C and AC region was identical. As a result, the AC region can be used to estimate the environmental steady-state background present in the C region. The subtraction of C-AC therefore only includes beam-related events (section~\ref{section:sns-analysis:cenns-observation}).\par

\begin{figure}[!htbp]
\begin{center}
\includegraphics[scale=1]{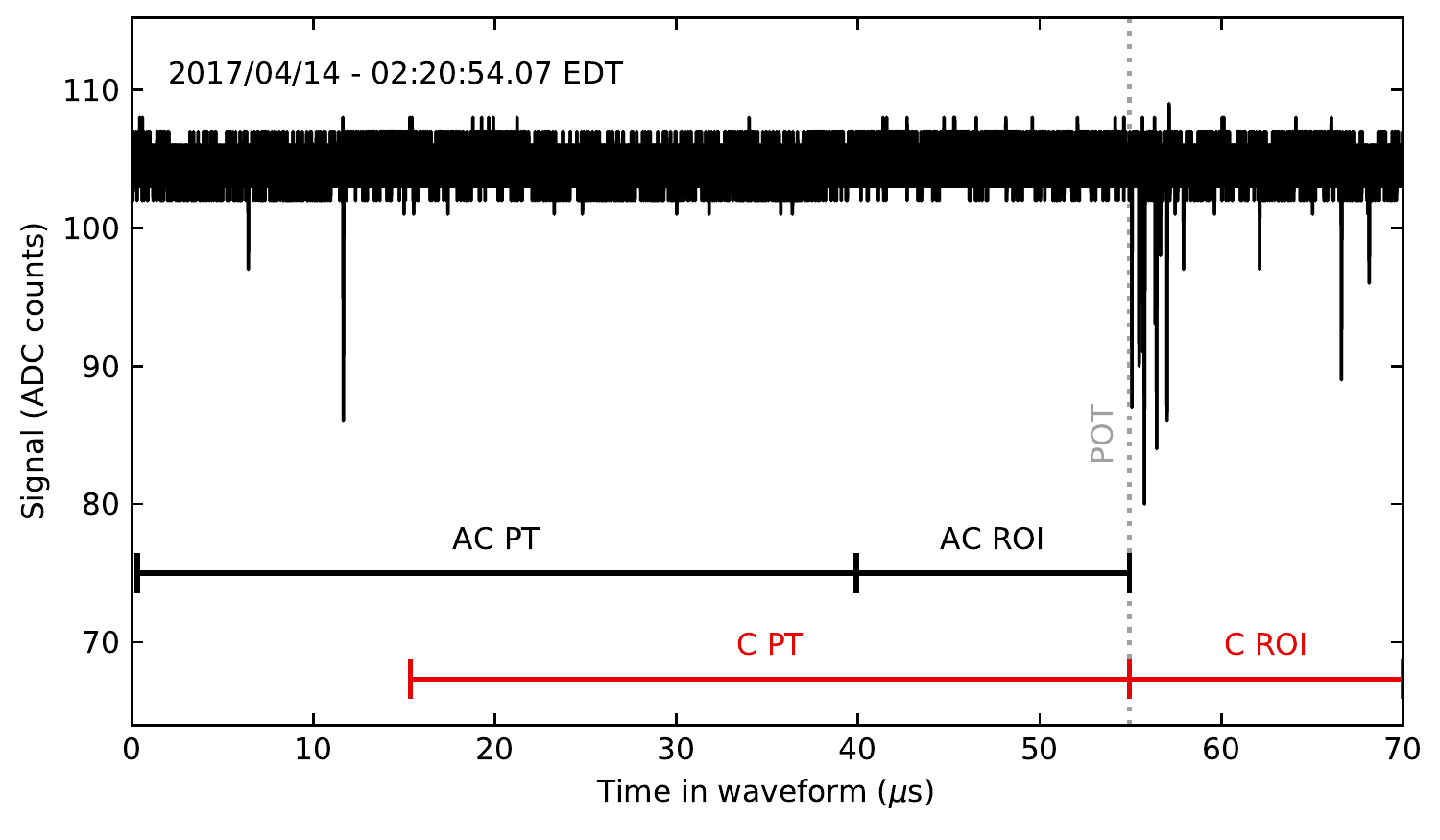}
\end{center}
\caption[Example \csi/ waveform highlighting the different analysis regions used in the CE$\nu$NS search data analysis]{Example \csi/ waveform highlighting the different analysis regions used in the CE$\nu$NS search data analysis. Also shown is the \ac{pot} trigger provided by the \acs*{sns}. The AC \ac{roi} cannot contain any beam-related events and was used to estimate the event rate from environmental backgrounds. The long \ac{pt} regions provided a cut against contamination from afterglow from preceding high-energy events.}
\label{fig:sns-analysis:example-waveform}
\end{figure}

Both C and AC are further sub-divided into a \ac{pt} and \ac{roi}. The positioning of AC and C was chosen such that the \acf{roi} of the former ends at the \acf{pot} trigger, whereas the \ac{roi} of the latter begins there. This choice guaranteed that no potential \ac{cenns} signal could be present in the AC \ac{roi}. The AC region can therefore be used to estimate the random, environmental background occurring before the \ac{pot} trigger. These random backgrounds include Cherenkov light emission in the PMT window, spurious \acp{spe} from afterglow, and random groupings of dark-current photoelectrons, among others. A waveform highlighting the different regions is shown in Fig.~\ref{fig:sns-analysis:example-waveform}. The exact start and end sample of each region are given in Table~\ref{tab:sns-analysis:pt-roi-ranges}.\par

\begin{table}[!htbp]
	\begin{center}
		\begin{tabular}{ccccc}
		\toprule
		& \multicolumn{2}{c}{AC region} & \multicolumn{2}{c}{C region}\\
		\cmidrule(lr){2-3}\cmidrule(lr){4-5}
		& Start & End & Start & End \\
		\midrule
		PT  &   \num{150} & \num{19950} &  \num{7675} & \num{27475} \\
		ROI & \num{19950} & \num{27475} & \num{27475} & \num{35000} \\
		\bottomrule
		\end{tabular}
	\end{center}
	\caption[Start and end sample number of each analysis region in the \acs*{cenns} search]{Start and end sample of each analysis region in the \ac{cenns} search. The starting sample was included in the region, whereas the ending sample was excluded.}
	\label{tab:sns-analysis:pt-roi-ranges}
\end{table}

For each waveform it was first determined whether it was fully contained within the digitizer range. An overflow ($f_\text{o}\,=\,1$) was recorded if at least one sample showed an amplitude of +127 or \SI{-128}{\adc} in either the \csi/ and/or the muon veto channel. Second, a linear gate flag ($f_\text{g}\,=\,1$) was recorded if the number of falling ($n_f$) and rising ($n_r$) threshold crossings of the \SI{18}{\adc} level were unequal in the \csi/ signal, i.e., $n_f \neq n_r$.\par

The global baseline $V_\text{median}$ of the \csi/ signal was estimated using the median of the first \num{20000} samples. The \csi/ signal $V$ was baseline shifted and inverted using
\begin{align}
\hat{V}_i = V_\text{median} - V_i\quad\text{for}\quad i\in[0,\num{35000}).
\end{align}

The location of each peak $p_i$ was determined using the same peak detection algorithm described in chapter~\ref{chapter:am-calibration}. A peak is defined as at least four consecutive samples with an amplitude of at least \SI{3}{\adc}. Both positive and negative threshold crossings were recorded for each peak and its charge calculated as described in chapter~\ref{chapter:am-calibration}. All charges are added to an \ac{spe} charge distribution on a \SI{10}{\minute} basis, i.e., a new distribution was created for each \SI{10}{\minute} time interval. Each charge distribution was fitted using Eq.~(\ref{eq:am-calibration:spe-polya}), providing an independent mean \ac{spe} charge \qspe/ for each time interval.\par

Once all peaks were integrated and added to the corresponding charge spectrum the individual C and AC regions were analyzed. The analysis for both of these regions was identical and as such no distinction is made in the following description. The analysis closely followed the that presented for the \isotope{Ba}{133} calibration (section~\ref{section:ba-calibration:waveform-analysis}), which is illustrated in Fig.~\ref{fig:ba-calibration:example-waveform-analysis}. The only difference between the waveform analysis for the \isotope{Ba}{133} calibration data acquired at the University of Chicago and the \ac{cenns} search data acquired at the \ac{sns}, is the exact onset of each analysis window (C and AC) and the length of the \acp{pt}. The analysis windows for the \isotope{Ba}{133} calibration are defined in Table~\ref{tab:ba-calibration:pt-roi-ranges}, whereas the analysis windows used for the \ac{cenns} search data are presented in Table~\ref{tab:sns-analysis:pt-roi-ranges}.\par

In a first step, the respective region (C or AC) were extracted from the waveform and the \ac{pt} and \ac{roi} sub-regions defined. The total number of peaks present in the \ac{pt} was recorded. The location of the first peak within the \ac{roi} was determined, which defined the onset of the \num{1500} sample =\SI{3}{\micro\second} long integration window. As already shown in section~\ref{section:quenching-calibration:fast-scintillator}, \csi/ represents a fast scintillator (Fig.~\ref{fig:csi-setup:csi-decay-times}). As a result, the onset of an event in the \csi/ is well defined as long as more than four peaks are produced (Fig.~\ref{fig:quenching:pe-arrival-times}). The \ac{cenns} search described in this chapter used a Cherenkov cut of \chc/=8, which demands the presence of at least eight peaks in the integration window. This ensured that the uncertainty on the onset is negligible.\par
A new baseline was determined for the integration window using the median of the \SI{1}{\micro\second} immediately preceding the window. The integration window was extracted from the \ac{roi} and shifted using the new baseline. The amplitude of any sample, that did not belong to any peak $p_i$ found by the peak detection algorithm, was set to zero. The scintillation curve $Q(t)$ was calculated integrating over the zero suppressed signal (Eq.~\ref{eq:ba-calibration:integrated-scintillation-curve}). The total charge $Q_\text{total}$ was recorded and the $T_{10}$, $T_{50}$ and $T_{90}$ rise-times were determined as described in chapter~\ref{chapter:am-calibration} and shown in Fig.~\ref{fig:am-calibration:spe-and-integration}. The parameters extracted from each \csi/ waveform are identical to the ones determined in the \isotope{Ba}{133} calibration (chapter~\ref{chapter:ba-calibration}) and are given in Table~\ref{tab:ba-calibration:parameters-extracted-per-wf}.\par

In contrast to the data acquired for the \isotope{Ba}{133} calibration the \ac{cenns} search data set also included the muon veto channel. For each trigger the location of all muon veto events present in the trace was determined, using the standard peak finding algorithm described in chapter~\ref{chapter:am-calibration}. The logical signal of the discriminated muon veto output was detected if at least ten samples showed a minimum amplitude of \SI{10}{\adc} above baseline. If a muon is present in either C or AC a muon veto flag ($f_\text{m}=1$) was raised for the trigger. If there was any muon present in the waveform, the position of the first muon event was recorded. A muon was recorded in approximately $\SI{1}{\percent}$ of all triggers recorded (panel \textbf{E} of Fig.~\ref{fig:sns-analysis:stability}).\par

All parameters were recorded in a separate and independent n-tuple for the AC and C region. The procedure described above was repeated for each waveform, where all n-tuples calculated based on the AC region were added to the \acDS/ data set and all n-tuples calculated for the C region were added to the \cDS/ data set. Even though the analysis program was optimized to extract all necessary parameters in two passes of each waveform, it still required several weeks on a computer cluster to analyze all of the $\sim2.8$ billion triggers acquired. The analysis was performed on the HCDATA cluster provided by the Physics Division at \ac{ornl}. The analysis program fully made use of the multi-core processing capabilities provided by the cluster. Approximately 20-30 analysis jobs were continuously running in parallel on HCDATA, using the same amount of CPUs. The full analysis code used on HCDATA is available at \cite{scholz-01}.\par 

\section{Detector stability performance over the full data-taking period}
\label{section:sns-analysis:stability}
\begin{figure}[tbhp]
\begin{center}
\includegraphics[scale=1]{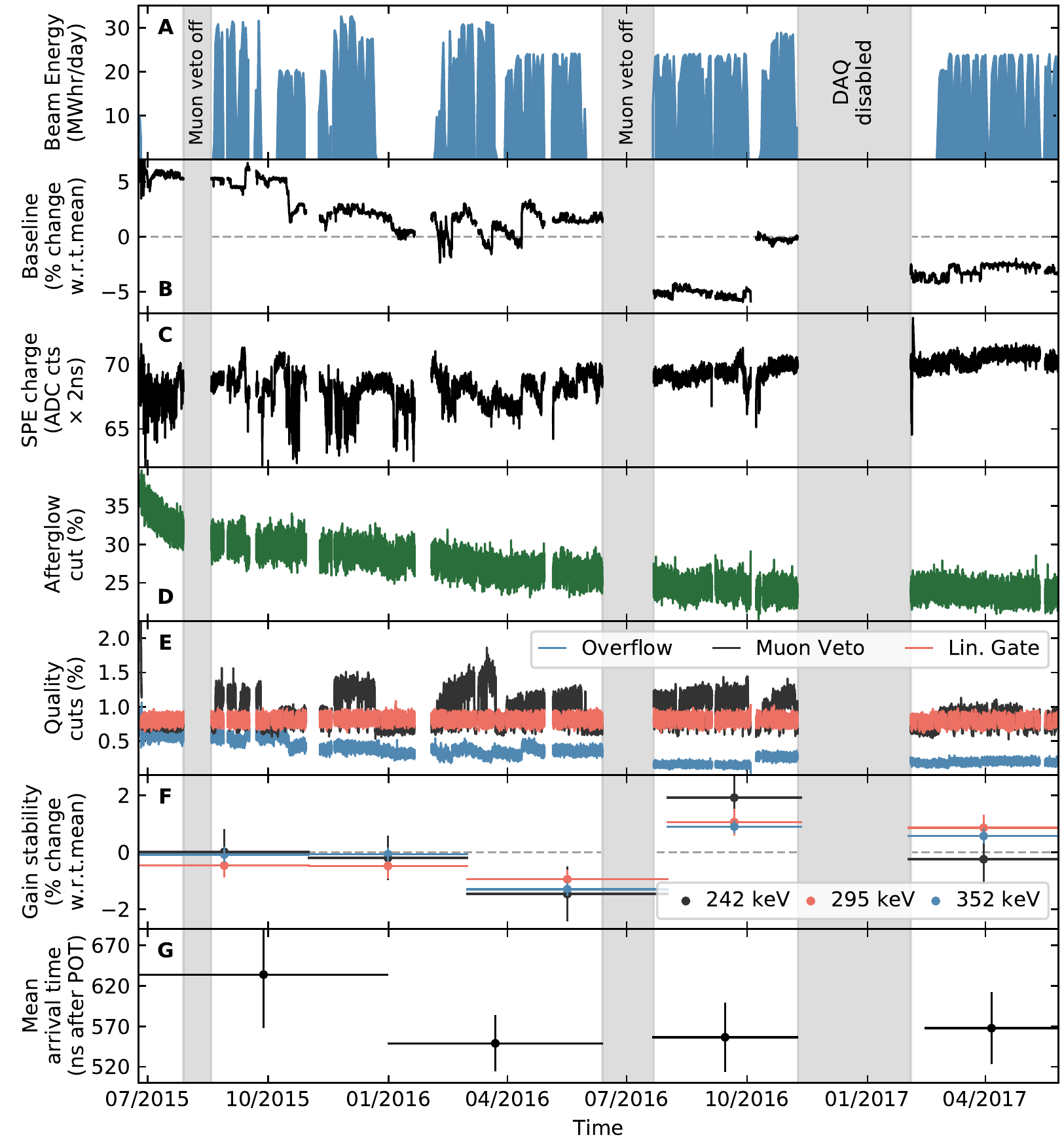}
\end{center}
\caption[Detector stability tests performed during the two years of \csi/ data-taking at the SNS]{Detector stability tests performed during the two years of \csi/ data-taking at the SNS. Panel \textbf{A} is described in section~\ref{section:sns-analysis:beam-power}. Panel \textbf{B} and \textbf{C} are examined in section~\ref{section:sns-analysis:baseline-spe-charge}. An in depth discussion of panel \textbf{D} is provided in section~\ref{section:sns-analysis:afterglow-cut}. Panel \textbf{E} is discussed in detail in section~\ref{section:sns-analysis:quality-cuts}. Lastly, panel \textbf{F} and \textbf{G} are examined in section~\ref{section:sns-analysis:light-yield-trigger-position}.}
\label{fig:sns-analysis:stability}
\end{figure}

Once all data was processed, the detector stability over the two years of operation was investigated in order to ensure a high data quality for all time periods used in this analysis. Several different metrics were calculated using the n-tuples created for each waveform, all of which are shown in Fig.~\ref{fig:sns-analysis:stability}. In the following sections each panel is discussed in detail.\par

\subsection{Beam energy delivered on the mercury target}
\label{section:sns-analysis:beam-power}
Panel \textbf{A} of Fig.~\ref{fig:sns-analysis:stability} shows the total daily beam energy delivered by the \ac{sns} on the mercury target. Several distinct features are apparent. First, three time periods were excluded from the analysis, shown in shaded gray. The first two were excluded, as the muon veto was not operational during these dates. Both of these incidents were linked to power outages at the \ac{sns} after which the high voltage supply for the muon veto did not properly restart. The last gap, in contrast, is linked to a failure of the digitizer used in the data acquisition system occurring after the full data acquisition system was restarted. The \si{NI} 5153 was sent off to the manufacturer for repairs. Once this issue was resolved the data acquisition was restarted and no further issues were found. Second, gaps in beam energy delivered are apparent, which are not linked to any issues within the experimental setup. These represent times during which the \ac{sns} underwent long-term maintenance and no \acp{pot} were delivered. Third, dips within beam windows are apparent, which are caused by outages that last less than a full day, e.g., short-term maintenance every Tuesday morning.\par

\begin{figure}[thb]
\begin{center}
\includegraphics[scale=1]{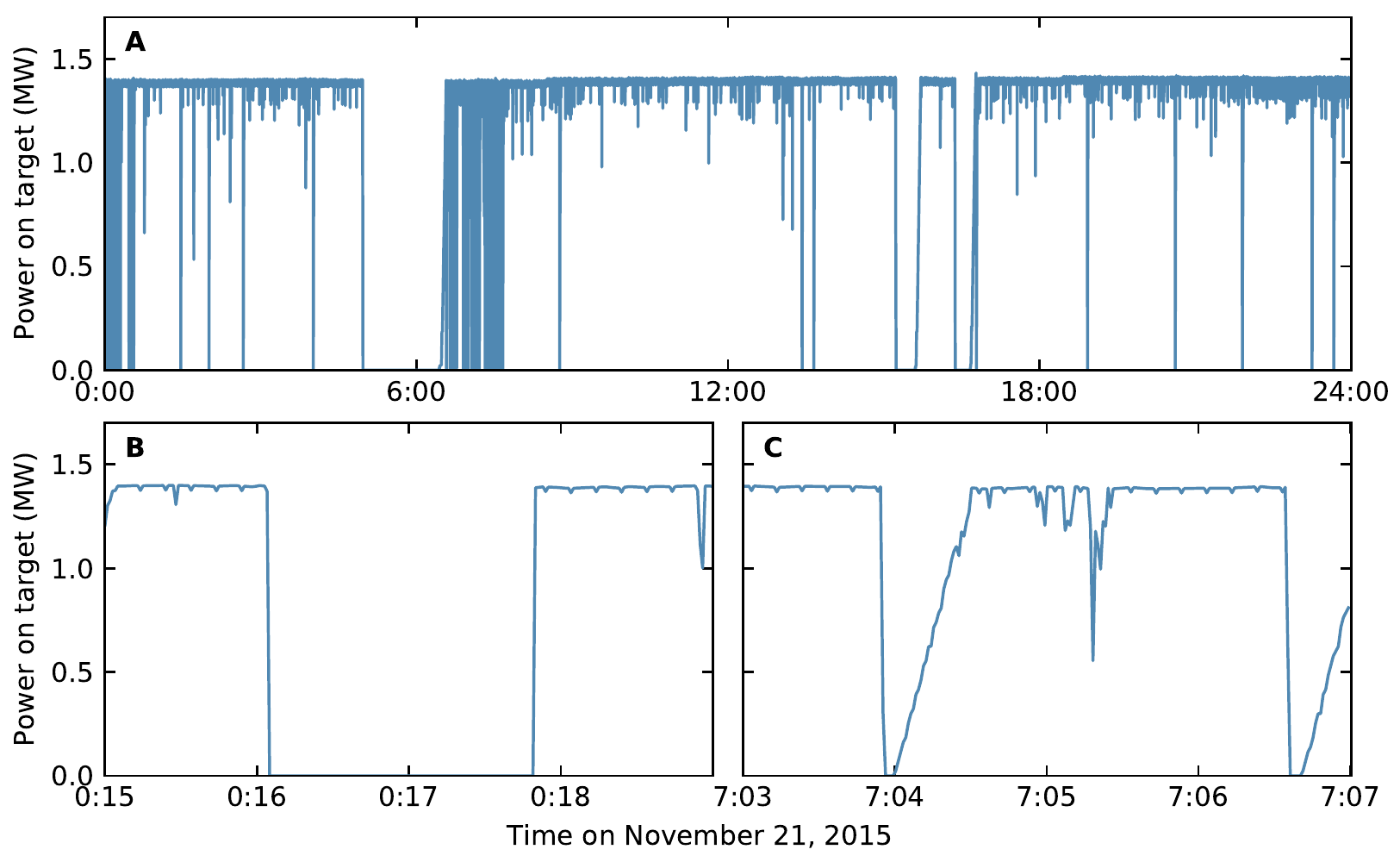}
\end{center}
\caption[Exemple beam power retrieved from the \acs*{sns} archives for November 21, 2015]{Exemplary beam power retrieved from the \acs*{sns} archives for November 21, 2015. Panel \textbf{A} shows the beam power on target for the full day. Short and long drops in beam power are visible. Panel \textbf{B} shows a section of bad beam power data, i.e., the recorded data instantaneously jumps from 0 to \SI{1.4}{\mega\watt} without a proper ramp-up process. SNS personnel pointed out that this behavior is unphysical. Panel \textbf{C}, in contrast, shows a beam power drop followed by a proper ramp-up process. This drop was therefore deemed physical and included in the analysis.}
\label{fig:sns-analysis:example-beam-power}
\end{figure}

The \ac{sns} delivered an energy of up to \SI{30}{\MWh} on the \ac{sns} target, per day in 2015 and early 2016. However, due to several target failures the beam power was reduced to provide a more stable operation at $\sim$\SI{24}{\MWh}. The beam power information is available from the \ac{sns} archive server with a timing resolution of \SI{1}{\second}. A post-processing step was applied to the retrieved beam power information data, which is described in the following.\par

Panel \textbf{A} of Fig.~\ref{fig:sns-analysis:example-beam-power} shows an example beam power time series retrieved from the archive for November 21st, 2015. The recorded power occasionally drops to zero and instantly recovers (panel \textbf{B}). Such a behavior is unphysical for an accelerator and to be addressed. For each drop in beam power the corresponding rise in beam power at the end of the gap is analyzed. Whenever an instantaneous rise back to full operational power is visible the gap was flagged as unphysical and exclude from the analysis. In contrast, if the beam was found to be properly ramped back up to full power (panel \textbf{C}) the period in question was flagged as physical and a zero power delivered on target was recorded for the gap duration. Less than $\SI{0.1}{\percent}$ of the triggers acquired in the \ac{cenns} search data set had to be excluded from the analysis.\par

Once these beam power drops were properly addressed, the \csi/ data acquired at the \ac{sns} was split into \onDS/ and \offDS/ data sets on a second by second basis. Using this information the total beam energy delivered on the mercury target was calculated for the whole \onDS/ data set. It is
\begin{align}
E_\text{beam}=\SI{7475}{\MWh}.
\end{align}
This information is used in section~\ref{section:sns-analysis:optimizing-cuts} to scale the \ac{cenns} interaction rate prediction in the \csi/.\par

\subsection{Baseline and mean \acs*{spe} charge stability}
\label{section:sns-analysis:baseline-spe-charge}
Panel \textbf{B} of Fig.~\ref{fig:sns-analysis:stability} shows the change in the \csi/ baseline over the two years of data-taking. The data shown is based on \SI{10}{\minute} averages of the baseline determined using the first \SI{40}{\micro\second} of each waveform. The dashed line represents the the average for the full two years of data-taking.\par

\begin{figure}[htbp]
\begin{center}
\includegraphics[scale=1]{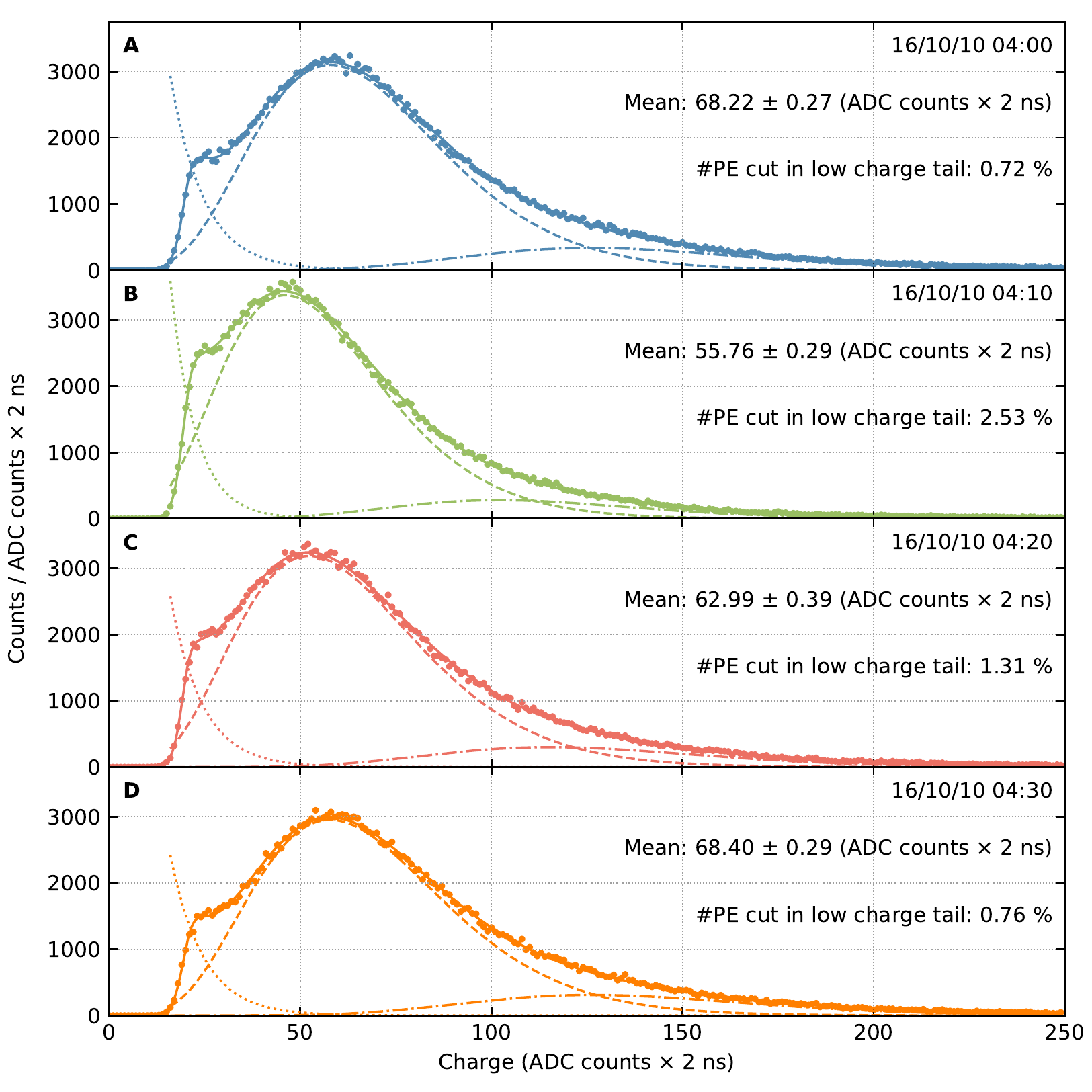}
\end{center}
\caption[Example charge spectra for an occasional drop in the mean \acs*{spe} charge]{Example charge spectra for an occasional drop in the mean \acs*{spe} charge. Panel \textbf{A} shows the charge distribution immediately preceding a recorded drop. Panel \textbf{B} and \textbf{C} show the distribution during the drop, whereas panel \textbf{D} shows the distribution once the charge has fully recovered. These drops are therefore not the result of badly converged \acs*{spe} charge fits. However, as shown in the text, these drops in \ac{spe} charge happened only infrequently and were later attributed to transients in the Ortec 556 power supply providing the high voltage for the R877-100 PMT.}
\label{fig:sns-analysis:speq-drift}
\end{figure}

Panel \textbf{C} of Fig.~\ref{fig:sns-analysis:stability} shows the time evolution of the mean \ac{spe} charge in \SI{10}{\minute} intervals. Several short term drops in the \ac{spe} charge are apparent. These drops are not correlated with a drop in baseline. Fig.~\ref{fig:sns-analysis:speq-drift} shows the charge spectra recorded during such a drop. The charge spectra were fitted using Eq.~(\ref{eq:am-calibration:spe-polya}). Panel \textbf{A} shows the \SI{10}{\minute} interval directly preceding the \ac{spe} charge drop. The average \ac{spe} is large enough such that less than \SI{1}{\percent} of \acp{spe} are missed using the standard peak finding algorithm. In addition these \ac{pe} only carry a charge less than one third of \qspe/. The bias introduced is therefore negligible.\par

Panel \textbf{B} highlights the spectrum calculated during the interval containing the \ac{spe} charge drop. The full distribution is shifted towards lower charge values, i.e., the reduced charge is not the result of any fit error. The mean charge dropped by almost \SI{20}{\percent}. A much higher number of \acp{spe} are missed in the low charge tail.\par

Panel \textbf{C} shows the slow recovery after the initial sharp drop and panel \textbf{D} shows the full recovery of the mean \ac{spe} charge. For almost all cases a drop in charge only lasts for a singular \SI{10}{\minute} interval, after which the charge recovers as shown. The cause of this behavior was not determined, however, one possibility was a faulty power supply. The original Ortec 556 power supply which provides the high voltage for the \csi/ detector was replaced with another unit of the same kind in 2017.  No more drops in the mean \ac{spe} charge were recorded since then. As these events happened infrequently enough, all periods with a mean \ac{spe} charge of $<\SI{62}{\adcq}$ were excluded from the data analysis.\par

\subsection {Afterglow cut and the decay of cosmogenics}
\label{section:sns-analysis:afterglow-cut}
Panel \textbf{D} of Fig.~\ref{fig:sns-analysis:stability} shows the percentage of events rejected by an afterglow cut of $\agc/=3$. The overall behavior is the same for all choices of $\agc/\in[1,10]$. The rejection percentage decays over time. This decline consists of two different time scales, one on a much shorter than the other. Both decay times were determined by fitting the afterglow acceptance time evolution with
\begin{align}
f(t,a_s,\tau_s,a_l,\tau_l,f_0) = a_s\text{e}^{-\nicefrac{t}{\tau_s}} + a_l\text{e}^{-\nicefrac{t}{\tau_l}} + f_0.\label{eq:sns-analysis:decay-time-fit}
\end{align}

\begin{figure}[thbp]
\begin{center}
\includegraphics[scale=1]{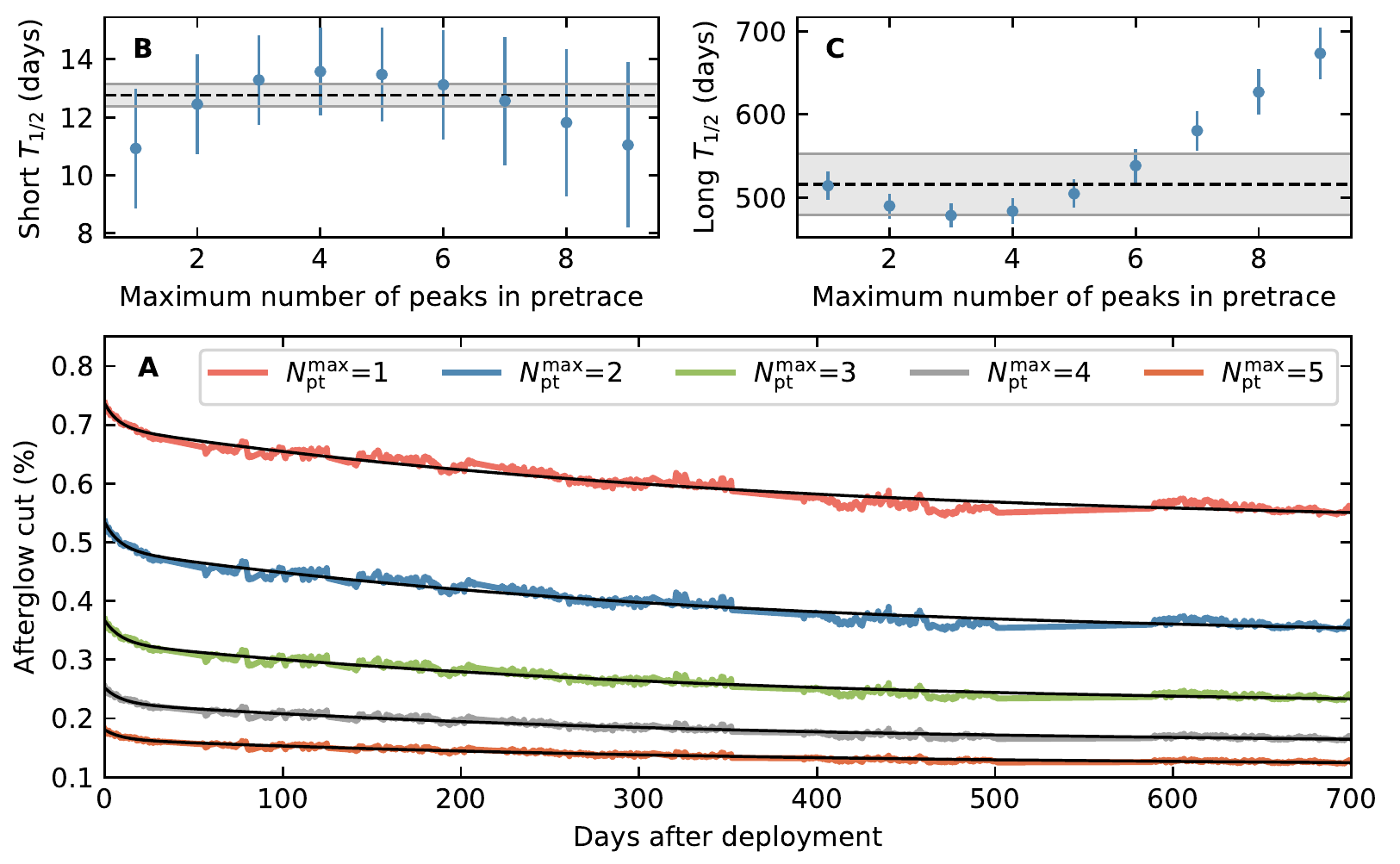}
\end{center}
\caption[Determining the decay times apparent in the afterglow cut]{Determining the decay times apparent in the afterglow cut. Panel \textbf{A} shows the percentage of events rejected by the afterglow cut for different choices of \agc/ (color). The rejection percentage decays over time for all choices of \agc/. Shown in black are fits of Eq.~(\ref{eq:sns-analysis:decay-time-fit}) to the respective data. Panel \textbf{B} shows the short half-life for all choices of \agc/ The uncertainty weighted average is shown as dashed black line. The $1\sigma$ error is shown as a shaded gray band. Panel \textbf{C} shows the same for the long half-life.}
\label{fig:sns-analysis:decay-times-in-afterglow}
\end{figure}

Panel \textbf{A} of Fig.~\ref{fig:sns-analysis:decay-times-in-afterglow} shows the time evolution of the afterglow cut for multiple choices of \agc/ (colored). The black lines represent the corresponding fits of Eq.~(\ref{eq:sns-analysis:decay-time-fit}). Panel \textbf{B} (\textbf{C}) shows the short (long) half-life determined for different choices of \agc/. The dashed black line shows the uncertainty weighted average of the data. The shaded gray region represent the associated $1\sigma$ uncertainty. The decay times were found to be
\begin{align}
T^\text{short}_{\nicefrac{1}{2}} & = 12.77 \pm 0.39\,\text{days}\nonumber\\
T^\text{long}_{\nicefrac{1}{2}}  & = 515.83 \pm 36.92\,\text{days}\nonumber
\end{align}

As the \csi/ detector was driven from the underground laboratory at the University of Chicago to \ac{ornl} there was a period of approximately one day over which the detector was above ground and during which cosmogenics could were created. The short decay time is compatible with cosmogenic \isotope{I}{126} production as the half-life of this isotope is given by $T_{\nicefrac{1}{2}}(\isotope{I}{126}) = 12.93\pm0.05\,$days \cite{amare-01}. The long decay time is probably dominated by cosmogenic-origin neutron capture on \isotope{Cs}{133}, as the half-life of the associated isotope is given by $T_{\nicefrac{1}{2}}(\isotope{Cs}{134}) = 2.0648\,$years \cite{chu-01,lee-01}.\par

This calculation did not use the decaying event rate from a distinct peak in the recorded energy spectrum to determine the half-life of potential cosmogenics, but rather relies on the random coincidences between the afterglow from a potential \isotope{Cs}{134}- or \isotope{I}{126}-decay and the \ac{sns} trigger. This might explain the difference between the measured half-life and the literature value. However, after approximately two years of operation under an overburden of \SI{8}{\mwe}, the level of radiation present in the crystal has stabilized as can be seen in panel \textbf{D} of Fig.~\ref{fig:sns-analysis:stability}.\par

The average afterglow cut acceptance $\eta_\text{afterglow}$ was calculated for different choices of \agc/. This acceptance is used in section~\ref{section:sns-analysis:optimizing-cuts} to adjust the predicted \ac{cenns} signal rate in the \csi/. The overall average afterglow cut acceptance is calculated from the afterglow cut acceptances found for each \SI{10}{\minute} interval shown in Fig.~\ref{fig:sns-analysis:stability}. Since the \ac{cenns} event rate is directly correlated to the beam energy delivered on the \ac{sns} target, the afterglow cut acceptances found for each \SI{10}{\minute} interval are weighted by the beam energy delivered in that interval. It is 
\begin{align}
\eta_\text{afterglow} = \frac{\sum\limits_t E_B(t)\,\eta_\text{afterglow}(t)}{\sum\limits_t E_B(t)}\label{eq:sns-analysis:power-weighted-cut}
\end{align}
where $\eta_\text{afterglow}(t)$ denotes the fraction of events accepted in the \SI{10}{\minute} interval $t$ and $E_B(t)$ represents the total integrated power on target delivered in the same window. This only includes time periods within the \onDS/. The exact beam energy averaged acceptances for different \agc/ is given in Table~\ref{tab:sns-analysis:afterglow-cut-acceptance}

\begin{table}[!htbp]
	\begin{center}
		\begin{tabular}{cccccccc}
		\toprule
		\agc/ & $\eta_\text{afterglow}$ && \agc/ & $\eta_\text{afterglow}$ && \agc/ & $\eta_\text{afterglow}$ \\
		\midrule
		0 & 0.167 && 4 & 0.817 && 7 & 0.910\\
		1 & 0.405 && 5 & 0.863 && 8 & 0.923\\
		2 & 0.606 && 6 & 0.891 && 9 & 0.932\\
		3 & 0.738 \\
		\bottomrule
		\end{tabular}
	\end{center}
	\caption[Beam energy-weighted afterglow cut acceptances for varying \agc/]{Beam energy-weighted afterglow cut acceptances for a number of different choices of \agc/.}
	\label{tab:sns-analysis:afterglow-cut-acceptance}
\end{table}

\subsection{Quality cuts}
\label{section:sns-analysis:quality-cuts}
Panel \textbf{E} of Fig.~\ref{fig:sns-analysis:stability} shows the time evolution of events rejected by the different quality cuts. The red line represents the percentage removed due to a linear gate being present in the \csi/ signal (section~\ref{section:csi-setup:wiring}). The rejection rate remained constant over the full data acquisition period. This agrees with the assumption that most of the high energy events being cut by the linear gate were induced by cosmic ray muons traversing the crystal.\par

The black line represents the percentage of events rejected due to a coincident event in the muon veto waveform. A clear correlation between the energy delivered on the mercury target (panel \textbf{A}) and the muon veto rejection rate is visible . This correlation arises from the close proximity of the \ac{mots} exhaust pipe to the muon veto panels, which was described in section~\ref{sec:csi-setup:muon-veto}.\par

Finally, the blue line represents the percentage of events rejected due to digitizer overflows. A correlation between the overflow rate and the average baseline level (panel \textbf{B}) is apparent. The closer the baseline is to the upper digitizer range the more triggers are rejected due to overflows.\par

A beam energy-weighted acceptance fraction is calculated for each quality cut. $\eta_\text{muon-veto}$, $\eta_\text{linear-gate}$ and $\eta_\text{overflow}$ were defined following Eq.~(\ref{eq:sns-analysis:power-weighted-cut}). The beam energy weighted acceptance fractions are given by
\begin{align}
\eta_\text{muon-veto}\,=\,0.989\quad\qquad\eta_\text{linear-gate}\,=\,0.992\quad\qquad\eta_\text{overflow}\,=\,0.997
\end{align}

\subsection{Light yield and trigger position}
\label{section:sns-analysis:light-yield-trigger-position}
The data presented in panels \textbf{F} and \textbf{G} of Fig.~\ref{fig:sns-analysis:stability} were both calculated by Alexey Konovalov at the National Research Nuclear University MEPhI. Panel \textbf{F} shows the gain stability of the \csi/ detector using $\gamma$-lines from backgrounds internal of the crystal, namely \isotope{Pb}{212} and \isotope{Pb}{214}~\cite{collar-02}. These $\gamma$-rays carry energies beyond the digitizer range. Alexey estimated the total energy of each event exhibiting an overflow by removing the clipped part of the scintillation curve and measuring the charge in the tail only. The gain remained stable within $\sim\SI{1.5}{\percent}$ over the course of the two year period.\par

Panel \textbf{G} shows the stability of the \ac{pot} trigger provided by the \ac{sns} by searching for prompt neutron interactions in the muon veto panels. Even though these panels cover a large area, their neutron efficiency is severely limited due to the high discriminator threshold. Only a small excess of events caused by prompt neutrons is visible in the muon veto over the steady state background. The low count rate is largely responsible for the large uncertainty on the measurement. The arrival times of the prompt neutrons following the \ac{pot} trigger derived using this method are similar to those determined in chapter~\ref{chapter:background-studies}.\par

\begin{figure}[htb]
\begin{center}
\includegraphics[scale=1]{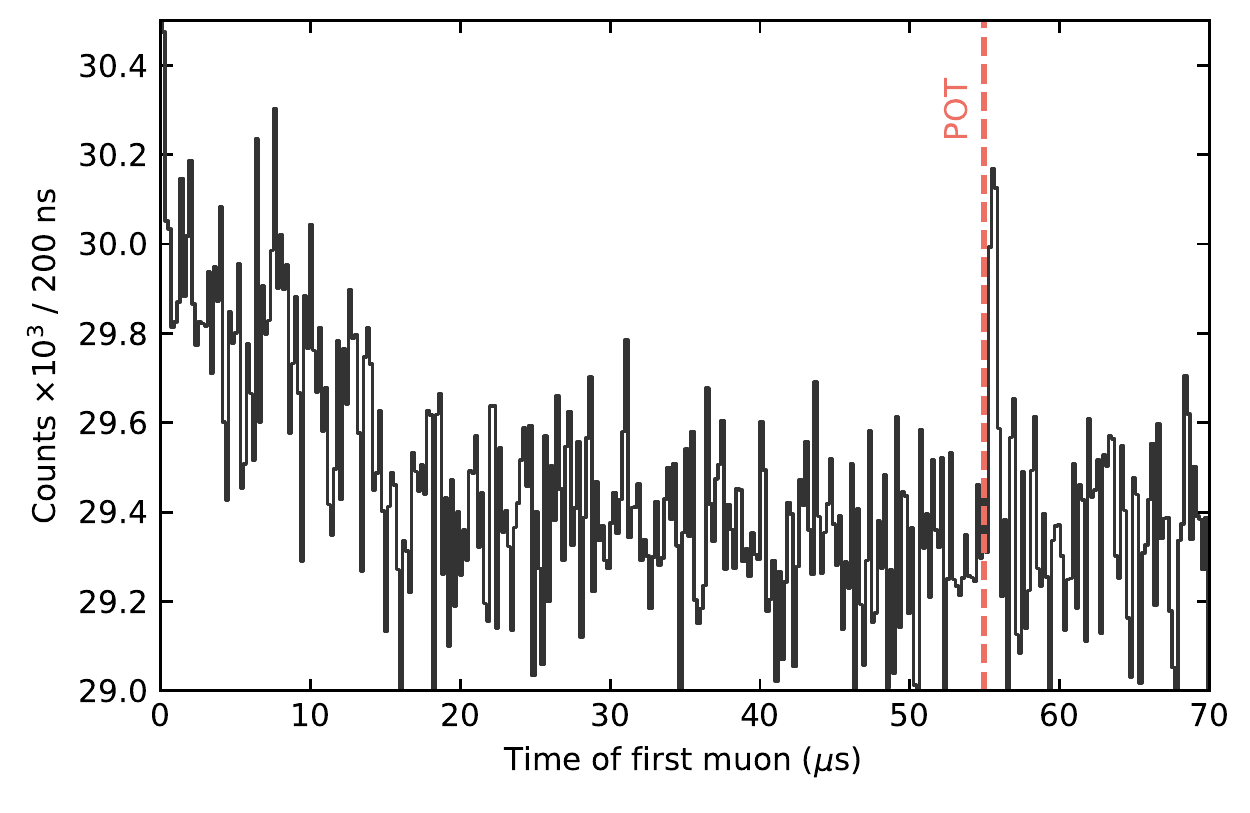}
\end{center}
\caption[Muon veto hit distribution showing a prompt neutron component shortly following the \acs*{pot} trigger in the \acs*{cenns} search data]{Muon veto hit distribution showing a prompt neutron component shortly following the \acs*{pot} trigger in the \acs*{cenns} search data. The excess at the beginning of the trace represents a bias in the analysis as only the onset of the first muon veto hit within each waveform was recorded. However, this bias was deemed unimportant as the purpose of this measurement was to detect beam-related prompt neutrons.}
\label{fig:sns-analysis:muon-hit-dist}
\end{figure}

A replica of Alexey's analysis is shown in Fig.~\ref{fig:sns-analysis:muon-hit-dist}, which included a total of \SI{4651}{\hour} of \onDS/ data. An excess from prompt neutrons shortly following the \ac{pot} trigger is apparent. As only the arrival time of the first muon was recorded, Fig.~\ref{fig:sns-analysis:muon-hit-dist} shows an excess for low muon arrival times.

\section{Monitoring prompt neutrons using the inelastic scattering \isotope{I}{127}(n,n'$\gamma$) reaction}
In the previous section the arrival time after the \ac{pot} trigger was measured for beam-related, prompt neutrons using the muon veto event distribution. The event rate from these prompt neutrons in the \csi/ was found to be small in chapter~\ref{chapter:background-studies}. However, these calculations heavily depended on neutron transport simulations using \mcnp/. In this section a test of the accuracy of these simulations is provided using the inelastic scattering \isotope{I}{127}(n,n'$\gamma$) reaction. Due to its large \xs/~\cite{collar-02,fields-01}, this process provides an excellent monitoring tool for prompt neutrons following the \ac{pot}, that were able to penetrate the $\sim\SI{20}{\meter}$ of shielding between the \ac{sns} target and the \csi/ detector (Fig.~\ref{fig:sns:neutrino-alley}). Using the \isotope{Am}{241} source to calibrate the \csi/ detector (chapter~\ref{chapter:am-calibration}) ensured that the main gamma emission of the inelastic scattering reaction at \SI{57.6}{\keV} was initially contained within the digitizer range. However, due to a drift in the DC baseline of the \csi/ detector signal it is no longer fully contained. In the following sections a calibration measurement using a \isotope{Cf}{252} neutron source, and a search for the inelastic scattering \isotope{I}{127}(n,n'$\gamma$) reaction in the \csi/ caused by prompt neutrons associated with the \ac{sns} \acl{pot} are described.

\subsection{Validating neutron transport simulations using a \isotope{Cf}{252} source}
\label{section:sns-analysis:cf-calibration}
The prompt neutron event rate presented in chapter~\ref{chapter:background-studies} strongly depended on \mcnp/ simulations. It is therefore crucial to confirm that these simulations are accurate. One way to test the accuracy of these simulations is to use an external neutron source and to compare the simulated prediction to the experimental measurement. As such, a \isotope{Cf}{252} neutron source with a yield of $\sim8600$ neutrons per second was placed on the outside of \csi/ shielding, i.e., outside of the water tanks. The emission spectrum above \SI{1}{\MeV} for this neutron source shows a comparable hardness to the prompt neutron spectrum found in chapter~\ref{chapter:background-studies} \cite{smith-01}. Its emitted neutrons carry enough energy to penetrate the neutron moderator and are still able to undergo inelastic scattering within the detector.\par

However, such a neutron source could potentially activate the detector or shielding material. This calibration measurement was therefore performed at the beginning of a scheduled, month-long \ac{sns} maintenance so that any potentially activated materials could decay away before new \onDS/ data was acquired. In addition to this, the full \ac{cenns} search data set described in this thesis was acquired prior to this calibration measurement to avoid any possible contamination.\par

One of the potential long-lived backgrounds that could be produced in this \isotope{Cf}{252} calibration is \isotope{Cs}{134} due to neutron capture on \isotope{Cs}{133}. As already discussed in section~\ref{section:sns-analysis:afterglow-cut}, the half-life of this isotope is approximately two years. However, a rough estimation shows that this potential background is negligible: Assuming that all of the neutrons emitted from the \isotope{Cf}{252} source actually thermalize within the moderator between the source and the \csi/ detector and that no neutron is captured in between a total of $\sim$160 neutrons per second would traverse the \csi/ detector. All of the neutrons are assumed to capture as the capture \xs/ is $\mathcal{O}(\SI{50}{\barn})$~\cite{jendl-01}. For a total calibration period of \SI{2}{\hour} this would result in the creation of \num{1.1e6} \isotope{Cs}{134} isotopes. Approximating a linear decay of half of these isotopes over a two year period results in 0.01 decays per second. Assuming each of these events contaminates a total of \SI{1}{\milli\second} due to its afterglow a rate of random coincidences between the decaying \isotope{Cs}{134} and the \ac{pot} trigger is expected on the order of
\begin{align}
\Gamma_c =\,\Delta t\,\Gamma_\text{sns}\,\Gamma_\text{Cs-134}\,=\,\left(\SI{70}{\micro\second} + \SI{1}{\milli\second}\right)\times\SI{60}{\hertz}\times\SI{0.01}{\hertz} = \SI{6.42e-4}{\hertz},
\end{align}
which is negligible.\par

The \isotope{Cf}{252} source was placed on the outside of the \csi/ shielding, in the middle of the side of water tanks facing the \ac{mots} pipe at a height of \SI{60}{\cm} measured from the floor (Fig.~\ref{fig:csi-setup:csi-shielding-sketch}). The data acquisition was set to trigger directly on the \csi/ channel, similar to what is described in chapter~\ref{chapter:am-calibration}. Data was acquired for a total of \SI{107}{\minute}. The overall data analysis closely follows the light yield calibration presented in chapter~\ref{chapter:am-calibration}.\par

In contrast to the light yield calibration no digitizer overflows were excluded from the data set. This was necessary as a significant part of the events in the inelastic peak actually exceeded the digitizer range. As a result events showing a digitizer overflow had to be included in the data analysis to correctly determine the total number of inelastic scattering events in the \csi/. The energy for events exceeding the digitizer range was corrected using the following approach.\par

\begin{figure}[htbp]
\begin{center}
\includegraphics[scale=1]{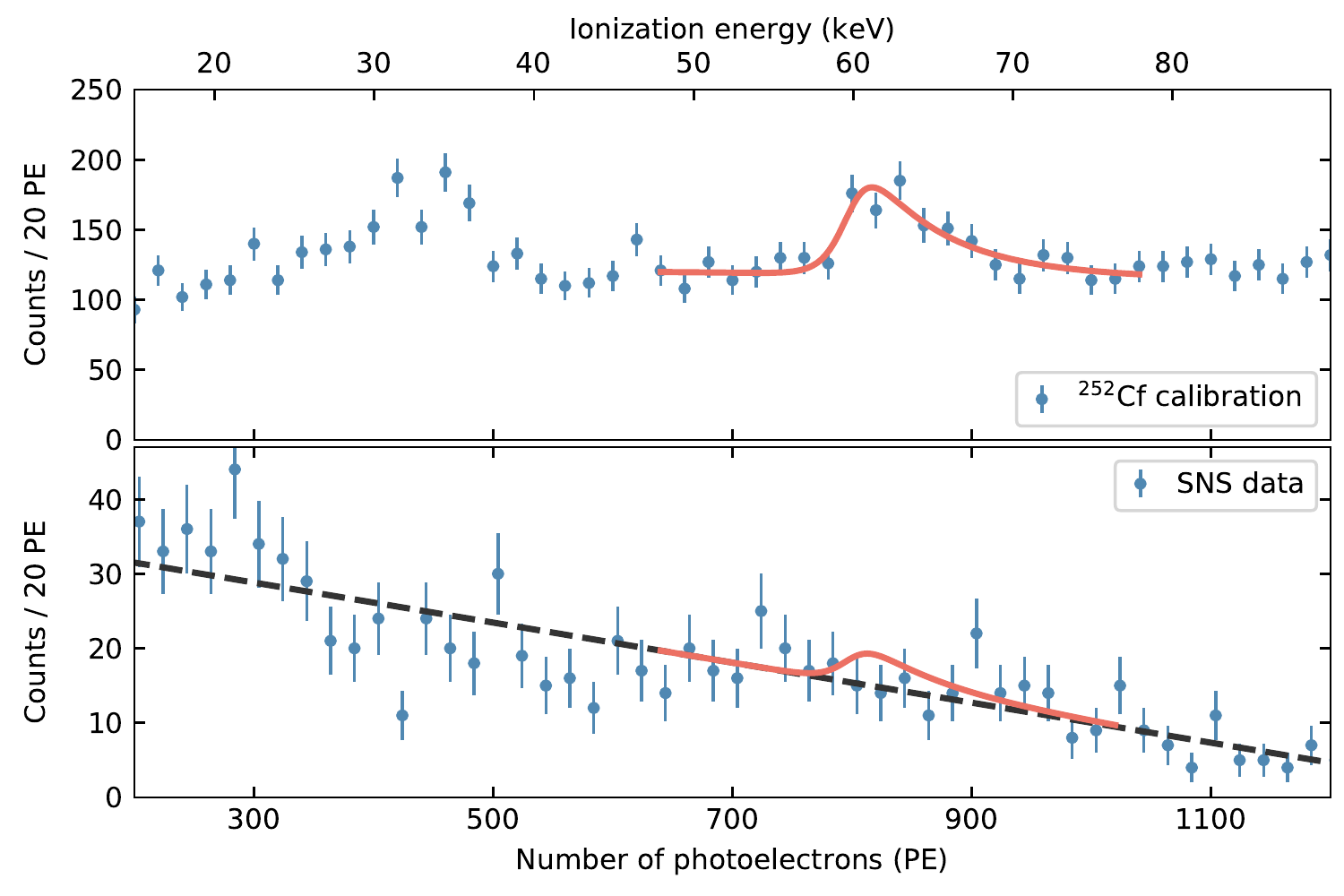}
\end{center}
\caption[Validating neutron transport simulations using the inelastic scattering \isotope{I}{127}(n,n'$\gamma$) reaction]{\textbf{Top}: Energy spectrum recorded in the \csi/ during the \isotope{Cf}{252} calibration measurement. The \emph{shark-tooth} shaped peak at approximately \SI{60}{\keV} is due to inelastic neutron scattering \isotope{I}{127}(n,n'$\gamma$). The unique shape and the shift towards higher energies is caused by the addition of the energy from the recoiling nucleus as well as $\gamma$ de-excitation~\cite{jovancevic-01,collar-04}. The second peak at approximately \SI{30}{\keV} is caused by the electron capture decay of \isotope{I}{128}. The red curve represents the fit of an \emph{ad hoc} peak template (Eq.~(\ref{eq:sns-analysis:peak-template})) to the data. \textbf{Bottom}: Energy spectrum recorded in the \csi/ for events with an arrival time of 200-\SI{1100}{\ns}, i.e., the arrival time window for prompt neutrons (chapter~\ref{chapter:background-studies}). Shown are only events taken from \cDS/ in the full \onDS/ data set. A linear background model was fitted to the whole spectrum, shown in dashed gray. The fit was used as background in the \emph{ad hoc} peak template, which was fitted to the data. The only free parameter in this fit was the peak amplitude. Shown in red is the \SI{90}{\percent} confidence limit for this fit.}
\label{fig:sns-analysis:inelastic-fit}
\end{figure}

\begin{figure}[htb]
\begin{center}
\includegraphics[scale=1]{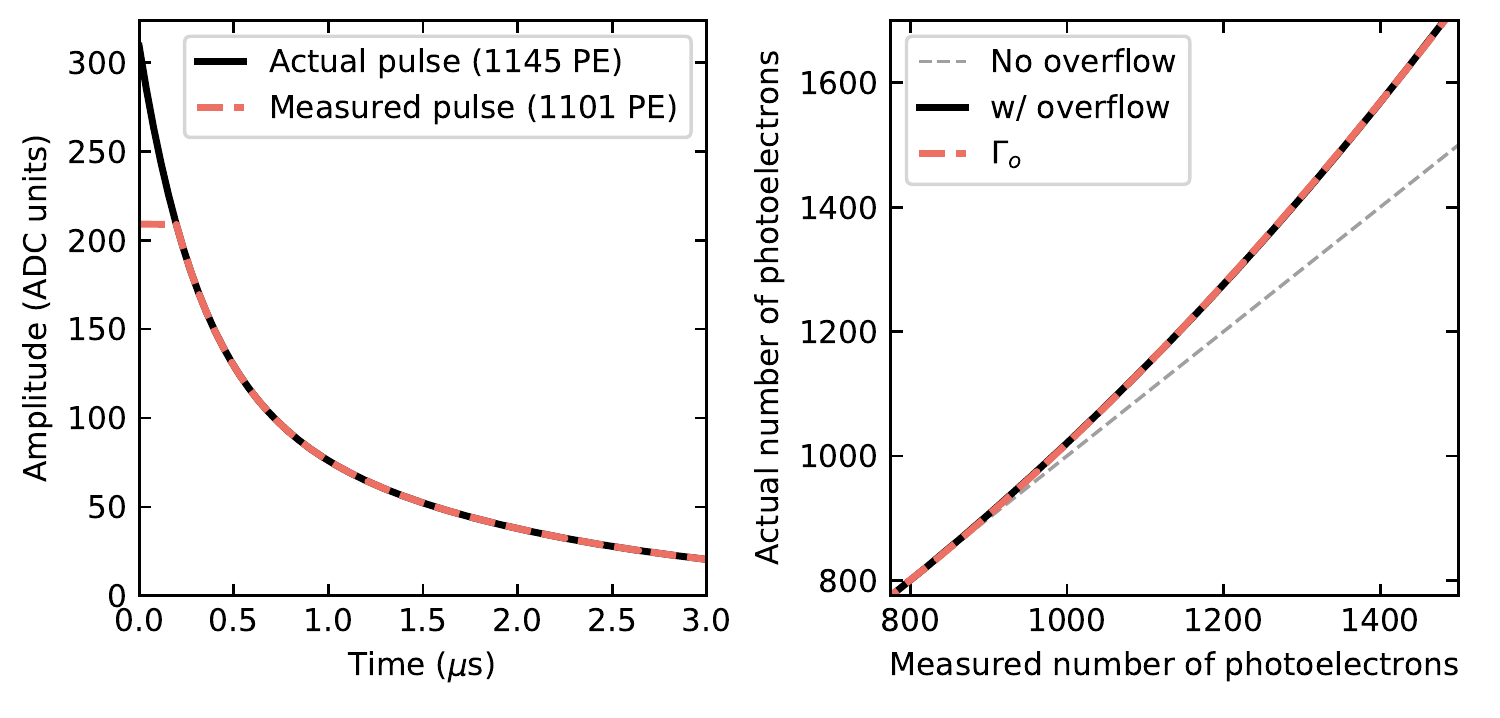}
\end{center}
\caption[Correcting the energy of events experiencing digitizer overflows]{Correcting the energy of events experiencing digitizer overflows. \textbf{Left}: The simulated light curve for an event with an energy of \SI{1145}{\pe} is shown in black. The corresponding truncated waveform due to digitizer range overflow is shown in dashed red. Only the first $\sim$\SI{400}{\ns} exceed the digitizer range. The charge cut by the truncation (44 PE = 1145 PE - 1101 PE) is only a small fraction of the full integrated charge. \textbf{Right}: Comparison of the total number of photoelectrons in an event and the number of photoelectrons measured from integrating a truncated event. The deviation between both measures below $\sim\SI{1000}{\pe}$ is negligible. The red dashed line represents a fit of Eq.~(\ref{eq:sns-analysis:overflow-conversion}) to the relationship.}
\label{fig:sns-analysis:overflow-conversion}
\end{figure}

First, events were simulated for different energies, i.e., $N_\text{pe}$, using the light yield calibration (chapter~\ref{chapter:am-calibration}) and light emission profile as shown in Eq~(\ref{eq:quenching:pe-arrival}). The light emission profile for \csi/ slightly changes with energy. Using the \isotope{Am}{241} calibration data the following parameters were found for Eq~(\ref{eq:quenching:pe-arrival}) for an energy of approximately \SI{60}{\keVee}
\begin{align}
r=0.691 \quad\quad \tau_\text{fast}=\SI{309}{\ns} \quad\quad \tau_\text{slow}=\SI{1649}{\ns}.
\end{align}
The simulated events were clipped at \SI{209}{\adc} to mimic the digitizer range overflow. This particular value represented the total range available given the average baseline value during the \isotope{Cf}{252} calibration.\par

The left panel of Fig.~\ref{fig:sns-analysis:overflow-conversion} shows a simulated light emission profile in black and the corresponding truncated profile in dashed red. Both light curves are integrated and the total number of photoelectrons \npe/ measured in both cases was compared. The evolution of the ratio between true and measured energy is shown in the right panel of Fig.~\ref{fig:sns-analysis:overflow-conversion}. The following empirical model $\Gamma_o$ was fitted to the data
\begin{align}
\Gamma_o(N_\text{pe},a_1,a_2) = N_\text{pe} + \Theta_\text{H}(N_\text{pe} - 775)\left[a_1\left(N_\text{pe} - 775\right) + a_2\left(N_\text{pe} - 775\right)^2\right].\label{eq:sns-analysis:overflow-conversion}
\end{align}
The best fit parameters are given by
\begin{align}
a_1 = -0.014 \quad\quad a_2=4.5\times 10^{-4}.
\end{align}

The difference between the \npe/ calculated for the full light emission and the \npe/ calculated for the truncated emission was found to be negligible up to $\sim\SI{1000}{\pe}$. However, the visible energy of all events showing a digitizer range overflow was adjusted using $\Gamma_o$.\par

The resulting energy spectrum for the \isotope{Cf}{252} calibration is shown in the top panel of Fig.~\ref{fig:sns-analysis:inelastic-fit}. A peak from inelastic  scattering \isotope{I}{127}(n,n'$\gamma$) at around \SI{60}{\keV} is apparent. The small shift towards higher energies as well as the \emph{shark tooth} shape can be explained by the addition of the energy of the recoiling nucleus and additional $\gamma$ de-excitation \cite{jovancevic-01,collar-04}. The second peak at $\sim\SI{30}{\keV}$ is caused by the electron capture decay of \isotope{I}{128} ($E_\gamma\sim\SI{31}{\keV}$) which is produced by thermal neutron capture \cite{chu-01}.\par

The total number of inelastic scattering events was calculated by fitting an \emph{ad hoc} peak template to the peak region. This template is given by
\begin{align}
\zeta(N_\text{pe}) &= I(N_\text{pe},a,N_0,N_1,\delta_N + B(N_\text{pe},m,c) \label{eq:sns-analysis:peak-template}\\
I(N_\text{pe},a,N_0,N_1,\delta_N) & = a\left[ 1 + \text{e}^{-0.1\left(N_\text{pe} - N_0\right)}\right]^{-1} \text{e}^{-{\nicefrac{(N_\text{pe}-N_1)}{\delta_N}}}\\
B(N_\text{pe},m,c) & = m N_\text{pe} + c
\end{align}
where $I$ represents the inelastic peak and $B$ a linear background. The exact fit parameters are given in Table~\ref{tab:sns-analysis:ad-hoc-template-fit}. The total number of inelastic scattering events was determined by integrate the contribution from $I$ alone, which yielded
\begin{align}
N^\text{exp}_\text{is}=589\pm68
\end{align}

\begin{table}[tb]
	\begin{center}
		\begin{tabular}{cclccl}
		\toprule
		\multicolumn{3}{c}{$I(N_\text{pe},a,N_0,N_1,\delta_N)$} & \multicolumn{3}{c}{$B(N_\text{pe},m,c)$}\\
		\midrule
		$a$   &=& $69.1 \pm 7.1$                 & $m$ &=& $(-1.4 \pm 4.6) \times 10^{-2}$ \\
		$N_0$ &=& $(7.99 \pm 0.10)\times 10^{2}$ & $c$ &=& $(1.29 \pm 0.34) \times 10^{2}$ \\
		$N_1$ &=& $(8.20 \pm 0.38)\times 10^{2}$ \\
		$\delta_N$ &=& $68 \pm 39$\\
		\bottomrule
		\end{tabular}
	\end{center}
	\caption[\emph{Ad hoc} template fit parameters for the californium calibration]{\emph{Ad hoc} template fit parameters for the californium calibration. The fit parameters found for $N_0,\,N_1$ and $\delta_N$ were used to fix the shape of the \isotope{I}{127}(n,n'$\gamma$) peak within the \ac{cenns} search data set.}
	\label{tab:sns-analysis:ad-hoc-template-fit}
\end{table}	

The number of inelastic scattering events calculated above were further compared to the prediction based on an \mcnp/ simulation. The number of inelastic scattering events expected was calculated as follows: First, the total number of neutrons emitted from the \isotope{Cf}{252} source was calculated. The source was acquired two months prior to this measurement and its activity was given as $A=\SI{2}{\micro\curie}\pm\SI{10}{\percent}$ by the manufacturer. The branching ratio of spontaneous fission is given by \SI{3.092}{\percent} \cite{chu-01} and the number of neutrons per fission is 3.7692 \cite{carlson-01}. The half-life of \isotope{Cs}{252} is given by $T_{\nicefrac{1}{2}}=\SI{2.645}{\year}$ \cite{chu-01}. The total neutron yield $\Gamma_n$ can therefore be written as 
\begin{align}
\Gamma_n = \SI{2}{\micro\curie} \times 0.03092 \times 3.7692 \times \text{Exp}\left[-\frac{\SI{0.167}{\year}}{\SI{2.645}{\year}\,\text{ln}(2)}\right]= 7873\frac{\text{n}}{\text{s}}\pm\SI{10}{\percent}
\end{align}

Second, the efficiency of a single neutron emitted by the \isotope{Cf}{252} source to actually undergo inelastic scattering within the \csi/ detector was calculated using an \mcnp/ simulation. A comprehensive \mcnp/ simulation was performed and this efficiency was found to be $\eta_\text{is}=\num{1.31e-5}$. The total number $N^\text{sim}_\text{is}$ of inelastic scattering \isotope{I}{127}(n,n'$\gamma$) events expected based on these simulations is given by
\begin{align}
N^\text{sim}_\text{is} = \Delta T\times\Gamma_n\times\eta_\text{is} = \SI{6420}{\second} \times 7873 \frac{\text{n}}{\text{s}} \times \num{1.31e-5} = 662 \pm \SI{10}{\percent}.
\end{align}
The number of experimental counts $N^\text{exp}_\text{is}=589\pm68$ is compatible with the predicted number $N^\text{sim}_\text{is}=662\pm66$ within their respective errors. The \emph{shark-tooth} shape of the inelastic scattering \isotope{I}{127}(n,n'$\gamma$) peak was also correctly predicted by the simulations. This exercise confirmed the validity of the neutron transport simulations used in chapter~\ref{chapter:background-studies}.

\subsection{Inelastic \isotope{I}{127}(n,n' $\gamma$) scattering in the CE$\nu\!$NS search data}
Using the inelastic scattering \isotope{I}{127}(n,n'$\gamma$) reaction, bounds on the maximum background rate caused by prompt neutron interactions in the \csi/ can be determined in addition to the ones derived in chapter~\ref{chapter:background-studies}. The high energy spectrum, i.e., 200 to \SI{1200}{\pe}, is calculated for the full \cDS/ data in the \onDS/ data set. Based on the arrival time spectrum of prompt neutrons measured in chapter~\ref{chapter:background-studies}, only events with an arrival time of $T_\text{arr}\in[200,1100]\,$ns are included in this analysis. The energy of events exceeding the digitizer range were adjusted using Eq.~(\ref{eq:sns-analysis:overflow-conversion}). The resulting spectrum is shown in the bottom panel of Fig.~\ref{fig:sns-analysis:inelastic-fit}.\par

No peak is apparent at the location predicted by the \isotope{Cf}{252} calibration for the inelastic scattering \isotope{I}{127}(n,n'$\gamma$) reaction. To constrain the total number of inelastic events the following fit procedure was used. First a linearly decaying background was fitted to the full energy range, shown in dashed black. Second, the \emph{ad hoc} peak template, which is shown in Eq.~(\ref{eq:sns-analysis:peak-template}), was fitted to the peak region. However, most of the fit parameters of the \emph{ad hoc} template were predetermined. The parameters $m$ and $c$ are taken from the fit of the linearly decaying background. $N_0$, $N_1$, and $\delta_N$ were taken directly from the californium calibration (Table~\ref{tab:sns-analysis:ad-hoc-template-fit}). The only remaining free parameter was given by the peak amplitude $a$.\par

Once the template was fitted to the data the corresponding $I$ was integrated to yield the total number of inelastic scatter events. It is
\begin{align}
N^\text{exp}_\text{is} = 3.9 \pm 11.1,\label{eq:sns-analysis:nexp-is}
\end{align}
which is compatible with zero. The \SI{90}{\percent} confidence level of the upper limit is given by 22.2 counts and its corresponding fit is shown in red in the bottom panel of Fig.~\ref{fig:sns-analysis:inelastic-fit}. The fit uncertainty was mainly driven by the overall background achieved and the total exposure recorded. This uncertainty will diminish with more statistics in the future. A \mcnp/ simulation of prompt neutrons was conducted. The spectral hardness and a total flux of these prompt neutrons was set to the values measured in chapter~\ref{chapter:background-studies}. The neutrons coming from the \ac{sns} target unidirectionally bathed the \csi/ setup. The total number of inelastic scattering events induced by these prompt neutrons is 
\begin{align}
N^\text{sim}_\text{is} = 1.2 \pm 0.2,
\end{align}
for the full \onDS/ data set, i.e., a total beam energy on target of \SI{7.475}{\GWh}. This prediction is compatible with the value found in Eq.~(\ref{eq:sns-analysis:nexp-is}) at the $1\sigma$ level.\par

The excellent agreement between simulated predictions and measurements confirms the accuracy of the simulation results presented in chapter~\ref{chapter:background-studies}.

\section{\acs*{cenns} analysis}
In section~\ref{section:sns-analysis:stability} the \csi/ detector stability over the course of the two year data-taking period was examined. Time periods were identified and excluded from the analysis for which no valid information is available regarding the beam energy delivered on the \ac{sns} target. Rare time intervals showing a significant decrease in \ac{spe} charge were also excluded. This section provides a detailed description of the post-processing of the n-tuples calculated for the remaining time periods.\par

As discussed in chapter~\ref{chapter:ba-calibration} the analysis focuses on the residual \rDS/ between the \acDS/ and the \cDS/ data sets. The former precedes the \ac{pot} trigger and as such can not contain any beam-related contributions. It only contains steady-state backgrounds due to environmental radiation. \cDS/ in contrast contains both beam-related events and random coincidences. By calculating the residual spectra in energy and arrival time (\rDS/=\cDS/-\acDS/), all contributions from steady-state backgrounds are removed. As a result, \rDS/ only contains contributions from beam-related events, be those \ac{cenns}, prompt neutrons, or \acp{nin} (chapter~\ref{chapter:background-studies}).\par

The theoretical \ac{cenns} prediction can be compared to the residual spectrum \rDS/ to determine the observed level of agreement. However, both \rDS/ and the \ac{cenns} signal prediction depend on the exact choice of cut parameters used in the post-processing. The following section therefore discusses the analysis used to determine optimized cut parameters, which yield the highest ratio of the expected \ac{cenns} signal to the steady-state environmental backgrounds.

\subsection{Optimizing cut parameters}
\label{section:sns-analysis:optimizing-cuts}
All data cuts used in the analysis can be classified into three main categories. The first category consists of the quality cuts (section~\ref{section:sns-analysis:quality-cuts}), which cannot be adjusted. The three cuts in this category are the overflow, linear gate and muon veto cut. These cuts reject a fixed percentage of events and contribute an energy independent signal acceptance fraction. The second category consists of cuts for which the \ac{cenns} search data was used in order to determine their signal acceptance. The first of these is the afterglow cut (section~\ref{section:sns-analysis:afterglow-cut}), the second is the diagonal rise-time cut (its definition is analog to that in section~\ref{section:ba-calibration:calculating-acceptances}), both of which also provide an energy independent signal acceptance fraction. The third category contains cuts for which the \isotope{Ba}{133} calibration data was used to calculate a proper signal acceptance (chapter~\ref{chapter:ba-calibration}). This category includes the Cherenkov cut (section~\ref{section:ba-calibration:cherenkov-cut}), as well as the different orthogonal rise-time cuts (section~\ref{section:ba-calibration:rt-cuts}). The signal acceptance obtained from these cuts is given by an energy-dependent acceptance fraction, as described by Eq.~(\ref{eq:ba-calibration:acceptance-function}).\par

In order to establish the optimal choice of cut parameters, a \ac{fom} was defined which quantifies how well a certain set of parameters performs in maximizing the \ac{cenns} signal to steady-state background ratio. In order to prevent introducing any bias, the following approach was chosen: First, the quantification was limited to the \offDS/ data set, i.e., periods for which there can not be any beam related events, neither in \acDS/ nor \cDS/. This represents a form of a blind analysis. All data cuts were applied to the \acDS/ data, and the corresponding energy and arrival time spectra of the stead-state background events were determined. Only events with an arrival time of $T_\text{arr}\in[0,6]\SI{}{\micro\second}$ were included in the energy projection. The arrival time projection only included events with an energy of $N_\text{pe}\in[0,40]$. These choices are based on the expected \ac{cenns} energy and arrival time distributions, which show a negligible \ac{cenns} rate outside of these parameter bounds.\par

A total of \num{778634460} and \num{1558323928} triggers were recorded for the \offDS/ and the \onDS/ data sets, respectively. As discussed earlier, the corresponding beam energy on target delivered for the \onDS/ data is given by $E_\text{beam}=\SI{7475}{\MWh}$. In order to provide a real background estimate for the \ac{cenns} prediction, the exposure between the \offDS/ and \onDS/ data-taking periods needs to be matched. To this end, both energy and arrival time spectra determined from the \acDS/ data set, were multiplied by a factor of 2.00135. An example of the resulting, scaled spectra for different choices of the Cherenkov cut(\chc/, section~\ref{section:ba-calibration:cherenkov-cut}) is shown in panel \textbf{A} and \textbf{B} of Fig.~\ref{fig:sns-analysis:signal-vs-background}. 
The number of events passing all cuts decreases significantly with more stringent Cherenkov cut choices, i.e., increasing \chc/. However, it is also apparent that most of the events rejected carry energies below \SI{20}{\pe}, which is consistent with the energy expected from Cherenkov radiation in the \ac{pmt} window. In contrast, the arrival time distribution shows an overall decrease in event rate, consistent with what is expected from random coincidences. The slight increase towards earlier arrival times is caused by afterglow. Even though a cut based on \agc/ was applied to the data set, the arrival time of \ac{spe} from phosphorescence from high-energy depositions is still slightly biased towards earlier times in the waveform. This will further be discussed in section~\ref{section:sns-analysis:cenns-observation}.

\begin{figure}[htbp]
\begin{center}
\includegraphics[scale=1]{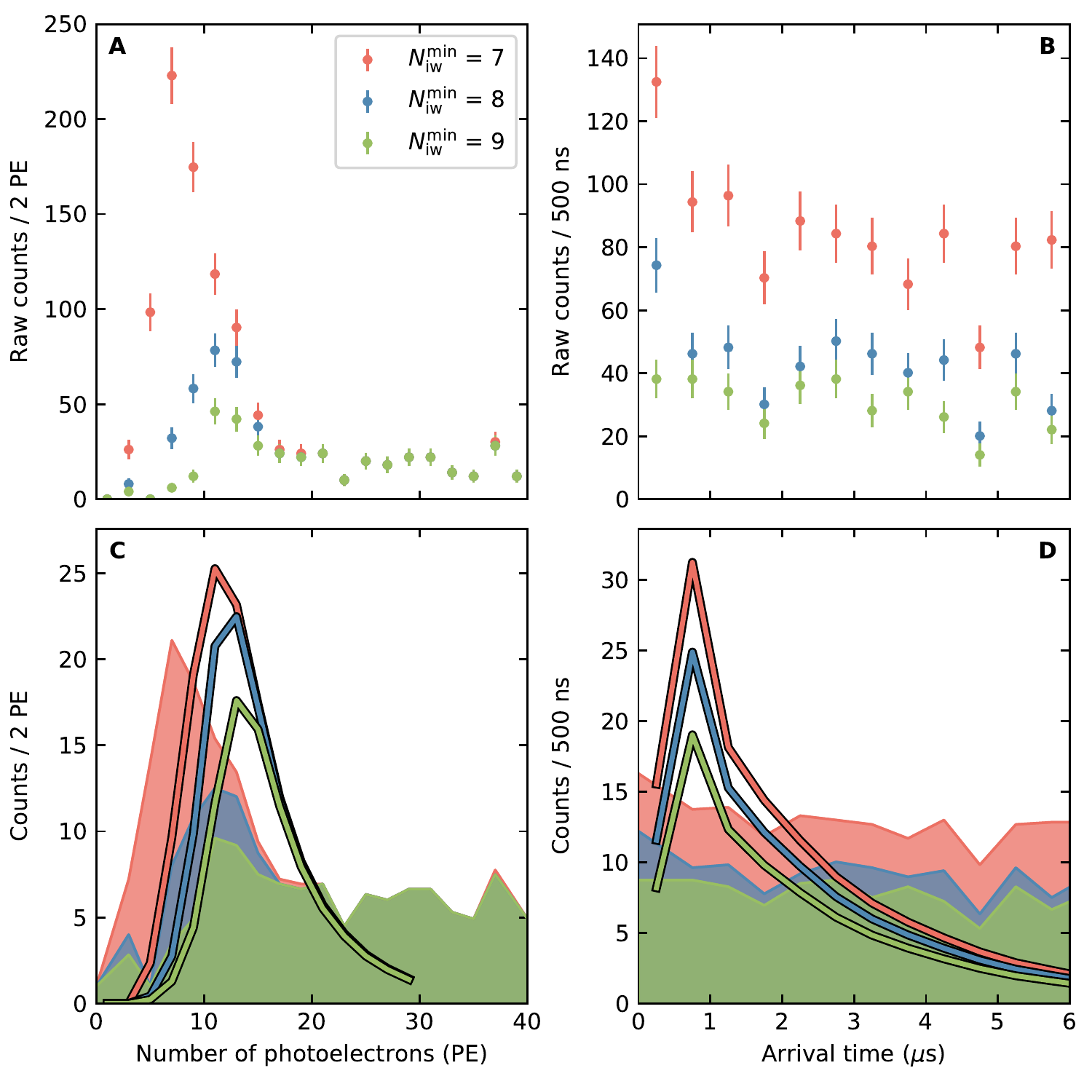}
\end{center}
\caption[Estimating the signal-to-background ratio using only \acDS/ data from the \offDS/ data set]{Panel \textbf{A} and \textbf{B} show the energy and arrival time spectra for three different choices of the Cherenkov cut \chc/, calculated from \acDS/ of the \offDS/ data set.  The data was scaled to match the number of triggers found in the \onDS/ data set. A significant reduction in the number of events passing the data cuts for increasing \chc/ can be observed. Panel \textbf{A} highlights that mostly events with an energy of less than \SI{20}{\pe} are rejected, whereas panel \textbf{B} shows an overall reduction independent of the arrival time. This is consistent with the assumption that most of these events consist of random coincidences of Cherenkov spikes with the \acs*{pot} trigger. 
The trend in time projections towards monotonically decreasing rates with increasing arrival time originates in events with a small afterglow component, able to pass all cuts.}
\label{fig:sns-analysis:signal-vs-background}
\end{figure}

\begin{figure}[htbp]
\begin{center}
\ContinuedFloat
\end{center}
\caption[]{(Continued) The afterglow cut needed to be chosen small enough to reduce this trend to a level at which the \offDS/ residual shows no systematic deviation from zero in the energy and arrival time spectra. However, softening the afterglow cut increases the number of steady-state backgrounds passing all cuts. This effect is more pronounced in \acDS/ than in \cDS/. As as result, the residual spectra show a systematic deviation opposite to a potential \ac{cenns} signal and can therefore not be mistaken for a \ac{cenns} signal. Panels \textbf{C} and \textbf{D} show the statistical fluctuations $\sigma_\text{bg}$ around zero for the corresponding residual \rDS/, caused by random coincidences, as colored bands. The colors correspond to the data shown in panels \textbf{A} and \textbf{B}. Only the positive fluctuation is shown. The solid lines represent the predicted \ac{cenns} spectra for the different cut values. Comparing the expected \ac{cenns} spectrum to the expected steady-state background allows the definition of a \acf*{fom} (Eq.~(\ref{eq:sns-analysis:fom}) which is used to find an optimal set of cut parameters that maximizes the \ac{cenns} signal to steady-state background ratio.}
\end{figure}

To quantify the steady-state background in the \ac{cenns} search, the fluctuations $\sigma_\text{bg}$ around zero expected for the residual \rDS/ in the absence of any beam-related signal (\ac{cenns}, prompt neutrons, or \acp{nin}) were estimated. As there is no difference in the steady-state background rate between \acDS/ and \cDS/ for the \offDS/ data set, these fluctuations can be estimated using
\begin{align}
\sigma^i_\text{bg}=\sqrt{2 N^i_\mathcal{AC}},\label{eq:sns-analysis:residual-background}
\end{align}
where $i$ denotes the $i$-th bin of either energy or arrival time distribution and $N^i_\mathcal{AC}$ the number of events found with the corresponding energy or arrival time.\par

The magnitude of these fluctuations is shown in panel \textbf{C} and \textbf{D} of Fig.~\ref{fig:sns-analysis:signal-vs-background} as colored areas, with only the positive band being shown. It can be observed that the size of the fluctuations naturally decreases with increasing \chc/ as fewer and fewer events pass the cuts, i.e., $N^i_\mathcal{AC}$ becomes smaller and smaller.\par

To estimate a proper signal-to-background ratio, the expected \ac{cenns} signal for different cut parameter combinations needs to be calculated. For this purpose, the uncut \ac{cenns} induced nuclear recoil spectrum was calculated as it would be expected in the \csi/ detector at the \ac{sns}. Combining Eq.~(\ref{eq:cenns-theory:form-factor-inclusion}), (\ref{eq:cenns-theory:diff-cross-section-2}) and (\ref{eq:cenns-theory:form-factor}) yields the differential \xs/ for both cesium and iodine and with its dependence on the incoming neutrino energy. The differential \xs/ was convolved with the three different neutrino emission spectra, Eq.~(\ref{eq:sns:nmu-emission},~\ref{eq:sns:ne-emission},~\ref{eq:sns:anmu-emission}), to extract a differential recoil spectrum for each neutrino type and isotope. The resulting differential recoil spectrum was further scaled to match a reference beam energy delivered on the \ac{sns} target of \SI{1}{\giga\watt\hour}.\par

\geant/ simulations regarding the neutrino production rate were conducted by the University of Florida group within the \coherent/ collaboration. The full target geometry, the neutron moderators, as well as the beryllium reflector surrounding the target were incorporated in the simulation \cite{collar-04}. The QGSP\textunderscore BERT physics list was chosen, which uses the Bertini model~\cite{bertini-01} to simulate the intra-nuclear cascade. A total production rate of $0.08\pm\SI{10}{\percent}$ neutrinos per flavor per proton was found, which amounts to \num{1.98e21} per \SI{}{\GWh} per flavor as a reference for a proton energy of $\sim\SI{960}{\MeV}$.\par

\begin{figure}[htb]
\begin{center}
\includegraphics[scale=1]{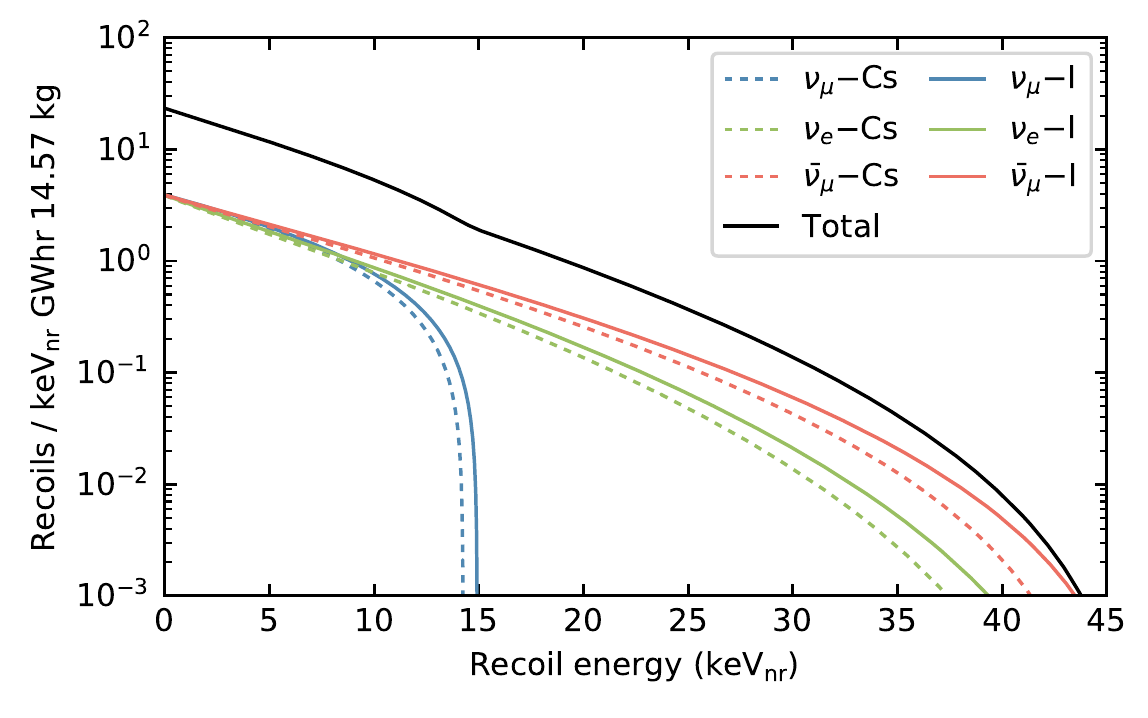}
\end{center}
\caption[Nuclear recoil rate from \acs*{cenns} interactions in the \csi/ detector at the \acs*{sns}]{Nuclear recoil rate from \acs*{cenns} interactions in the \csi/ detector at the \acs*{sns}. Iodine (cesium) recoils are shown as solid (dotted) lines. The recoil spectra for \isotope{I}{127} and \isotope{Cs}{133} are very similar as already discussed in earlier chapters. The total recoil rate is shown in black.}
\label{fig:sns-analysis:sns-recoil-spectrum}
\end{figure}

With a detector mass of \SI{14.57}{\kg} and a distance between target and detector of \SI{19.3}{\m} (Fig.~\ref{fig:sns:neutrino-alley}), the final recoil rate for each flavor and isotope can be determined (Fig.~\ref{fig:sns-analysis:sns-recoil-spectrum}) for a beam energy delivered on the mercury target, of \SI{1}{\GWh}. The total recoil rate is shown in black.\par

The individual recoil spectra con be integrated to extract the total number of \ac{cenns} events $N^\alpha_{\nu_x}$ that are expected in a zero-threshold detector for each neutrino flavor $x$ and isotope $\alpha$. The total number of events expected for a perfect detector is given by $N_\text{total} = \sum\limits_\alpha\sum\limits_x N^\alpha_{\nu_x} = 1146$ events per \SI{1}{\giga\watt\hour}.\par

This estimate does not include any threshold effects and other data quality cuts. To properly account for these, the energy deposited in an event needs to be converted into the number of \ac{spe} produced. A \acl{mc} approach was used to convert the nuclear recoil energy deposited in the \csi/ into a corresponding number of photoelectrons produced \npe/, which is described next.\par

\begin{figure}[tb]
\begin{center}
\includegraphics[scale=1]{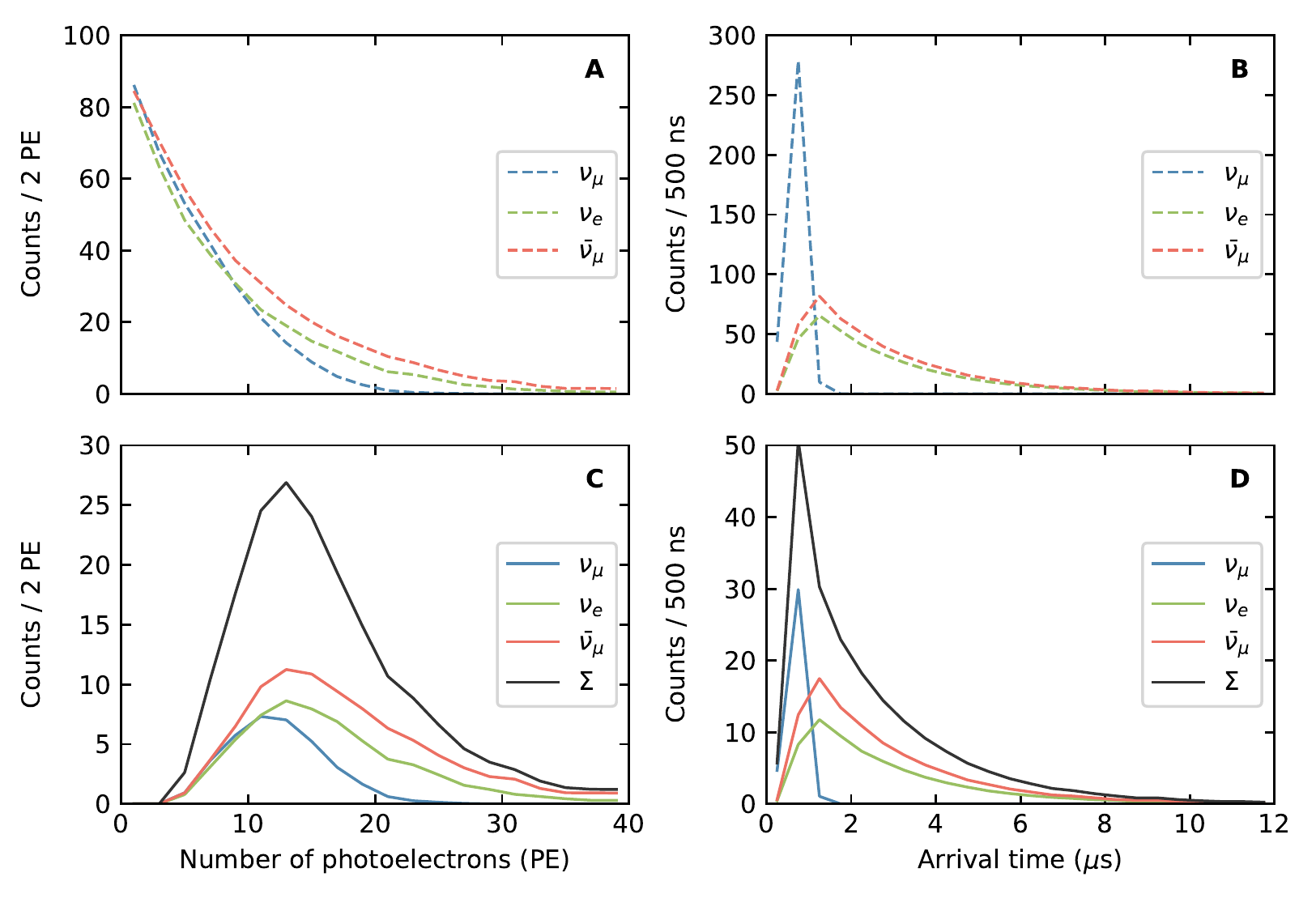}
\end{center}
\caption[Comparison between \acs*{cenns} spectra before and after applying the signal acceptance function to the data]{Panels \textbf{A} and \textbf{B} show the uncut \ac{cenns} energy and arrival time spectra for each neutrino flavor expected in the \csi/ crystal respectively. The spectra were scaled to match the total total neutrino production expected from a beam energy delivered on target of \SI{7475}{\MWh}. Panels \textbf{C} and \textbf{D} show the corresponding \ac{cenns} spectra after the signal acceptance function was applied.}
\label{fig:sns-analysis:cenns-prediction-cut-uncut}
\end{figure}

In order to extract a smooth energy spectrum, a total of $N^\alpha_{\nu_x}\times 10000$ recoil events were simulated for each isotope $\alpha$ and neutrino flavor $x$. Each event energy was drawn from the corresponding differential recoil spectrum shown in Fig.~\ref{fig:sns-analysis:sns-recoil-spectrum}. The energy of each event was converted from \SI{}{\keVnr} to \SI{}{\keVee} using the quenching factor calculated in chapter~\ref{chapter:quenching-calibration}, i.e., $Q = 8.78\pm\SI{1.66}{\percent}$. Using the light yield determined in chapter~\ref{chapter:am-calibration}, i.e., $\mathcal{L}_\text{CsI} = 13.348 \pm 0.019\,\frac{\text{PE}}{\si{\keVee}}$, the ionization energy was converted to \npe/. Poisson fluctuations were applied to the number of photoelectrons. For each event an arrival time was drawn from the distributions shown in the right panel of Fig.~\ref{fig:cenns:sns-neutrino-production-mechanism}. These results were combined to calculate the uncut \ac{cenns} distributions for all neutrino flavors (Fig.~\ref{fig:sns-analysis:cenns-prediction-cut-uncut}, panels \textbf{A} and \textbf{B}), where the spectra were divided by a factor of \num{10000} to account for the \ac{mc} smoothing factor.\par

\begin{figure}[htbp]
\begin{center}
\includegraphics[scale=1]{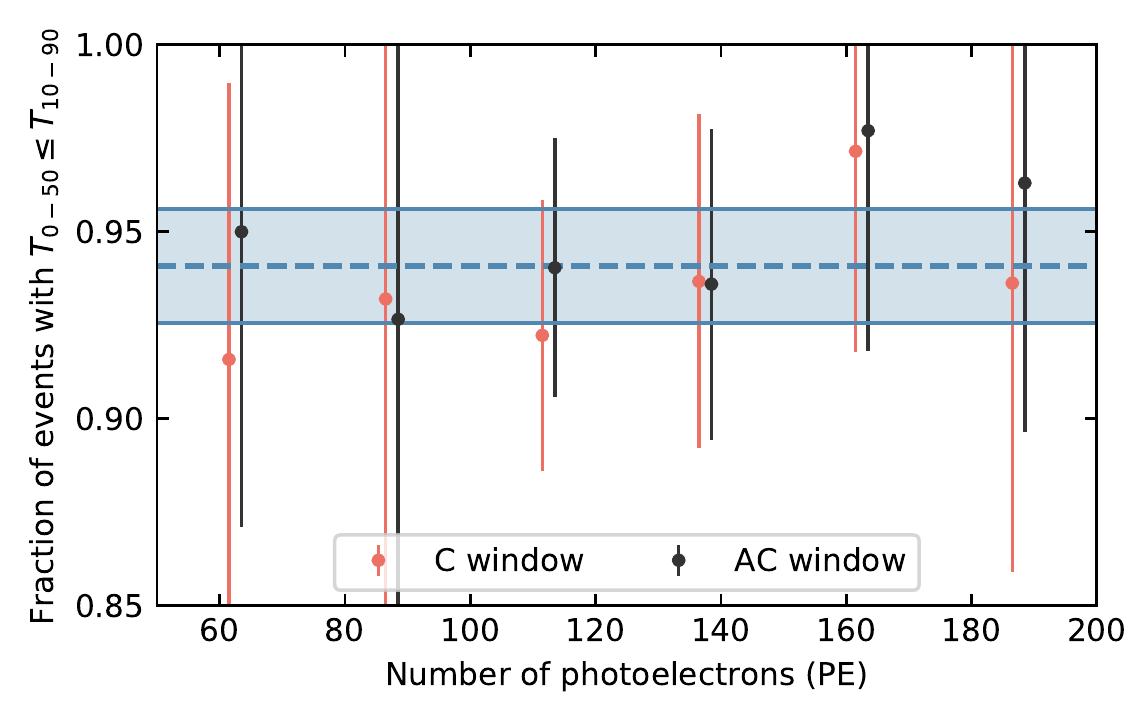}
\end{center}
\caption[Determination of the signal acceptance fraction for the diagonal rise-time cut]{Determination of the signal acceptance fraction for the diagonal rise-time cut using high-energy events in \acDS/ and \cDS/. The \acDS/ and \cDS/ are offset from their respective bin center to improve readability. The fraction of triggers with a misidentified onset is independent of the energy deposited in an event. As a result one can use high-energy events, which are unaffected by the  Cherenkov and the orthogonal rise-time cuts, to determine $\eta_\text{diag-rise-time}$.}
\label{fig:sns-analysis:diag-rt-cut}
\end{figure}

The \ac{cenns} prediction needs to be properly scaled by incorporating the signal acceptance fractions from the different cuts employed in this analysis. An overall signal acceptance function is defined that combines all different cut fractions. This overall acceptance function can be written as
\begin{align}
\eta(N_\text{pe}) = \eta_\text{Ba}(N_\text{pe},a,k,x_0)\;\eta_\text{afterglow}\;\eta_\text{muon-veto}\;\eta_\text{linear-gate}\;\eta_\text{diag-rise-time}\label{eq:sns-analysis:acceptance-function}
\end{align}
where $\eta_\text{Ba}(N_\text{pe},a,k,x_0)$ is discussed in detail in chapter~\ref{chapter:ba-calibration}, whereas $\eta_\text{afterglow},\;\eta_\text{muon-veto}$ and $\eta_\text{linear-gate}$ are discussed in section~\ref{section:sns-analysis:stability}. The following approach was used to calculate $\eta_\text{diag-rise-time}$, which was previously motivated and described in section~\ref{section:ba-calibration:calculating-acceptances}. To determine the acceptance of the diagonal rise-time cut all other cuts had to be applied to the \acDS/ and \cDS/ data of the \onDS/ first. Next, the percentage of events with $T_{0-50}\leq T_{10-90}$ was calculated for energies between 50 - 200 PE. As the chance to misidentify the onset of an event is independent of its energy this percentage directly represents the acceptance fraction for the diagonal rise-time cut (section~\ref{section:ba-calibration:calculating-acceptances}). Fig.~\ref{fig:sns-analysis:diag-rt-cut} shows the percentage of events satisfying $T_{0-50}\leq T_{10-90}$ as a function of energy for a choice of $\agc/\,=\,3$ and $\chc/\,=\,8$. The dashed blue line represents the uncertainty weighted average of the percentages shown. The $1\sigma$ uncertainty is shown as shaded blue region. The acceptance fraction found for this particular choice of afterglow and Cherenkov cut is
\begin{align}
\eta_\text{diag-rise-time}\,=\,0.941\,\pm\,0.015,
\end{align}

\begin{figure}[tb]
\begin{center}
\includegraphics[scale=1]{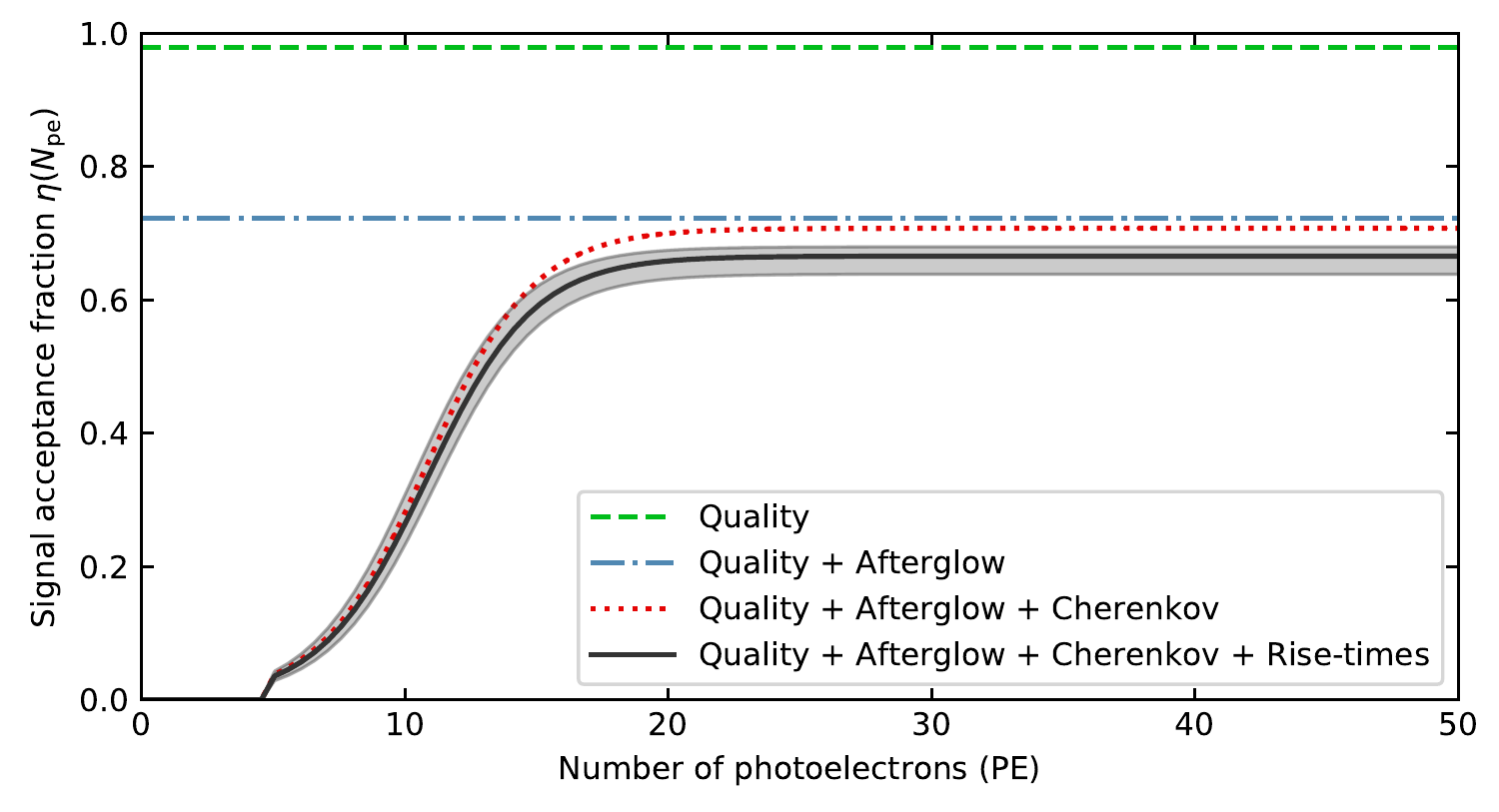}
\end{center}
\caption[Overall signal acceptance function as calculated for the optimized cut parameters]{The overall signal acceptance function as calculated for the optimized cut parameters, given in by Eq.~(\ref{eq:sns-analysis:best-cut-parameters}), is shown in black. The associated uncertainty is shown as a gray band and dominated by the available \isotope{Ba}{133} statistics, that was used to quantify the acceptances of the Cherenkov and orthogonal rise-time cuts. The cumulative evolution of this signal acceptance function upon inclusion of all individual cuts is also shown.}
\label{fig:sns-analysis:acceptance-curve}
\end{figure}

As an example, Fig.~\ref{fig:sns-analysis:acceptance-curve} shows the overall signal acceptance function corresponding to the optimal choice of cut parameters as calculated later in this section (Eq.~(\ref{eq:sns-analysis:best-cut-parameters})). Shown is the evolution of the acceptance function as different cuts are added. As such, every curve includes all cuts already shown above in addition to the one shown in the label. The black curve represents the full overall signal acceptance function as given in Eq.~(\ref{eq:sns-analysis:acceptance-function}).\par

The overall acceptance function $\eta(N_\text{pe})$ was applied to the predicted \ac{cenns} energy spectrum which is shown in panel \textbf{A} of Fig.~\ref{fig:sns-analysis:cenns-prediction-cut-uncut}. The fraction of rejected events was recorded for each individual neutrino flavor and the corresponding arrival time spectra for each flavor were scaled by this fraction. For the resulting arrival time spectrum the fraction of events with an arrival time of $T_\text{arr}<\SI{6}{\micro\second}$ was computed and the energy spectra were scaled accordingly. Last, the arrival time spectrum was adjusted to only include events with an energy of $\leq\SI{30}{\pe}$. The resulting \ac{cenns} spectra are shown in panels \textbf{C} and \textbf{D} of Fig.~\ref{fig:sns-analysis:cenns-prediction-cut-uncut}.\par

It is apparent from panel \textbf{C} of Fig.~\ref{fig:sns-analysis:cenns-prediction-cut-uncut}, that a majority of \ac{cenns} events was cut due to overall signal acceptance function. However, for the choice of cuts shown in this example plot (afterglow and rise-time cuts are given in Eq.~(\ref{eq:sns-analysis:best-cut-parameters}), Cherenkov cut \chc/ as given in the labeling of the figure) one would still expect to observe $\mathcal{O}(100-200)$ events in the energy spectrum.\par

A simple \acf{fom} can be defined to maximize the \ac{cenns} signal to steady-state background ratio, by demanding that the optimized cut parameters should minimize the background $\sigma_\text{i,bg}$ while maximizing $N_\text{i,signal}$ for each bin $i$ at the same time. A straight forward choice for the \ac{fom} is to minimize the likelihood of a residual with zero counts and a dispersion of $\sigma_\text{i,bg}$ to actually mimic a corresponding \ac{cenns} signal. This directly leads to
\begin{align}
\text{FOM} = \sum_i\frac{N_\text{i,signal}^2}{\sigma_\text{i,bg}^2},\label{eq:sns-analysis:fom}
\end{align}
which is a $\chi^2$ goodness-of-fit test between the predicted signal and a residual fluctuating around zero counts. In contrast to a maximum likelihood fit, the goal is to minimize the likelihood that a residual fluctuating around zero is able to reproduce the potential \ac{cenns} signal, which means the \ac{fom} was maximized.
\par

To identify the optimal cuts, the \ac{fom} needs to be calculated for each possible combination of parameters. This can not be done in a sequential manner, i.e, finding the best parameter choice for one cut, since these cuts interplay on a complex manner. To efficiently find the set of parameters that jointly maximize the \ac{fom}, the range of the parameters explored in this search was limited to

\begin{equation}
\begin{aligned}
\text{Afterglow cut:}\quad\agc/ &\in [0;9]\\
\text{Cherenkov cut:}\quad\chc/ &\in [6,7,8,9]\\
\text{Rise-time cuts:}\quad T_{0-50}^\text{min} &\in [0,50,100,150,200]\,\SI{}{\nano\second}\\
                                       T_{0-50}^\text{max} &\in [1.50,2.00,2.50,3.00]\,\SI{}{\micro\second}\\
                                       T_{10-90}^\text{min} &\in [0.00,0.25,0.50,0.75,1.00]\,\SI{}{\micro\second}\\
                                       T_{10-90}^\text{max} &\in [2.25,2.50,2.75,2.85,2.95,3.00]\,\SI{}{\micro\second}.
\end{aligned}
\end{equation}
This amounts to a total of $10 \times 4 \times 5 \times 4 \times 5 \times 6 = \num{24000}$ potential cut parameter combinations. The rise-time cut parameter ranges were chosen by looking at Fig.~\ref{fig:ba-calibration:rt-distribution-02} and estimating an equal percentage of signal events being cut in each step.\par

For all cut combinations the energy and arrival time spectra were calculated for \acDS/ using the \offDS/ data set in order to determine $\sigma_\text{i,bg}$ using Eq.~(\ref{eq:sns-analysis:residual-background}). The acceptance function of the Cherenkov and orthogonal rise-time cuts was calculated using the barium calibration (chapter~\ref{chapter:ba-calibration}). By combining all data cuts into a single overall signal acceptance function, (Eq.~(\ref{eq:sns-analysis:acceptance-function}), the \ac{cenns} prediction (Fig.~\ref{fig:sns-analysis:cenns-prediction-cut-uncut}) was properly adjusted. Finally, the \ac{fom} was calculated using Eq.~(\ref{eq:sns-analysis:fom}). The parameter choice which maximizes the \ac{fom} is given by
\begin{equation}
\begin{aligned}
\label{eq:sns-analysis:best-cut-parameters}
\text{Afterglow cut:}\quad\agc/ & = 3\\
\text{Cherenkov cut:}\quad\chc/ & = 8\\
\text{Rise-time cuts:}\quad T_{0-50}^\text{min} & = \SI{200}{\nano\second}\\
                                       T_{0-50}^\text{max} & = \SI{2.5}{\micro\second}\\
                                       T_{10-90}^\text{min} & = \SI{0.5}{\micro\second}\\
                                       T_{10-90}^\text{max} & = \SI{2.85}{\micro\second}.       
\end{aligned}
\end{equation}

\begin{figure}[htbp]
\begin{center}
\includegraphics[scale=1]{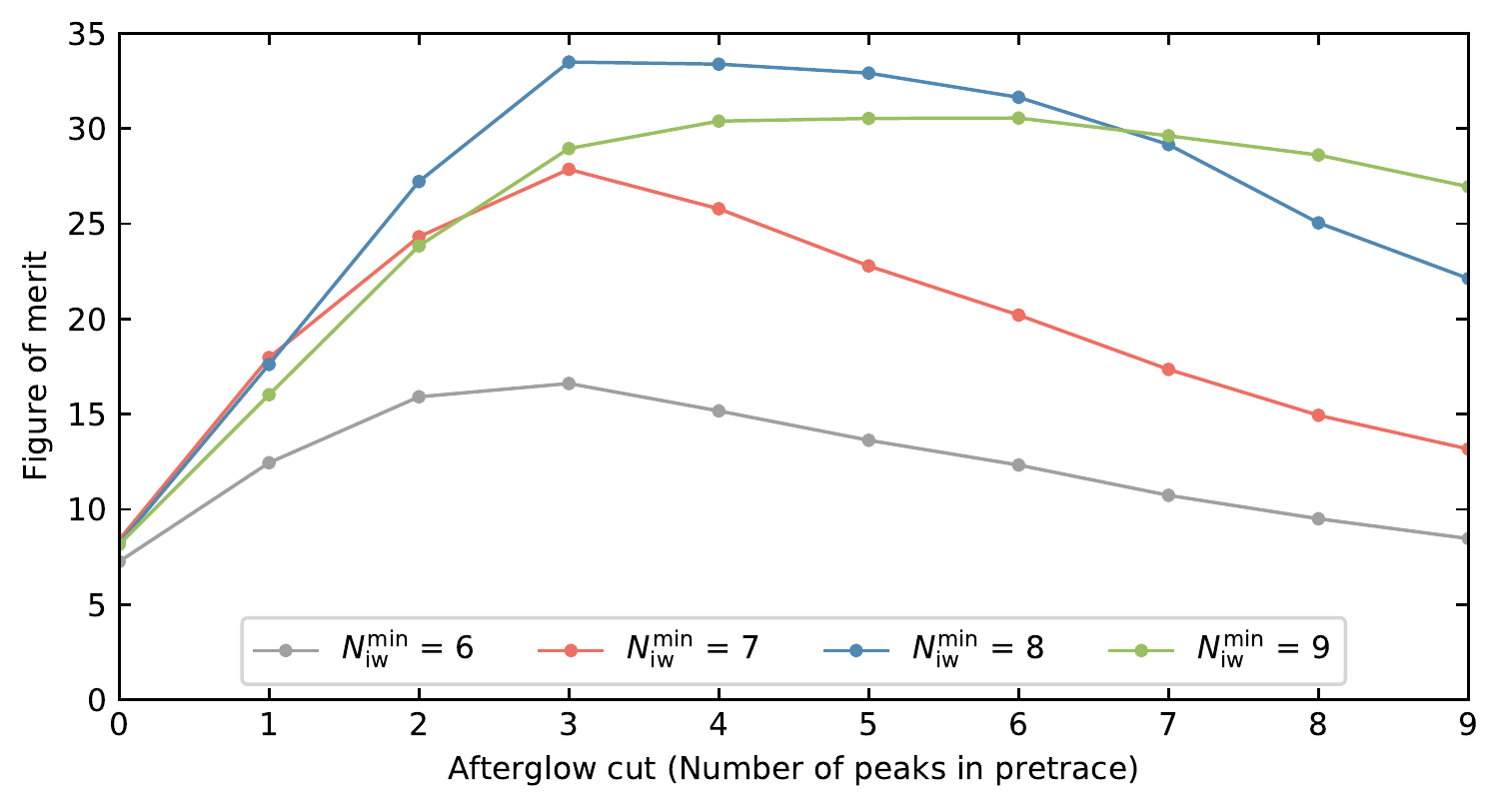}
\end{center}
\caption[Figure of merit, i.e., a measure of the signal-to-background ratio, for different afterglow and Cherenkov cuts in the \acs*{cenns} search]{Shown is the \ac{fom} calculated using Eq.~(\ref{eq:sns-analysis:fom}) for different afterglow and Cherenkov cut choices. The orthogonal rise-time cuts were fixed to their optimized values as given in Eq.~(\ref{eq:sns-analysis:best-cut-parameters}). It is apparent that $\chc/=8$ yields the highest \ac{fom} for most choices of \agc/.}
\label{fig:sns-analysis:fom-vs-ppt}
\end{figure}

In order to ensure that this choice of cut parameters indeed represents the best choice and not simply an artifact in the analysis, the evolution of the \ac{fom} with respect to the choice of \chc/ and \agc/ was investigated. All orthogonal rise-time cuts were set to the values quoted in Eq.~(\ref{eq:sns-analysis:best-cut-parameters}) and the \ac{fom} was created for different \agc/ and \chc/. The evolution of the \ac{fom} is shown in Fig.~\ref{fig:sns-analysis:fom-vs-ppt}. It can be seen that for most choices of afterglow cuts \agc/ a Cherenkov cut of $\chc/ = 8$ results in the highest \ac{fom}. It can also be verified that the \ac{fom} peaks at an afterglow cut of 3 for most Cherenkov cuts. This additional test validates the choice of $\agc/= 3$ and $\chc/= 8$, which were found in Eq.~\ref{eq:sns-analysis:best-cut-parameters}.\par

The overall signal acceptance function associated with this choice of parameters can be calculated using Eq.~(\ref{eq:sns-analysis:acceptance-function}). Throughout this thesis these optimal cut parameters were used to illustrate the analysis. By incorporating all individual acceptances Eq.~(\ref{eq:sns-analysis:acceptance-function}) can be simplified to
\begin{align}
\eta_\text{best}\left(N_\text{pe}\right) =\frac{0.665}{1 + \text{e}^{-0.494\,(N_\text{pe}\,-\,10.85)}} \Theta_\text{H}\left(N_\text{pe} - 5\right),
\end{align}
where $\Theta_\text{H}$ again represents the Heaviside step function. This acceptance function is shown in black in Fig.~\ref{fig:sns-analysis:acceptance-curve}.

\subsection{First observation of CE$\nu$NS}
\label{section:sns-analysis:cenns-observation}
Once the cut parameters optimizing the signal-to-background ratio had been calculated, these cuts were applied to all of the data sets, i.e., \acDS/ and \cDS/ for both \onDS/ and \offDS/ data sets. Table~\ref{tab:sns-analysis:trigger-numbers} lists the the total number of events passing all cuts for all data set combinations. The large number of events rejected by the Cherenkov and the afterglow cut are apparent. The rise-time cuts only reject a small amount of the surviving events. The numbers presented in Table~\ref{tab:sns-analysis:trigger-numbers} include events of all energies. However, \ac{cenns}-induced recoils only deposit a small amount of energy in the crystal (lower left panel of Fig.~\ref{fig:sns-analysis:cenns-prediction-cut-uncut}). As such it is favorable to focus on events depositing an energy of $\npe/\leq\SI{50}{\pe}$ in the crystal. The total number of events passing all cuts in all data sets within this energy region is given by
\begin{equation}
\begin{aligned}
N^\offDS/_\acDS/ & = 518 \qquad\qquad N^\onDS/_\acDS/ = 1032\\
N^\offDS/_\cDS/ & = 507 \qquad\qquad N^\onDS/_\cDS/ = 1207.
\end{aligned}
\end{equation}
\begin{table}[tbp]
	\begin{center}
		\begin{tabular}{ccccc}
		\toprule
		Cumulative cuts & \multicolumn{4}{c}{Number of waveforms passing all cuts}\\
		\midrule
		Total	        & \multicolumn{4}{c}{\num{2825705648}} \\
		Time \& Power & \multicolumn{4}{c}{\num{2393035787}} \\
		Quality       & \multicolumn{4}{c}{\num{2336958388}} \\
								  \cmidrule(lr){2-3}\cmidrule(lr){4-5}
			            & \multicolumn{2}{c}{\onDS/}           & \multicolumn{2}{c}{\offDS/}\\
									\cmidrule(lr){2-3}\cmidrule(lr){4-5}
			            & \multicolumn{2}{c}{\num{1558323928}} & \multicolumn{2}{c}{\num{778634460}}\\
									\cmidrule(lr){2-2}\cmidrule(lr){3-3}\cmidrule(lr){4-4}\cmidrule(lr){5-5}
									& \cDS/ & \acDS/ & \cDS/ & \acDS/ \\
									\cmidrule(lr){2-2}\cmidrule(lr){3-3}\cmidrule(lr){4-4}\cmidrule(lr){5-5}
		Cherenkov	    & \num{7298862} & \num{7362478} & \num{3320220} & \num{3353320} \\
		Afterglow	    & \num{15779}   & \num{15713}   & \num{7692}    & \num{7740} \\
		Rise-times    & \num{12906}   & \num{12844}   & \num{6228}    & \num{6270} \\
		\bottomrule
		\end{tabular}
	\end{center}
	\caption[Number of events surviving the different cuts applied to the \acs*{cenns} search data]{Number of events surviving the different cuts applied to the \acs*{cenns} search data. The cuts are applied in a cumulative manner, as such each new row includes all cuts listed in the rows above.}
	\label{tab:sns-analysis:trigger-numbers}
\end{table}
\begin{figure}[htb]
\begin{center}
\includegraphics[scale=1]{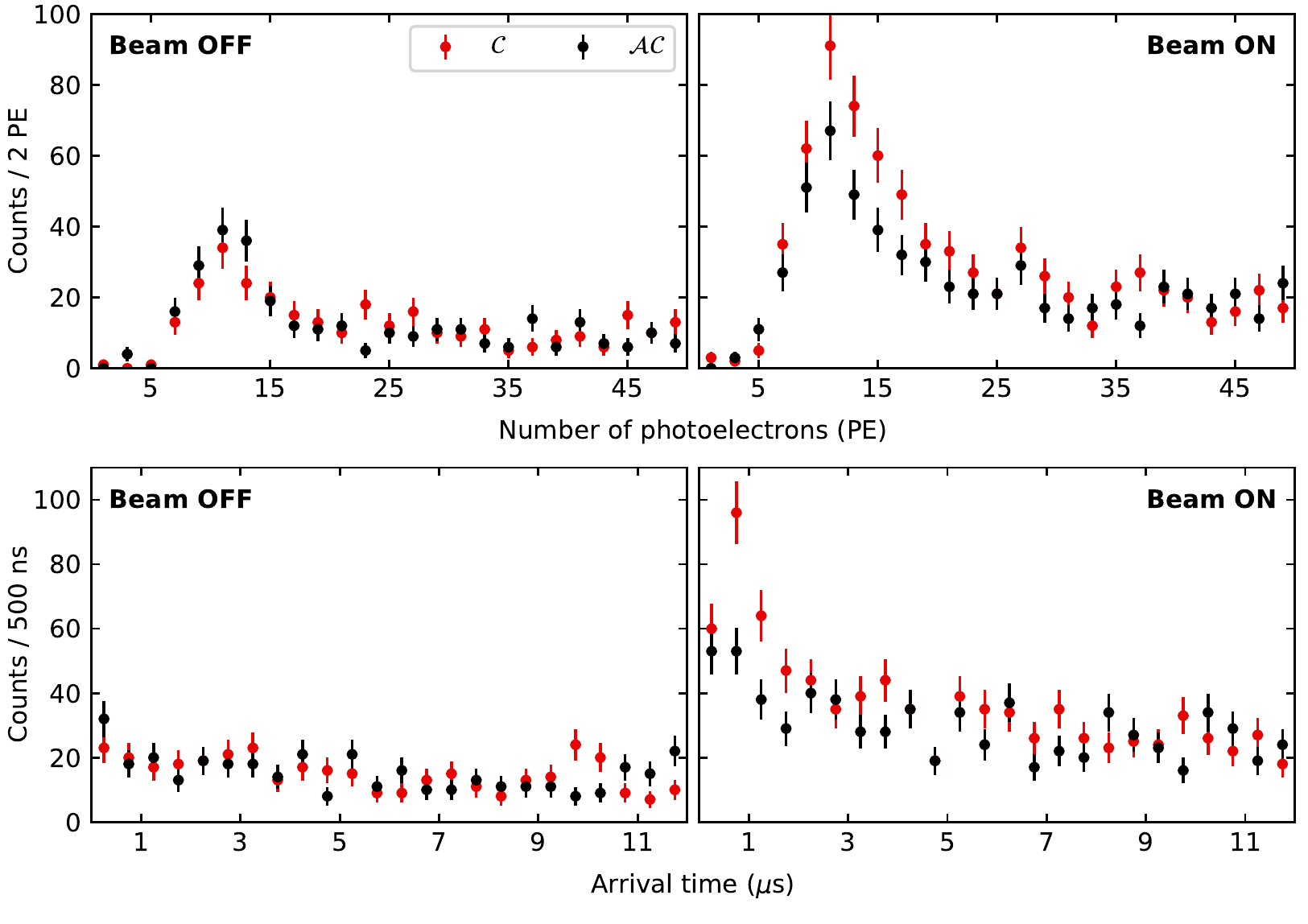}
\end{center}
\caption[Energy and arrival time spectra for all events passing the data cuts]{Energy (top) and arrival time (bottom) spectra for all events passing the optimized data cuts. The \offDS/ (\onDS/) data is shown on the left (right). For times during which the \acs*{sns} provided beam power on target, i.e., \onDS/, a clear excess of \cDS/ over \acDS/ is apparent, both in energy and arrival time. No such excess is visible for \offDS/.}
\label{fig:sns-analysis:raw-on-off-spectra}
\end{figure}

Fig.~\ref{fig:sns-analysis:raw-on-off-spectra} shows the energy and arrival time spectra for all events passing the optimized data cuts. Two additional constraints were used during the projection of the data onto the energy and arrival time spectra. First, only events with an arrival time of $T_\text{arr}\leq\SI{6}{\micro\second}$ were added to the energy spectrum. Second, the energy of events contributing to the arrival time spectrum was limited to $\npe/\leq\SI{30}{\pe}$, to maximize the contribution from \ac{cenns} induced events (Panel \textbf{C} of Fig.~\ref{fig:sns-analysis:signal-vs-background}). The \offDS/ data is shown on the left of Fig.~\ref{fig:sns-analysis:raw-on-off-spectra} and the \onDS/ data set is shown on the right. \par

In the top left panel of Fig.~\ref{fig:sns-analysis:raw-on-off-spectra} an excess below \SI{15}{\pe} is visible for both \acDS/ and \cDS/ in the \offDS/ data set. This feature is caused by steady-state environmental background in coincidence with the \ac{pot} trigger. Without data cuts this excess would monotonically increase for $\npe/\rightarrow 0$. However, a peak like feature arises once the Cherenkov and rise-time cuts are applied (Fig.~\ref{fig:sns-analysis:acceptance-curve}). Both \acDS/ and \cDS/ show a comparable number of events which leads to a residual fluctuating around zero. The corresponding arrival time spectrum is shown in the lower left of Fig.~\ref{fig:sns-analysis:raw-on-off-spectra}. The slight increase towards earlier arrival times is caused by a small afterglow component from preceding high-energy depositions, introducing a slight bias in the arrival time of the environmental background.\par

The right panels in Fig.~\ref{fig:sns-analysis:raw-on-off-spectra} show the energy (top) and arrival time spectra (bottom) corresponding to the \onDS/ data set. A clear excess of \cDS/ over \acDS/ is apparent in energy and time, demonstrating the presence of a beam-related signal in the data set. Comparing the raw number of counts found in \acDS/ for both \offDS/ and \onDS/ periods showed a comparable level of steady-state environmental background for both data sets. As a result, no degradation in the performance of the optimized cut parameters calculated in section~\ref{section:sns-analysis:optimizing-cuts} is to be expected.\par

\begin{figure}[htbp]
\begin{center}
\includegraphics[scale=1]{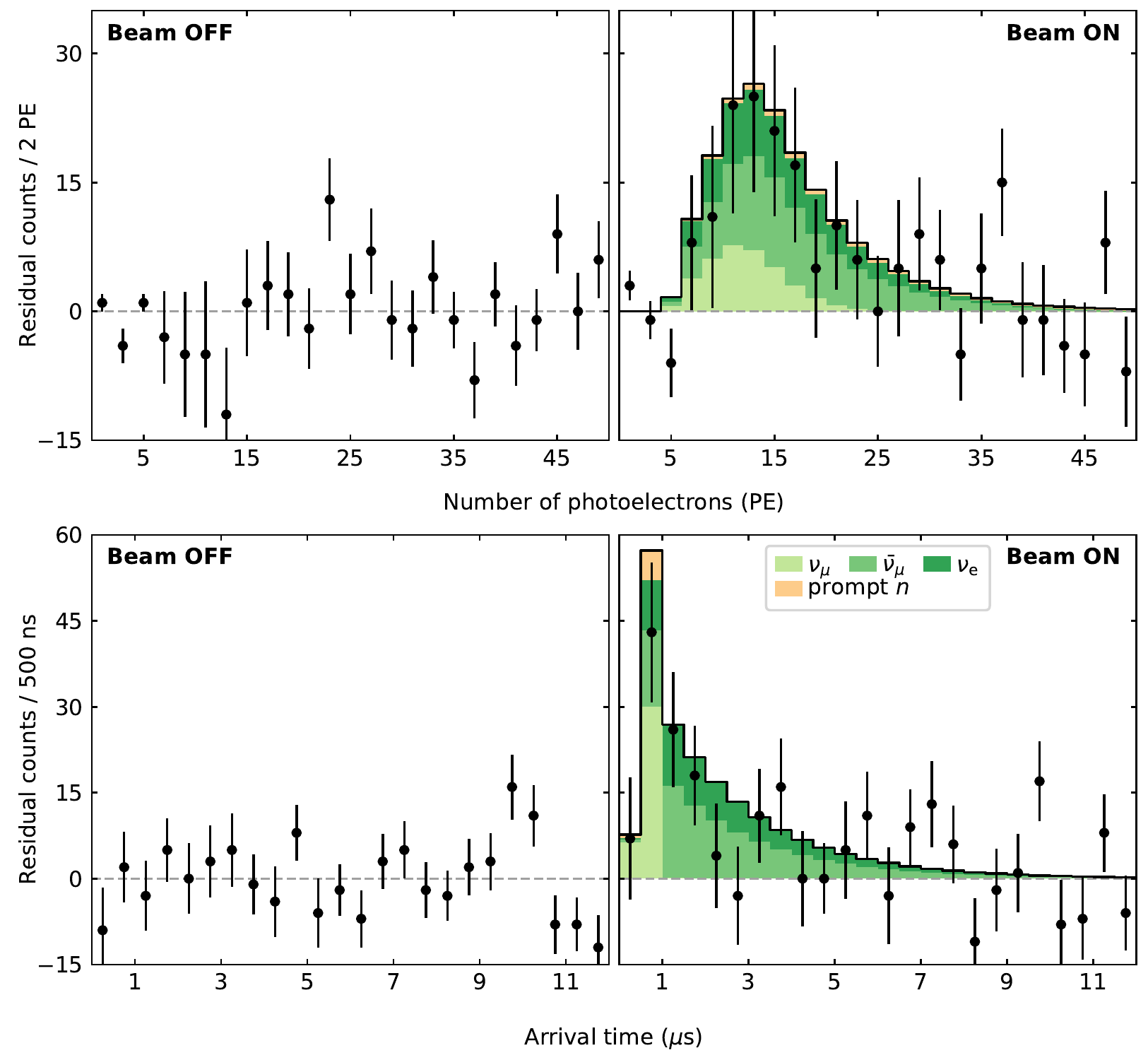}
\end{center}
\caption[First observation of \acs*{cenns}]{First observation of \acs*{cenns}.  Residual energy (top) and arrival time (bottom) spectra, i.e., \rDS/ = \cDS/ - \acDS/, for all events passing the optimized data cuts are shown in black. The \offDS/ (\onDS/) data is shown on the left (right). Error bars are statistical only. Both energy and arrival time residuals fluctuate around zero for the \offDS/ data set, confirming that steady-state environmental backgrounds contribute equally to \acDS/ and \cDS/. In contrast, the \onDS/ data shows a significant excess both in energy and arrival time. This excess is consistent with the Standard Model \ac{cenns} prediction, shown as stacked green histogram. Every neutrino flavor contributes to the recoil spectrum, as expected from a neutral-current interaction. The presence of a \ac{cenns} signal is favored at $6.7\sigma$ over its absence. A small background from prompt neutrons as derived in chapter~\ref{chapter:background-studies} is shown in orange. The \ac{nin} background was omitted as it consists of $\sim 4$ counts spread over an energy range of $\sim \SI{25}{\pe}$ and an arrival time range of $\sim\SI{6}{\micro\second}$. Similar results are obtained from an independent, secondary analysis \cite{konovalov-01}.}
\label{fig:sns-analysis:money-plot}
\end{figure}

The residual \rDS/=\cDS/-\acDS/ was calculated for all spectra shown in Fig.~\ref{fig:sns-analysis:raw-on-off-spectra}, which statistically removed all contribution from steady-state backgrounds. The resulting residual spectra are shown in Fig.~\ref{fig:sns-analysis:money-plot} in black, where the error bars are statistical only. The spectra on the left were calculated using \acDS/ and \cDS/ of the \offDS/ data set, whereas the \onDS/ data is shown on the right. The residuals of the \offDS/ data fluctuate around zero for both energy and arrival time. This was to be expected as steady-state environmental backgrounds contribute equally to \acDS/ and \cDS/ and vanish in the subtraction.\par

In contrast, the \onDS/ residual data shows a significant excess in both energy and arrival time. The \onDS/ panels also include the \ac{sm} \ac{cenns} prediction as calculated by Grayson Rich \cite{grayson-01} as stacked green histograms. Similar \ac{cenns} predictions were found by the author of this thesis, Juan Collar (University of Chicago) and Kate Scholberg (Duke University). The calculations made by Grayson Rich were adopted in this thesis to be consistent with the result published in \cite{collar-04}. The prediction was scaled to match the neutrino emission expected from a total beam energy delivered on the mercury target of $\SI{7.475}{\GWh}$ (section~\ref{section:sns-analysis:beam-power}). Rich included most of the secondary corrections that were omitted in chapter~\ref{chapter:cenns-theory}, such as axial vector couplings and strange quark contributions~\cite{grayson-01}. However, the small differences in the \ac{cenns} \xs/ for different neutrino flavors due to their different charge radii were neglected. A small contribution from prompt neutrons is shown in orange in Fig.~\ref{fig:sns-analysis:money-plot}. This beam-related background was calculated in chapter~\ref{chapter:background-studies}. Their arrival time is highly concentrated whereas their energy covers most of the range shown in the top panel. The contribution of prompt neutron induced events to each bin is barely discernible.\par

The beam-related background caused by \acp{nin} is omitted in the plot. The reason for this is first, the \ac{nin} event rate (Eq.~\ref{eq:background-studies:nin-rate}) is approximately 43 times smaller than the \ac{cenns} event rate (Eq.~\ref{eq:sns-analysis:cenns-rate}). Second, the arrival time of \ac{nin} induced events closely follows the \SI{2.2}{\micro\second} profile and is therefore spread out over the full arrival time range shown. Third, the corresponding events cover a large energy range. As a result the \ac{nin} background consists of just $\sim 4$ counts spread over an energy range of $\sim \SI{25}{\pe}$ and an arrival time range of $\sim\SI{6}{\micro\second}$. The \ac{nin} contribution to each individual bin is therefore negligible and omitted.\par

The excellent overall agreement between residual data and the predicted \ac{cenns} signal is apparent. It is further evident that events produced by all three neutrino flavors are necessary to reproduce the experimental data. As \ac{cenns} is a neutral-current process this was to be expected. The full prediction model including \ac{cenns}, prompt neutrons and \acp{nin} is described in more detail in the supplementary online materials of \cite{collar-04}.\par

The total number of events found within a two-dimensional energy-arrival time window bounded by $\npe/\in[6,30]$ and $T_\text{arr}\in[0,6]\SI{}{\micro\second}$ was calculated to be
\begin{align}
N^\text{sm}_{\text{ce}\nu\text{ns}} = 173 \pm 48.
\end{align}
This corresponds to a \ac{cenns} event rate of
\begin{align}
\label{eq:sns-analysis:cenns-rate}
\Gamma^\text{sm}_{\text{ce}\nu\text{ns}} = 23.1 \pm 6.4 \frac{\text{events}}{\si{\GWh}}
\end{align}
There are mainly four factors contributing to the uncertainty of this prediction. First, the uncertainty on the signal acceptance model contributes $\mathcal{O}(\SI{5}{\percent})$. Second, the form factor choice has a significant impact on the overall shape of the nuclear recoil spectrum, which leads to an additional $\mathcal{O}(\SI{5}{\percent})$ uncertainty. Third, a $\mathcal{O}(\SI{10}{\percent})$ uncertainty is associated with the total neutrino flux emitted by the \ac{sns} (section~\ref{section:sns-analysis:optimizing-cuts}). Last, the quenching factor was calculated in chapter~\ref{chapter:quenching-calibration} carrying a $\mathcal{O}(\SI{25}{\percent})$ uncertainty. Assuming all of these uncertainties are independent of one another, the total uncertainty on the \ac{cenns} count rate predicted by the Standard Model is $\sim\SI{28}{\percent}$.\par

The simple \ac{cenns} model presented in section~\ref{section:sns-analysis:optimizing-cuts} yields the identical number of events within the energy and arrival time boundaries given above. This further illustrates how well the \ac{cenns} \xs/ can be approximated by Eq.~(\ref{eq:cenns-theory:diff-cross-section-1}) and the relative insignificance of second-order corrections.\par

In a recent \coherent/ publication \cite{collar-04}, Rich further performed a binned, maximum likelihood analysis on the two-dimensional energy ($\npe/\in[6,30]\si{\pe}$) and arrival time ($T_\text{arr}\in[0,6]\SI{}{\micro\second}$) data of the coincidence region (\cDS/) in the \onDS/ data set presented in this thesis. The fit model included \acp{pdf} for \ac{cenns} signal, the prompt neutron background, and steady-state environmental backgrounds. No \acp{nin} were included in this analysis, due to their negligible contributon. The \ac{cenns} signal \ac{pdf} was calculated using the same approach as discussed in section~\ref{section:sns-analysis:optimizing-cuts}, with the inclusion of aforementioned second-order corrections to the \ac{cenns} \xs/. The \ac{pdf} for prompt neutron induced events was informed by the results presented in chapter~\ref{chapter:background-studies}. The steady-state environmental background \ac{pdf} was calculated using the \acDS/ \onDS/ data. Assuming the expected lack of correlation between energy and arrival time for this background, an analytical model was fit to the energy spectrum after the data was marginalized over arrival time. Likewise, the data marginalized over energy was fitted to an analytical model describing the arrival time. The two-dimensional background \ac{pdf} was constructed by the convolution of both models.\par

The uncertainties on the prompt neutron rate, the spectral hardness, and quenching factor, were included as a constraints in the likelihood fit. The amplitude of the steady-state background, and its constraint, were determined from the \acDS/ data. The amplitude of the \ac{cenns} signal was left unconstrained.

Rich performed this maximum likelihood fit of the \onDS/ residual data derived in this thesis using the RooFit analysis toolkit \cite{verkerke-01}. The best fit \ac{cenns} signal was found to be~\cite{collar-04}
\begin{align}
N^\text{fit}_{\text{ce}\nu\text{ns}} = 134\pm22
\end{align}
This value is approximately \SI{23}{\percent} smaller than the \ac{sm} prediction, however it is covered by the $1\sigma$ confidence level of $N^\text{sm}_{\text{ce}\nu\text{ns}} = 173 \pm 48$.\par

A more simplistic approach to extract the total number of \ac{cenns} induced events in the data set is to marginalize over the two-dimensional distributions and to subtract the known background contributions. The total number of events satisfying $\npe/\in[6,30]$ and $T_\text{arr}\in[0,6]\SI{}{\micro\second}$ are given by
\begin{equation}
\begin{aligned}
N^\offDS/_\acDS/ & = 209 \qquad\qquad N^\onDS/_\acDS/ = 405\\
N^\offDS/_\cDS/ & = 209 \qquad\qquad N^\onDS/_\cDS/ = 547
\end{aligned}
\end{equation}
In chapter~\ref{chapter:background-studies} the count rates for the two (prompt neutrons, \acp{nin}) main beam-related backgrounds were calculated. Recapitulating these rates were found to be
\begin{align}
\Gamma_\text{prompt} = 0.92\,\pm\,0.23\,\frac{\text{events}}{\SI{}{GWh}}\\
\Gamma_\text{nin}\,=0.54\,\pm\,0.18\,\frac{\text{events}}{\SI{}{GWh}}.
\end{align}
Scaling these rates with the total beam energy delivered on target (\SI{7.475}{\GWh}) yields a total of $6.9\pm1.7$ counts from prompt neutrons and $4.0\pm1.3$ counts from \acp{nin}. As was already discussed in the discussion of beam-related backgrounds (chapter~\ref{chapter:background-studies}) these backgrounds are $\sim 25$ (prompt neutrons) and $\sim 43$ (\acp{nin}) times smaller than the \ac{sm} prediction of \ac{cenns}.

The total number of \ac{cenns} events can therefore be approximated by
\begin{align}
N_{\text{ce}\nu\text{ns}} = N^\onDS/_\cDS/ - N^\onDS/_\acDS/ - N_\text{prompt} - N_\text{nin} = 131 \pm 31.
\end{align}
The total number of \ac{cenns} events determined using this approach agrees well with the number of \ac{cenns} events determined using the profile maximum likelihood fit. $N_{\text{ce}\nu\text{ns}}$ is in agreement with the \ac{sm} prediction at the $1\sigma$ level.\par

Rich and Barbeau further used the maximum likelihood fit to determine the significance of this observation \cite{collar-04}. The alternative hypothesis, i.e., a \ac{cenns} signal is present, is favored over the null hypothesis, i.e., \ac{cenns} is absent, at the 6.7-$\sigma$ level. This experiment therefore provides ample evidence for the first observation of \acl{cenns}.


\subsection{Correlation of residual excess and integrated beam energy}
\begin{figure}[htb]
\begin{center}
\includegraphics[scale=1]{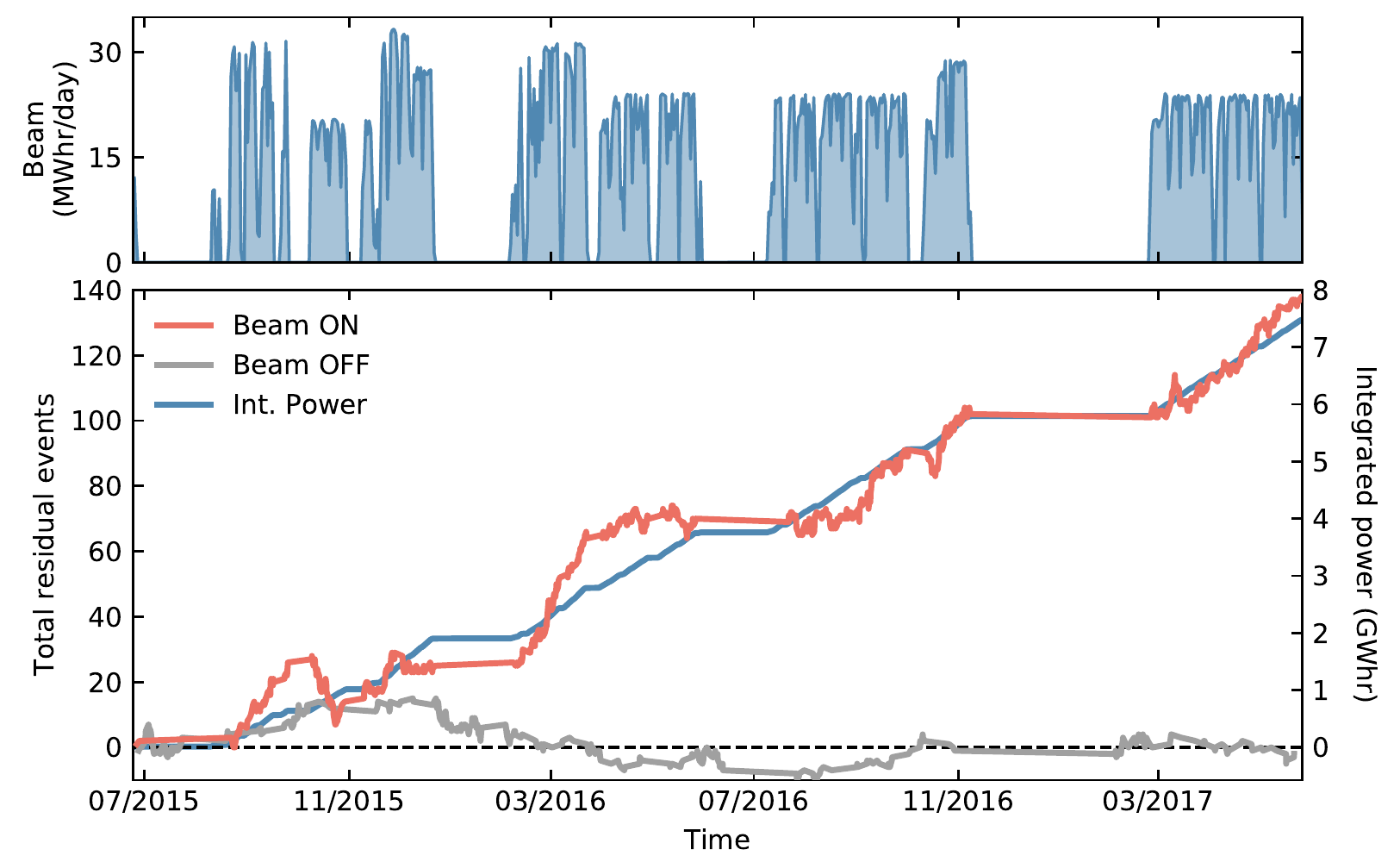}
\end{center}
\caption[Time evolution of residual counts for both \onDS/ and \offDS/ data]{\textbf{Top}: Daily integrated beam power delivered by the \acs*{sns} on the mercury target, which was discussed in section~\ref{section:sns-analysis:beam-power}. \textbf{Bottom}: Time evolution of the number of residual counts in both \onDS/ (red) and \offDS/ (black) data sets. The rate at which the \onDS/ residual grows is correlated to the total integrated beam power acquired (blue). In contrast, the \offDS/ residual fluctuates around zero. The \offDS/ residual excess changes continuously due to frequent, short, unplanned outages, not all visible in the top panel.}
\label{fig:sns-analysis:ks-plot}
\end{figure}

To further validate these results the time evolution of the number of residual events $N^\onDS/_\cDS/ - N^\onDS/_\acDS/$ was investigated by the author of this thesis. Fig.~\ref{fig:sns-analysis:ks-plot} shows the time evolution of the residual counts calculated for both \onDS/ (red) and \offDS/ (gray) data sets. The accumulated beam energy delivered on target is shown in blue which was normalized to the same vertical scale as the residual axis. The agreement between the increase in residual counts in the \onDS/ data and the total beam energy delivered on the mercury target is evident. In contrast, the \offDS/ residual only fluctuates around zero. The correlation between beam energy delivered per day and the rate at which \rDS/ increases for \onDS/ is also apparent.\par

The correlation between beam energy and \onDS/ residual excess was further investigated by Joshua Albert (Indiana University) using a Kolmogorov-Smirnov test \cite{chakravarti-01}. He created several test residual distributions using simulated event distributions. The \onDS/ residual as seen in Fig.~\ref{fig:sns-analysis:ks-plot} showed a stronger correlation with the integrated beam power than \SI{96}{\percent} of all \ac{mc} generated distributions. It is therefore most likely that the excess found in the \onDS/ residual is produced entirely by beam-related events. The correlation between the residual excess in the \onDS/ data set and the integrated beam energy is unmistakable in Fig.~\ref{fig:sns-analysis:ks-plot}.\par

\subsection{Future directions for the analysis using statistical discrimination between electronic and nuclear recoils}

\begin{figure}[htb]
\begin{center}
\includegraphics[scale=1]{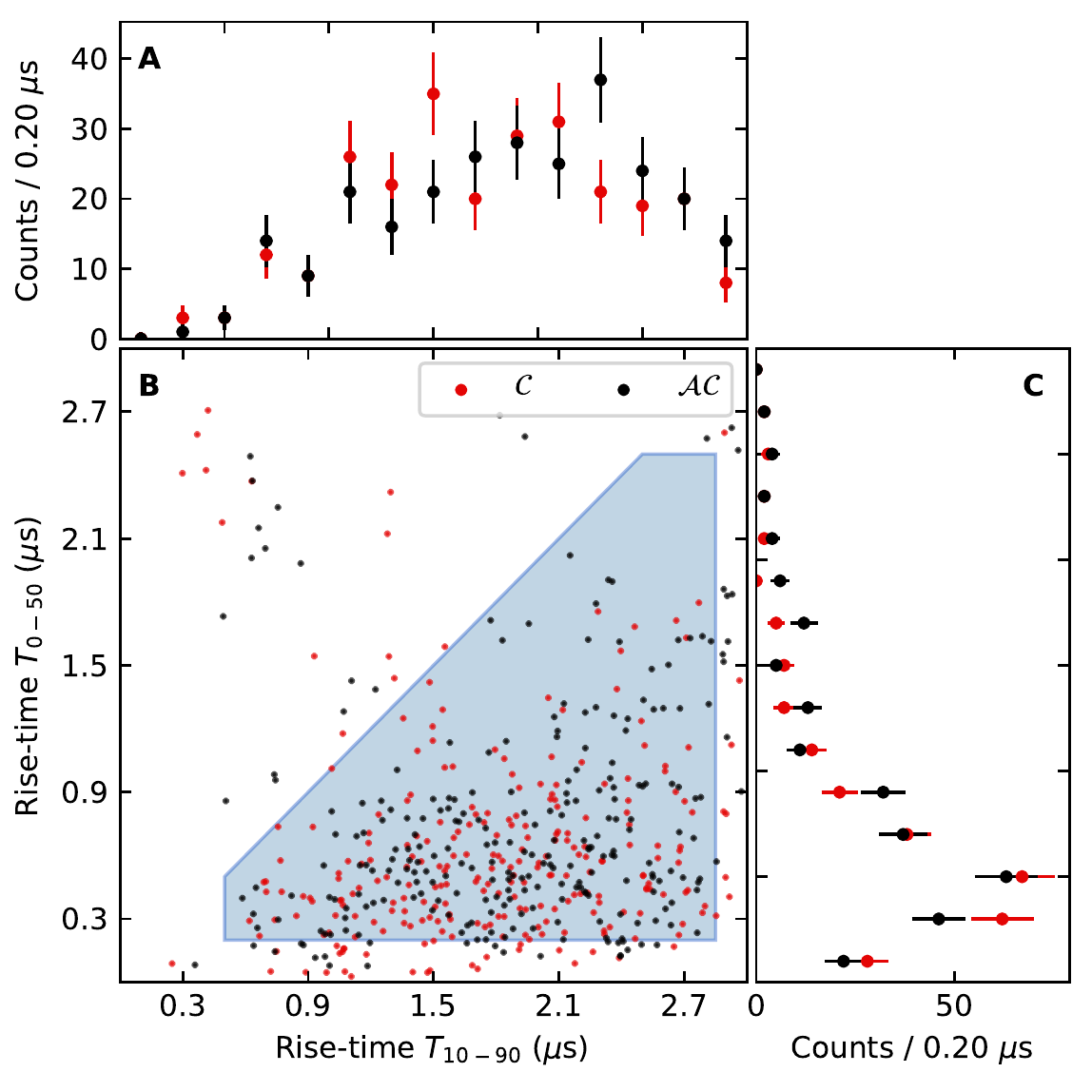}
\end{center}
\caption[Rise-time distributions of \offDS/ events passing all optimized data cuts]{Rise-time distributions of \offDS/ events passing all optimized data cuts during \ac{cenns} search runs. The \cDS/ (\acDS/) data is shown in red (black). The events of both data sets mainly cluster around the same rise-times seen in Fig.~\ref{fig:ba-calibration:rt-distribution-02}. The rise-time distributions for both \cDS/ and \acDS/ look mostly identical. A tail above the $T_{0-50}=T_{10-90}$ diagonal can be identified, that is caused by misidentified event onsets due to a preceding \ac{spe}. The shaded blue region shows the optimized rise-time cut window.}
\label{fig:sns-analysis:risetime-off}
\end{figure}

\begin{figure}[!ht]
\begin{center}
\includegraphics[scale=1]{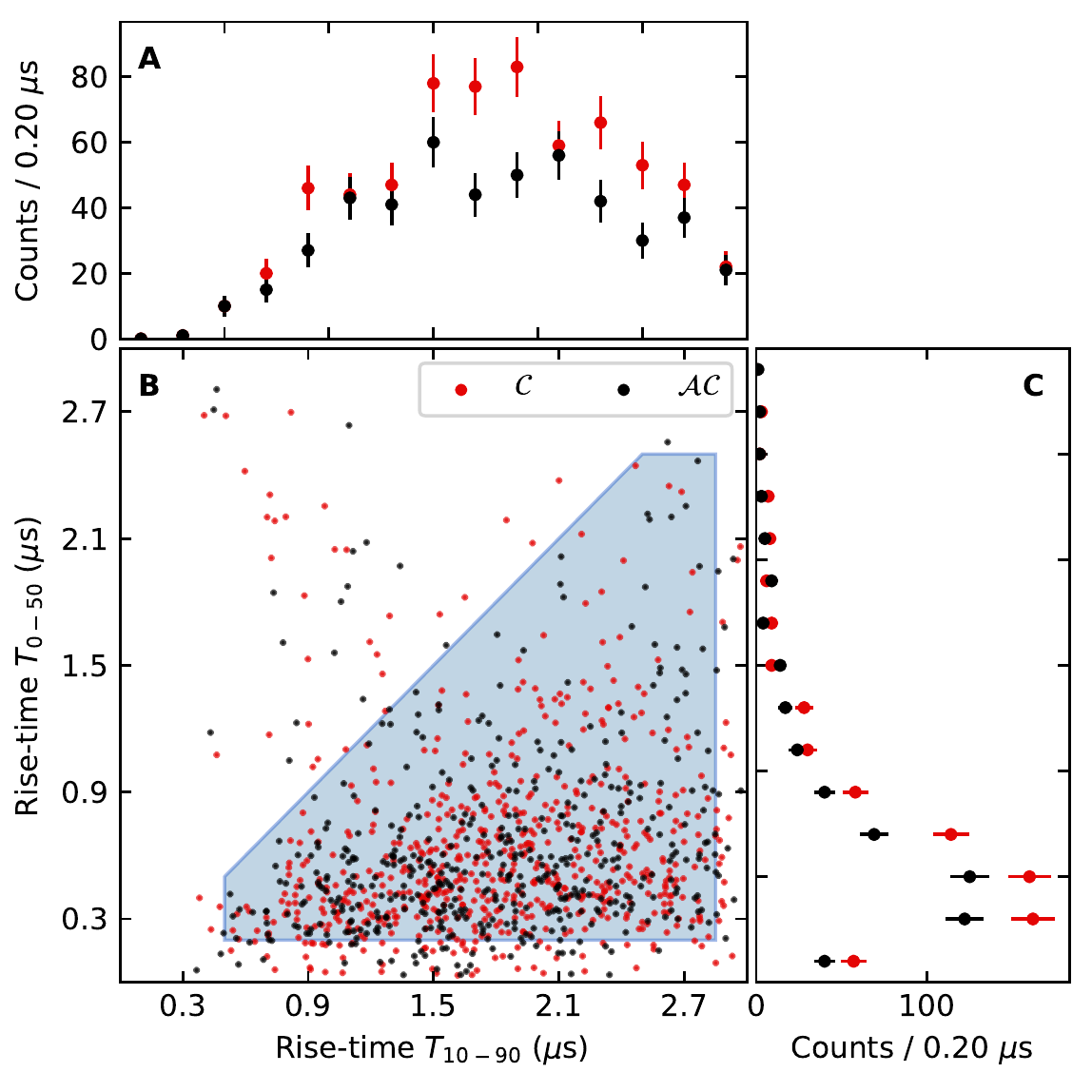}
\end{center}
\caption[Rise-time distributions of \onDS/ events passing all optimized data cuts]{Rise-time distributions of \onDS/ events passing all optimized data cuts during \ac{cenns} search runs. The \cDS/ (\acDS/) data is shown in red (black). The events of both data sets mainly cluster around the same rise-times seen in Fig.~\ref{fig:ba-calibration:rt-distribution-02}. In addition an excess is readily visible for the \cDS/ data over the \acDS/ data, which is caused by \ac{cenns}-induced events. This excess is therefore fully comprised of nuclear recoils. The residual \rDS/=\cDS/-\acDS/ is shown in Fig.~\ref{fig:sns-analysis:risetime-residual}. A tail above the $T_{0-50}=T_{10-90}$ diagonal can be identified, that is caused by misidentified event onsets due to a preceding \ac{spe}. The shaded blue region shows the optimized rise-time cut window.}
\label{fig:sns-analysis:risetime-on}
\end{figure}

In this section it is investigated whether there is any evidence in the \ac{cenns} search data that nuclear and electronic recoils can be statistically discriminated using the event rise-times~\cite{collar-02}. To this end, the rise-time distributions of all events that passed the optimized data cuts were analyzed by the author of this thesis. The distributions for the \offDS/ data is shown in Fig.~\ref{fig:sns-analysis:risetime-off}, whereas the \onDS/ data is shown in Fig.~\ref{fig:sns-analysis:risetime-on}. The \cDS/ (\acDS/) data is shown in red (black). The shaded blue region shows the acceptance window of the optimized rise-time cuts used in the \ac{cenns} search (section~\ref{section:sns-analysis:optimizing-cuts}).\par

The rise-time distributions for both \offDS/ and \onDS/ look similar to those extracted from the \isotope{Ba}{133} calibration data (chapter~\ref{chapter:ba-calibration}, Fig.~\ref{fig:ba-calibration:rt-distribution-02}). An excess in \cDS/ over \acDS/ for the \onDS/ data set can be identified, which is caused by \ac{cenns}-induced nuclear recoils.\par

\begin{figure}[thb]
\begin{center}
\includegraphics[scale=0.85]{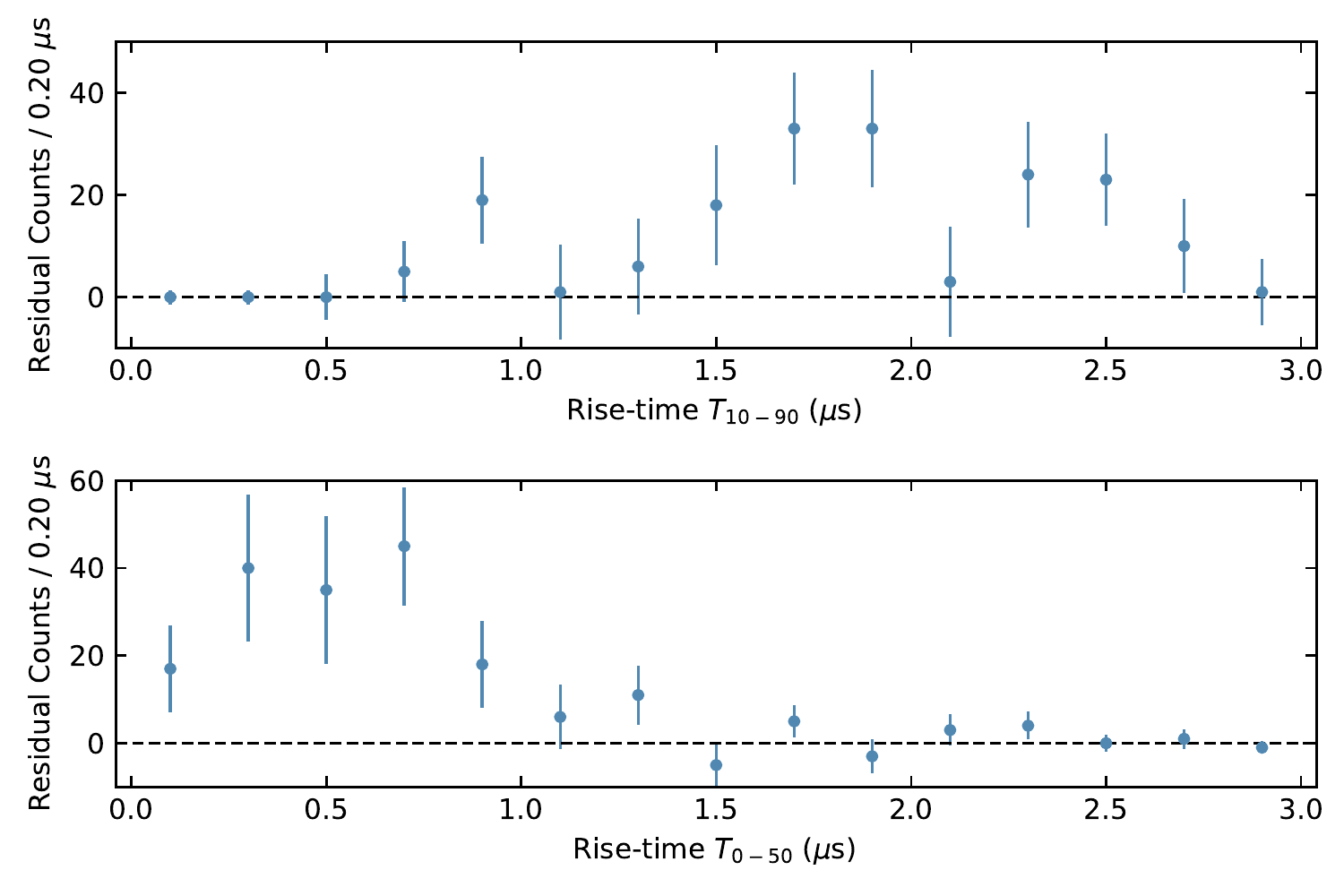}
\end{center}
\caption[Residual rise-time distributions of \onDS/ events passing all optimized data cuts]{Residual \rDS/ = \cDS/ - \acDS/ rise-time distributions of all events shown in Fig.~\ref{fig:sns-analysis:risetime-on}. The distributions show average rise-times differing from those calculated for electronic recoils, hinting at a possible nuclear and electronic recoil discrimination on a statistical basis.}
\label{fig:sns-analysis:risetime-residual}
\end{figure}

To investigate the rise-time distributions of the \ac{cenns}-induced events only, the residual spectra for $T_{0-50}$ and $T_{10-90}$ were calculated. These residual spectra only contain contributions from nuclear recoils. The resulting rise-time distributions for the \onDS/ data set are shown in Fig.~\ref{fig:sns-analysis:risetime-residual}. A broad excess in $T_{10-90}$, as well as a narrow excess in $T_{0-50}$ are evident.\par

The rise-time residuals for the \offDS/ data show no significant deviation from zero. The average rise-times for \ac{cenns}-induced nuclear recoils were calculated from the residual spectra. They are
\begin{equation}
\begin{aligned}
\left<T_{10-90}\right> & = \SI[separate-uncertainty = true]{1.84(10)}{\micro\second}\\
\left<T_{0-50}\right> & = \SI[separate-uncertainty = true]{0.68(4)}{\micro\second},
\end{aligned}
\end{equation}
where the error quoted is the standard error of the mean.\par

The expected rise-times for a nuclear and electronic recoil can be calculated using Eq.~(\ref{eq:ba-calibration:emission-curve}) with fast and slow scintillation decay times for nuclear and electronic recoils as measured in \cite{collar-02}. The rise-times from nuclear and electronic recoils according to Eq.~(\ref{eq:ba-calibration:emission-curve}) are
\begin{equation}
\begin{aligned}
\underline{\text{Nuclear recoils}} \qquad\qquad\qquad & \qquad\qquad \underline{\text{Electronic recoils}}\\
T_{10-90} = 1.94^{+0.02}_{-0.06}\,\SI{}{\micro\second}\qquad\qquad & \qquad\qquad T_{10-90} = 1.88^{+0.03}_{-0.05}\,\SI{}{\micro\second}\\
T_{0-50}  = 0.56^{+0.01}_{-0.02}\,\SI{}{\micro\second}\qquad\qquad & \qquad\qquad T_{0-50} = 0.51^{+0.01}_{-0.01}\,\SI{}{\micro\second}
\end{aligned}
\end{equation}
Comparing the rise-times calculated from the residual data with the rise-times calculated using Eq.~(\ref{eq:ba-calibration:emission-curve}) shows that the current level of statistics is insufficient to discriminate nuclear and electronic recoils. However, the rise-time $T_{0-50}$ measured for the residual spectrum hints at a possible nuclear-electronic recoil discrimination given larger statistics. The \csi/ detector is scheduled to continue acquiring data at the \ac{sns}. Consequently the statistics will improve and this might provide an opportunity to discriminate nuclear and electronic recoils on a statistical basis~\cite{konovalov-01}.

%% file: future-outlook.tex
%
%
\chapter{Conclusion}
\label{chapter:future-outlook}
Even though the \acf{sm} \xs/ predicted for \acf{cenns} is the largest of all low-energy neutrino couplings, it had eluded detection for over four decades. This thesis described the efforts of the \coherent/ collaboration that resulted in the first observation of \ac{cenns} at a $6.7$-$\sigma$ confidence level, using a low-background \csi/ detector with a total mass of \SI{14.57}{\kg}. The detector was located at the \ac{sns}, a stopped pion source at \ac{ornl}, where a pulsed proton beam impinges on a liquid mercury target producing \SI{}{\MeV} neutrinos. The \csi/ was deployed in a sub-basement approximately \SI{20}{\meter} away from the target. The location provides $\sim$\SI{19}{\meter} of continuous shielding against beam-related backgrounds and a total of \SI{8}{\mwe} overburden against cosmic-ray induced backgrounds.\par

Measurements of the beam-related background at the exact detector location found a background rate from prompt neutrons $\sim 25$ times smaller than the expected \ac{cenns} signal. The background caused by \acp{nin} originating in the lead surrounding the \csi/ detector was calculated to be $\sim 43$ times smaller than the expected \ac{cenns} rate. These low background rates allowed a \ac{cenns} search with a high signal-to-background level.\par

The \csi/ crystal and \ac{pmt} assembly showed a large light yield of $\mathcal{L}_\text{csi}=13.35 \frac{\text{SPE}}{\si{\keVee}}$, which is sufficient to achieve a low energy threshold. The light collection efficiency was further found to be uniform throughout the whole crystal, greatly simplifying the overall analysis.\par

To calibrate the analysis pipeline a library of low-energy radiation-induced events with a total energy below a few tens of \ac{pe} was recorded using a \isotope{Ba}{133} source. Given the small difference in the scintillation decay times of nuclear and electronic recoils in \csi/, this library provides close replicas of the detector response to \ac{cenns}-induced nuclear recoils. It enables calculating the signal acceptance fraction for several data cuts employed in the \ac{cenns} search. The purposes of these data cuts is to reject certain spurious backgrounds, such as Cherenkov light emission in the PMT window, or random groupings of dark-current photoelectrons.\par

In order to calculate the expected \ac{cenns} signal rate in the \csi/ detector, the total energy carried by a \ac{cenns}-induced nuclear recoil has to be properly converted into an electron equivalent energy. This conversion makes use of the quenching factor. To reduce the uncertainty associated with the quenching factor, additional measurements were performed at \ac{tunl}. These measurements focused on nuclear recoils with an energy of a few \SI{}{\keVnr} to a few tens of \SI{}{\keVnr}, covering the energy region expected for \ac{cenns}-induced nuclear recoils at the \ac{sns}.\par

The \csi/ detector was deployed at the \ac{sns} in June 2015. Since then data was acquired almost continuously at a \SI{60}{\hertz} acquisition rate. The \ac{cenns} search described in this thesis includes all data acquired between June 25th, 2015 and May 26th, 2017. Detector stability measurements show an excellent performance of the setup over the full two years of data taking, with only minor exceptions. Using data from time periods during which the \ac{sns} did not produce any neutrinos, an optimized set of data cuts was calculated that maximizes the \ac{cenns} signal-to-background ratio. Excesses in energy and event arrival time are observed for neutrino production periods only, which are in agreement with signatures predicted by the \ac{sm} for \ac{cenns}. The presence of a \ac{cenns} signal is favored at a $6.7$-$\sigma$ confidence level over the null hypothesis of environmental background only. The observed \ac{cenns} rate is further compatible with the \ac{sm} \ac{cenns} \xs/ at the $1$-$\sigma$ level. The signal rate was further found to be closely correlated to the integrated beam energy delivered on the mercury target, which itself determines the total number of neutrinos produced. This indicates that the measured excess is strictly beam-related.\par

This observation confirms the coherent enhancement of the neutrino-nucleus scattering \xs/ in cesium and iodine at low momentum transfers and substantiates an analogous enhancement in \ac{wimp}-nucleus scattering. It also provides confidence in the accuracy of \ac{cenns} \xs/s used in neutrino-floor calculations for \ac{wimp} searches as well as supernova calculations.\par

The \csi/ detector remains at the \ac{sns} and continues to acquire data. The \ac{sns} is expected to increase its power output in the near future. During August 2017 the \ac{sns} already operated at $\sim\SI{1.2}{\mega\watt}$ resulting in an increase in neutrino production of $\sim$\SI{20}{\percent} compared to earlier months of 2017. A higher beam power is beneficial as it directly increases the \ac{cenns} rate in the \csi/, whereas the environmental background rate remains the same. As such the expected signal-to-background ratio also increases with beam power. Combining the analysis presented in this thesis with new \csi/ data will therefore result in a \ac{cenns} \xs/ measurement with much lower uncertainties.\par

The \coherent/ collaboration aims to measure \ac{cenns} for multiple targets. In addition to the \csi/, the collaboration currently operates a \SI{28}{\kg} single-phase \ac{lar} detector as well as \SI{185}{\kg} of NaI[Tl]. A planned expansion includes a 2 ton NaI[Tl] array, a $\sim$1 ton \ac{lar} detector and $\sim\SI{20}{\kg}$ of p-type point contact germanium detectors. The precision measurements expected from these detectors enable the \coherent/ collaboration to pursue new neutrino physics opportunities in the future.\par